\documentclass[a4paper]{article}

\usepackage[pages=all, color=black, position={current page.south}, placement=bottom, scale=1, opacity=1, vshift=5mm]{background}
\SetBgContents{
}      % copyright

\usepackage[margin=1in]{geometry} % full-width

\usepackage{amsmath}
\usepackage{amsthm}
\usepackage{amssymb}
\usepackage{bbold}

\usepackage[utf8]{inputenc}
\usepackage[colorlinks,citecolor=blue,urlcolor=blue,filecolor=blue,backref=page]{hyperref}
\hypersetup{
	unicode,
	pdfauthor={Author One, Author Two, Author Three},
	pdftitle={A simple article template},
	pdfsubject={A simple article template},
	pdfkeywords={article, template, simple},
	pdfproducer={LaTeX},
	pdfcreator={pdflatex}
}

\usepackage{amsthm}
\usepackage{amsmath}
\usepackage{natbib}
\usepackage{graphicx}

\usepackage{xcolor} % SB
\usepackage{subfigure} % SB
\usepackage{lscape} % SB

\title{\textbf{Straightforward Phase I Dose-Finding Design for Healthy Volunteers Accounting for Surrogate Activity Biomarkers}}
\author{Sandrine Boulet$^{1,2}$, Emmanuelle Comets$^{3,4}$ , Antoine Guillon$^{5,6}$,  Linda B.S. Aulin$^7$, \\ Robin Michelet$^7$,  Charlotte Kloft$^7$, Sarah Zohar$^{1,2,*,**}$ and Moreno Ursino$^{1,2,*}$}

\date{
	$^1$Inserm, Centre de Recherche des Cordeliers, Sorbonne Université, Université Paris Cité, F-75006 Paris, France\\%
	$^2$Inria, HeKA, F-75015 Paris, France\\%
 	$^3$Inserm, Univ Rennes, EHESP, Irset (Institut de recherche en santé, environnement et travail) - UMRS 1085, F-35000 Rennes, France\\%
	$^4$Inserm, Université Paris Cité, IAME, F-75018 Paris, France\\%
    $^5$Inserm, Centre d'\'Etude des Pathologies Respiratoires (CEPR), UMR 1100, Tours, France; Université de Tours, Tours, France \\
    $^6$Service de Médecine Intensive Réanimation, CHRU de Tours, Tours, France\\
	$^7$Department of Clinical Pharmacy \& Biochemistry, Institute of Pharmacy, Freie Universitaet Berlin, 12169 Berlin, Germany\\%
	$^{*}$Sarah Zohar and Moreno Ursino made equal contributions and are co-last authors.\\%
	$^{**}$Corresponding author.\\%
}

\begin{document}
\maketitle

\begin{abstract} % 200 words
Conventionally, a first-in-human phase I trial in healthy volunteers aims to confirm the safety of a drug in humans. In such situations, volunteers should not suffer from any safety issues and simple algorithm-based dose-escalation schemes are often used. However, to avoid too many clinical trials in the future, it might be appealing to design these trials to accumulate information on the link between dose and efficacy / activity under strict safety constraints. Furthermore, an increasing number of molecules for which the increasing dose-activity curve reaches a plateau are emerging. 
\\In a phase I dose-finding trial context, our objective is to determine, under safety constraints, among a set of doses, the lowest dose whose probability of activity is closest to a given target. For this purpose, we propose a two-stage dose-finding design. The first stage is a typical algorithm dose escalation phase that can both check the safety of the doses and accumulate activity information. The second stage is a model-based dose-finding phase that involves selecting the best dose-activity model according to the plateau location. 
\\ Our simulation study shows that our proposed method performs better than the common Bayesian logistic regression model in selecting the optimal dose.

\noindent\textbf{Keywords:} Bayesian; first-in-human;  plateau.
\end{abstract}

\section{Introduction}
\label{sec:intro}

The goal of a first-in-human (FIH) phase I trial in healthy volunteers is first to assess the safety, then tolerability, pharmacokinetics (PK) and pharmacodynamics (PD) of a drug and thus estimate the maximum tolerated dose (MTD). This phase typically follows  preclinical studies, e.g., \textit{in vitro} and/or \textit{in vivo} studies \citep{european_medicines_agency_2017}. 

Usually, after defining the safety endpoint of the trial as a binary variable, algorithm-based dose-escalation schemes, such as ``A+B" designs, in which the dose is gradually increased, are used \citep{storer_1989}. Alternatively, when continuous endpoints could be used, such as PK measures of exposure, a model-based design may be adopted, and the aim of the trial becomes finding the dose that gives, on average, the desired PK measure value or that does not exceed a threshold value~\citep{whitehead_2001, whitehead_2006}.

When PK measurements cannot be obtained, for example, because they are below the limit of quantification or they cannot be performed, in some cases, PD endpoints can be used as surrogates. Indeed, if the PD changes, it could be expected that it is due to the PK of the drug, which could not be observed. Furthermore, when considering PD measurements, one should take into account the plateau effect in the increasing dose-activity curve, i.e. from a certain dose onwards, the activity no longer increases and remains almost constant. In this context, the objective becomes to determine, under a safety main endpoint, the lowest dose whose probability of activity is closest to a given activity target among a set of doses. If there is a plateau in the dose-activity relationship, the dose to be selected may be either the first dose in the plateau or a lower dose depending on the intended target activity. Otherwise, the target dose is simply the dose corresponding to the target activity. Therefore, the estimated dose is the minimum activity dose (MAD) under safety constraints. For instance, increasing dose-activity curve with plateau effect is observed within the receptor occupancy theoretical framework, i.e. at some dose saturation of the target occurs and maximal response is reached. However, for newer modalities such as bispecific antibodies other type of exposure/dose-response curves (e.g. bell-shapes) can be seen due to their binding to different sites.

Some work has already been done on modeling a plateau in the context of phase I/II dose-finding trials in cancer trials. For instance, \citet{riviere_2018} considered molecularly targeted agents, such as monoclonal antibodies, for which efficacy often increases initially with the dose and then plateaus. They developed a Bayesian phase I/II dose-finding design to find the optimal dose, defined as the lowest safe dose that achieves the highest efficacy. More precisely, they employed a logistic model with a plateau parameter to capture the increasing-then-plateau feature of the dose-response (efficacy) relationship. Similarly, \citet{altzerinakou_2021} implemented a change point linear mixed effects skeleton model for biomarker measurements for a phase I/II trial design for targeted therapies and immunotherapies. In the field of oncology drug development, the US Food and Drug Administration (FDA) set up Project Optimus to reform the dose optimization and dose selection paradigm  (\href{https://www.fda.gov/about-fda/oncology-center-excellence/project-optimus}{https://www.fda.gov/about-fda/oncology-center-excellence/project-optimus})\citep{shah_2021, fourie_zirkelbach_2022}. In this context, for oncology trials with targeted drugs, \cite{guo_2023} proposed a dose-ranging approach to optimizing dose (DROID) as a way to escape the ``more is better" paradigm. Indeed, the approach focused first on targeting a therapeutic dose range rather than a single dose such as the MTD. Their approach is based jointly on three endpoints: a toxicity endpoint that is quickly available, an efficacy endpoint that takes more time to collect, and an efficacy surrogate endpoint such as a PD biomarker. In their modeling, they also proposed a Bayesian Emax model for the dose-PD relationship and thus assumed a plateau in this relationship.

However, these designs have been developed in the context of clinical trials using ill volunteer patients, in which the expected toxicity is higher than that for healthy volunteers for whom high toxicity is often unacceptable. For healthy volunteers, the search for the lowest dose that is closest to a target activity is justified by the need to minimize the risk of safety issues, which can be expected to increase as the dose increases.

To meet our objectives, we propose a straightforward two-stage dose-finding design. The first stage, called the ``start-up phase", consists of a dose-escalation algorithm to ensure the safety of the drug being tested in humans while allowing the gathering of sufficient information on the molecule activity to move on to the second stage. The second stage is a model-based dose-finding phase. It involves estimating several dose-activity models that differ in the location of the plateau and then either selecting the one that best fits the data by looking at the largest \textit{posterior} probability of the models or by weighting and combining the models.

This paper is organized as follows. The motivation is summarized in Section \ref{sec:motivation}. The methods are presented in Section \ref{sec:methods}. Section \ref{sec:simulation_settings} describes the simulation settings used to assess our methodology, and the simulation results are given in Section \ref{sec:simulation_results}. A discussion closes the paper in Section \ref{sec:discussion}.

\section{Motivation}\label{sec:motivation}

Our motivation is based on the European collaborative project FAIR (Flagellin Aerosol therapy as an Immunomodulatory adjunct to the antibiotic treatment of drug-Resistant bacterial pneumonia)  (\href{https://fair-flagellin.eu/}{https://fair-flagellin.eu/}). It aims to evaluate the activation of the innate immune system in airway aerosols by the delivery of immunomodulator flagellin as an alternative adjunct strategy to standard-of-care antibiotics for treating pneumonia caused by antibiotic-resistant bacteria. In this project, the dose-response-exposure nebulization of flagellin to treat respiratory infections is studied in a FIH phase 1 trial. Since the flagellin molecule is inhaled, PK measures of exposure cannot be measured via blood sampling since it is expected that the drug concentration would be under the limit of quantification. Thus, to study the relationship between the dose administered and the concentration of the molecule in the lungs over time, PD surrogates were used. Indeed, in preclinical (\textit{in vitro} and \textit{in vivo}) studies, good surrogates for molecule activity were found. Then, clinically relevant thresholds were used to dichotomize the PD surrogates and define the binary activity response for the clinical trial. Therefore, we propose a phase I dose-finding approach following these objectives, i.e., (1) to investigate the clinical safety/tolerability of the administration of a single dose of nebulized flagellin administered via the Aerogen\textregistered Solo nebulizer in spontaneously breathing healthy volunteers and (2) to estimate the dose-activity relationship of nebulized flagellin in humans under safety constraints. In this work, for the sake of illustration, the MAD is defined as the lowest dose showing the probability closest to the target activity of 50\%, which means that activity has to be reached for 50\% of volunteers.

\section{Methods}\label{sec:methods}

Let $n \in \mathbb N^*$ be the maximum number of healthy volunteers that can be included in the trial.
Let $\boldsymbol{\mathcal D} = \{d_1, ..., d_L\}$ be the set of doses that can be administered to healthy volunteers, where $d_l < d_{l+1}$, for $l \in \{1, ..., L-1\}$. 
 
\subsection{Dose-Activity Model}

Let $\tau \in \{1, ..., L\}$ be the plateau parameter that indicates at which dose level the dose-activity curve reaches the plateau and $d_{l_{\xi}}$ be a reference dose that we initially guess to have an activity probability equal to the target activity rate $\xi$.

The relationship between the binary activity response $Z_i$ of volunteer $i${, $i \in \{1,..., n\}$, and his or her dose $x_i \in \boldsymbol{\mathcal D}$ can be modeled using one of the following equations \citep{neuenschwander_2008, riviere_2018}:
\begin{equation}\label{eq:dose_act_model}
\left\{
\begin{array}{ll}
\mathcal{M}_{1}: \text{logit}\{\mathbb P(Z_i=1)\} =  \gamma_0  & \text{if~} \tau = 1, \\ 
\mathcal{M}_{\tau}: \text{logit}\{\mathbb P(Z_i=1)\} =  \gamma_0+ \gamma_1\left(\log\left(\dfrac{x_i}{d_{l_{\xi}}}\right) \mathbb{1}_{x_i < d_{\tau}} + \log\left(\dfrac{d_{\tau}}{d_{l_{\xi}}}\right)\mathbb{1}_{x_i \ge d_{\tau}}\right) & \text{otherwise.}
\end{array}
  \right.
\end{equation}
where $\boldsymbol \gamma = (\gamma_{0}, \gamma_1) \in \mathbb R \times \mathbb R^*_+$. These equations assume that the relationship between the dose and the probability of activity is non-decreasing, potentially reaching a plateau at a higher dose. More precisely, the logistic distribution imparts an S shape to the dose-response curve. This curve can be truncated before reaching the maximum probability of one due to the plateau effect. Indeed, when the dose level is less than $\tau$, the activity increases with the dose. Otherwise, the dose-activity relationship reaches a plateau with a constant dose effect $\gamma_1 \log(d_{\tau}/d_{l_{\xi}})$. Furthermore, since our activity response is binary, the probability scale and a logistic function are used. Nonetheless, for other endpoints, other functions, like a linear or an exponential function with a plateau parameter, could be used. Such change-point models allow to determine the location of a plateau parameter encompassing more shapes than the only $E_{\max}$ model. 

For $d \in \boldsymbol{\mathcal D}$,  let $\phi_d = \mathbb P(Z = 1|d)$ denote the activity probability of the dose $d$. 

\subsection{Dose-Finding Design}

Our dose-finding design, largely inspired by \cite{riviere_2018}, is composed of two stages. In the first stage, an algorithm dose escalation process is used to check the safety of the drug while accumulating information on the drug activity that is necessary to move to the second stage. The second stage is the model-based dose-finding phase, which involves estimation of several dose-activity models and the selection of the one that best fits the data or the weighted combination of all the models. 

\subsubsection{First Stage: Start-up Phase}

For the start-up phase, let the cohort size be equal to $K_{\text{start}}$. The lowest dose level is administered to the first cohort of volunteers, and if no volunteer suffers from safety issues, the next dose level is given to the second cohort. The dose escalation is continued until the first safety issue is observed or the highest dose level is reached. Thereafter, the start-up phase is stopped, and we switch to the model-based dose-finding phase, as follows.

In this work, placebo or diluent is not taken into account. Nevertheless, it may be  needed, for instance, in PD modeling, to estimate some baseline effects or for go/no-go decisions. In this case, it can be recommended, for example, either (1) to assign a few volunteers to this product at the beginning of the trial or (2) to randomly attribute this product to one of the volunteers of each cohort in the start-up phase. 

\subsubsection{Second Stage: Model-Based Dose-Finding}

A dose $d \in \boldsymbol{\mathcal D}$ is said to be admissible if the following activity requirement is satisfied:
\begin{equation}\label{eq:act_constraint}
\mathbb P(\phi_d > \xi) \ge C_F
\end{equation}
where $C_F$ is a chosen threshold on the \textit{posterior} probability of activity. 

Let $\mathcal B = \{l; 1 \le l \le L'\}$ be the set of dose levels that are not higher than the highest dose level for which no observed safety issues have yet been observed $L'$, and $\mathcal A$ be the set of admissible dose levels in $\mathcal B$, that satisfy \ref{eq:act_constraint}. 

After obtaining the activity outcomes of the enrolled volunteers, for each possible value $l \in \{1, ..., L\}$ of the plateau parameter $\tau$, the dose-activity model $\mathcal{M}_{l}$ given by equation \ref{eq:dose_act_model} is estimated. For $l \in \{1, ..., L\}$, let $\pi_l$ denote the \textit{posterior} probability of the $l$th dose level being the plateau point, i.e., the \textit{posterior} probability of the $l$th dose-activity model, $\mathcal{M}_l$,
$$\pi_l = \mathbb P (\mathcal{M}_l|\text{data})=\frac{\mathbb P (\text{data}|\mathcal{M}_l)\mathbb P (\mathcal{M}_l)}{\sum_{t=1}^L \mathbb P (\text{data}|\mathcal{M}_{t})\mathbb P (\mathcal{M}_{t})},$$
where the marginal likelihood of model $\mathcal{M}_l$ is
$$\mathbb P (\text{data}|\mathcal{M}_l)=\int \mathbb P (\text{data}|\phi_{d_l},\mathcal{M}_l) \mathbb P (\phi_{d_l}|\mathcal{M}_l)d\phi_{d_l},$$
$\mathbb P (\text{data}|\phi_{d_l},\mathcal{M}_l)$ is the likelihood and $\mathbb P (\phi_{d_l}|\mathcal{M}_l)$ is the \textit{prior} distribution.

Thereafter, two methods are proposed. In the first, which we later call the "selection" method, the dose-activity model $\mathcal{M}_l$ linked to the largest value of $\pi_l$ is selected. In the second, we use Bayesian model averaging (BMA) to combine the dose-activity models $\mathcal{M}_t$ resulting from all possible values $t \in \{1, ..., L\}$ for $\tau$. In this method, the predictive distribution of $\phi_{d_l}$, $l \in \{1, ..., L\}$,  is a weighted average of \textit{posterior} distributions of $\phi_{d_l}$ using our model $\mathcal{M}_{t}$ for each possible value $t \in \{1, ..., L\}$ of $\tau$ \citep{raftery_1997}:
$$\mathbb P (\phi_{d_l}|\text{data})=\sum_{t=1}^L \pi_t \mathbb P (\phi_{d_l}|\mathcal{M}_t,\text{data}).$$

The expected value and variance of $\phi_{d_l}$ are given by
\begin{equation*}
\left\{
\begin{array}{l}
\mathbb E (\phi_{d_l}|\text{data})=\sum_{t=1}^L \hat{\phi}_{d_l}^{(t)} \pi_t \\
\\
\mathbb V (\phi_{d_l}|\text{data})=\sum_{t=1}^L [\mathbb V (\phi_{d_l}|\text{data},\mathcal{M}_t)+(\hat \phi_{d_l}^{(t)})^2]\pi_t-\mathbb E (\phi_{d_l}|\text{data})^2
\end{array}
  \right.
\end{equation*}
where $\hat{\phi}_{d_l}^{(t)} = \mathbb E (\phi_{d_l}|\text{data},\mathcal{M}_t)$.

Then, the MAD is estimated as the dose with the estimated activity probability closest to the target activity rate $\xi$:
\begin{equation}\label{eq:MAD}
\widehat{\text{MAD}} = \underset{d \in \boldsymbol{\mathcal D}}{\arg \min} |\hat \phi_d-\xi|.
\end{equation}

If the estimate of the plateau parameter is equal to the highest dose level  $L$, that is, $\hat{\tau} = \underset{1 \le l \le L}{\arg \max}(\pi_l) = L$, there is probably no plateau point (at least, among the dose range studied); otherwise, there is probably a plateau point.
\begin{itemize}
\item If the estimate of the plateau parameter is greater than the dose level corresponding to the estimated MAD, the next cohort of $K_{\text{model}}$ volunteers is assigned to the dose level in $\mathcal A$ that is closest to the dose level corresponding to the estimated MAD.
\item Now, if the estimate of the plateau parameter is less than or equal to the dose level corresponding to the estimated MAD, let $\mathcal R$ be the set of dose levels whose \textit{posterior} probabilities of being the plateau point were close to the largest one with a difference less than a positive threshold $s$, i.e.,
\begin{equation}\label{eq:R}
\mathcal R = \{ l' : |\underset{1 \le l \le L}{\max}(\pi_l)-\pi_{l'}| \le s ; 1 \le l' \le L\}.
\end{equation}
This means that $\mathcal R$ is a set of dose levels that are most likely to be the plateau point. The next cohort is randomized to dose level $l \in \mathcal R$ with a probability $\pi_l / \sum_{l' \in \mathcal R} \pi_{l'}$. If the randomly selected dose level is not admissible (that is, not in $\mathcal A$), then the cohort receives the dose level in $\mathcal A$ that is closest to that dose level. The above randomization procedure for dose assignment avoids dose finding becoming stuck at suboptimal doses due to high estimation uncertainty.
\end{itemize}
When the maximum sample size $n$ is reached, this dose assignment process ends. The optimal dose is selected as the lowest dose that is admissible and whose probability of activity is closest to the given activity target $\xi$. At any time, if no dose is admissible, the trial is stopped to protect volunteers from futile doses.

Our method is summarized in Figure \ref{fig:methods}. The main differences with \cite{riviere_2018}'s work lie in the construction of the final dose-activity model and the introduction of the target activity. In fact, Riviere combines the estimated dose-activity models (for each value of the plateau parameter) using a method similar to the BMA. On the contrary, because we want to preserve the plateau in the estimation of the final model, we propose the “selection” method that selects the dose-activity model that best fits the data according to the \textit{posterior} probability of the models. What's more, for healthy volunteers, it could be interesting to find the lowest dose that is closest to a target activity, since this dose could be seen as the starting point of an activity window.

\begin{landscape}
\begin{figure} 
\begin{center}
\includegraphics[scale=0.76]{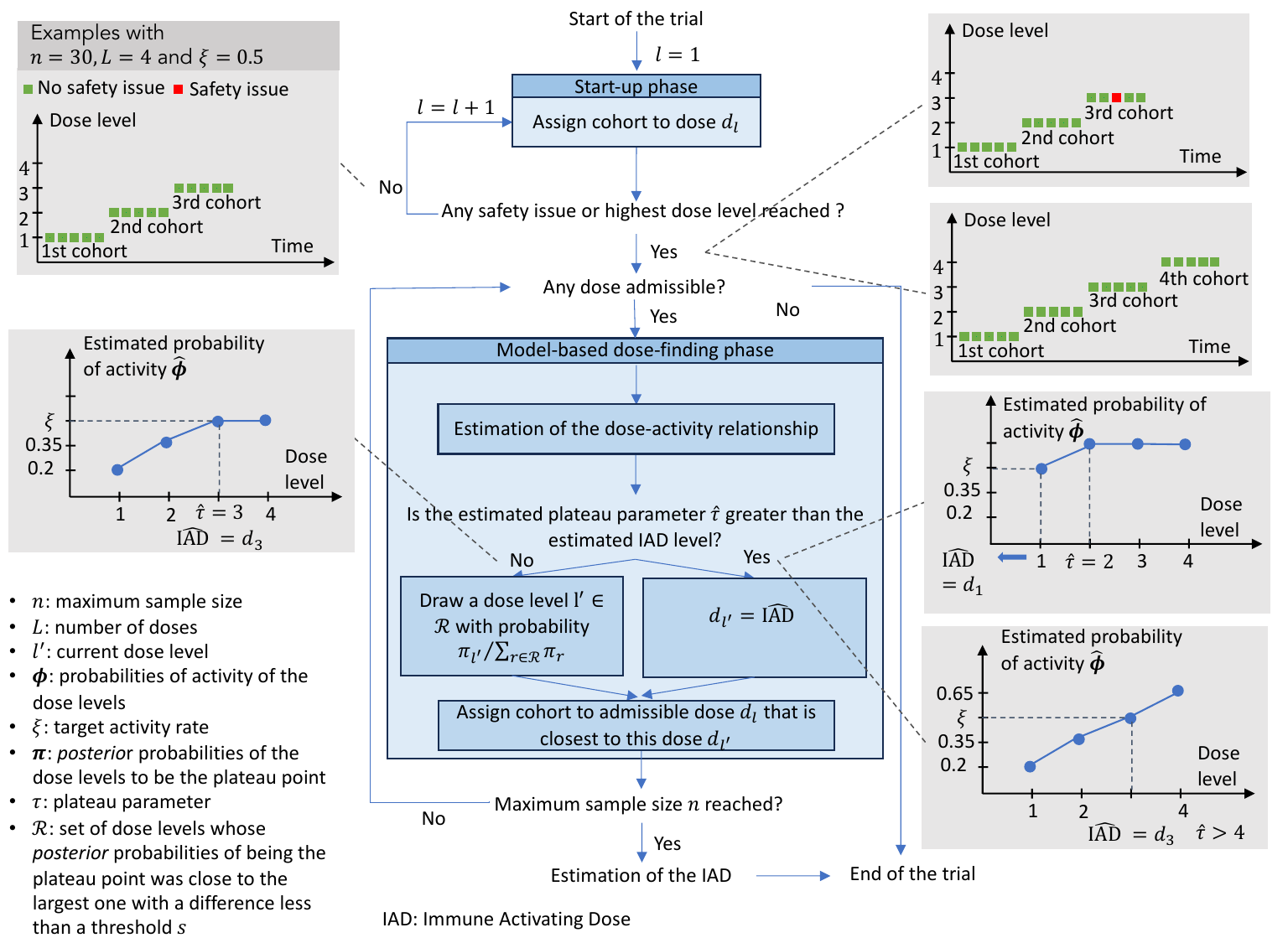}
\end{center}
\caption{
The dose-finding design in two stages: (1) start-up phase and (2) model-based dose-finding phase.
This figure appears in color in the electronic version of this article.
\label{fig:methods}}
\end{figure}
\end{landscape}

\section{Simulation Settings} \label{sec:simulation_settings}

To test the characteristics of our design, we conduct a simulation study including several scenarios. For these scenarios, toxicities and activities are simulated using Bernoulli distributions.

\subsection{Scenarios}

For $d \in \boldsymbol{\mathcal D}$, let $\psi_d$ denote the safety issue probability of dose $d$. 

We propose 8 scenarios that  differ according to the activity probabilities $\boldsymbol{\phi}$ associated with each dose and for which we vary the number $L$ of doses tested from 3 to 5 (see Table \ref{tab:simu_parameters}). The  target activity rate $\xi$ is set to 0.5, the value for which the variance is greatest and which therefore corresponds to the most difficult situation.
For instance, in Scenario 1, when $L = 3$ doses are tested, the 3 doses are respectively linked to activity probabilities of 0.5, 0.65 and 0.8. 
In Scenarios 1 to 4, no plateau is involved. For Scenarios 1, 2 and 4, the target dose is the dose associated to the target activity 0.5, which is one of the first two for Scenario 1, one of the last two for Scenario 2 and one of the middle for Scenario 4, depending on the number of doses tested. In fact, this means that, in Scenario 1, almost all doses are active while, for Scenario 2, only the last doses are active.
Scenarios 5 to 8 involve a plateau. In Scenarios 5 and 6, the target dose corresponds to both the first dose of the plateau and the target activity.
In Scenarios 3 and 7, no dose is effective. 
All doses of all scenarios are linked to very low safety issues, with probabilities from 0 to 0.004.
For each scenario, 1000 datasets are simulated. 

\begin{table}[t]
\small
\caption{Simulation parameters. $L$: number of doses tested; $\boldsymbol{\phi}$: activity probabilities; $\boldsymbol{\psi}$: safety issue probabilities; $n$: maximum number of healthy volunteers; $K_{\text{model}}$: cohort size for the model-based dose-finding phase; $\xi$: target activity rate; $C_F$: threshold on the \textit{posterior} probability of activity.\label{tab:simu_parameters}}
\begin{center}
\begin{tabular}{l|c|c|c}
\hline
 &  $L=3$ & $L=4$ & $L=5$   \\ 
\hline
\hline
&\multicolumn{3}{c}{$\boldsymbol{\phi}$} \\ 
\textbf{Scenario 1} & (\textbf{0.5}, 0.65, 0.8) & (0.35, \textbf{0.5}, 0.65, 0.8) & (0.35, \textbf{0.5}, 0.65, 0.8, 0.95) \\
\textbf{Scenario 2} & (0.2, 0.35, \textbf{0.5}) & (0.05, 0.2, 0.35, \textbf{0.5}) & (0.05, 0.2, 0.35, \textbf{0.5}, 0.65) \\ 
\textbf{Scenario 3} & (0.05, 0.2, 0.35)  & (0.01, 0.05, 0.2, 0.35) & (0.01, 0.05, 0.15, 0.25, 0.35) \\
\textbf{Scenario 4} & (0.35, \textbf{0.5}, 0.65)  & (0.2, 0.35, \textbf{0.5}, 0.65) & (0.2, 0.35, \textbf{0.5}, 0.65, 0.8) \\
\textbf{Scenario 5} & (0.35, \textbf{0.5}, 0.5)  & (0.2, 0.35, \textbf{0.5}, 0.5) &(0.05, 0.2, 0.35, \textbf{0.5}, 0.5) \\ 
\textbf{Scenario 6} & (\textbf{0.5}, 0.5, 0.5)  & (0.35, \textbf{0.5}, 0.5, 0.5) &(0.35, \textbf{0.5}, 0.5, 0.5, 0.5) \\ 
\textbf{Scenario 7} & (0.35, 0.35, 0.35) & (0.2, 0.35, 0.35, 0.35) & (0.2, 0.35, 0.35, 0.35, 0.35) \\ 
\textbf{Scenario 8} & (\textbf{0.65}, 0.65, 0.65) & (\textbf{0.5}, 0.65, 0.65, 0.65) & (\textbf{0.5}, 0.65, 0.65, 0.65, 0.65) \\ 
\hline
$\boldsymbol{\psi}$ & (0.0005, 0.001, 0.002) & (0, 0.0005, 0.001, 0.002) & (0, 0.0005, 0.001, 0.002, 0.004)  \\ 
\hline
$n$ & \multicolumn{3}{c}{18 or 24 or 30 or 40}  \\
$K_{\text{model}}$ & \multicolumn{3}{c}{2}  \\
$\xi$ & \multicolumn{3}{c}{0.5}  \\ 
$C_F$ & \multicolumn{3}{c}{0.05} \\
\hline
\end{tabular}
\end{center}
\end{table}

The cohort size for the start-up phase is chosen equal to 
\begin{equation*}
K_{\text{start}} =
\left\{
\begin{array}{ll}
[n/L-K_{\text{model}}]  & \text{if~} n \text{~is divisible by~} L \text{~or~} L \text{~is~an~even~number}, \\ 
\Big [ \dfrac{[n/L-K_{\text{model}}]}{2} \Big ] \times 2& \text{otherwise,}
\end{array}
  \right.
\end{equation*}
where the square brackets indicate the integer part. Note that this formula can only be used if $n$ is an even number.
We choose to use $s = 0.05 (1-N/n)$ \citep{riviere_2018}. Since this threshold is a function of the current sample size $N$, $s$ is larger when the trial begins (that is, when there is high uncertainty on model estimates) and decreases as the trial progresses (that is, when more data are available for the model estimation).

\subsection{\textit{Prior} Distributions}

Regarding \textit{prior} distributions, a normal distribution is used for the intercept, $\gamma_0 \sim \mathcal N(\overline{\gamma_0},\sigma_{\gamma_0}^2)$, where $\overline{\gamma_0} = \text{logit}(\xi)$ is the mean and $\sigma_{\gamma_0} = 2$ is the standard deviation, and a gamma distribution is considered for the slope to ensure positivity, $\gamma_1 \sim \gamma(\alpha_{\gamma}, \dfrac{\alpha_{\gamma}}{\overline{\gamma_1}})$, where $\alpha_{\gamma}=5$ is the shape parameter and $\overline{\gamma_1}$ is the mean. We choose 
$\overline{\gamma_1} = (\text{logit}(\tilde \phi_2)-\overline{\gamma_0})/\log(2/d_{l_{\xi}})$ 
when $d_{l_{\xi}} \in \{1, L\}$, and 
$\overline{\gamma_1} = (\text{logit}(\tilde \phi_1)-\overline{\gamma_0})/\log(1/d_{l_{\xi}})$ otherwise.
For the plateau parameter $\tau$, we assign a uniform distribution, $\tau \sim \mathcal U(\{1, ..., L\})$.
The values of the initial guesses of the activity probabilities for several values of the number of dose levels $L$ are given in Table \ref{tab:initial_guesses}.

\begin{table}[bt]
\caption{Initial guesses of the activity probabilities $\boldsymbol{\tilde \phi}$ for several values of the number of dose levels $L$.\label{tab:initial_guesses}}
\begin{center}
\begin{tabular}{l|c|c|c}
\hline
 &  $L=3$ & $L=4$ & $L=5$   \\ 
 \hline
 \hline
$\boldsymbol{\tilde \phi}$ & (0.5, 0.65, 0.8) & (0.35, 0.5, 0.65, 0.8) & (0.35, 0.5, 0.65, 0.8, 0.95) \\
\hline
\end{tabular}
\end{center}
\end{table}

All analyses were performed using R software version 4.2.1 with the Jags package \textit{R2jags} version 0.7-1. In \textit{R2jags}, 3 chains, a burn-in of 3000 and 6000 other iterations are used, and \cite{gelman_1992}'s potential scale reduction factor is implemented as a convergence criterion.

\section{Simulation Results}\label{sec:simulation_results}

Our proposed "selection" and BMA methods are compared with a method using a simple Bayesian logistic regression model (BLRM) that does not take into account a potential activity plateau. 

Figures \ref{fig:boxplot_nb_volunt_by_dose_n30_L4} and \ref{fig:boxplot_nb_volunt_by_dose_n40_L4} show the distributions of the number of volunteers when $L = 4$ dose levels are used for the $n \in \{30, 40\}$ maximum number of volunteers. Similarly, Figures \ref{fig:barplot_MAD_hat_n30_L4} and \ref{fig:barplot_MAD_hat_n40_L4} show the percentages of dose selection. In these last figures, the dose-activity link ("true activity") and the activity target horizontal line are also drawn. As a reminder, our aim is to select, under safety constraints, the MAD defined as the lowest dose whose probability of activity is closest to a given activity target among a set of doses. If no dose is sufficiently active, none should be selected (see Equation \ref{eq:act_constraint} for activity requirement). Otherwise, if there is a plateau in the dose-activity relationship, the MAD may be either the first dose in the plateau or a lower dose depending on the intended target activity. In the absence of a plateau, the MAD is simply the dose corresponding to the target activity. For instance, in Figure \ref{fig:barplot_MAD_hat_n30_L4}, in the absence of a plateau (Scenarios 1 to 4), the MAD, enclosed in square brackets, is the dose for which the "true activity" and "activity target" lines cross (doses 2, 4, "none" and 3, respectively). In the presence of a plateau (scenarios 5 to 8), the MAD is the first dose of the plateau if it is sufficiently active (dose 3 in Scenario 5 and dose 2 in Scenario 6), a lower dose if the target activity rate is reached before the plateau (dose 1 in Scenario 8), or no dose if the activity is too low compared to the target (Scenario 7). The MAD selection percentage must be as high as possible.

\begin{figure}[h!]
\begin{center}
\subfigure[Scenario 1]{\includegraphics[scale=0.29]{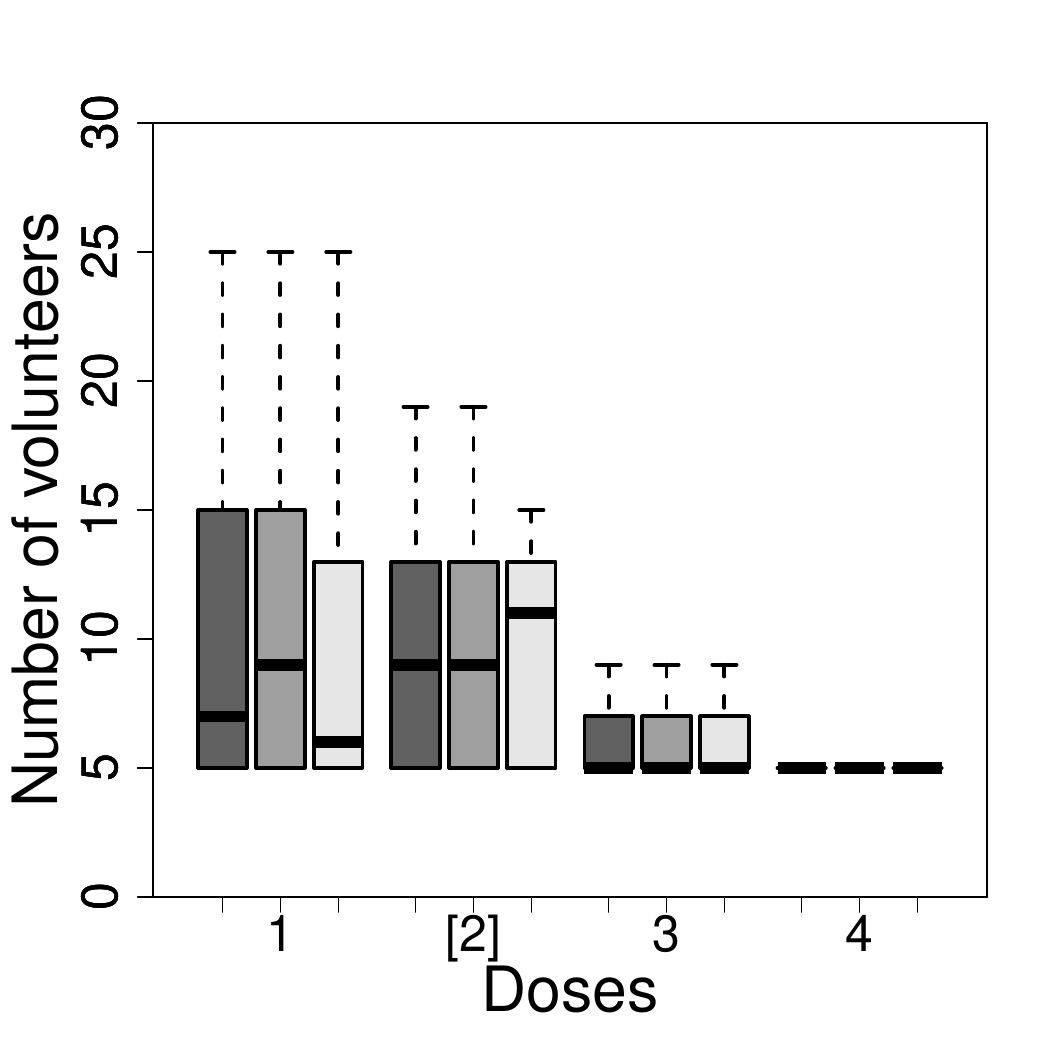}} 
\subfigure[Scenario 2]{\includegraphics[scale=0.29]{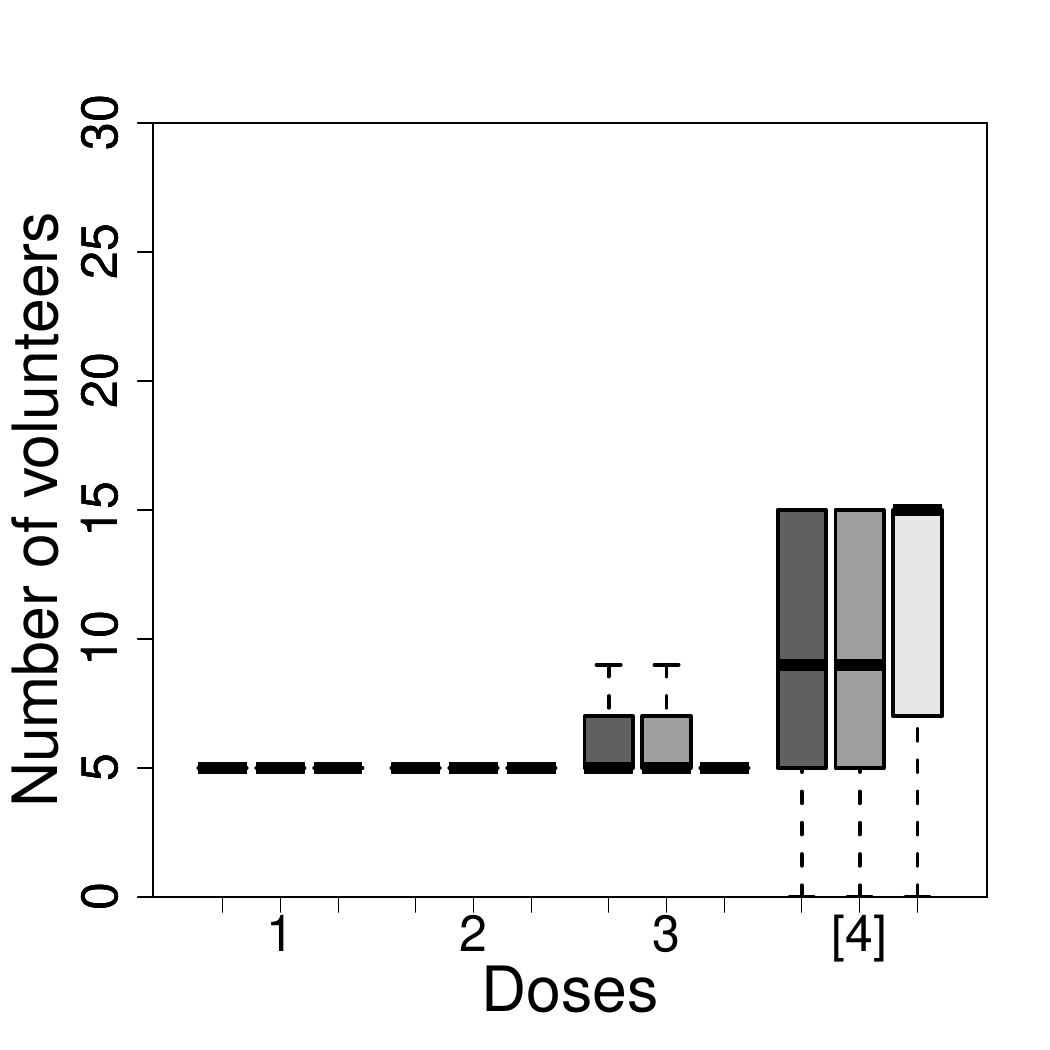}} 
\subfigure[Scenario 3]{\includegraphics[scale=0.29]{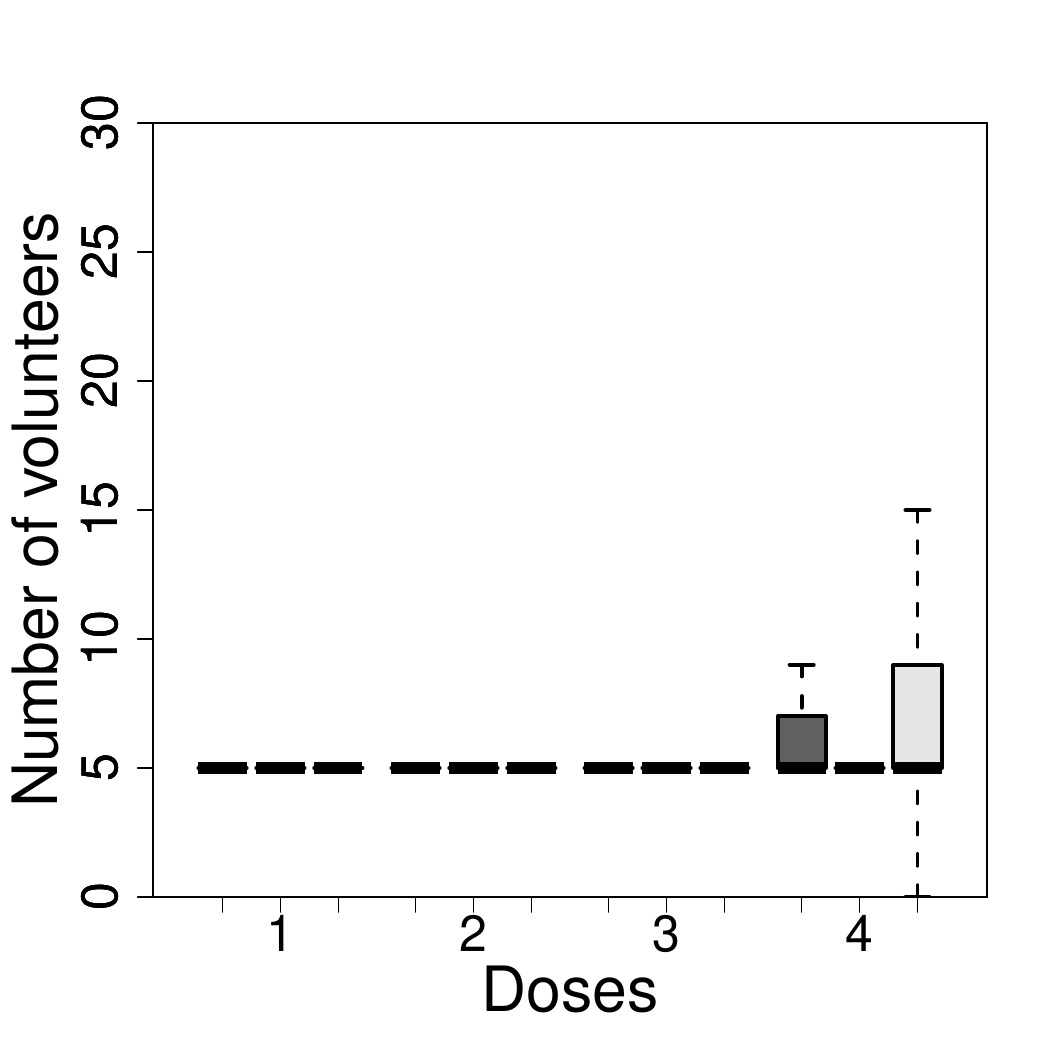}} \\ 
\subfigure[Scenario 4]{\includegraphics[scale=0.29]{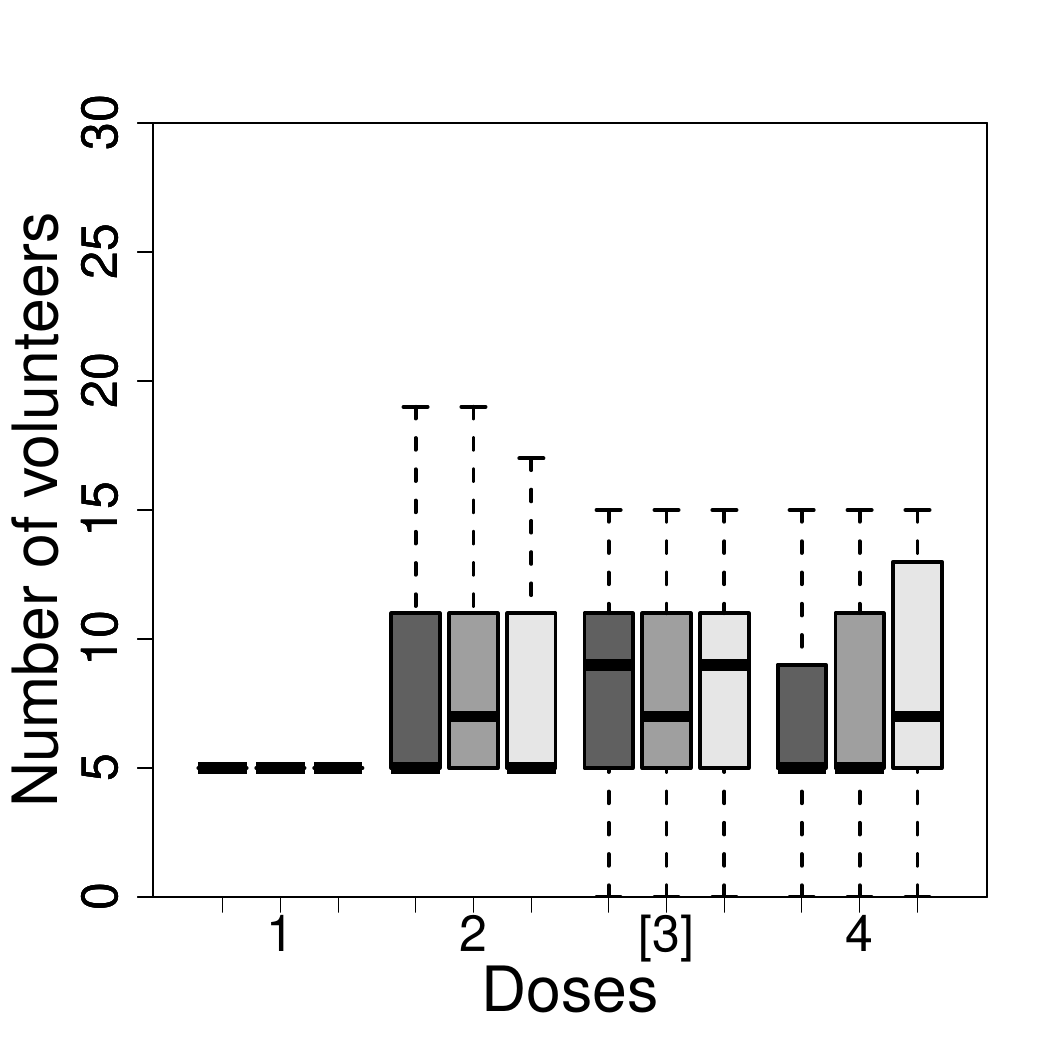}}
\subfigure[Scenario 5]{\includegraphics[scale=0.29]{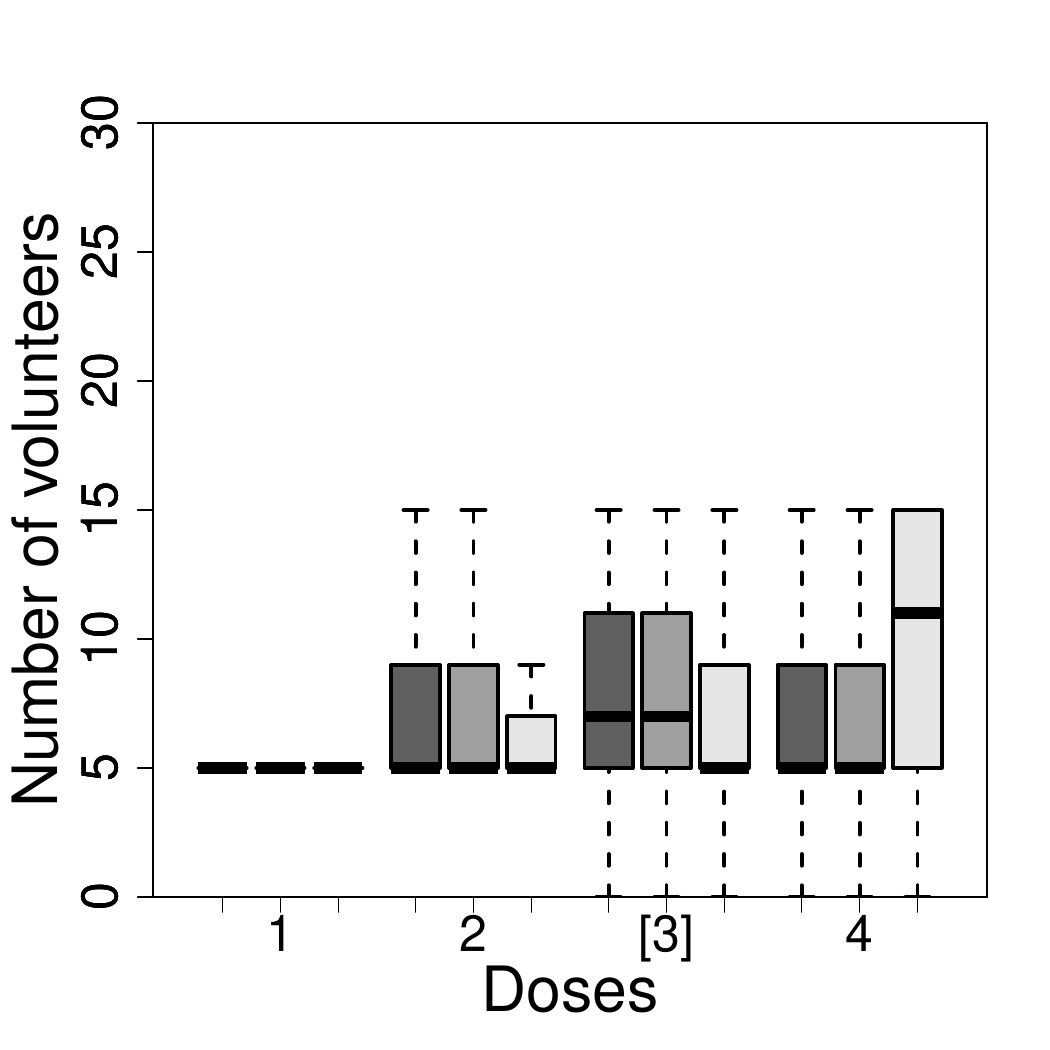}} 
\subfigure[Scenario 6]{\includegraphics[scale=0.29]{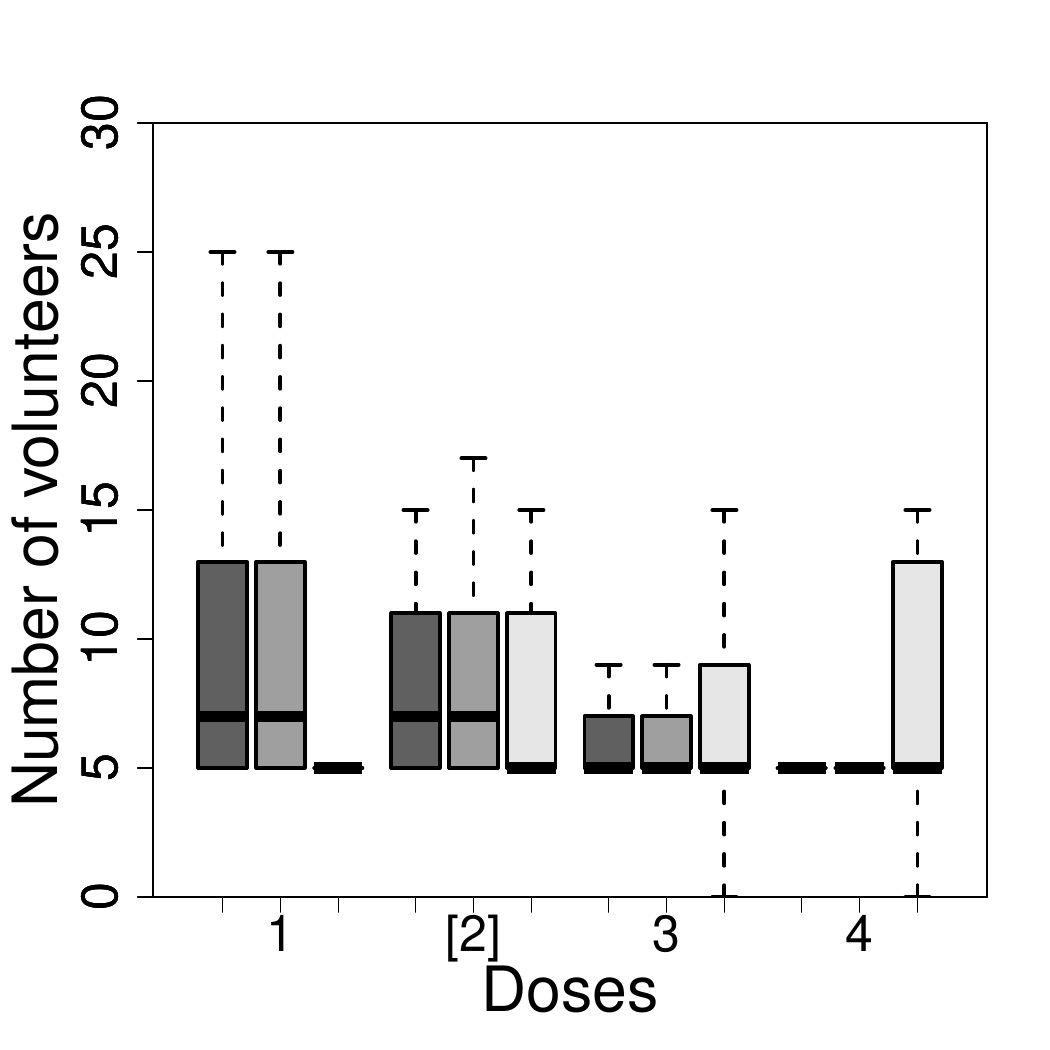}} \\ 
\subfigure[Scenario 7]{\includegraphics[scale=0.29]{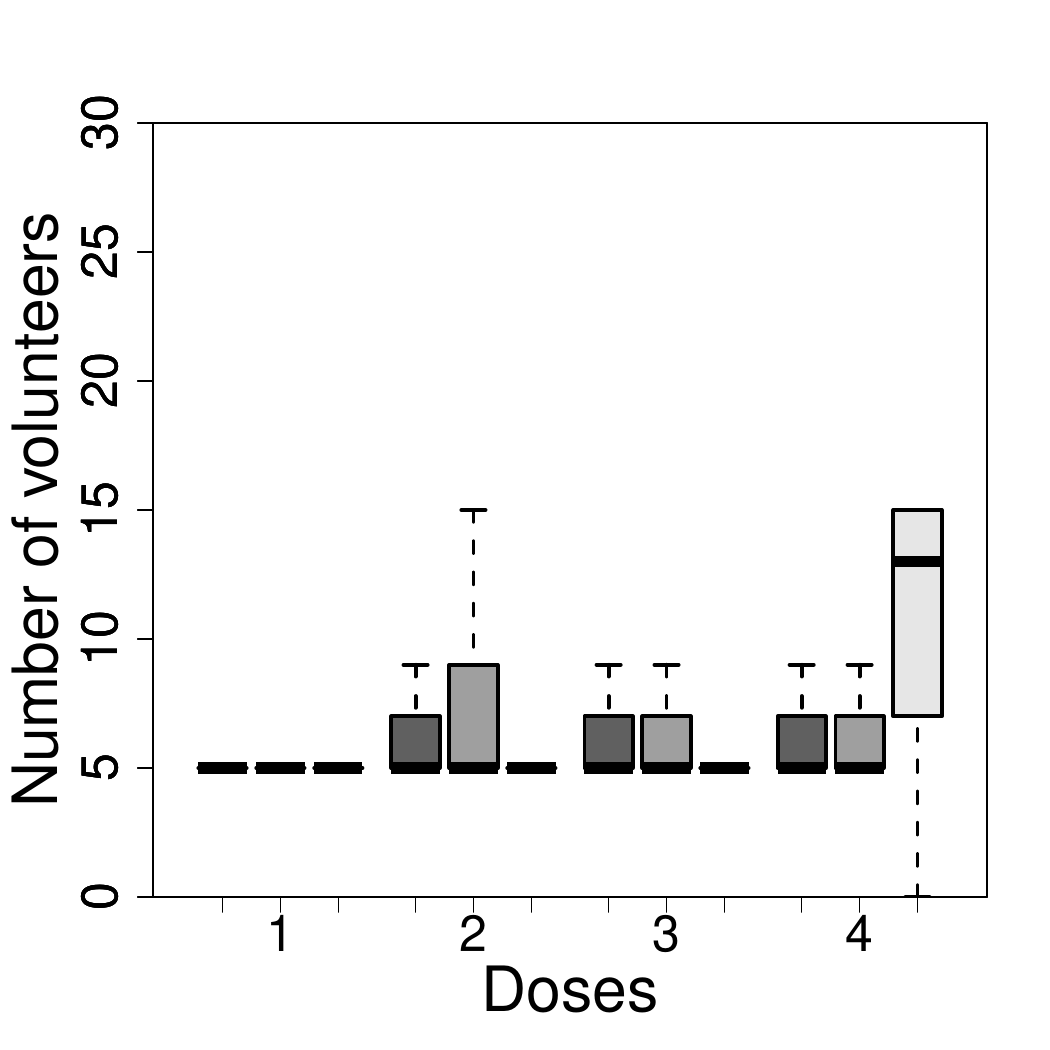}} 
\subfigure[Scenario 8]{\includegraphics[scale=0.29]{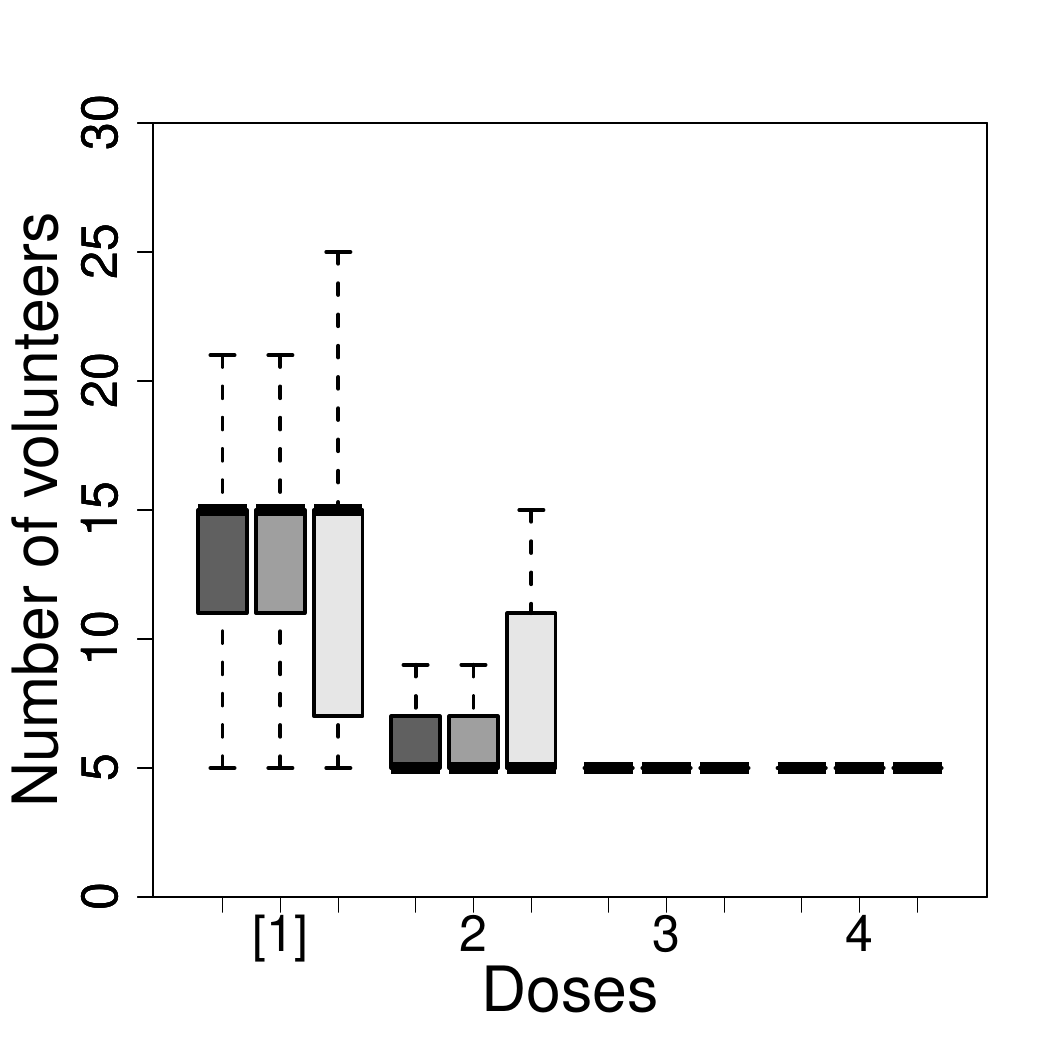}} 
\subfigure{\includegraphics[scale=0.29]{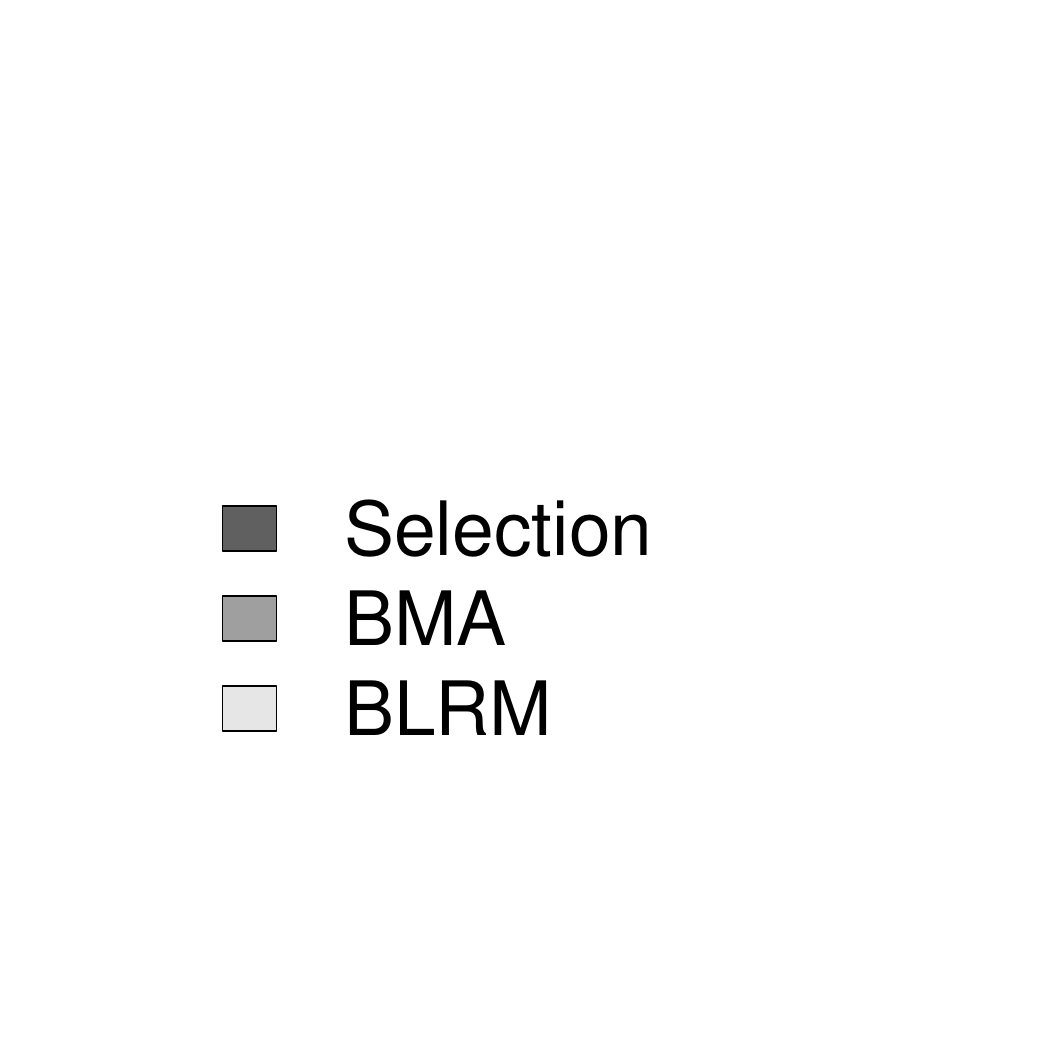}}
\end{center}
\caption{Comparison of the number of volunteers by dose for the selection method, BMA method  and BLRM for a maximum number of volunteers of $n = 30$, when $L = 4$ dose levels are used. 
\label{fig:boxplot_nb_volunt_by_dose_n30_L4}
}
\end{figure}

\begin{figure}[h!]
\begin{center}
\subfigure[Scenario 1]{\includegraphics[scale=0.29]{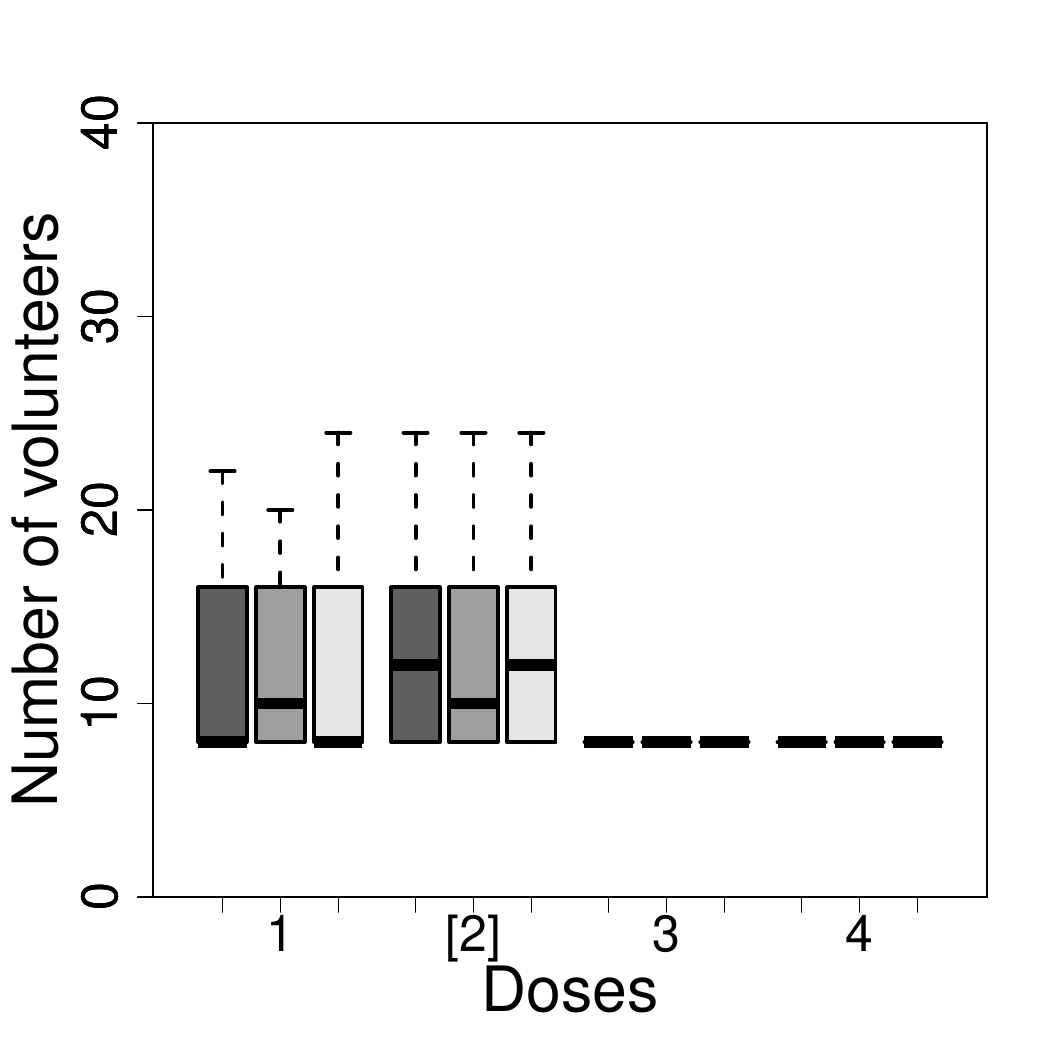}} 
\subfigure[Scenario 2]{\includegraphics[scale=0.29]{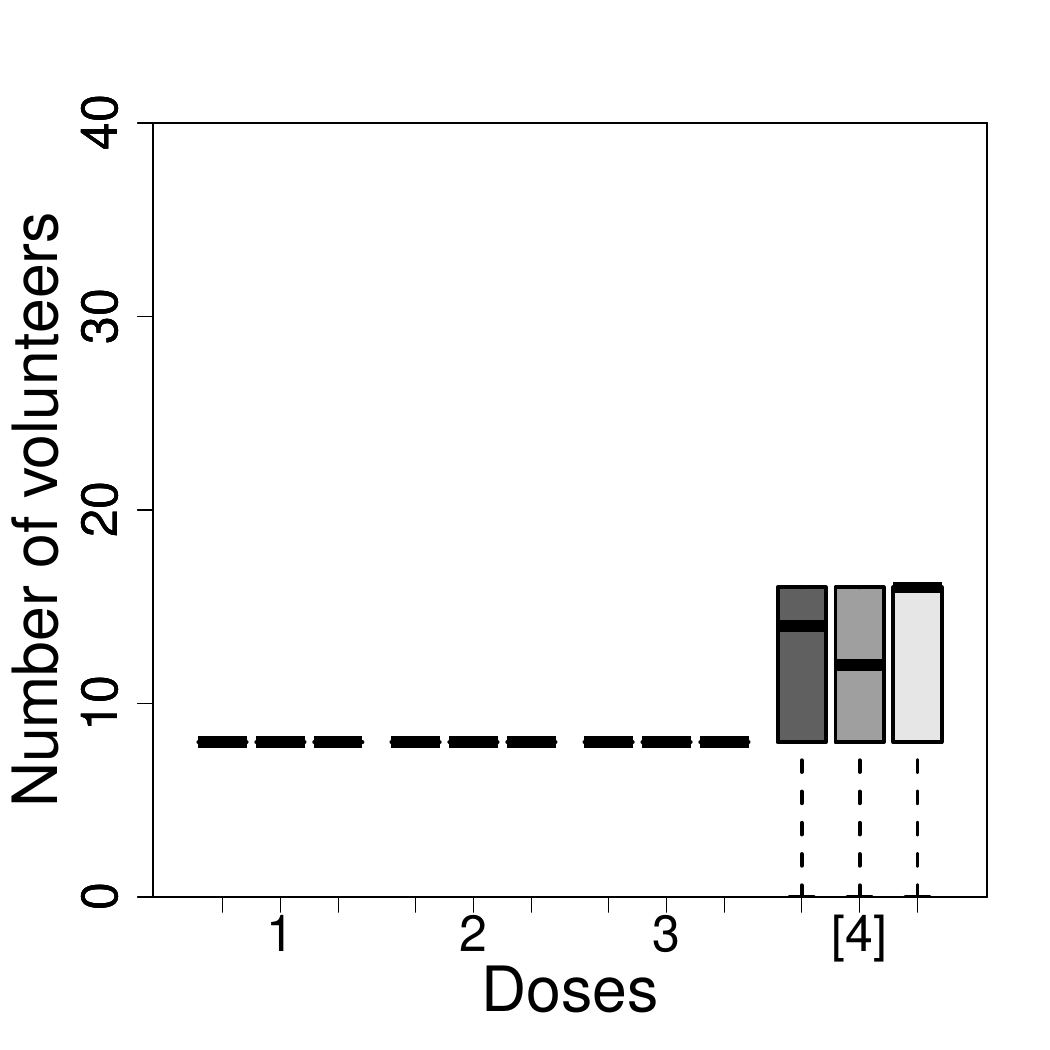}} 
\subfigure[Scenario 3]{\includegraphics[scale=0.29]{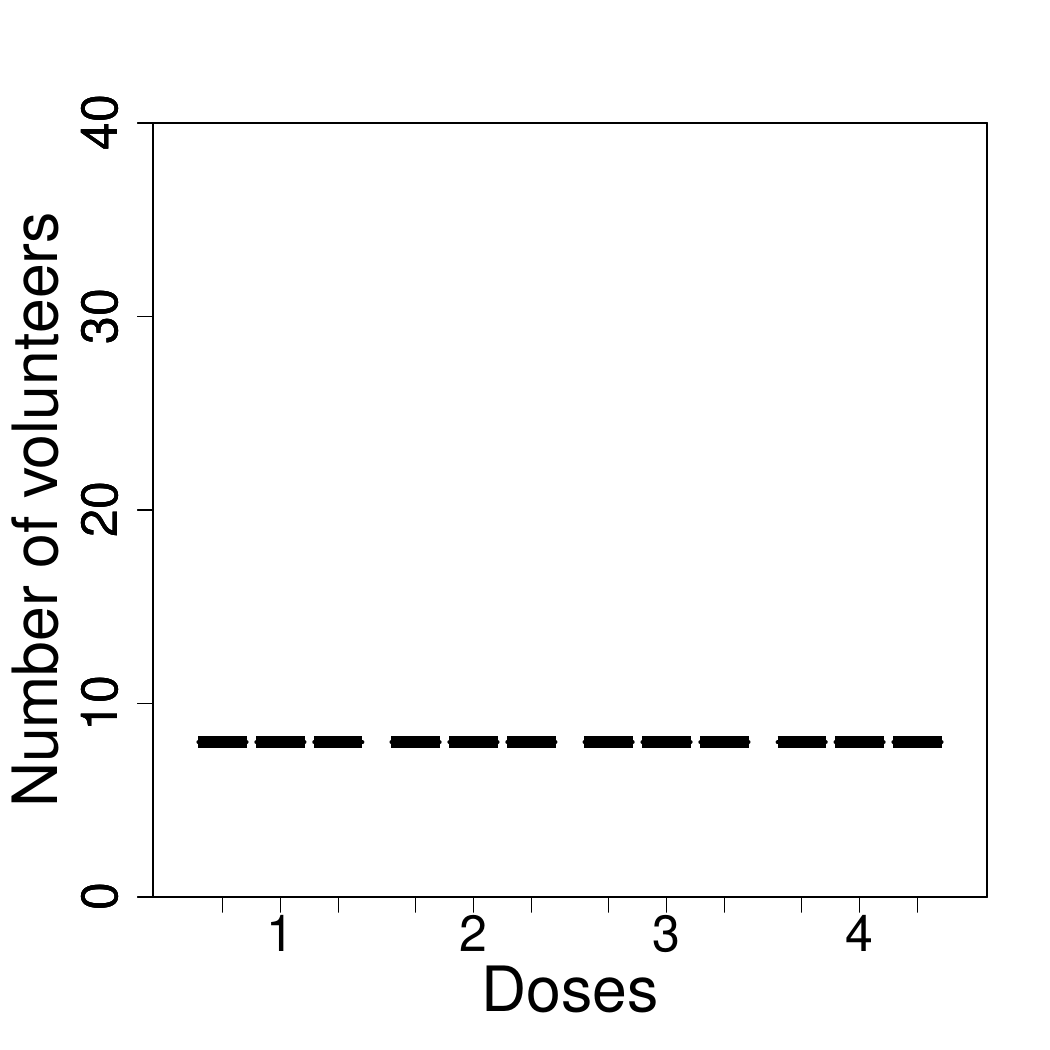}} \\ 
\subfigure[Scenario 4]{\includegraphics[scale=0.29]{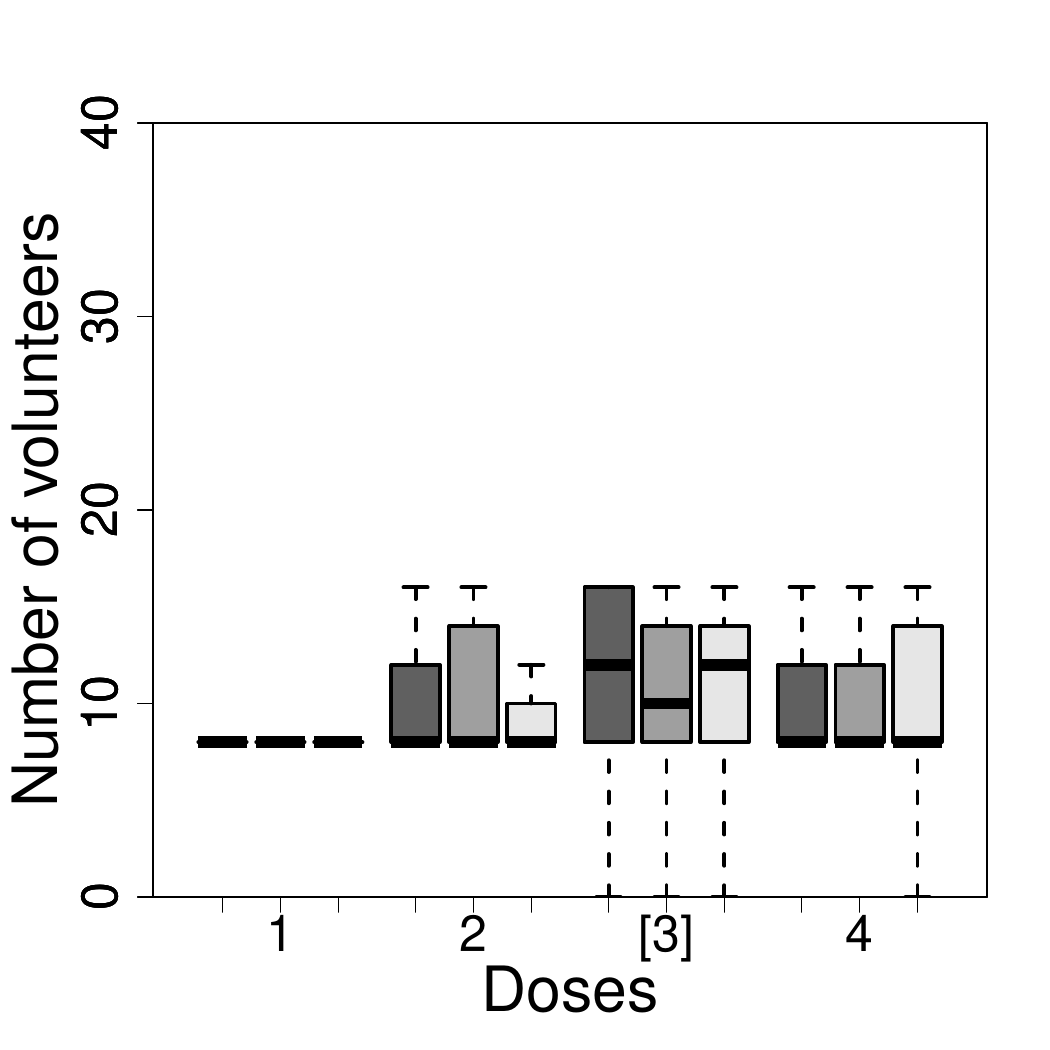}}
\subfigure[Scenario 5]{\includegraphics[scale=0.29]{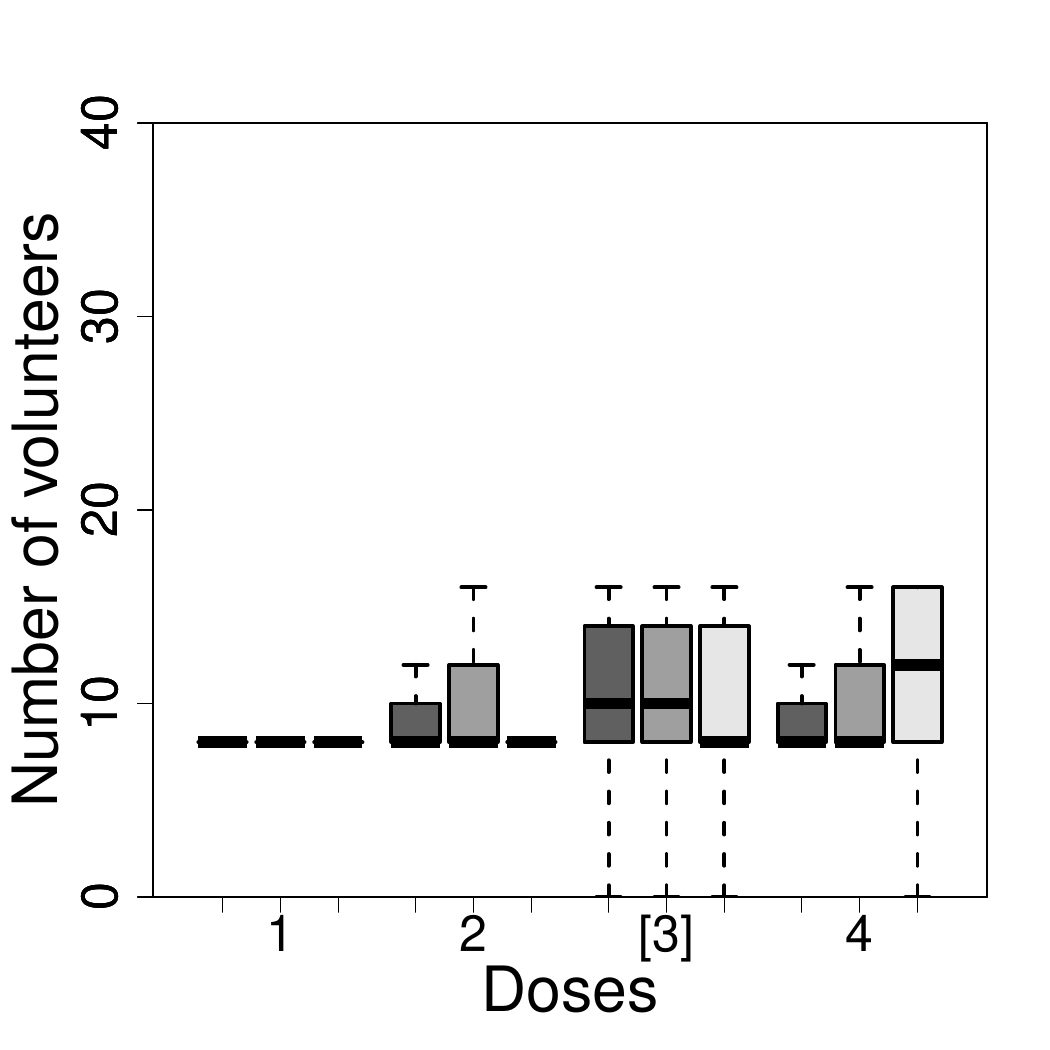}} 
\subfigure[Scenario 6]{\includegraphics[scale=0.29]{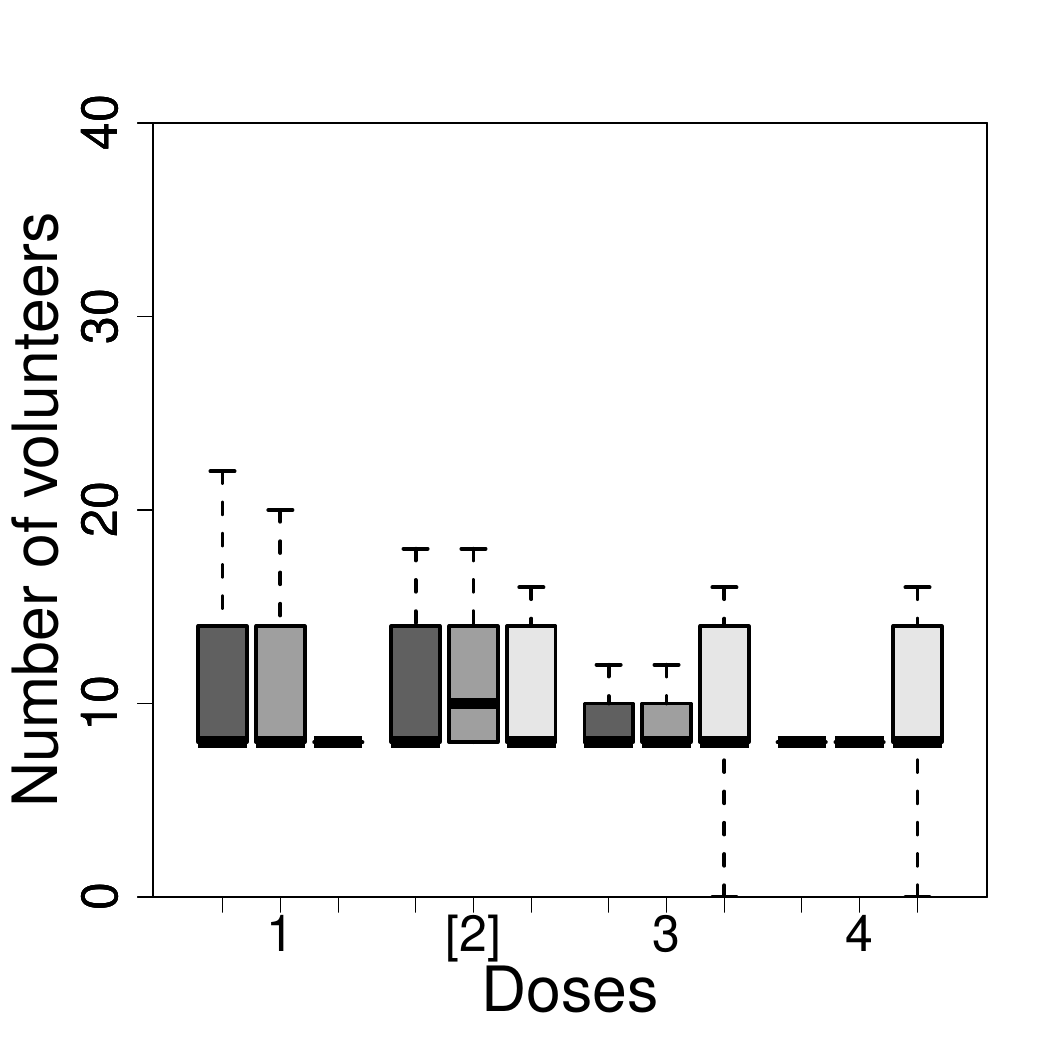}} \\ 
\subfigure[Scenario 7]{\includegraphics[scale=0.29]{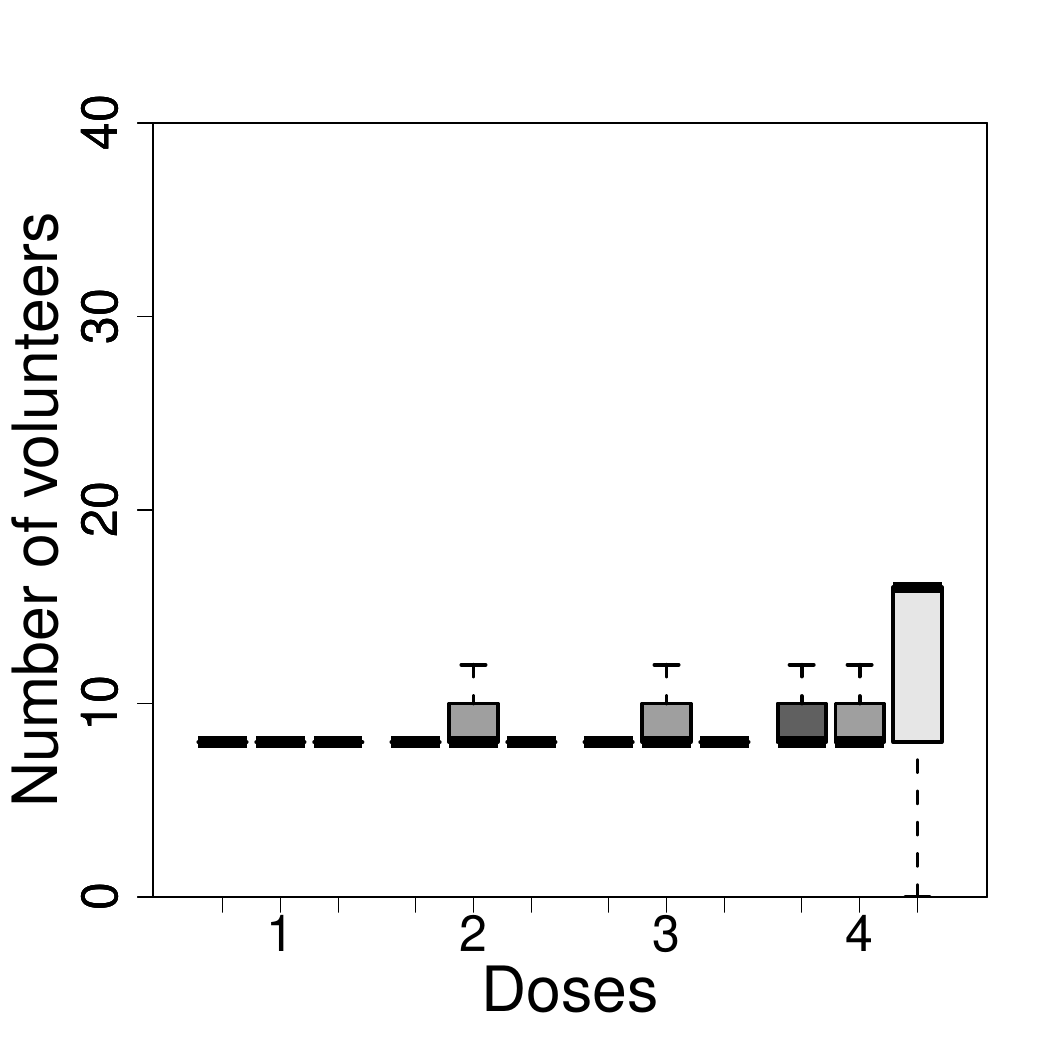}} 
\subfigure[Scenario 8]{\includegraphics[scale=0.29]{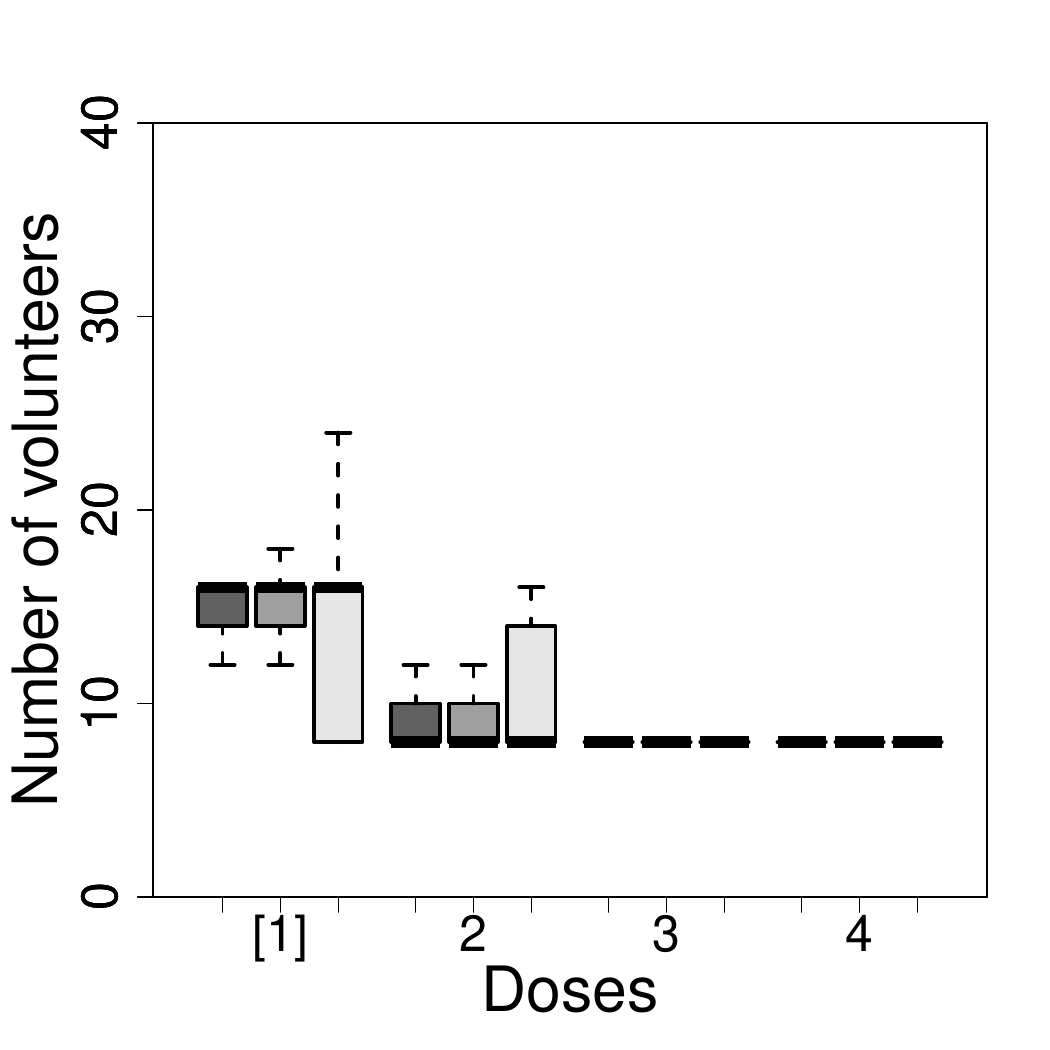}} 
\subfigure{\includegraphics[scale=0.29]{legend_boxplot_nb_volunt_by_dose.pdf}}
\end{center}
\caption{Comparison of the number of volunteers by dose for the selection method, BMA method  and BLRM for a maximum number of volunteers of $n = 40$, when $L = 4$ dose levels are used.  \label{fig:boxplot_nb_volunt_by_dose_n40_L4}}
\end{figure}

\begin{figure}[h!]
\begin{center}
\subfigure[Scenario 1]{\includegraphics[scale=0.29]{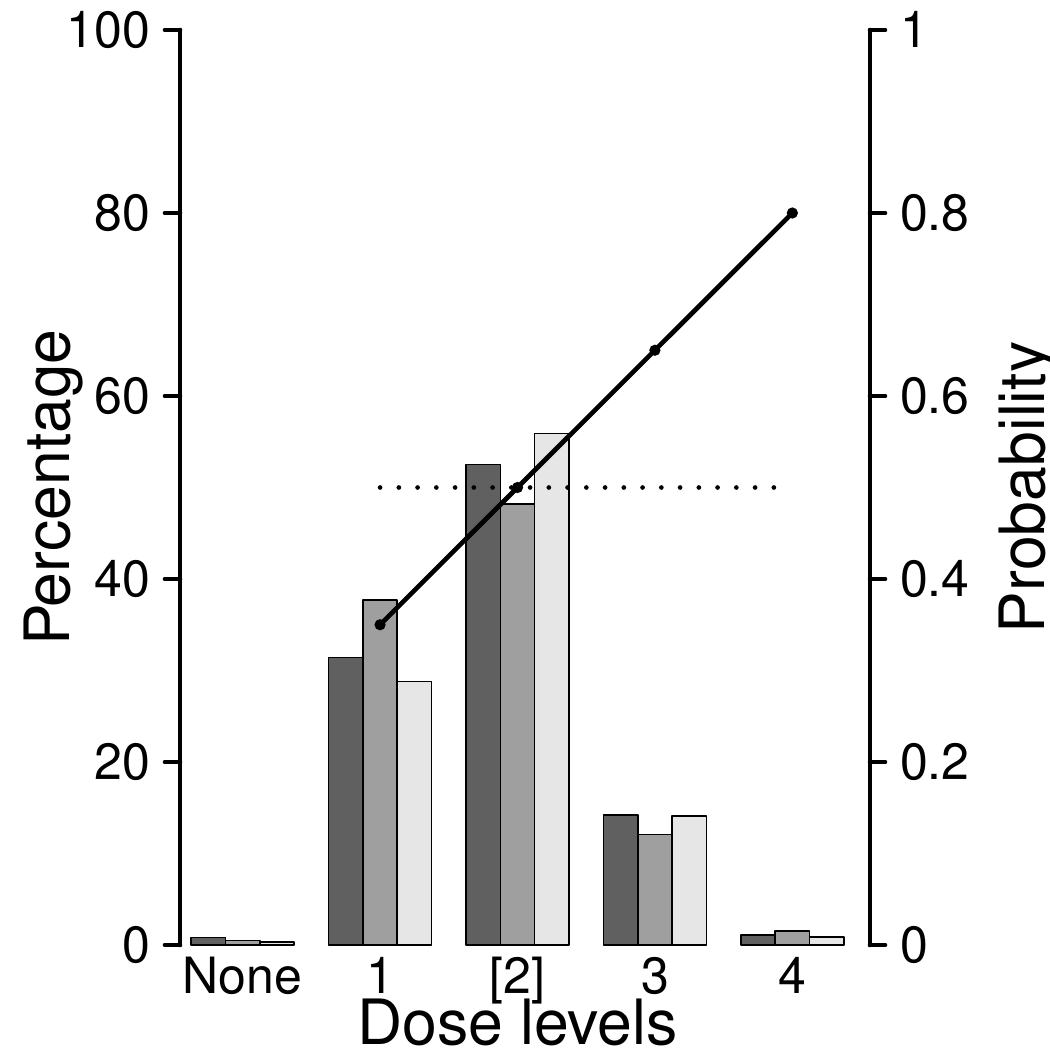}} 
\subfigure[Scenario 2]{\includegraphics[scale=0.29]{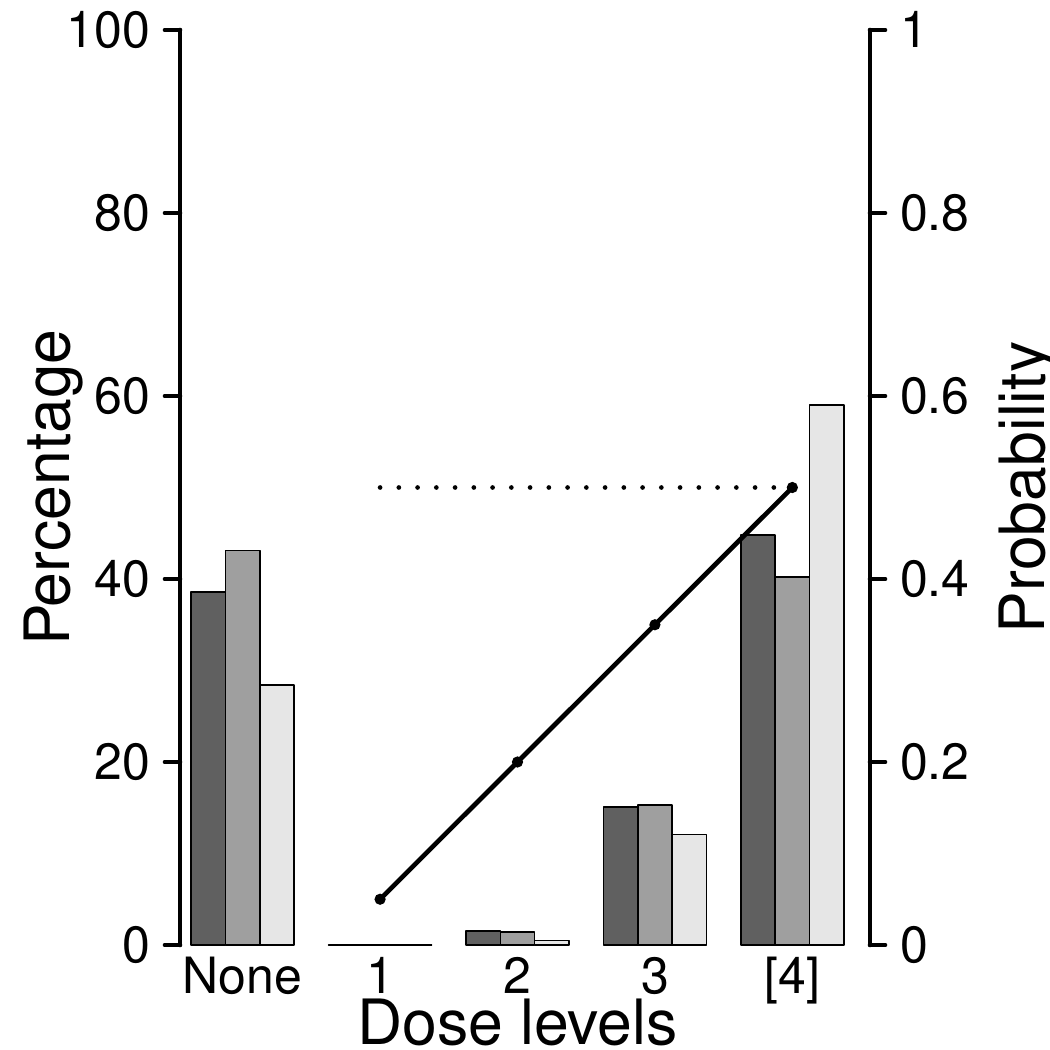}} 
\subfigure[Scenario 3]{\includegraphics[scale=0.29]{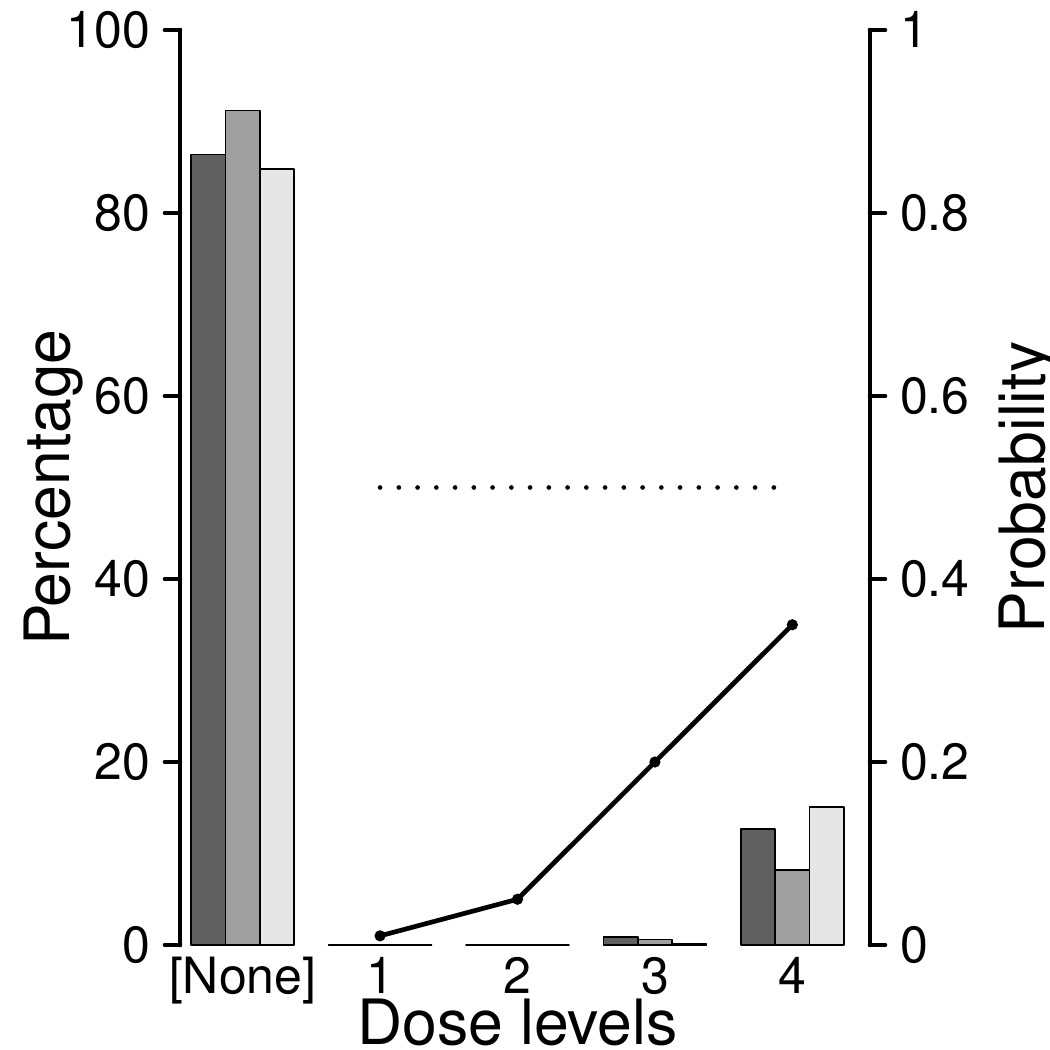}} \\ 
\subfigure[Scenario 4]{\includegraphics[scale=0.29]{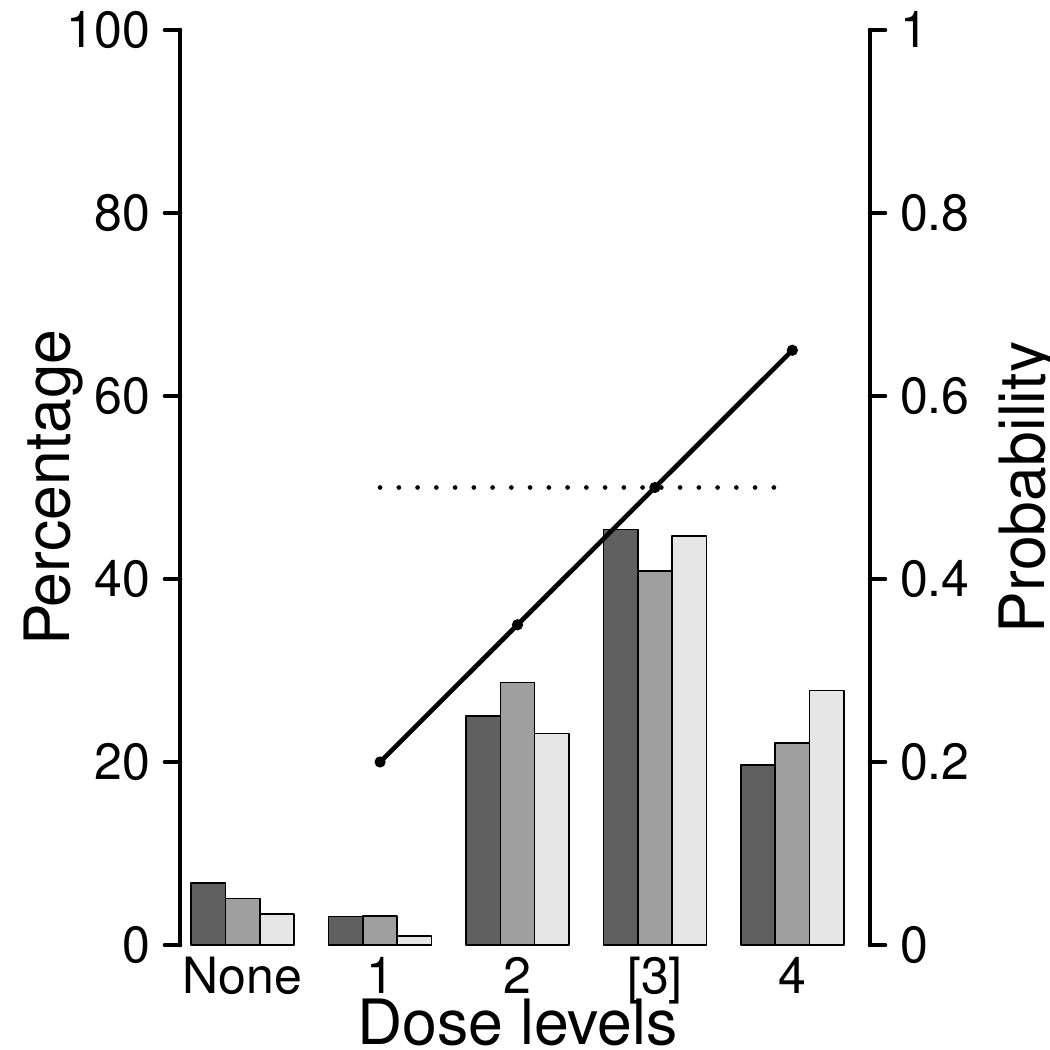}}
\subfigure[Scenario 5]{\includegraphics[scale=0.29]{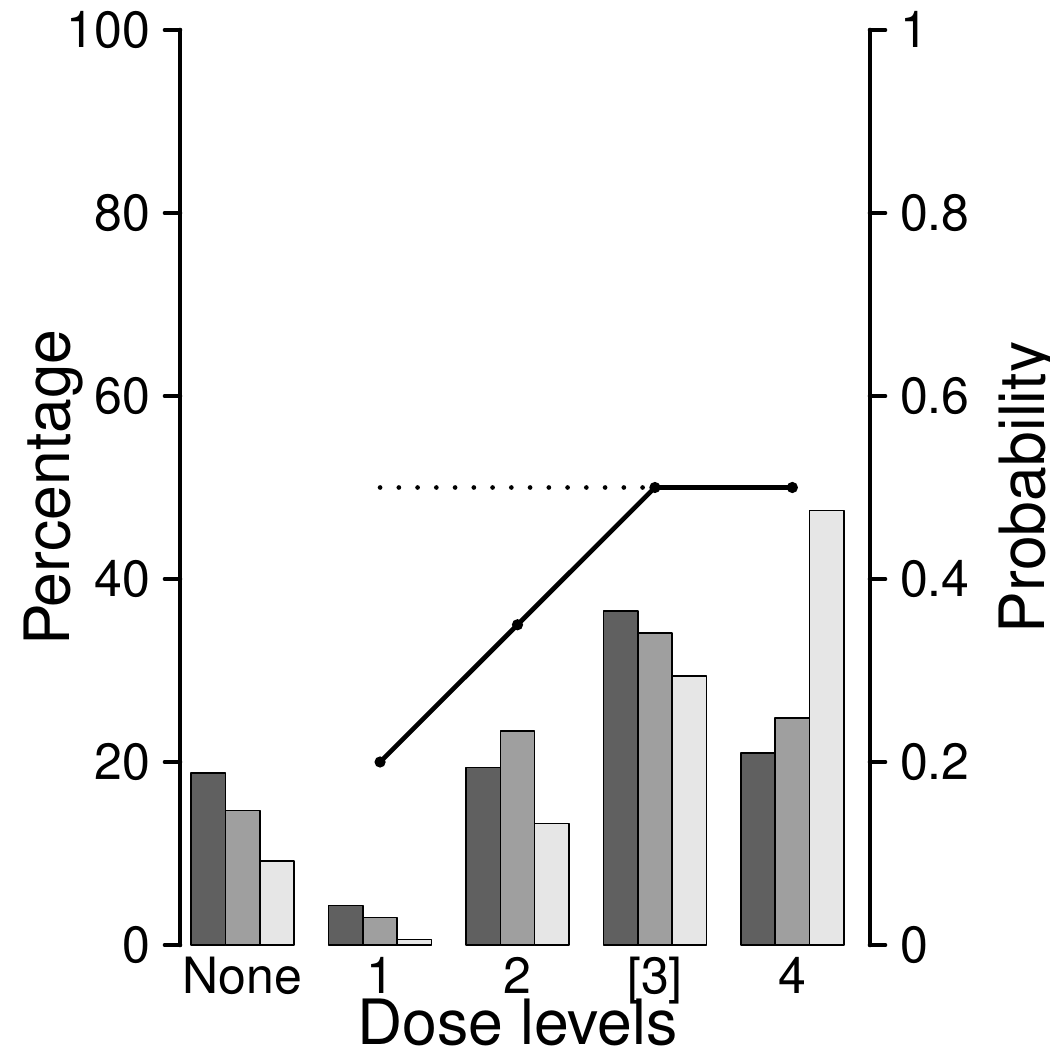}} 
\subfigure[Scenario 6]{\includegraphics[scale=0.29]{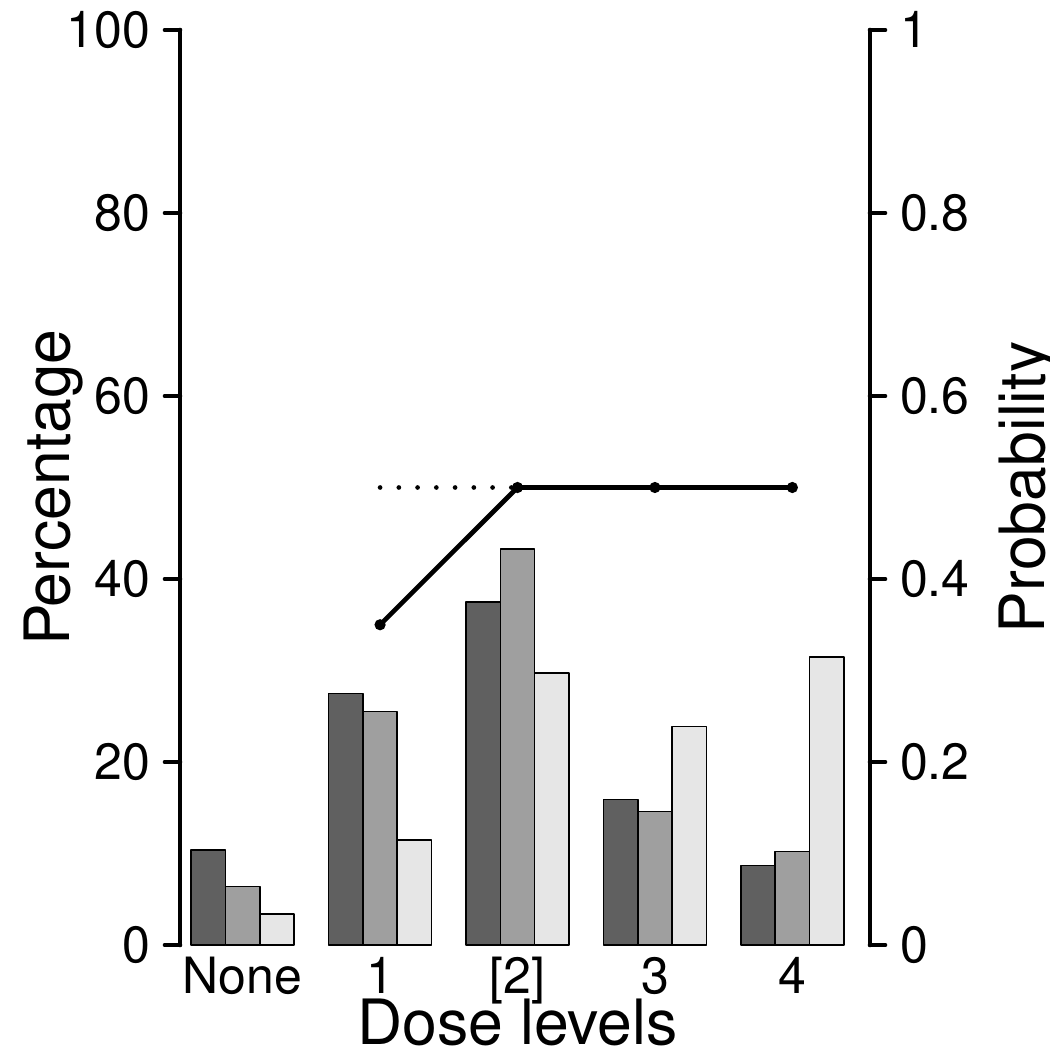}} \\ 
\subfigure[Scenario 7]{\includegraphics[scale=0.29]{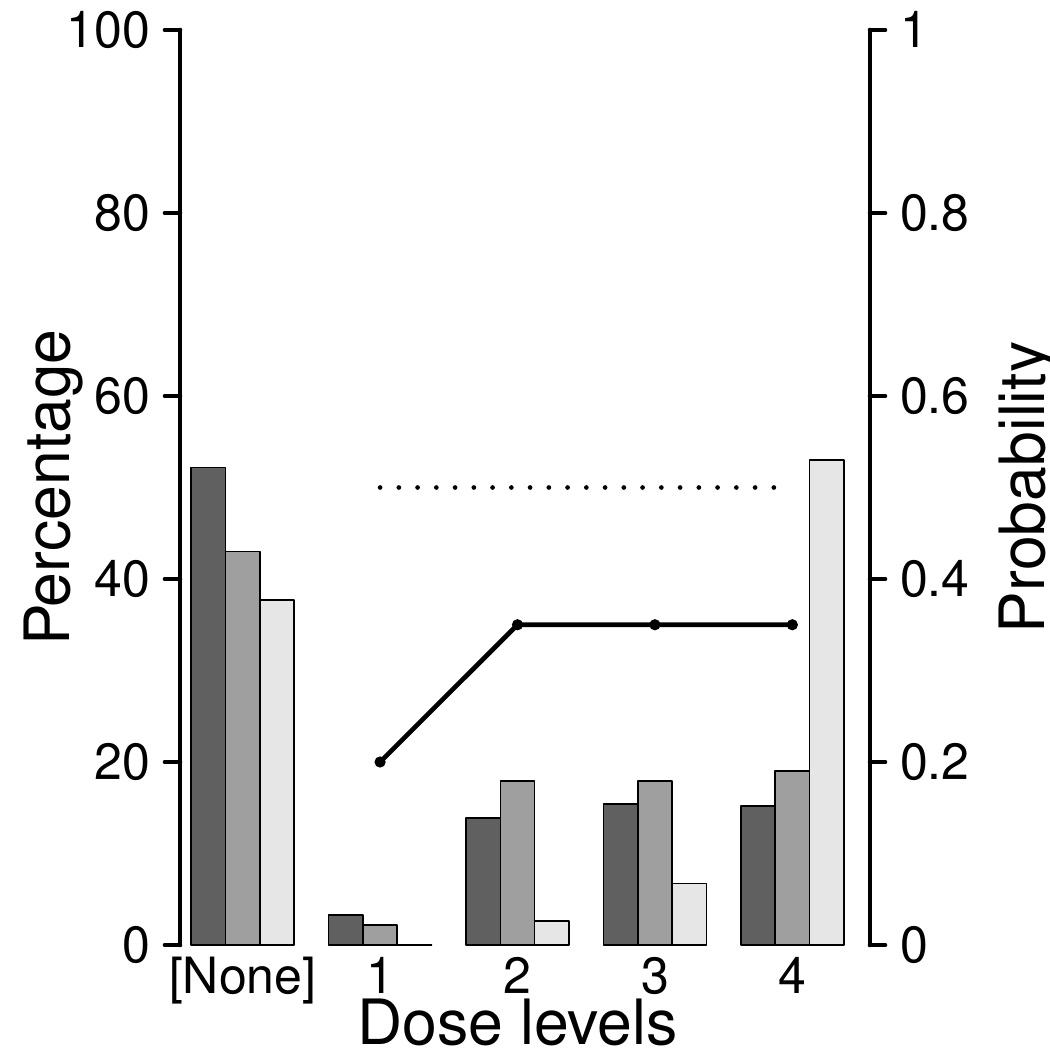}} 
\subfigure[Scenario 8]{\includegraphics[scale=0.29]{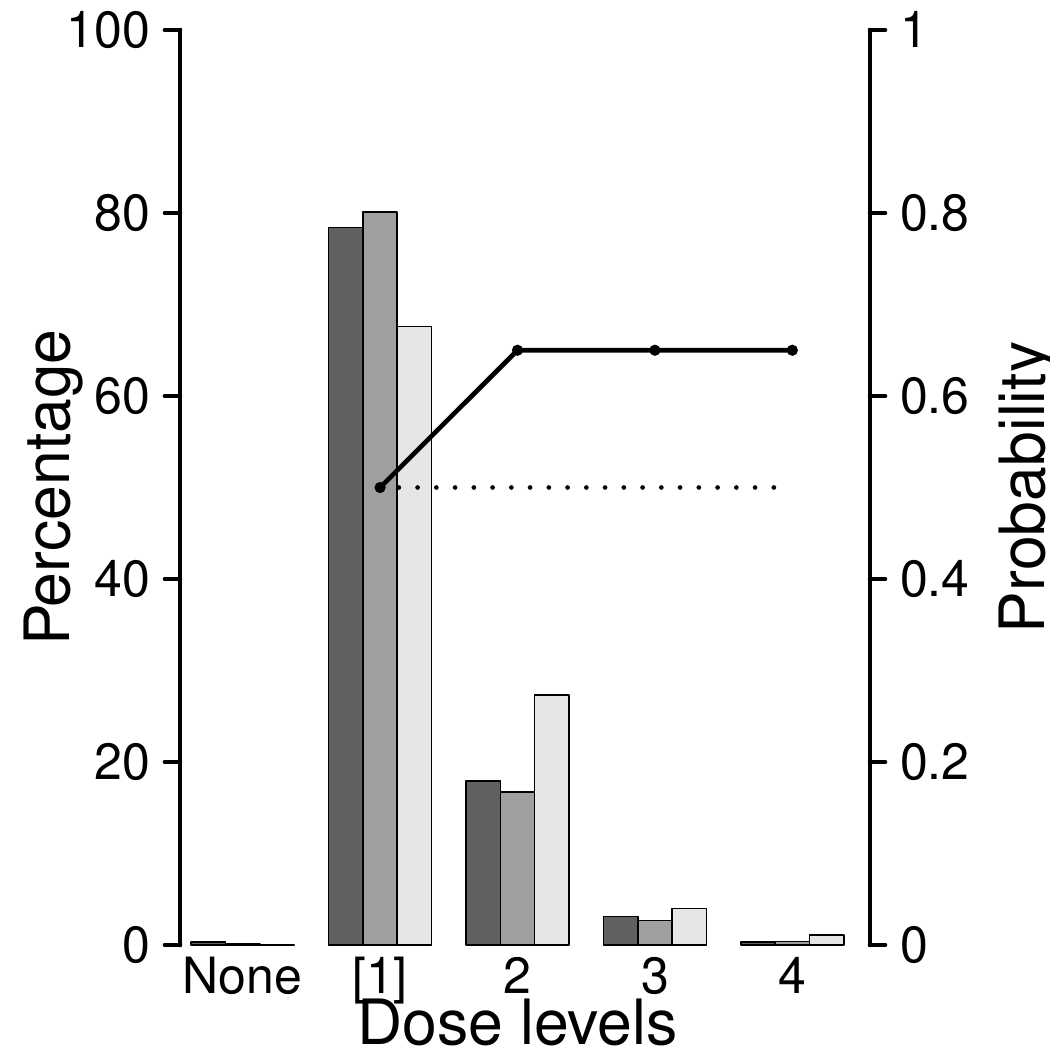}} 
\subfigure{\includegraphics[scale=0.29]{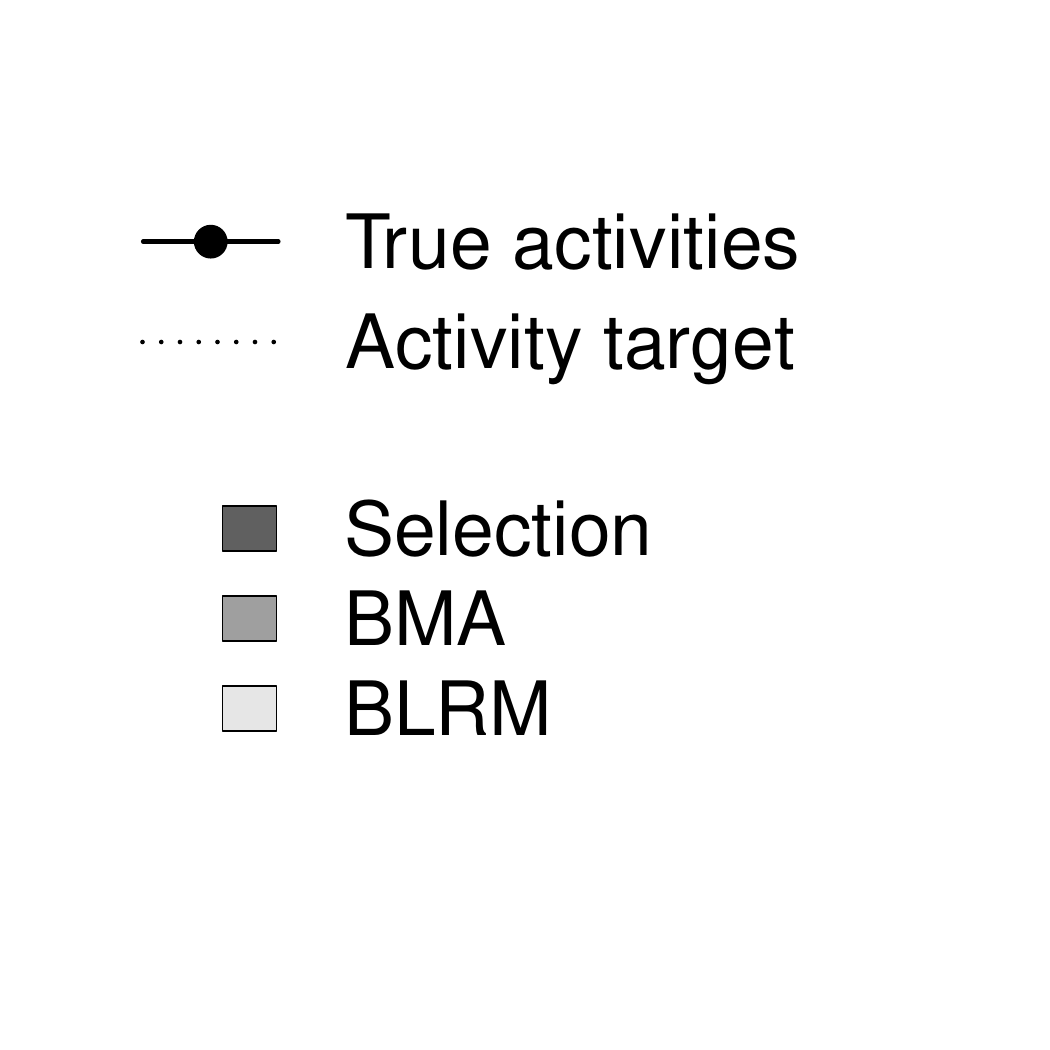}} 
\end{center}
\caption{Comparison of dose selection for the selection method, BMA method  and BLRM for a maximum number of volunteers of $n = 30$, when $L = 4$ dose levels are used. The solid line with circles represents the true dose-activity relationship. The horizontal dashed line is the activity probability target. \label{fig:barplot_MAD_hat_n30_L4}}
\end{figure}

\begin{figure}[h!]
\begin{center}
\subfigure[Scenario 1]{\includegraphics[scale=0.29]{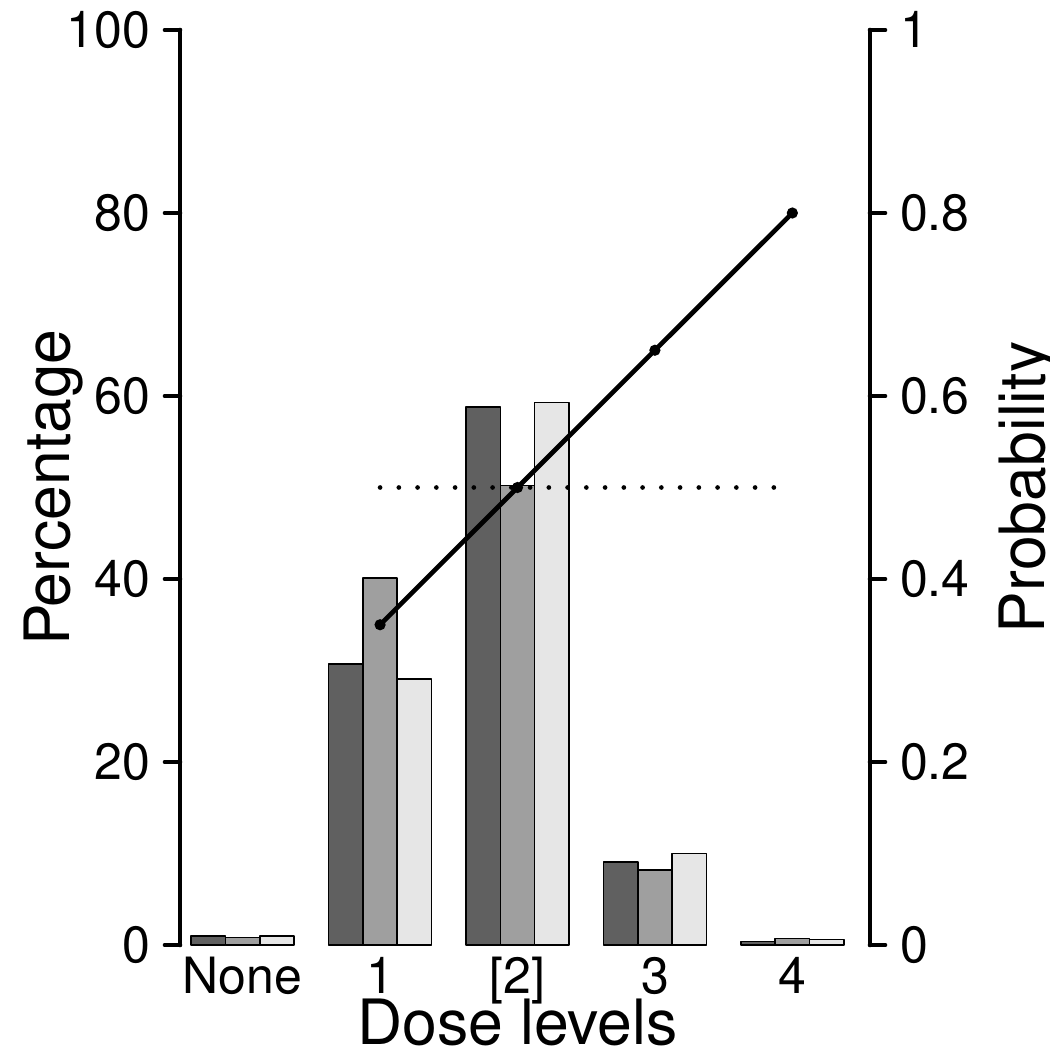}} 
\subfigure[Scenario 2]{\includegraphics[scale=0.29]{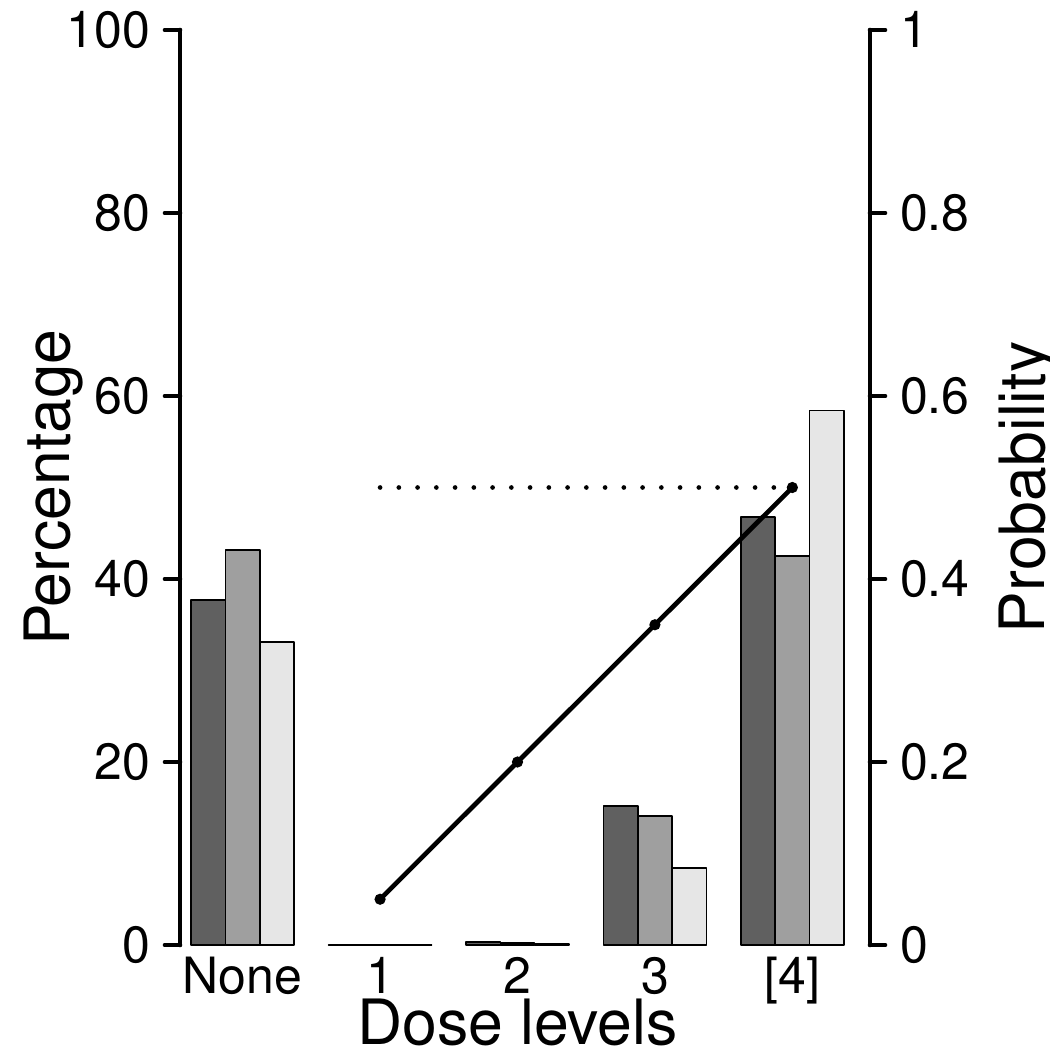}} 
\subfigure[Scenario 3]{\includegraphics[scale=0.29]{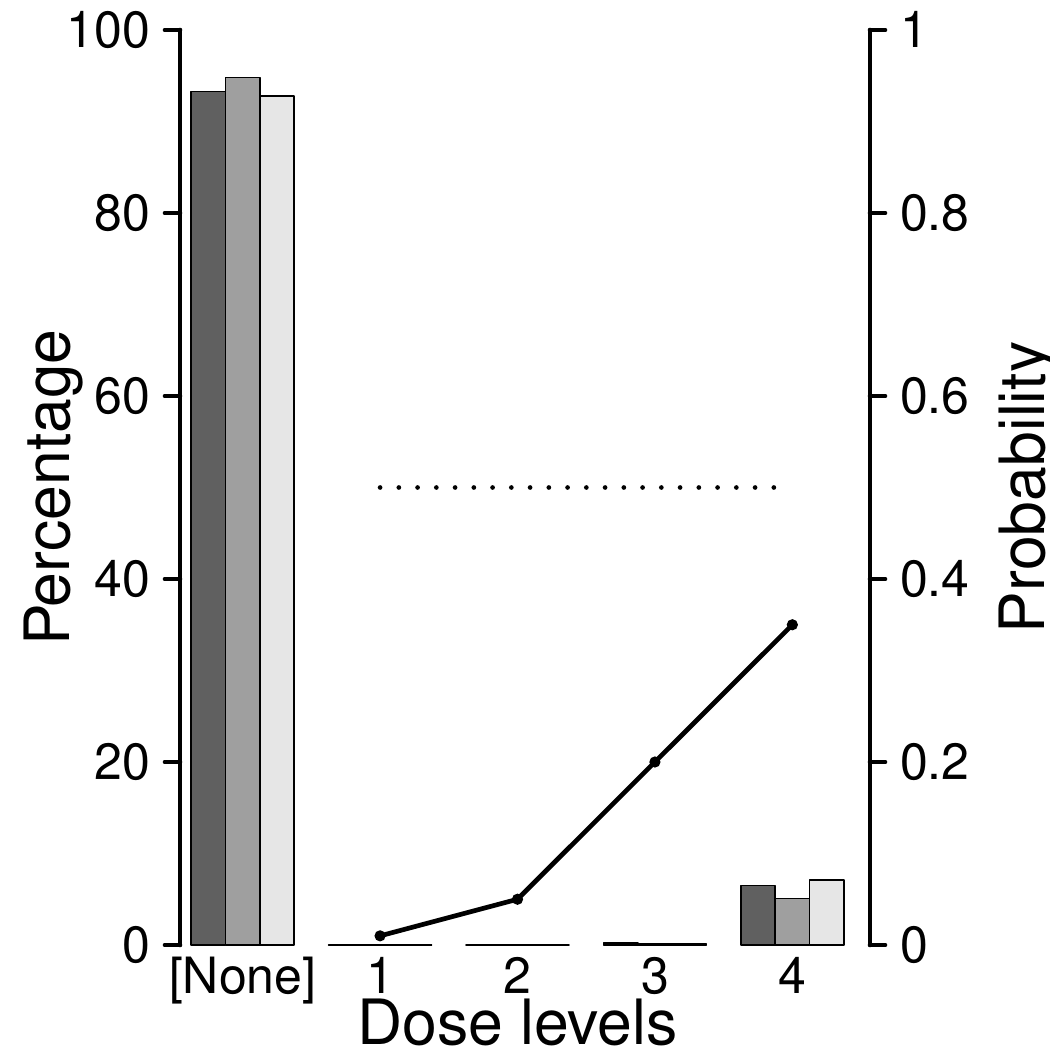}} \\ 
\subfigure[Scenario 4]{\includegraphics[scale=0.29]{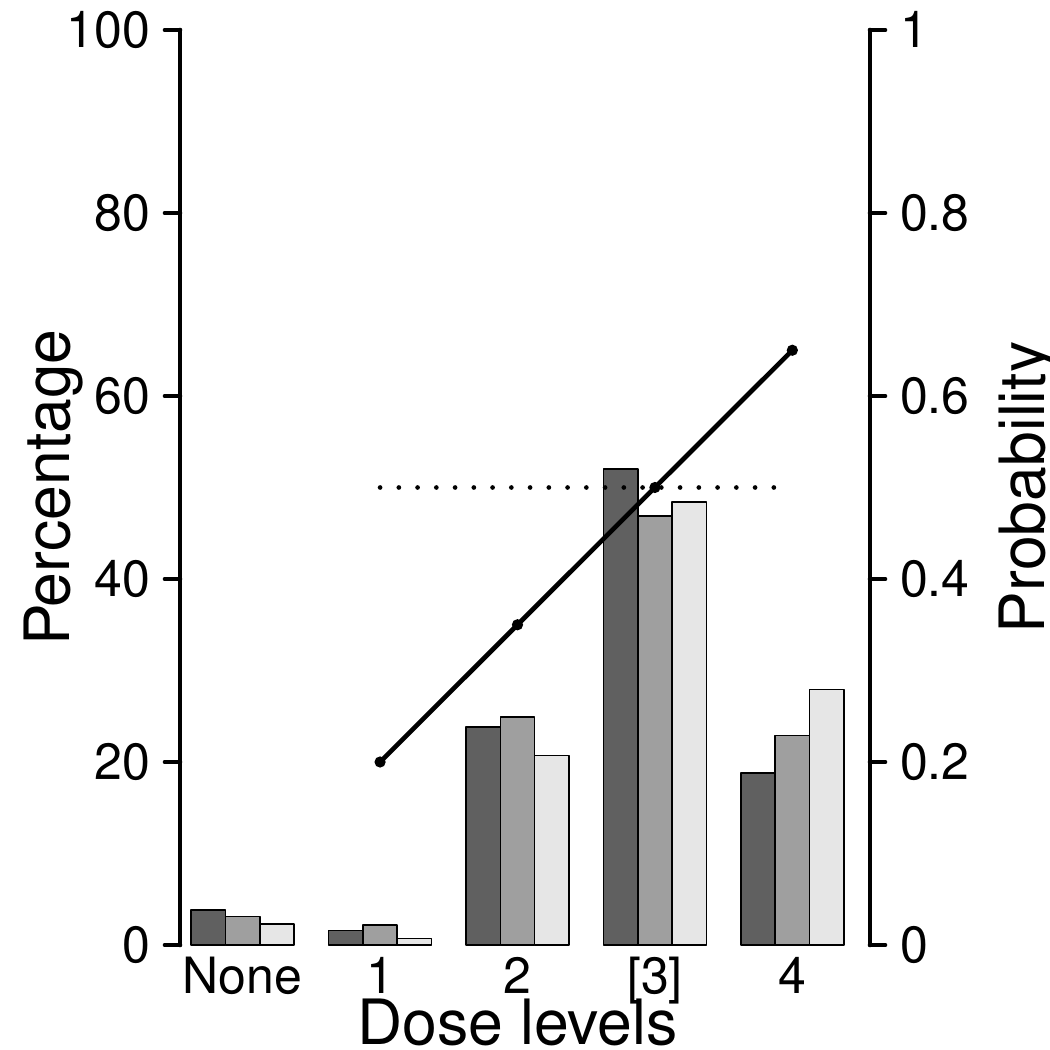}}
\subfigure[Scenario 5]{\includegraphics[scale=0.29]{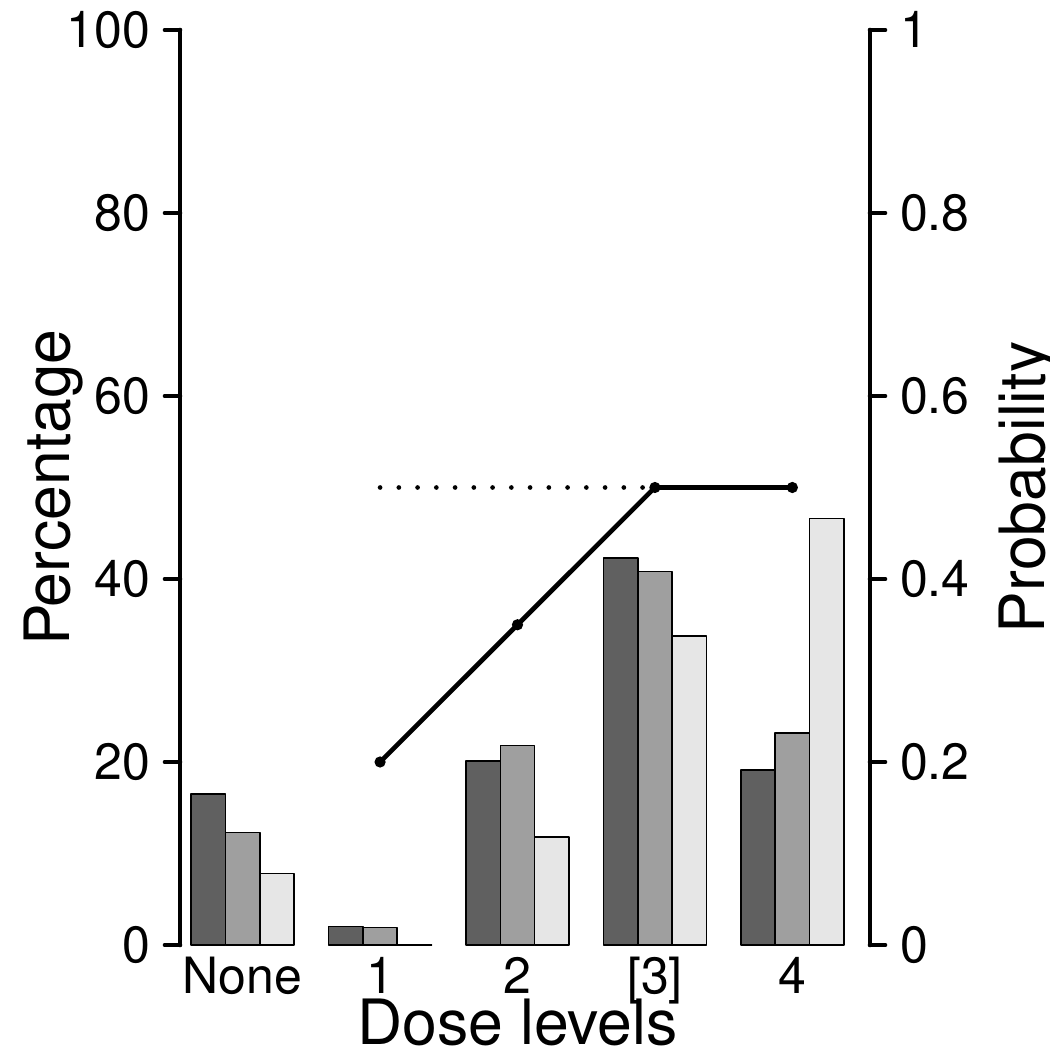}} 
\subfigure[Scenario 6]{\includegraphics[scale=0.29]{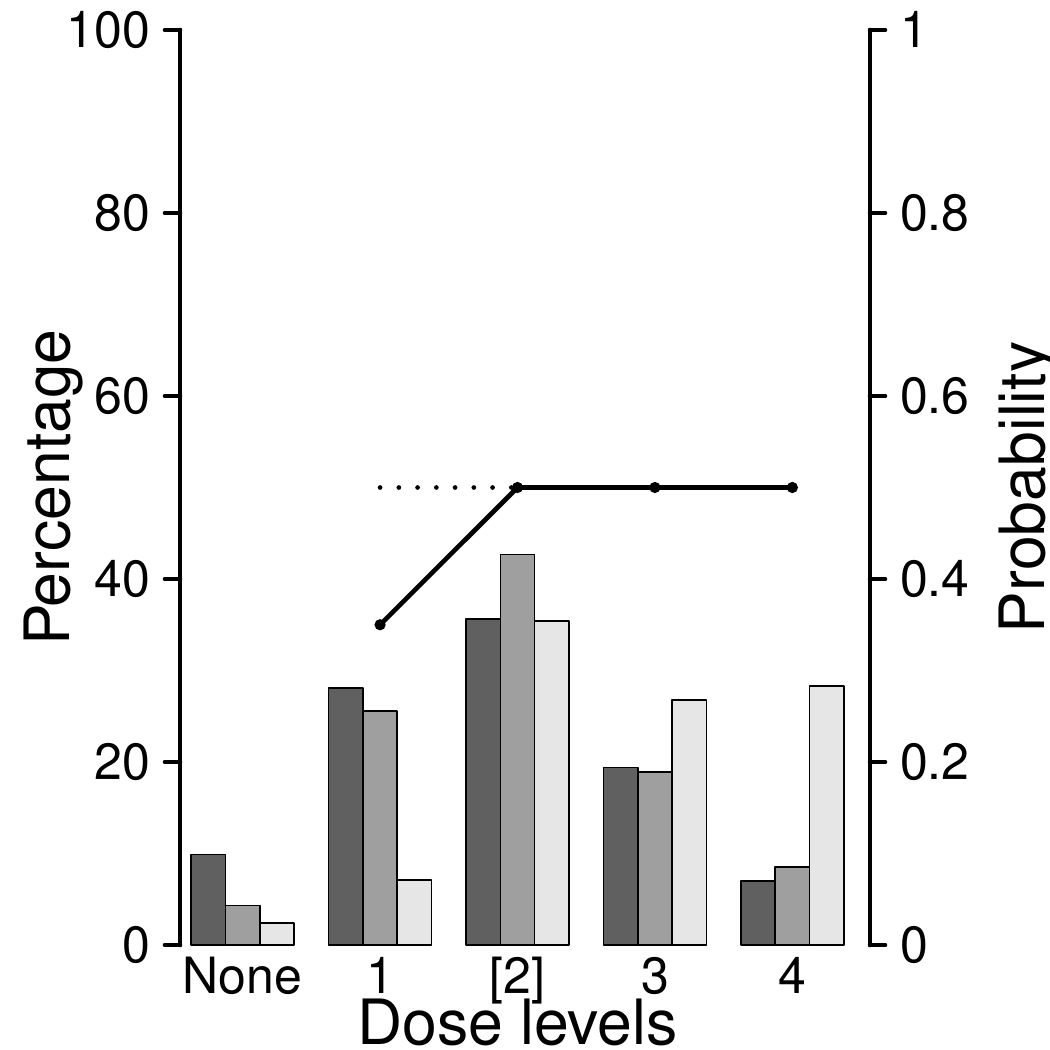}} \\ 
\subfigure[Scenario 7]{\includegraphics[scale=0.29]{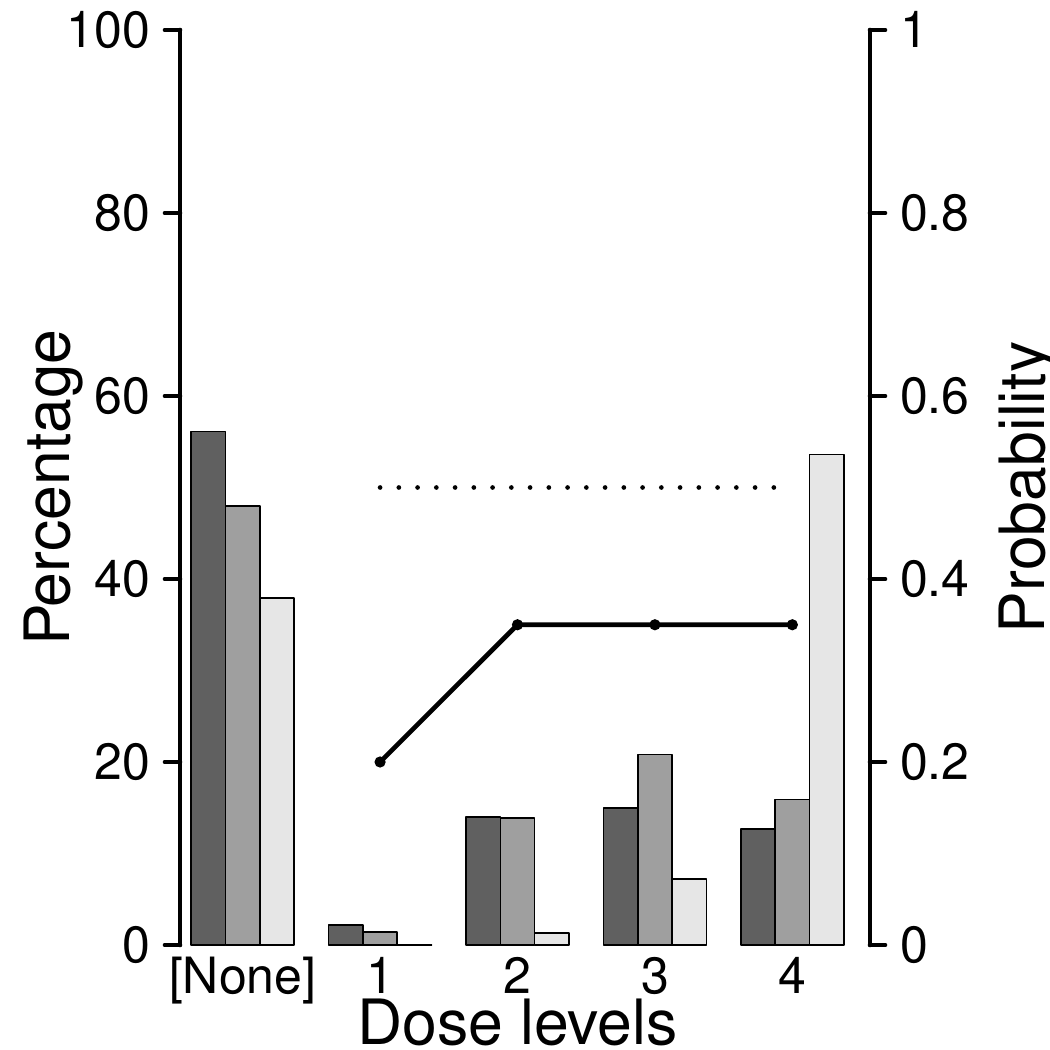}} 
\subfigure[Scenario 8]{\includegraphics[scale=0.29]{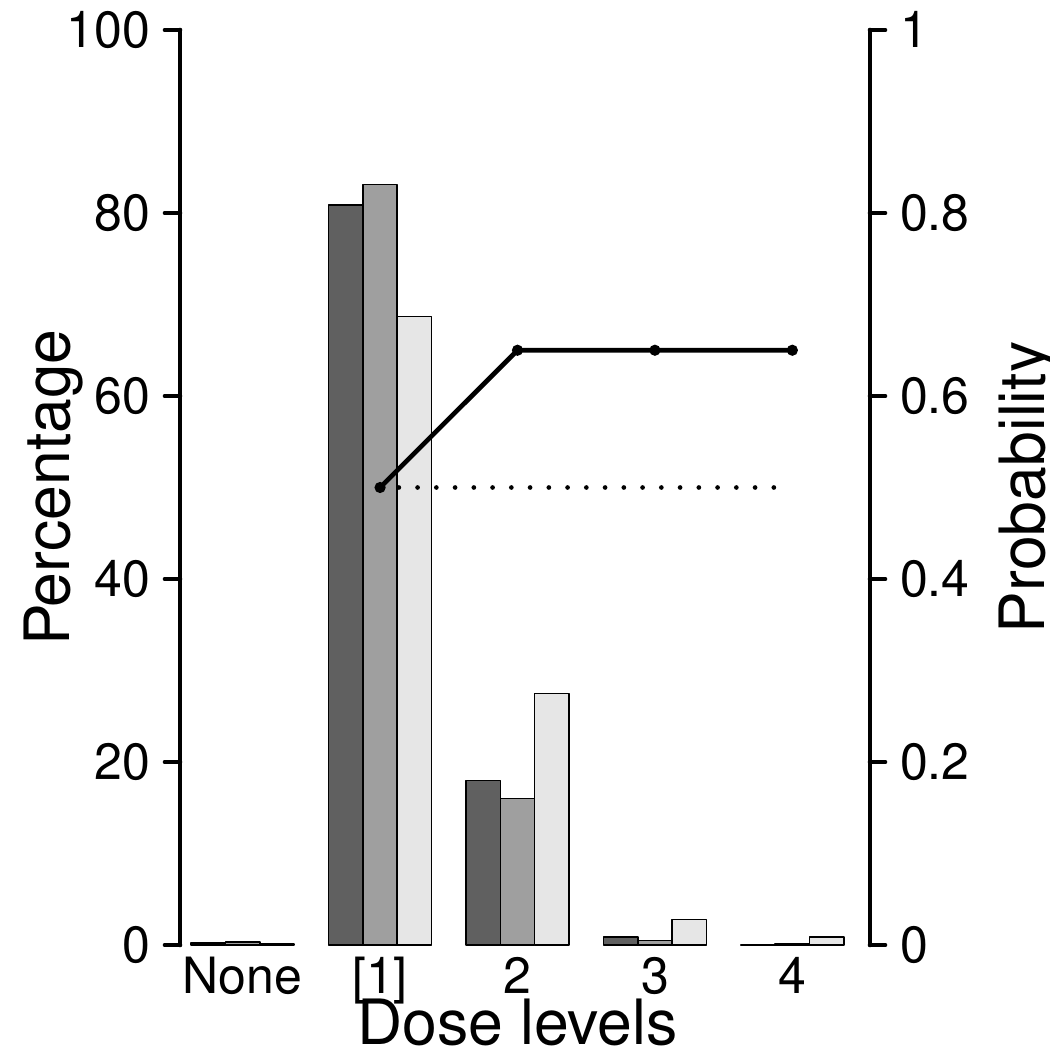}} 
\subfigure{\includegraphics[scale=0.29]{legend_barplot_OBD_hat}} 
\end{center}
\caption{Comparison of dose selection for the selection method, BMA method  and BLRM for a maximum number of volunteers of $n = 40$, when $L = 4$ dose levels are used. The solid line with circles represents the true dose-activity relationship. The horizontal dashed line is the activity probability target. \label{fig:barplot_MAD_hat_n40_L4}}
\end{figure}

More generally, for scenarios that do not involve a plateau, all three methods show similar performance except in Scenario 2. 
Indeed, when the MAD is the last dose, BLRM selects the optimal dose approximately 10\% to 20\% more often, and allocates a few more volunteers to this dose, than do our methods.

For scenarios that involve a plateau, BLRM more often selects a dose larger than the optimal dose than do our methods.
When the optimal dose is both the first dose of the plateau and the penultimate possible dose (Scenario 5), all the methods show similar performance in terms of the selection of the optimal dose but BLRM allocates a few more participants to the last dose. When the optimal dose is both the first dose of the plateau and the second possible dose (Scenario 6), BMA shows slightly better dose selection performance than the other methods, and BLRM tends to assign fewer participants to the first dose and more to the last. Finally, when the optimal dose is located before the plateau (Scenario 8), the selection method and BMA more often select the optimal dose (between 78.4\% and 83.1\%), and tend to assign more individuals to this dose, than does the BLRM method (67.6\% and 68.7\%). Moreover, in this scenario, BLRM allocates a few more volunteers to the second dose than our proposed methods. When no dose is sufficiently active (Scenario 7), the selection method and BMA select no dose (between 43.0\% and 56.1\%) more often than does BLRM (37.7\% and 37.9\%). In this last scenario, the number of volunteers assigned to the last dose is higher for BLRM than for our proposed methods.

Finally, in most scenarios, the selection method performs as well as the BMA method. Furthermore, the selection method and BMA stop the trial to protect volunteers from futile doses, usually more often than does BLRM. In other words, our methods enroll slightly fewer people in case of inactive drug.

In Appendix \ref{sec:app_simu_res}, in subsection \ref{subsec:app_nb_volunteers_res}, Figures \ref{fig:boxplot_nb_volunt_by_dose_n30_L3} and \ref{fig:boxplot_nb_volunt_by_dose_n40_L3} and Figures \ref{fig:boxplot_nb_volunt_by_dose_n30_L5} and \ref{fig:boxplot_nb_volunt_by_dose_n40_L5} show the distributions of the number of volunteers when $L = 3$ and $L =5$ dose levels are used, respectively, for the $n \in \{30, 40\}$ maximum number of volunteers. Similarly, in subsection \ref{subsec:app_MAD_hat_res}, Figures \ref{fig:barplot_MAD_hat_n30_L3} and \ref{fig:barplot_MAD_hat_n40_L3} and Figures \ref{fig:barplot_MAD_hat_n30_L5} and \ref{fig:barplot_MAD_hat_n40_L5} show the percentages of dose selection. In the Supplementary Material, supplementary tables and figures about dose selection percentages and numbers of volunteers allocated to each dose are provided, including results for $n \in \{18, 24\}$ maximum number of volunteers.

\clearpage

\section{Discussion}\label{sec:discussion}

We developed a dose-finding design in two steps that aims to determine, under safety constraints, among a set of doses, the optimal target dose when the increasing dose-activity curve is likely to reach a plateau. The first step is a dose escalation phase that enables us to guarantee the safety of the drug in volunteers and to collect information on the drug activity to prepare the second step. The second step is a model-based dose-finding phase. In this phase, several dose-activity models, which differ according to the possible location of a plateau in the dose-activity curve, are estimated. We proposed two different ways to choose the final dose-activity model. The first consists of simply selecting the model that best fits the data according to the \textit{posterior} probabilities of the models, while the second combines all the models using BMA. 

In the NEBUFLAG trial that motivates this work, PD markers are used as a surrogate for immunostimulation, assuming this coming from flagellin concentrations in the lung. Indeed, the concentration of the molecule in the lung over time is difficult to measure; consequently, no PK data are produced. Nevertheless, good surrogates for the biological activity of the drug can easily be  collected. Nonetheless, if scintigraphy is performed, more complex models based on PK data could be developed. Interested readers could be inspired by the work of \cite{gerard_2022}, who developed a Bayesian approach that uses a PD endpoint to estimate the relationship that links the dose regimen and toxicity at the end of the dose-escalation phase of a trial (once all PK/PD and toxicity data are collected). Another interesting work is that of \cite{su_2022}, who modelled the dose-response link using (1) a population PK model for the dose/schedule-concentration link, (2) a PD model for the concentration and latent pharmacologic effect intensity link and (3) a model for the link between the cumulative pharmacologic effect and clinical outcome. 

Since our trial is applied to healthy volunteers, no (or very little) toxicity is expected, so there is no toxicity model  and the toxicity is not estimated. The slightest toxicity that appears in a healthy volunteer leads to the elimination of the dose tested and higher doses.

Applying our design to a real study involves first to define the maximum number of healthy volunteers, the doses, the binary activity endpoint, the target activity rate and the initial guesses of the activity probabilities. Carry out a simulation study allows to select the last parameter values. Then, during clinical trials, there may be dropout and missing data. However, in our motivating trial, the healthy volunteers are paid, so usually a bit more compliant than patients, and the blood tests needed to define the activity endpoint are done in a short time window (one week). Hence, this problem is hardly expected. However, the majority of clinical trial designs encounter the same difficulties when such a problem arises, and simply does not use the data of the volunteer concerned.

In our method, none of the volunteers received a placebo or diluent. However, such control patient data can be  necessary in secondary analysis with continuous PD endpoint, for example to estimate the baseline effect $E_0$ in a $E_{\max}$ model or for other models that account for patient's variability and small sample size. Then, it is possible either (1) to assign three or four healthy volunteers to this control at the start of the trial or (2) for each new cohort of volunteers in the start-up phase to randomly assign one of the volunteers to the control. The first proposal ensures that enough volunteers receive the product, even if the trial is stopped early (without having tested all the proposed doses). The second proposal increases the probability that volunteers will be comparable between the treated and untreated groups. It is also standard practice in dose escalation designs in healthy volunteers to randomly assign subjects to placebo at each dose level, even in the second stage, in order to maintain study blinding.

We suggested to introduce a target activity rate to define the starting dose of an activity window. However, our models can be easily adapted to estimate the lowest dose showing an activity in a maximum number of volunteers.

The assumption of a plateau in the dose-activity relationship in humans is often derived from preclinical studies (\textit{in vitro} and \textit{in vivo}) where such a plateau is observed. However, preclinical studies can lead to very different results from those observed in humans, and the extrapolation methods used to go from an \textit{in vitro} or \textit{in vivo} model to a human model are questionable \citep{us_food_and_drug_administration_guidance_2005}. Nevertheless, if several preclinical studies agree on the presence of a plateau in the dose-activity relationship and these include the most relevant studies with respect to humans, it is reasonable to look for the presence of such a plateau in the FIH trial. In addition, our proposed model also considers the potential lack of a plateau.

When a plateau exists, the proposed methods outperform the usual BLRM in selecting the optimal dose and allocate fewer volunteers to futile doses. In the absence of a plateau, in the majority of cases, the proposed method shows similar performance to the usual BLRM. To sum up, the methods developed have good properties if the assumption about the non-decreasing relationship between the dose and the probability of activity is more or less respected. However, if this relationship is, for instance, bell-shaped (as in the case of bispecific antibodies), another method has to be used. A drawback of our method using BMA compared to our method using selection is that its estimated activity probabilities are not constant on the plateau. 

Our design does not stop volunteers enrollment even if the optimal dose has already been identified with sufficient precision. Indeed, in Phase I in healthy subjects and/or in FIH trial, it is preferable to allocate doses to as many healthy volunteers as possible, to extract more information with the final analyses. Moreover, in general regulatory authorities ask for these data. However, if required, a stopping rule for activity based on the length of the credibility interval, or similar to the stopping rule for futility (see equation \ref{eq:act_constraint}) with a threshold of 0.95 for example, could be used.

Finally, in this work, based on the inclusion and exclusion criteria of our motivating clinical trial, and as recommended by the \cite{ema_1998}, the patient population is assumed homogeneous. Nonetheless, different covariates, such as patient demographics (e.g. age, gender and ethnicity), baseline health conditions (e.g. pre-existing diseases, comorbidities and nutritional status), and other relevant factors (e.g. genetic factors, concomitant medications, lifestyle factors, adherence to treatment), might influence the activity response. That is why, usually, at the end of the trial, several exploratory analyses are performed to better understand trial results. Future work could introduce covariate effect in the dose allocation rules.

\paragraph{Acknowledgements} 
This work is part of the European FAIR project, which is supported by the European Union's Horizon 2020 research and innovation program under Grant number 847786.
The National Institute of Health and Medical Research (Inserm, France) will be the sponsor of the upcoming clinical trial (Inserm number: C20-48).
\\ The Version of Record of this manuscript has been published and is available in Statistics in Biopharmaceutical Research (02 Dec 2024) \url{http://www.tandfonline.com/10.1080/19466315.2024.2416410}.

\paragraph{Declaration of Interest}
The authors report that there are no competing interests to declare.

%% ** The bibliograhy **
%\bibliographystyle{ba}
\bibliography{FAIR_art2_main_manuscript_for_arXiv.bib}% place <bib-data-file> in ./bib folder

%%%%%%%%%%%%%%%%%%%%%%%%%%%%%%%%%%%%%%%%%%%%%%
%% Supplementary Material, if any, should   %%
%% be provided in {supplement} environment  %%
%% with title and short description.        %%
%%%%%%%%%%%%%%%%%%%%%%%%%%%%%%%%%%%%%%%%%%%%%%
\paragraph{Supplementary material}
Web Appendices, Tables and Figures, referenced in Section \ref{sec:simulation_results}}, are available as supplementary material.
%with this paper at the Statistics in Biopharmaceutical website on Taylors \& Francis Online. 

\appendix

\section{Figures Comparing Simulation Results}\label{sec:app_simu_res}

\subsection{Number of Volunteers by Dose}\label{subsec:app_nb_volunteers_res}

\begin{figure}[h!]
\begin{center}
\subfigure[Scenario 1]{\includegraphics[scale=0.29]{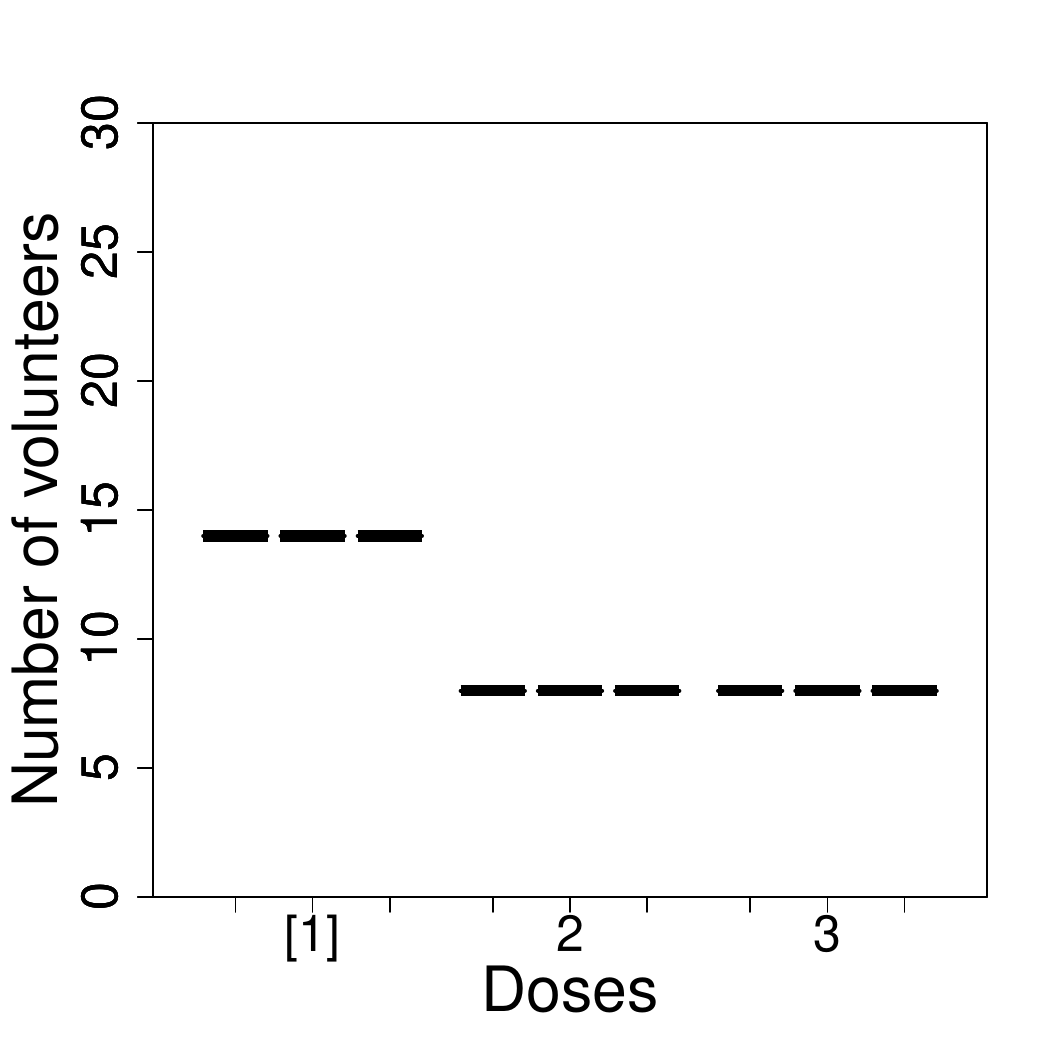}} 
\subfigure[Scenario 2]{\includegraphics[scale=0.29]{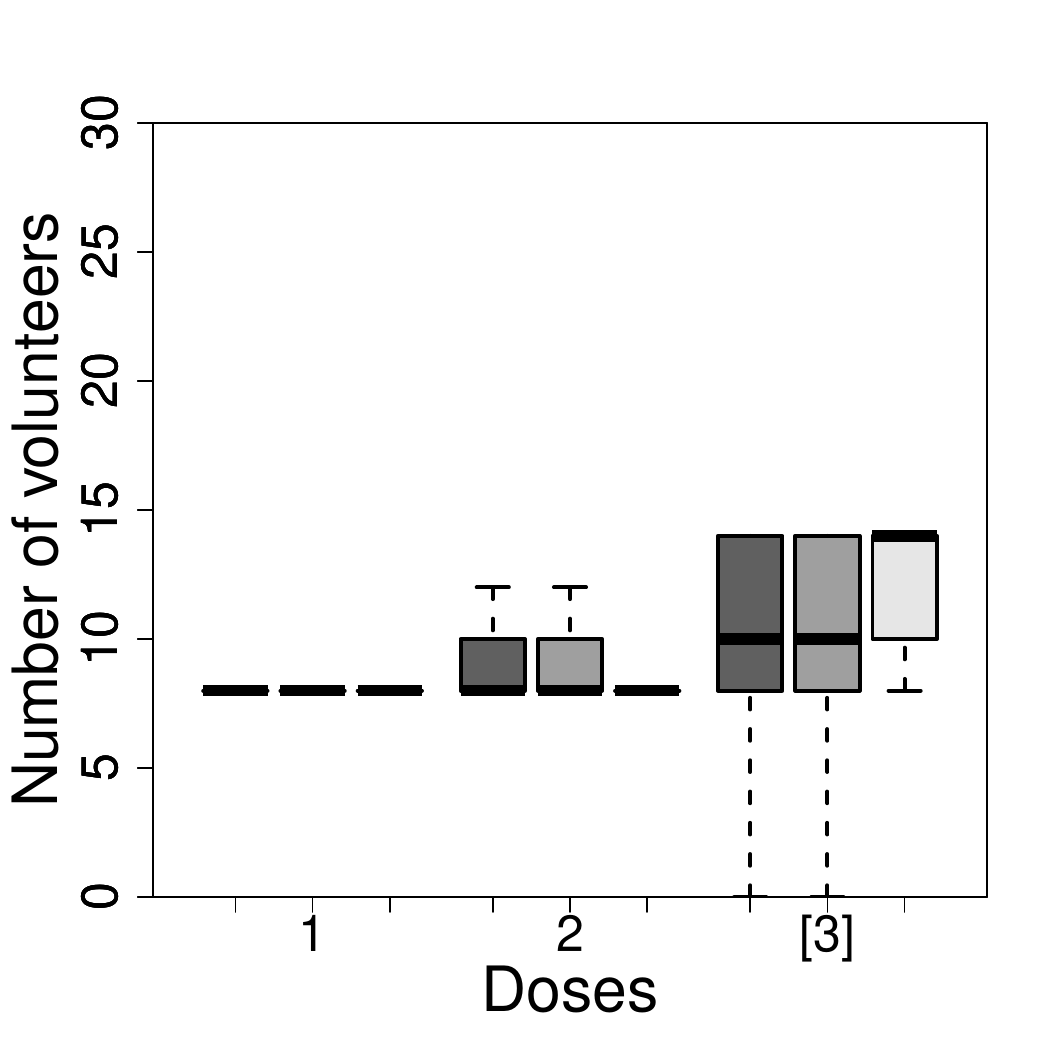}} 
\subfigure[Scenario 3]{\includegraphics[scale=0.29]{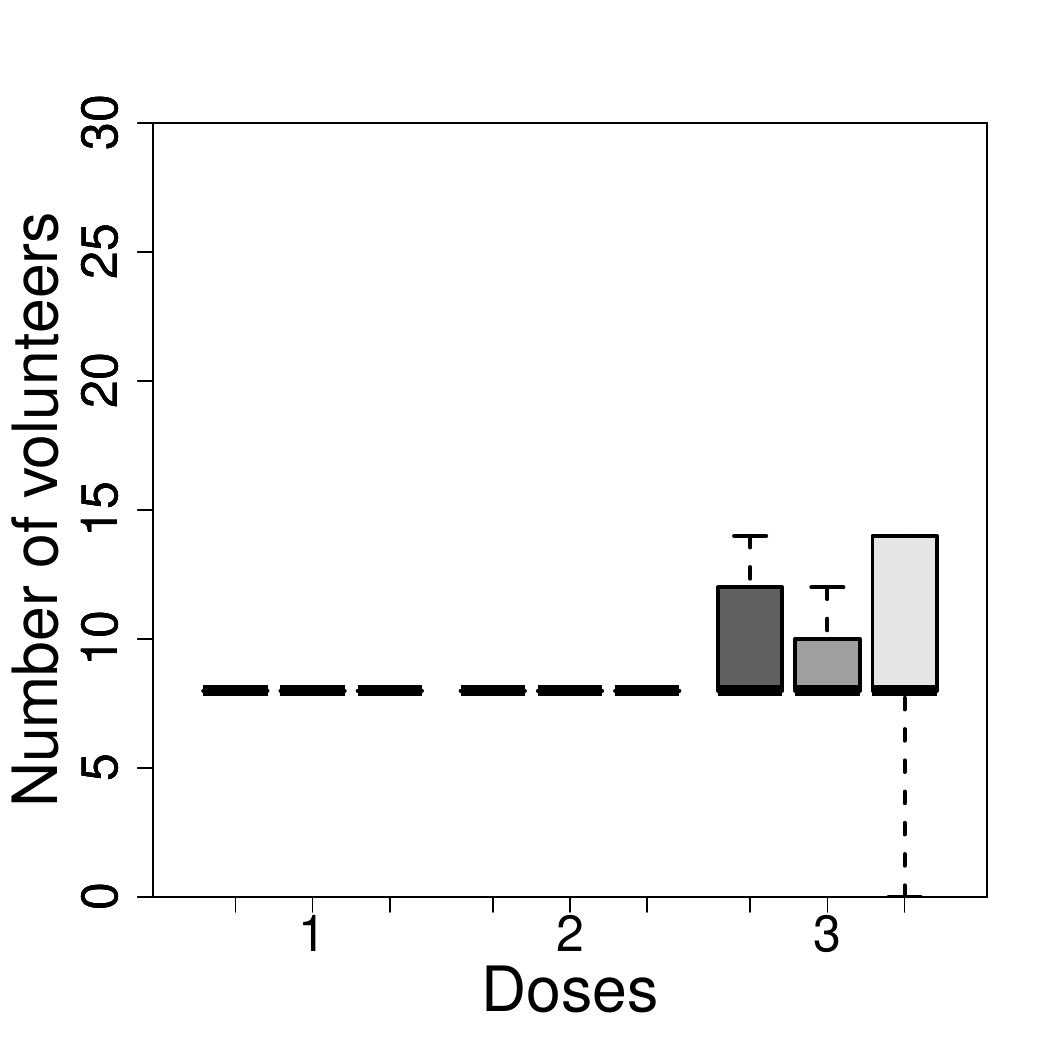}} \\
\subfigure[Scenario 4]{\includegraphics[scale=0.29]{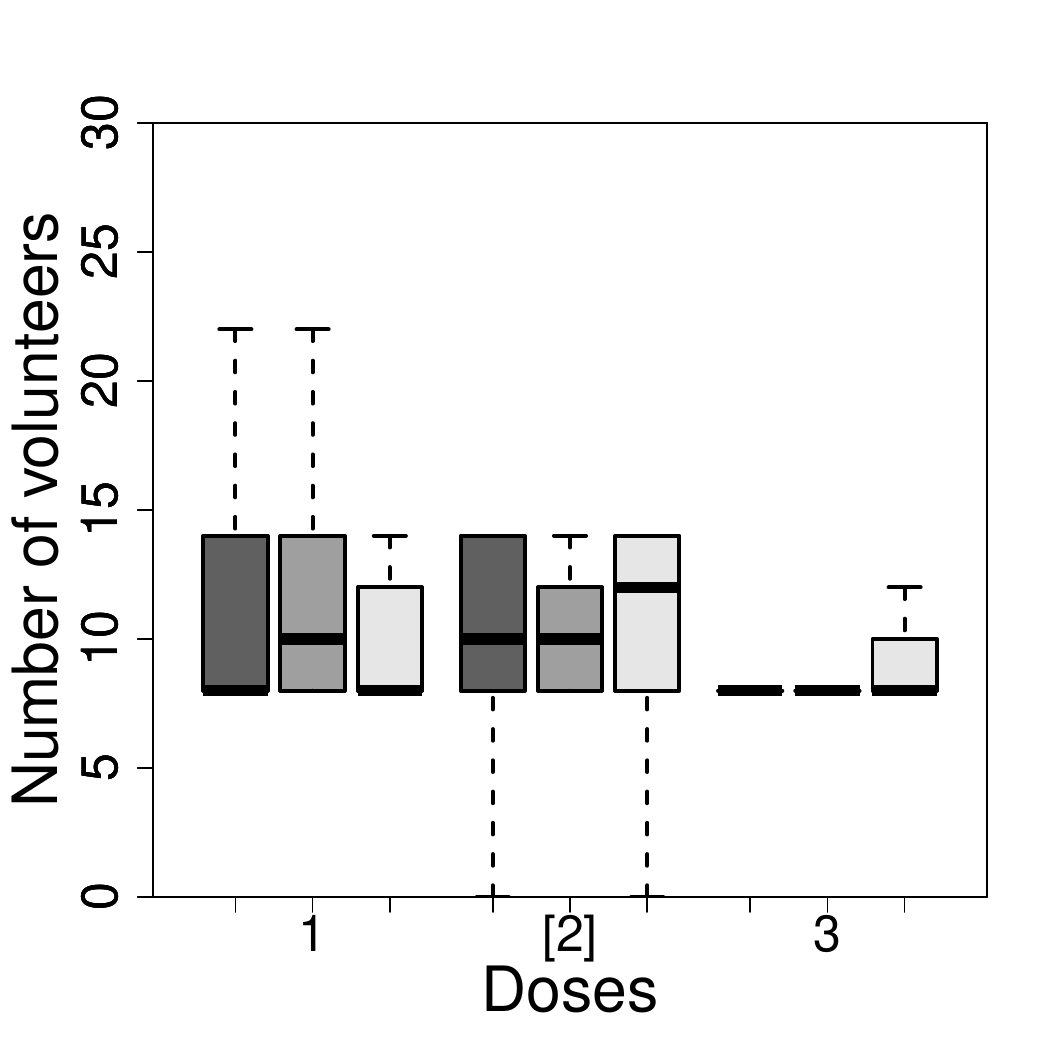}}
\subfigure[Scenario 5]{\includegraphics[scale=0.29]{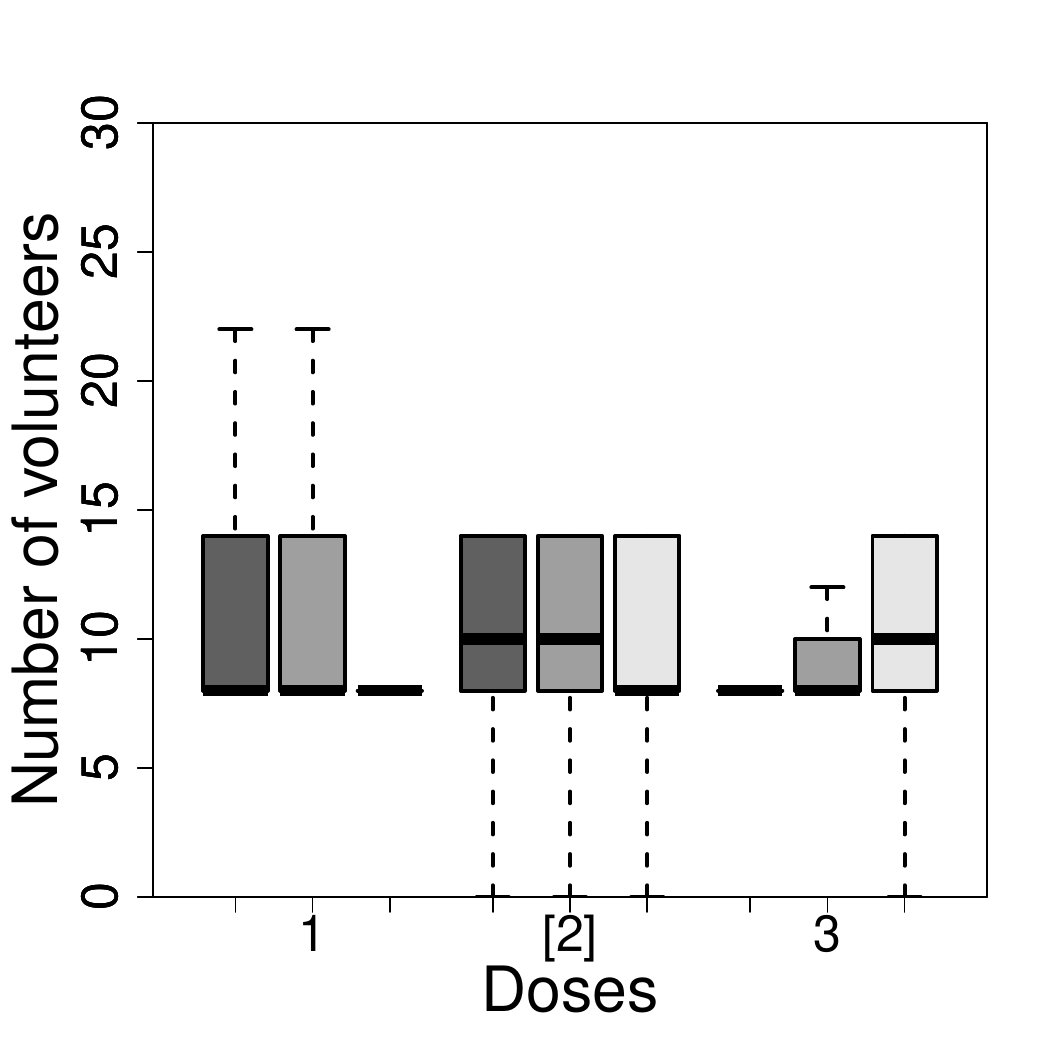}} 
\subfigure[Scenario 6]{\includegraphics[scale=0.29]{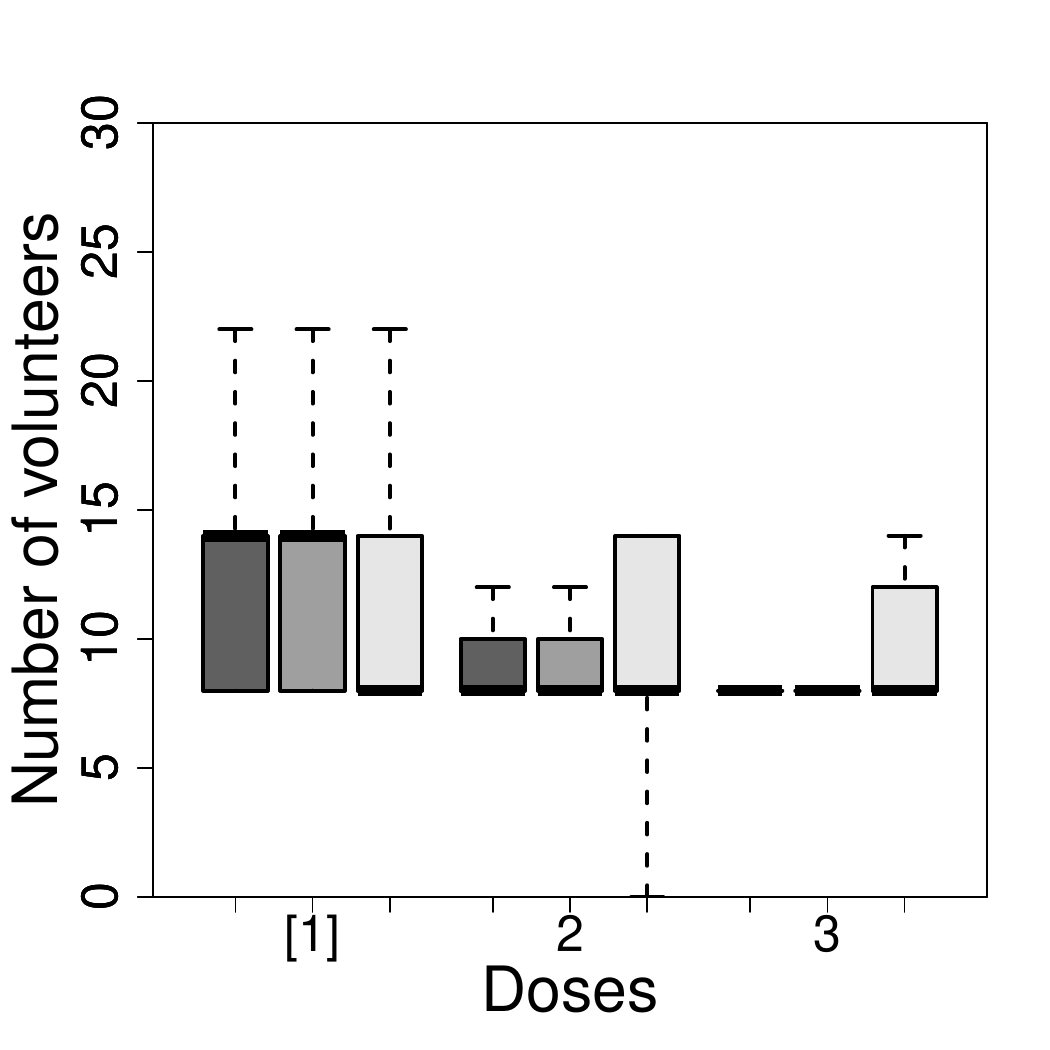}} \\ 
\subfigure[Scenario 7]{\includegraphics[scale=0.29]{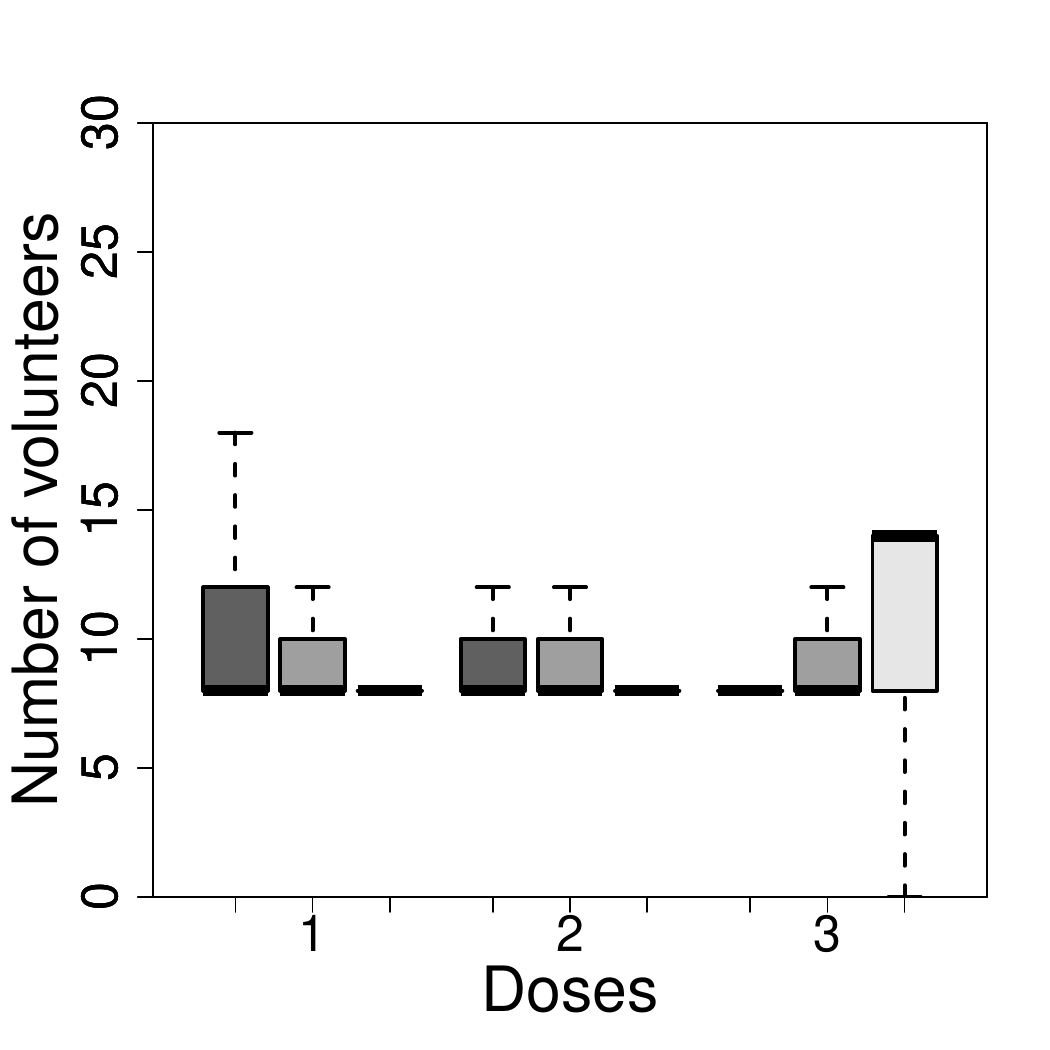}} 
\subfigure[Scenario 8]{\includegraphics[scale=0.29]{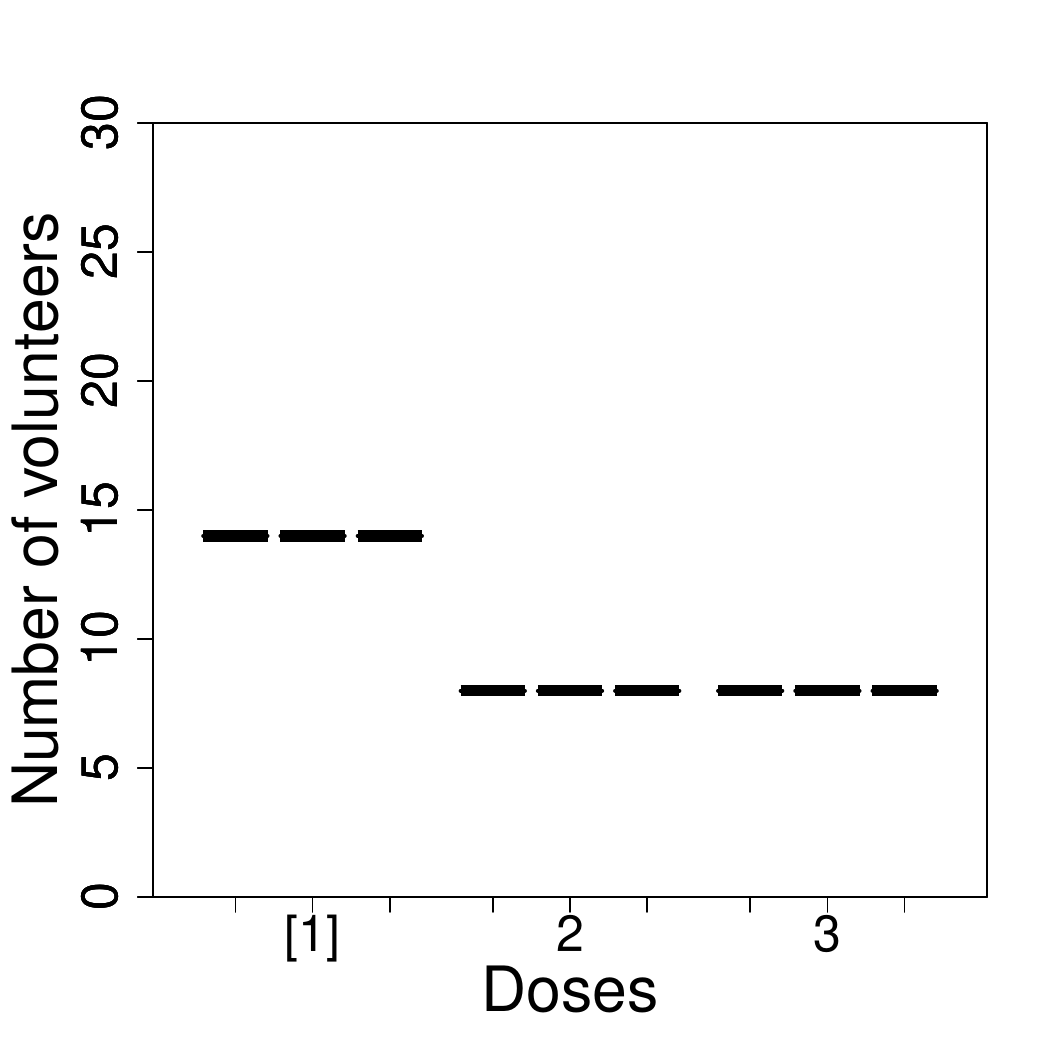}}
\subfigure{\includegraphics[scale=0.29]{legend_boxplot_nb_volunt_by_dose.pdf}}
\end{center}
\caption{Comparison of number of volunteers by dose for Selection method, BMA method  and BLRM for a maximum number of volunteers of $n = 30$, when $L = 3$ dose levels are used.  \label{fig:boxplot_nb_volunt_by_dose_n30_L3}}
\end{figure}

\begin{figure}[h!]
\begin{center}
\subfigure[Scenario 1]{\includegraphics[scale=0.29]{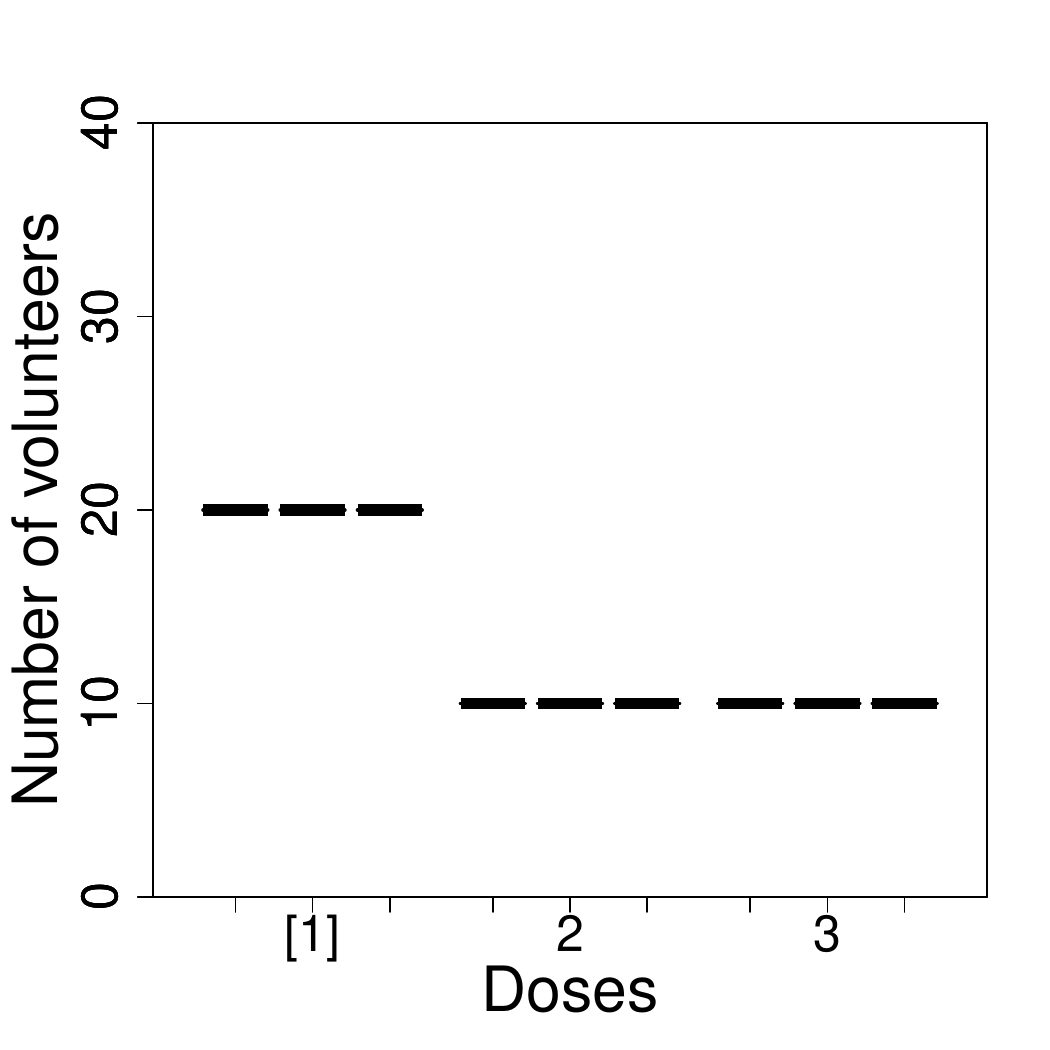}} 
\subfigure[Scenario 2]{\includegraphics[scale=0.29]{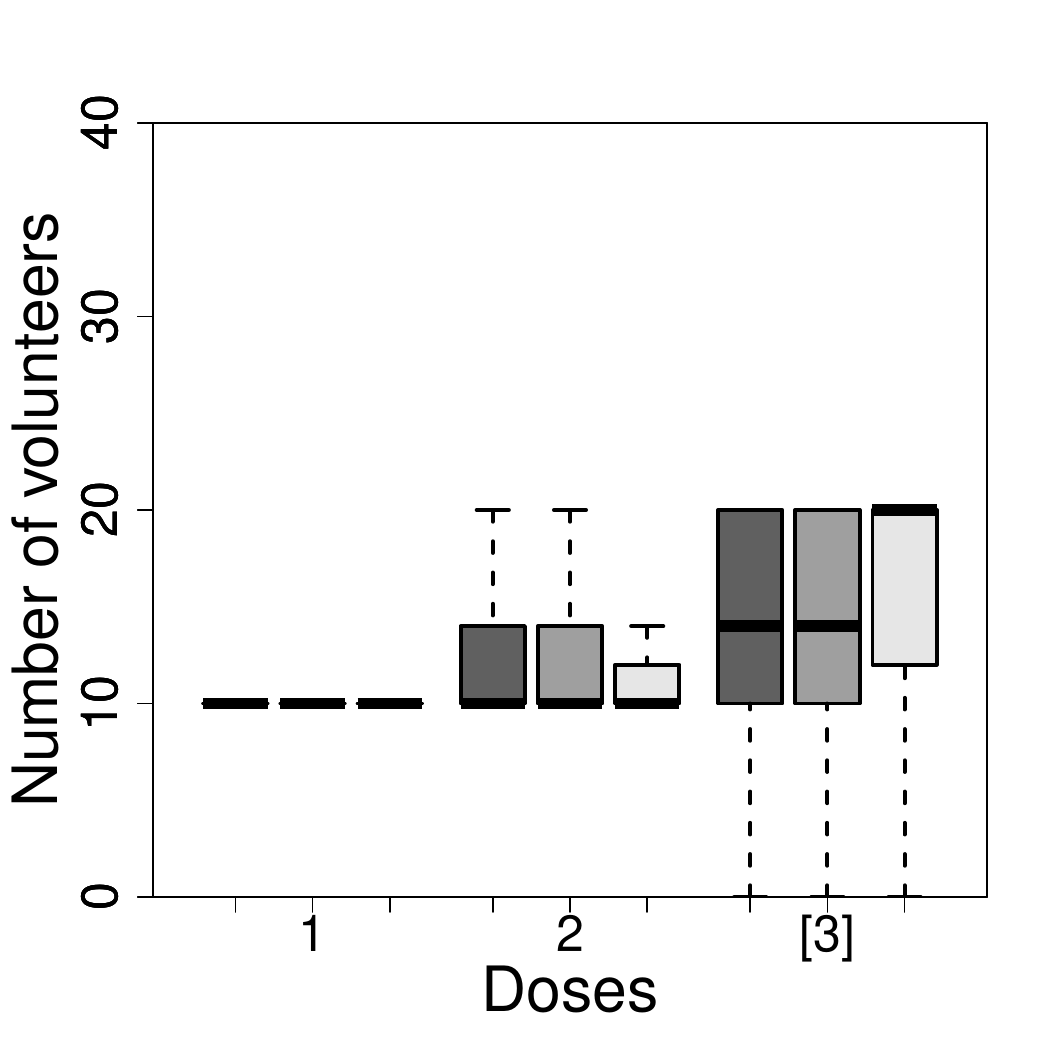}} 
\subfigure[Scenario 3]{\includegraphics[scale=0.29]{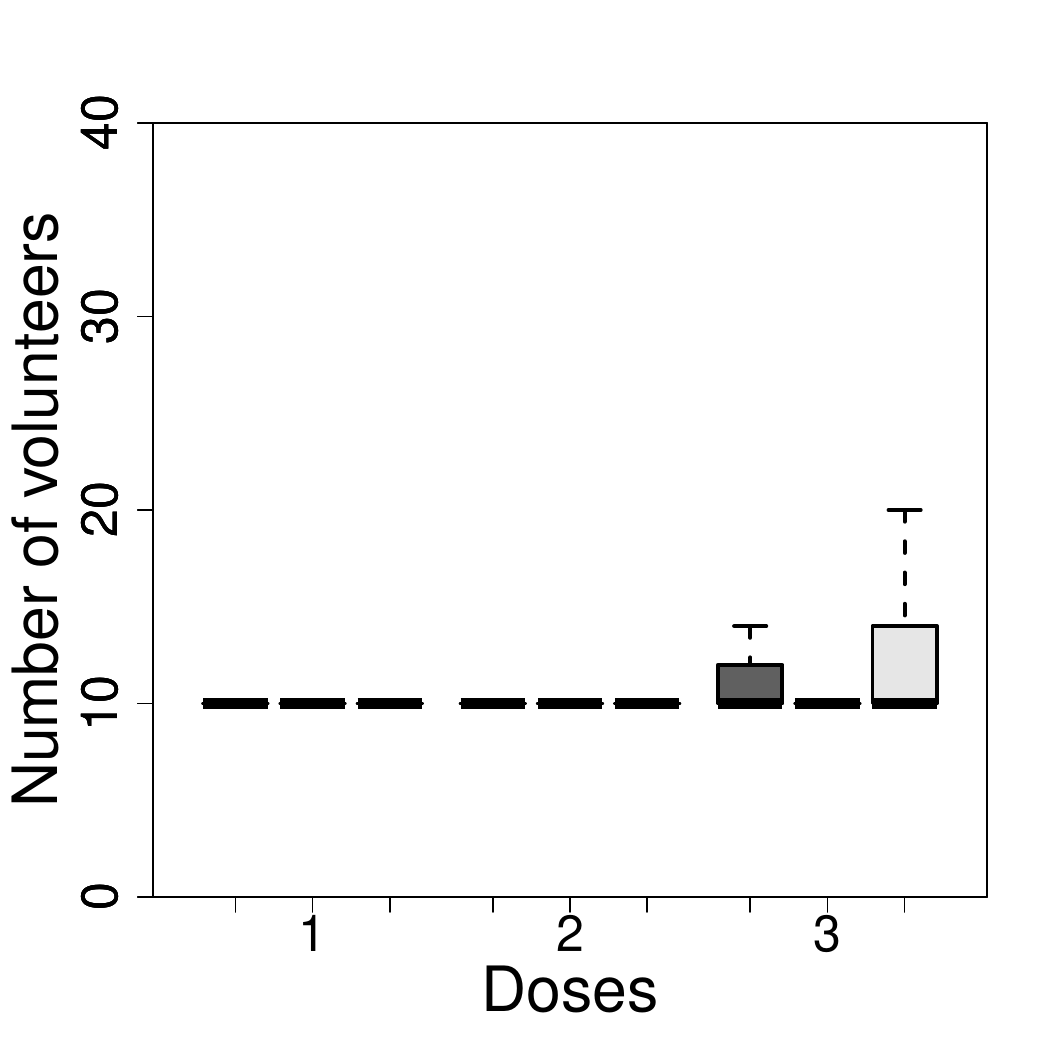}} \\
\subfigure[Scenario 4]{\includegraphics[scale=0.29]{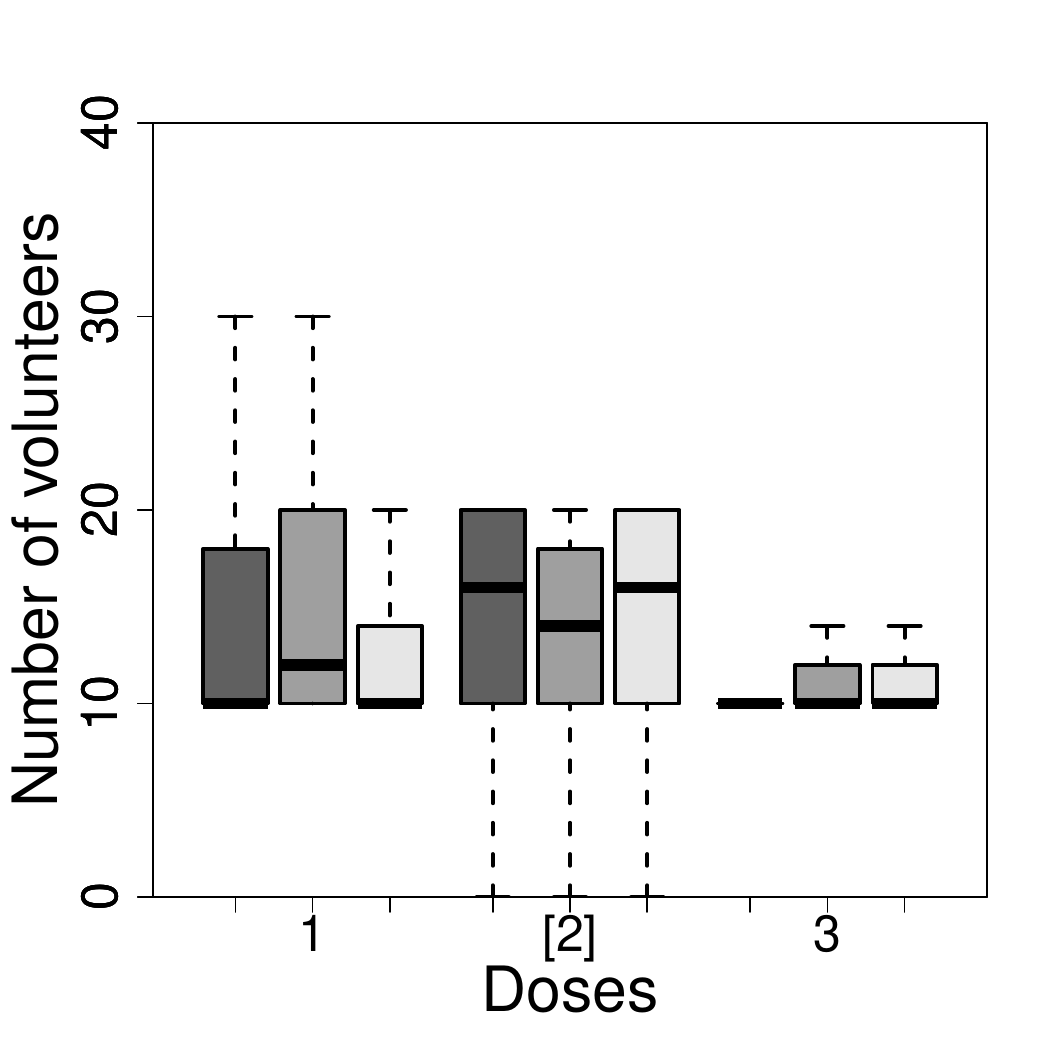}}
\subfigure[Scenario 5]{\includegraphics[scale=0.29]{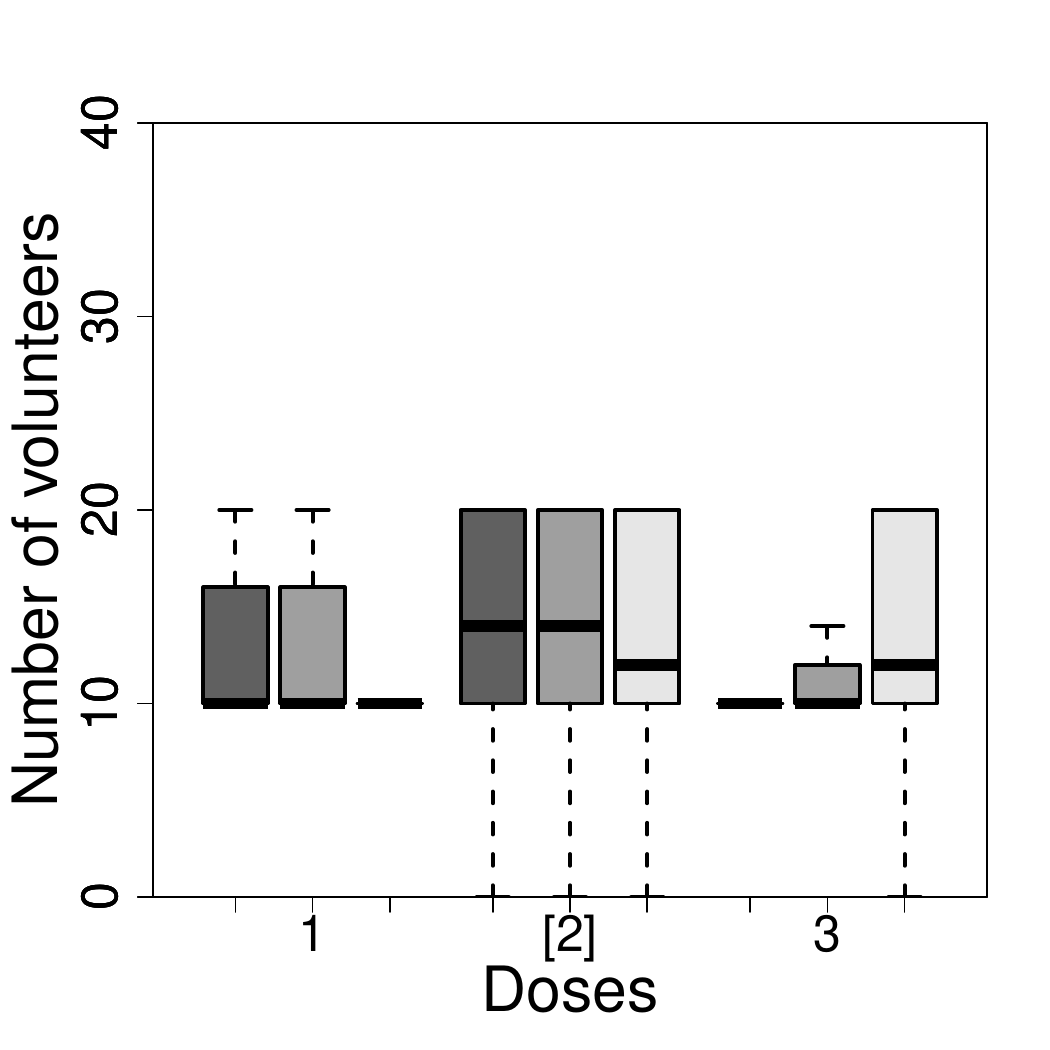}} 
\subfigure[Scenario 6]{\includegraphics[scale=0.29]{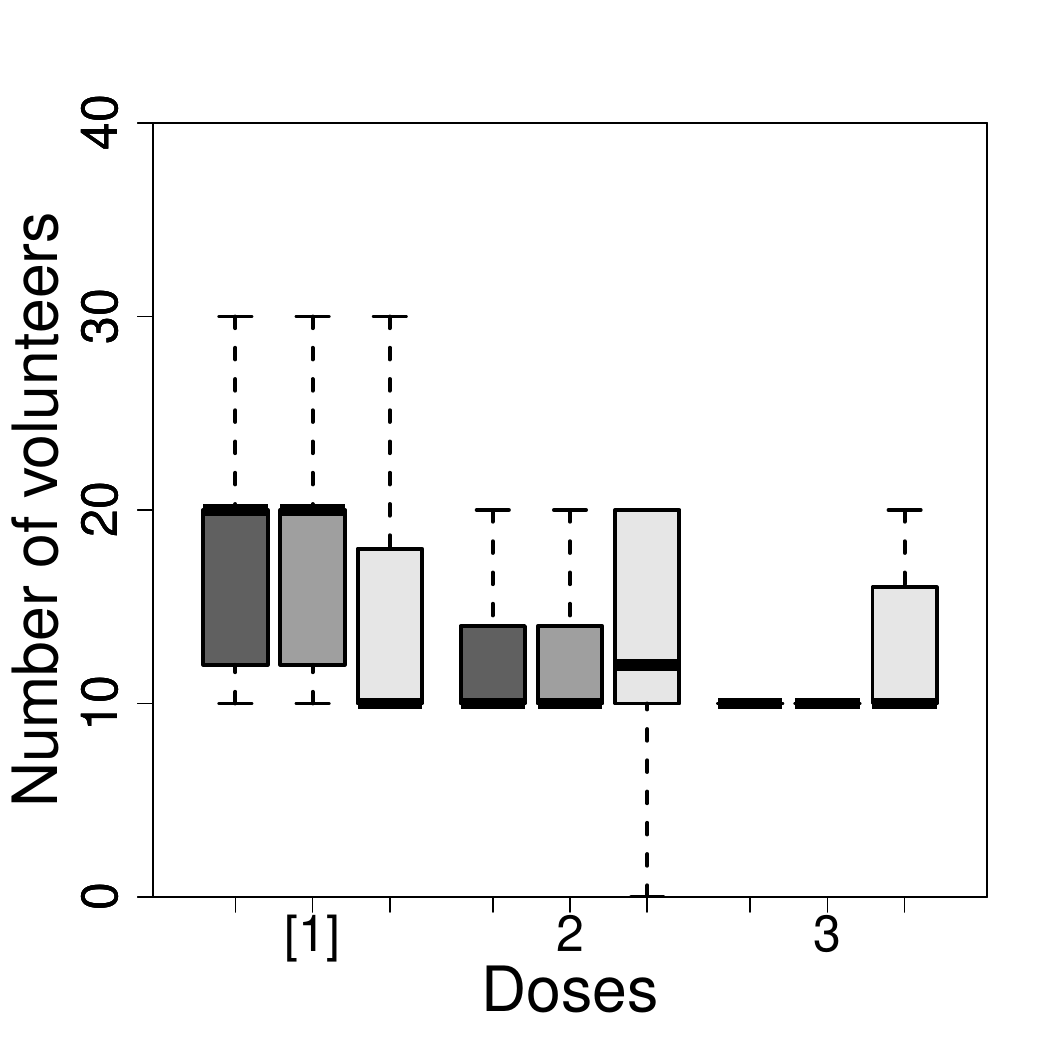}} \\ 
\subfigure[Scenario 7]{\includegraphics[scale=0.29]{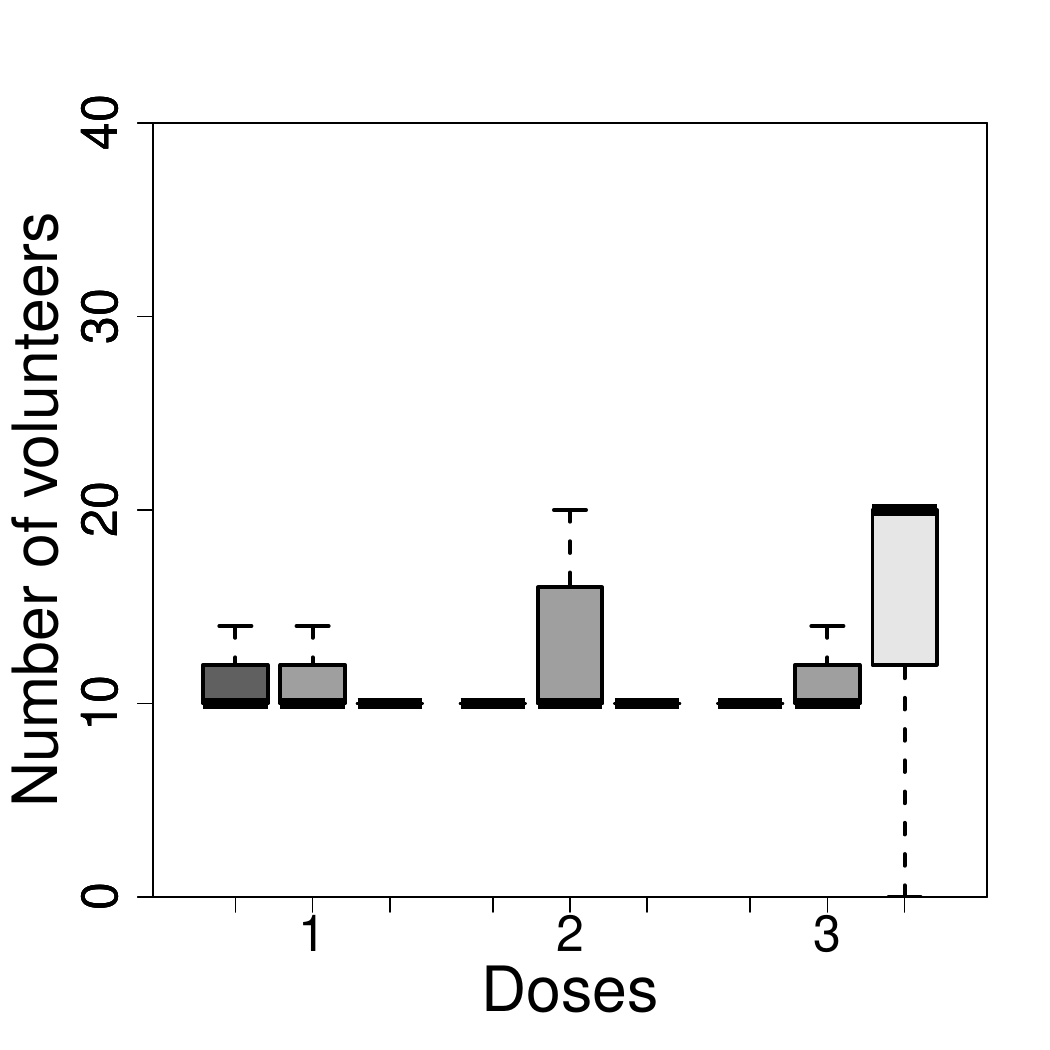}} 
\subfigure[Scenario 8]{\includegraphics[scale=0.29]{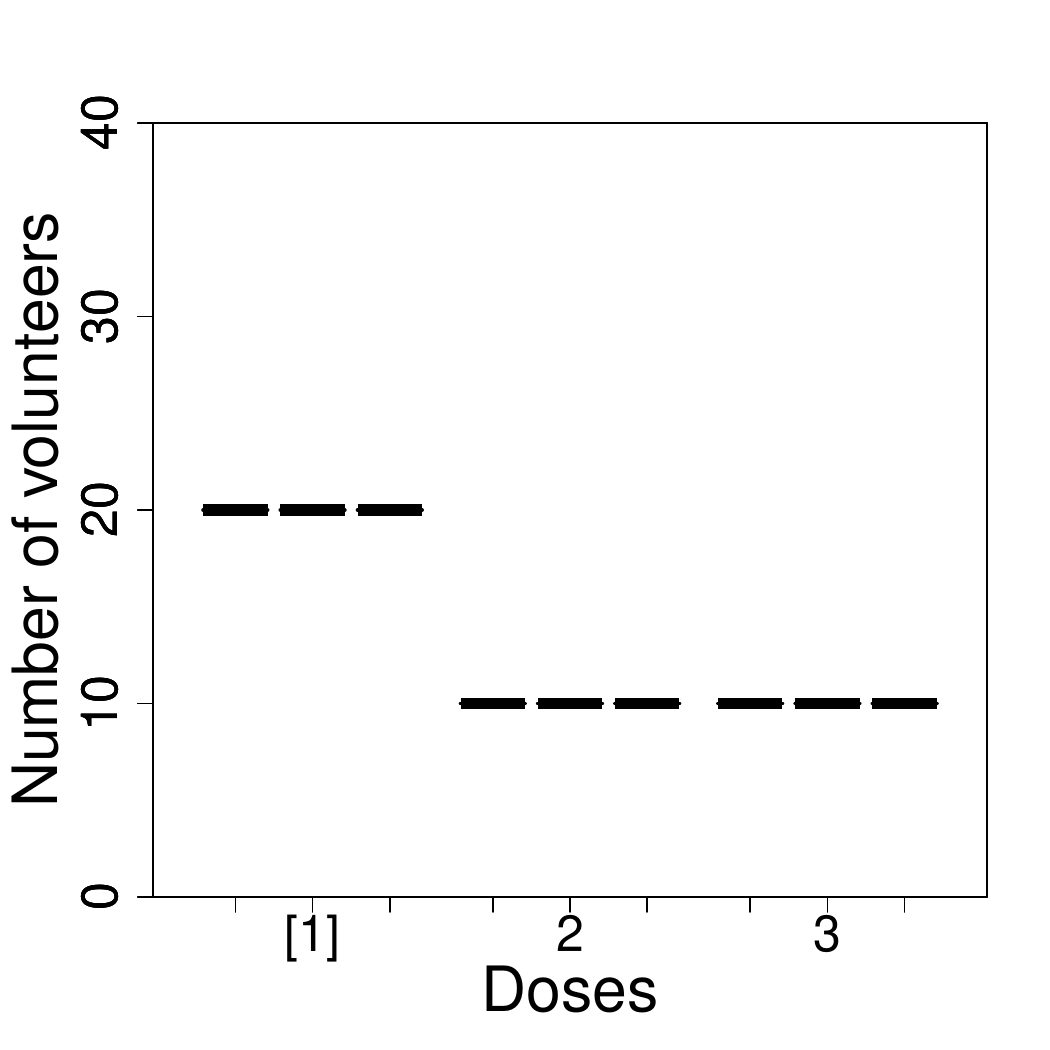}}
\subfigure{\includegraphics[scale=0.29]{legend_boxplot_nb_volunt_by_dose.pdf}}
\end{center}
\caption{Comparison of number of volunteers by dose for Selection method, BMA method  and BLRM for a maximum number of volunteers of $n = 40$, when $L = 3$ dose levels are used.  \label{fig:boxplot_nb_volunt_by_dose_n40_L3}}
\end{figure}

\begin{figure}[h!]
\begin{center}
\subfigure[Scenario 1]{\includegraphics[scale=0.29]{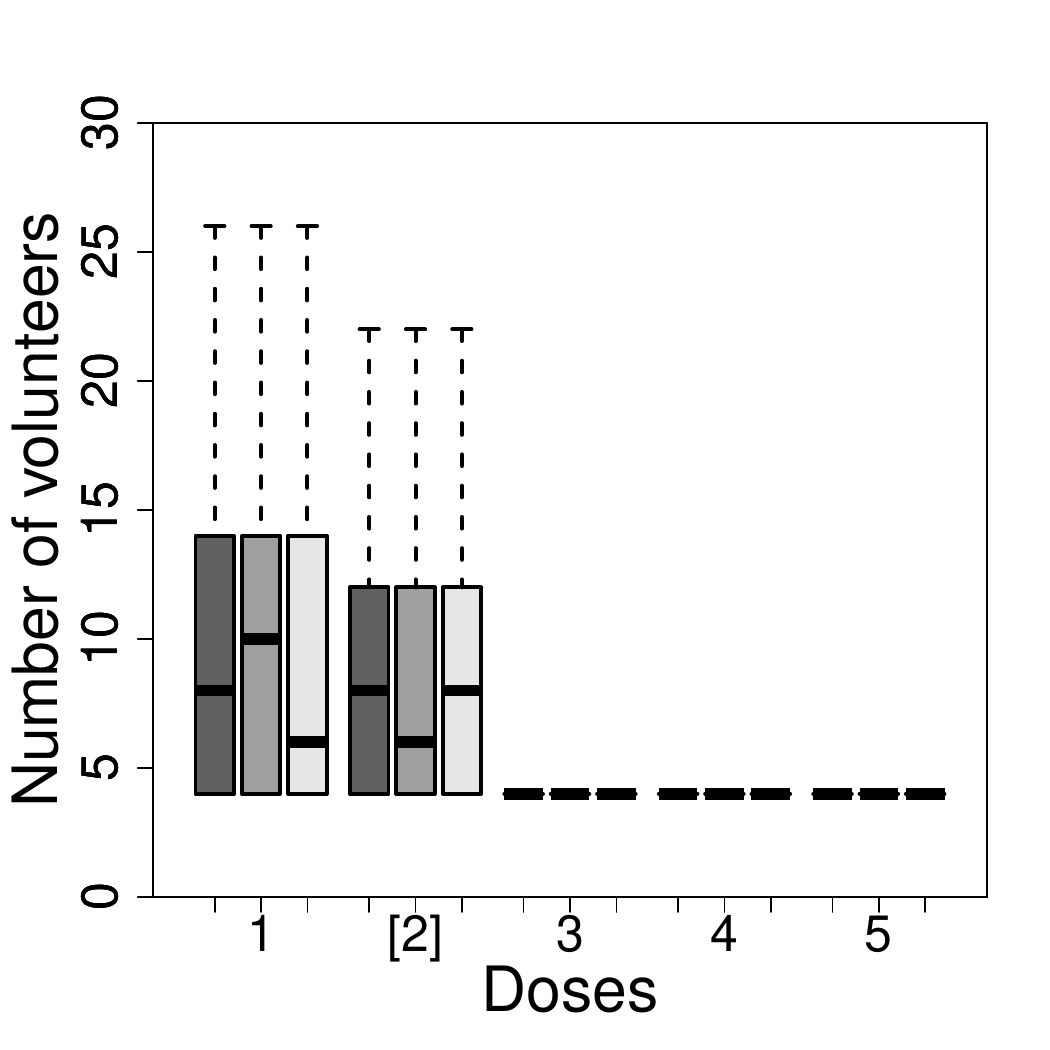}} 
\subfigure[Scenario 2]{\includegraphics[scale=0.29]{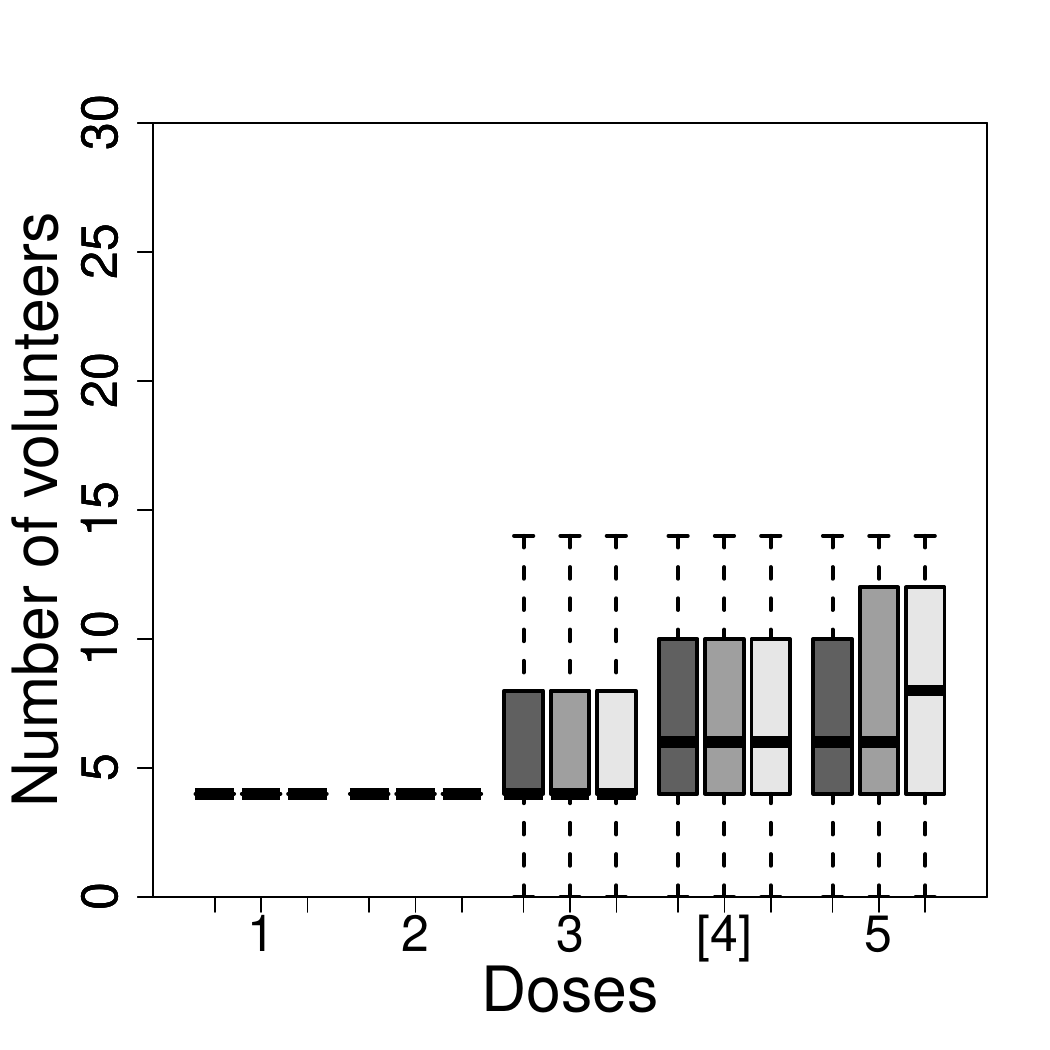}} 
\subfigure[Scenario 3]{\includegraphics[scale=0.29]{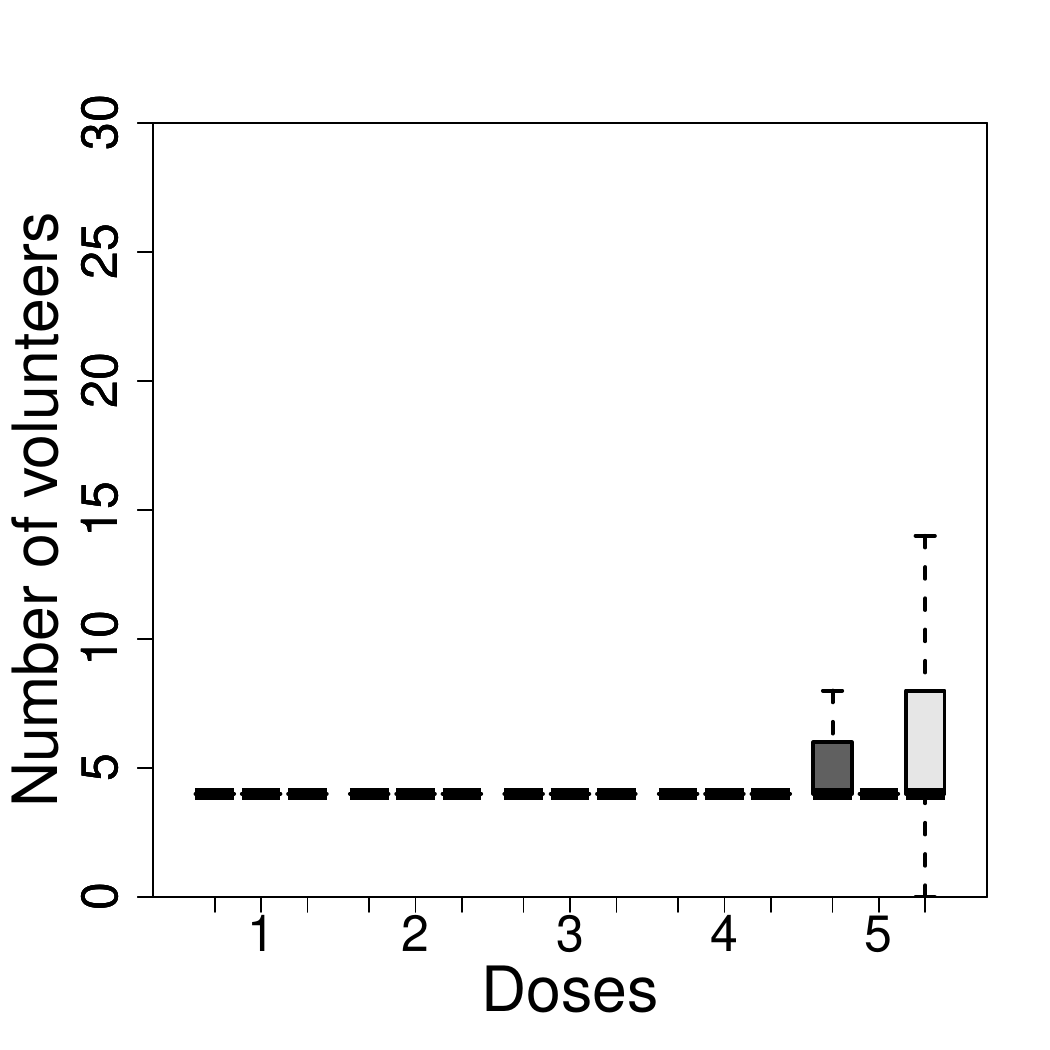}} \\ 
\subfigure[Scenario 4]{\includegraphics[scale=0.29]{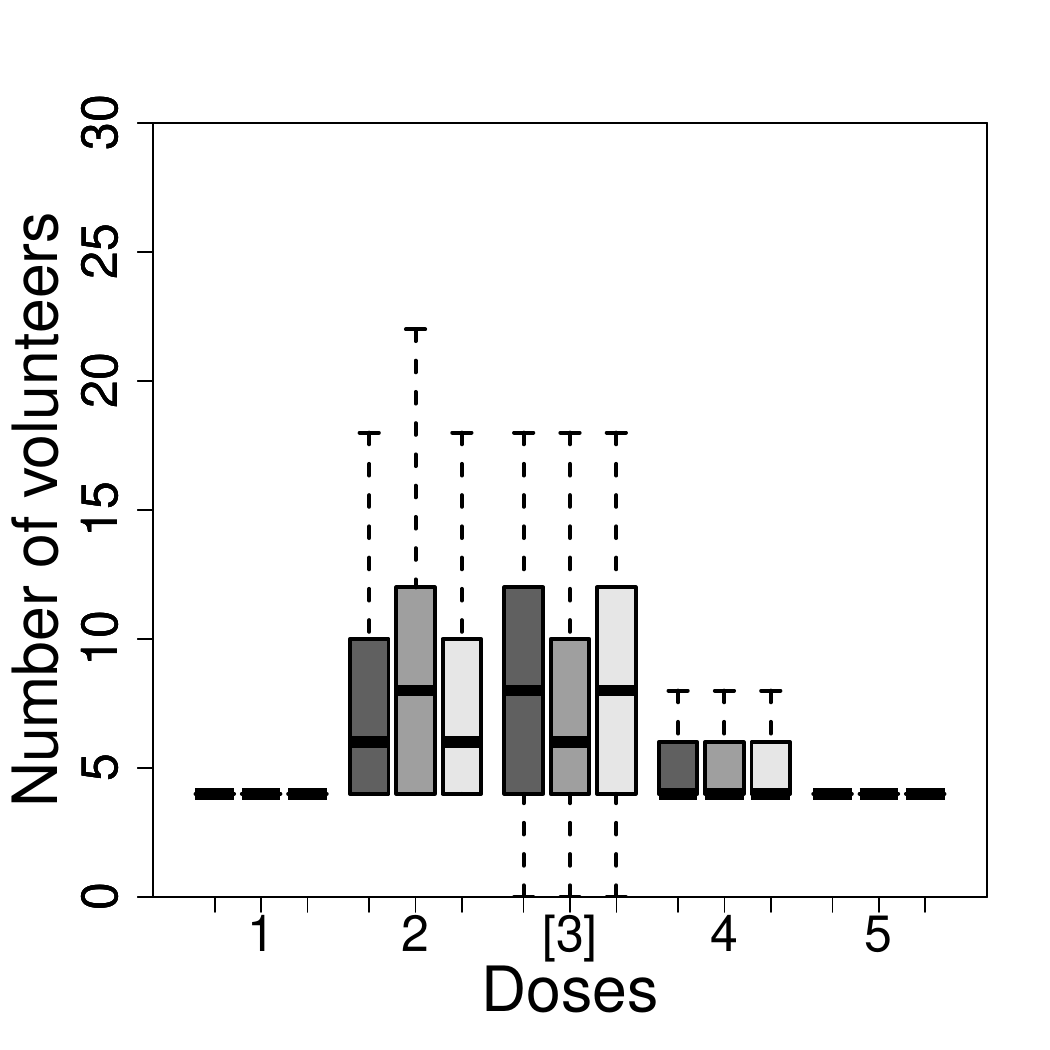}}
\subfigure[Scenario 5]{\includegraphics[scale=0.29]{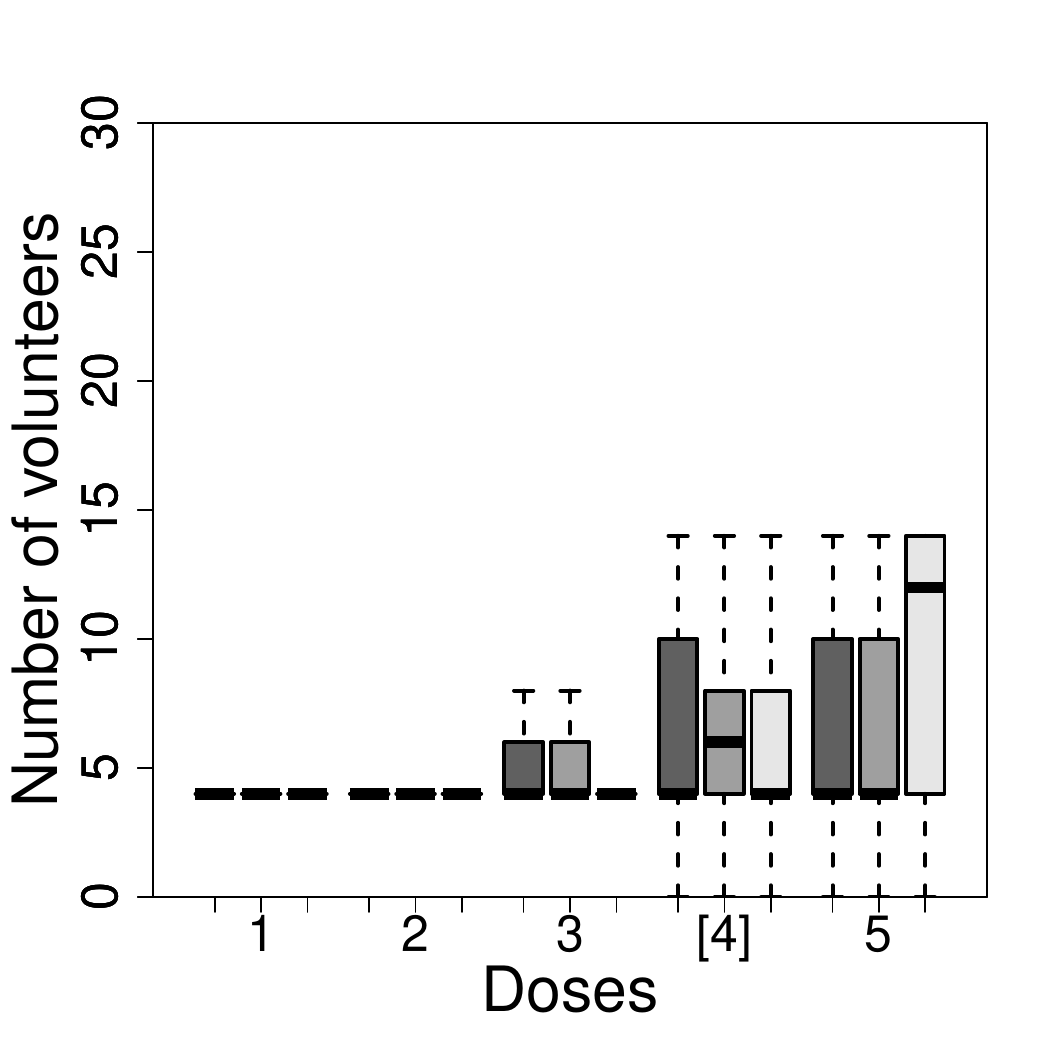}} 
\subfigure[Scenario 6]{\includegraphics[scale=0.29]{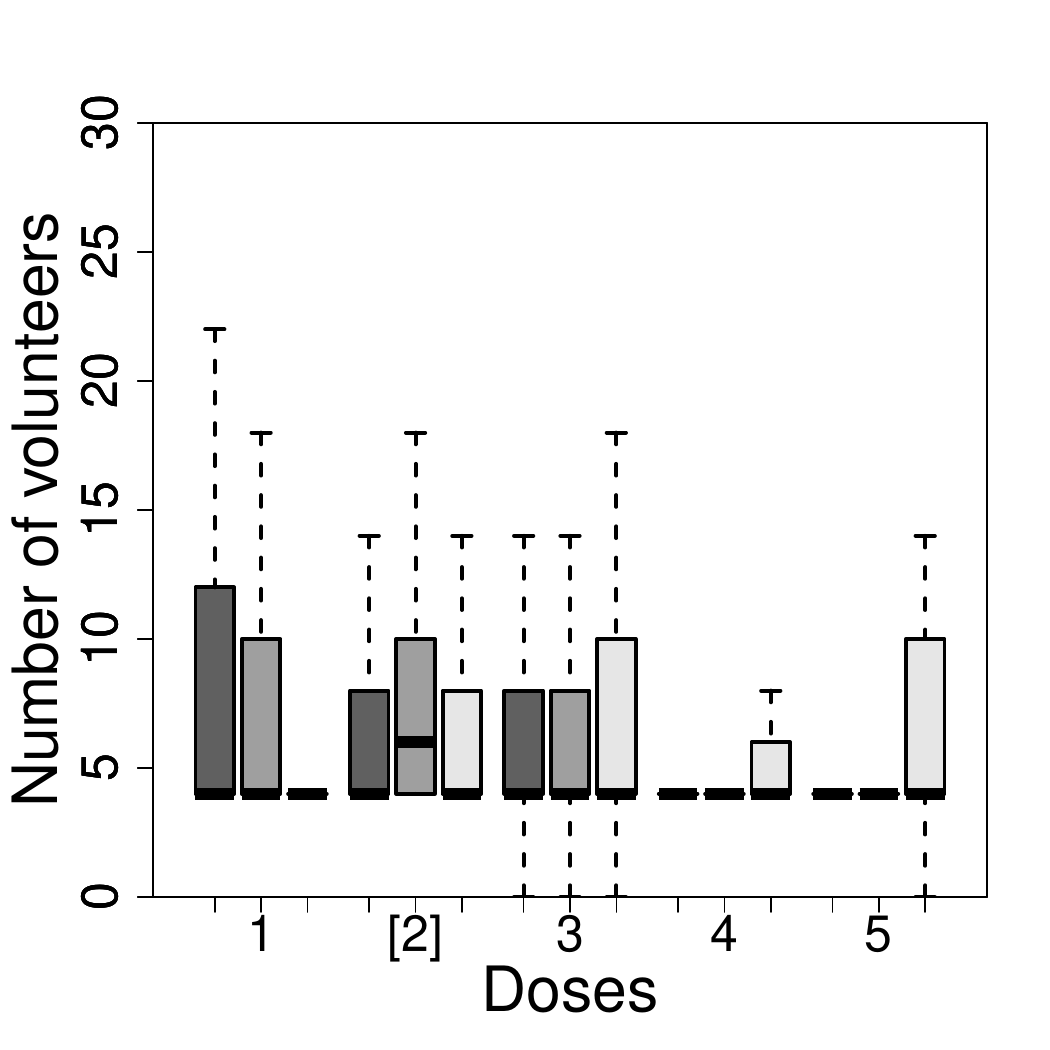}} \\ 
\subfigure[Scenario 7]{\includegraphics[scale=0.29]{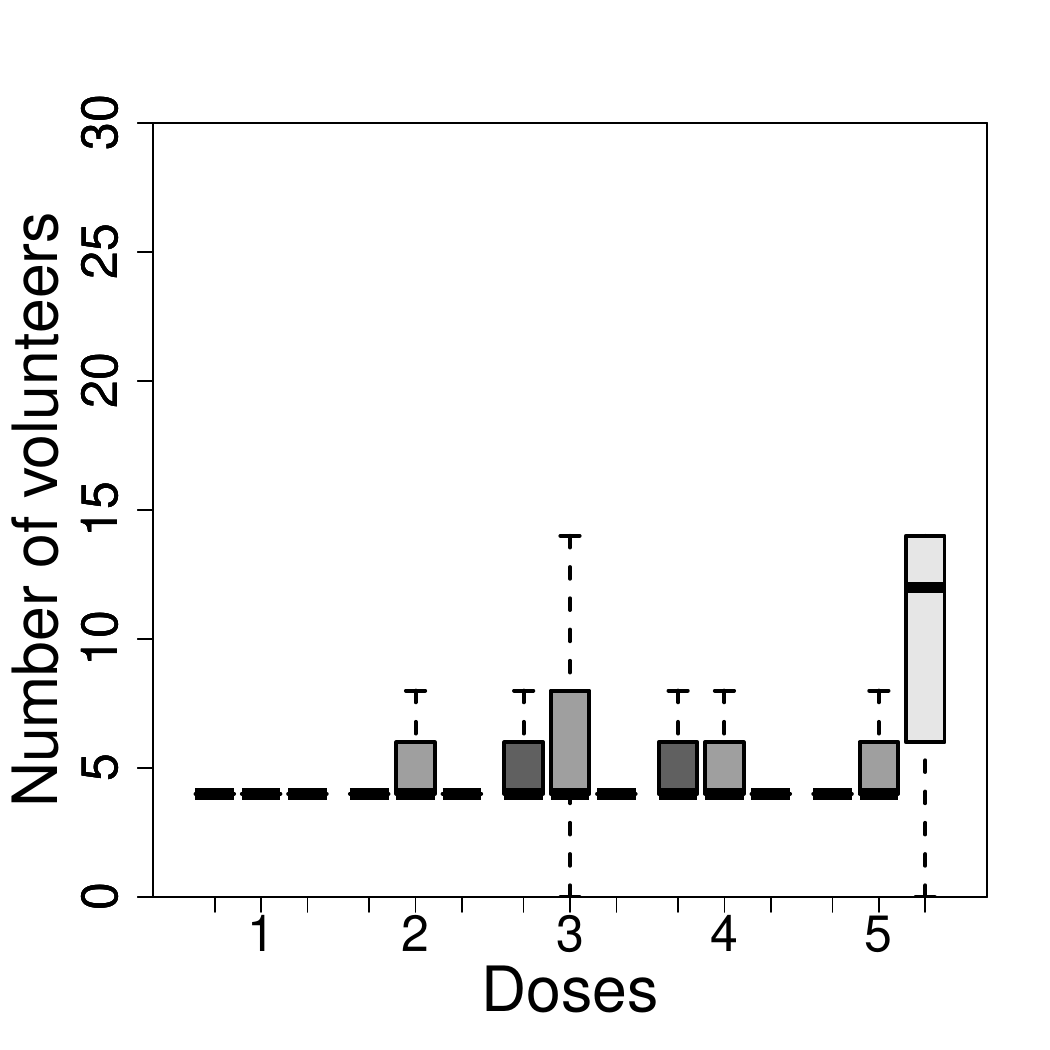}} 
\subfigure[Scenario 8]{\includegraphics[scale=0.29]{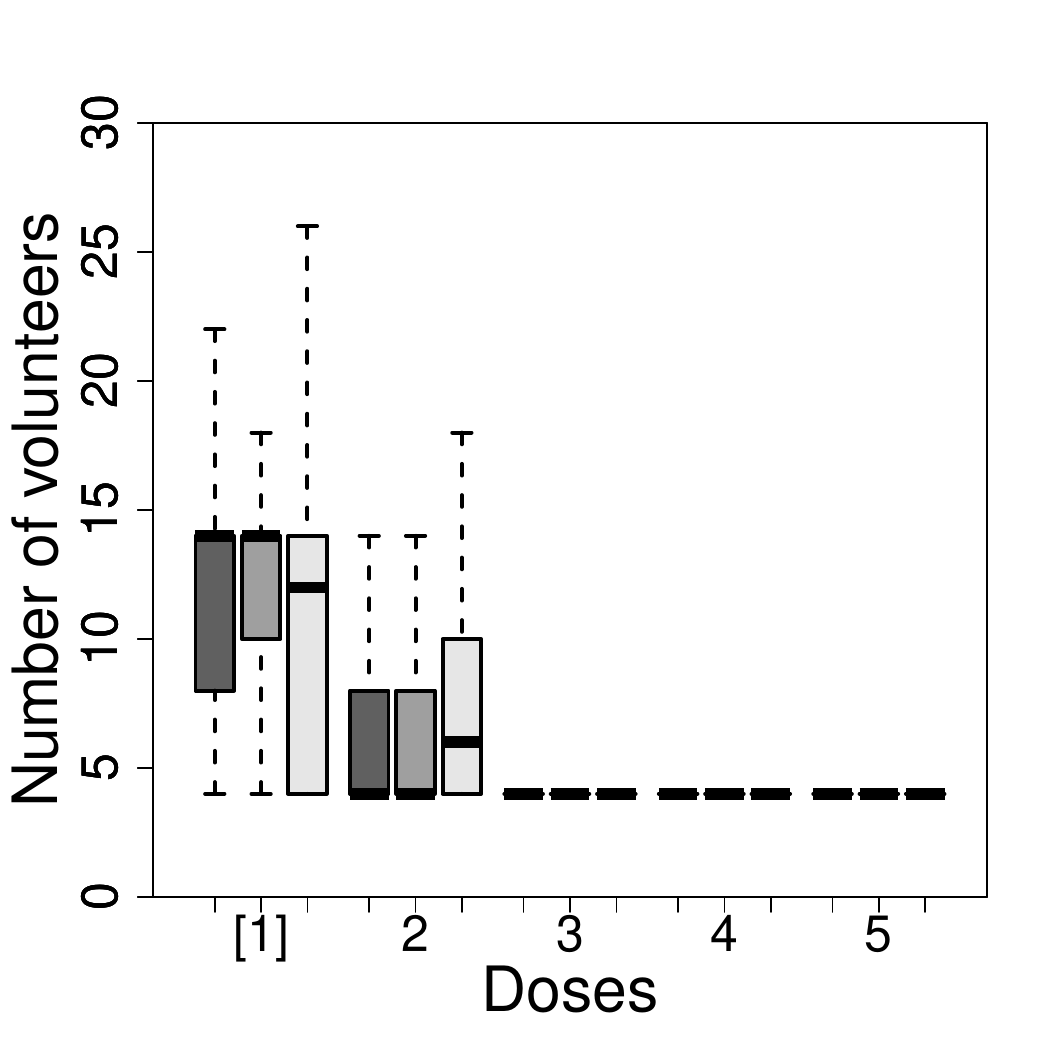}}
\subfigure{\includegraphics[scale=0.29]{legend_boxplot_nb_volunt_by_dose.pdf}}
\end{center}
\caption{Comparison of number of volunteers by dose for Selection method, BMA method and BLRM for a maximum number of volunteers of $n = 30$, when $L = 5$ dose levels are used.  \label{fig:boxplot_nb_volunt_by_dose_n30_L5}}
\end{figure}

\begin{figure}[h!]
\begin{center}
\subfigure[Scenario 1]{\includegraphics[scale=0.29]{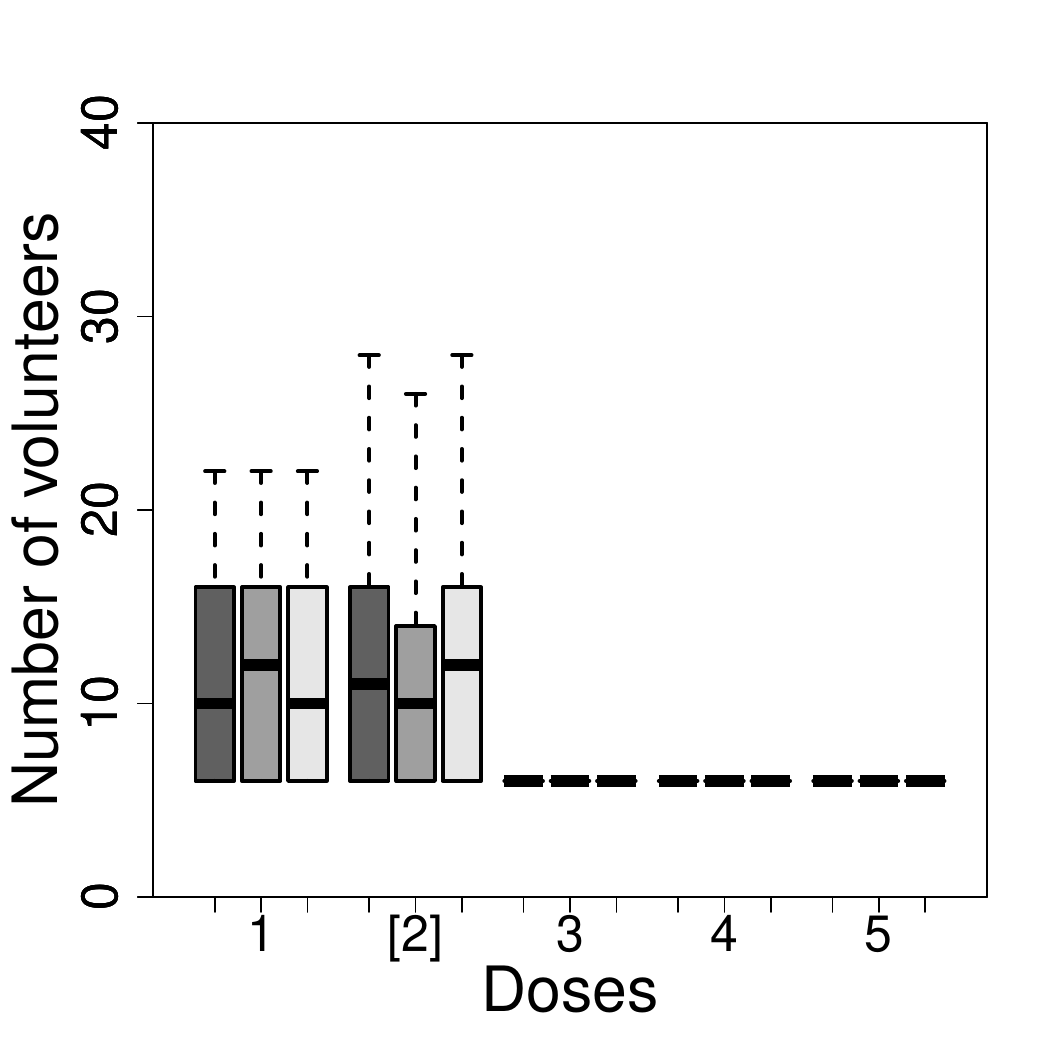}} 
\subfigure[Scenario 2]{\includegraphics[scale=0.29]{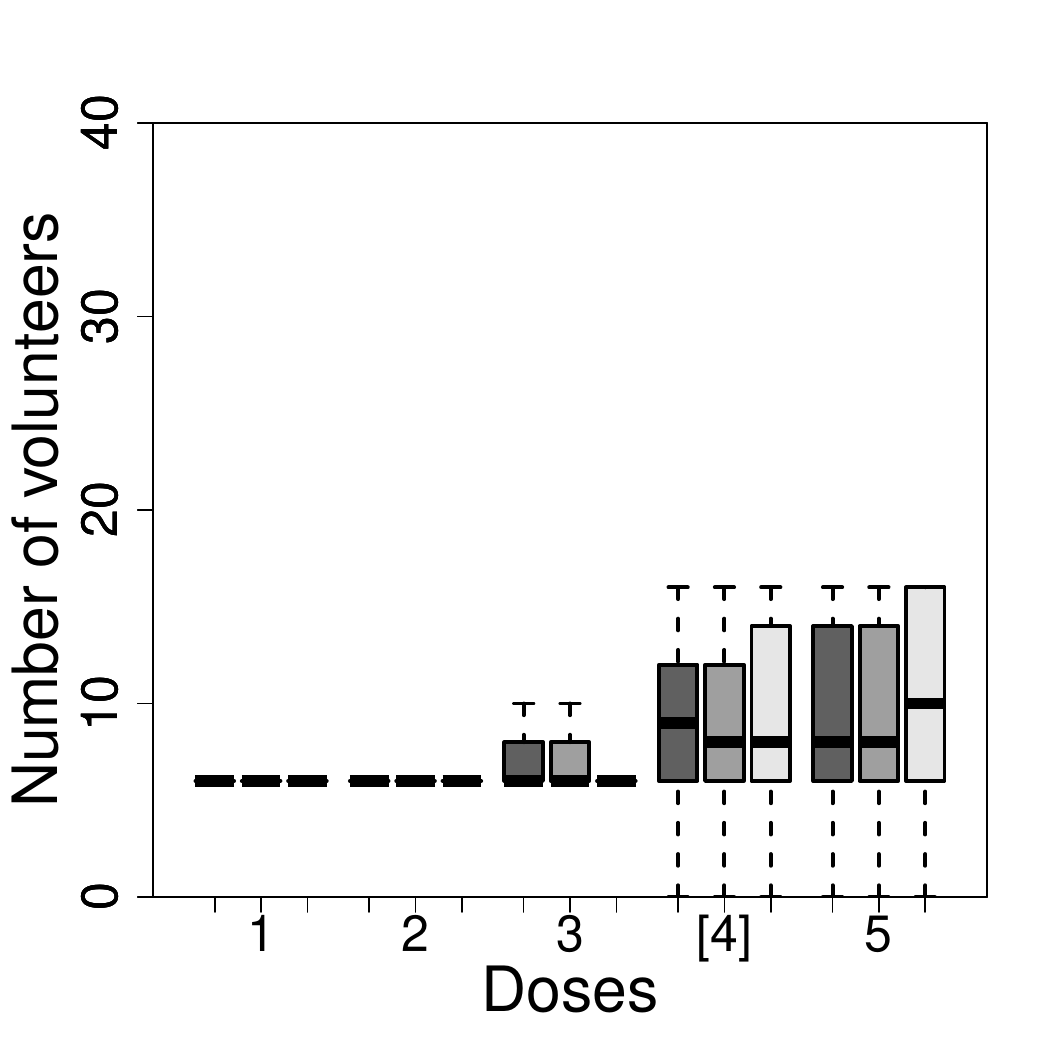}} 
\subfigure[Scenario 3]{\includegraphics[scale=0.29]{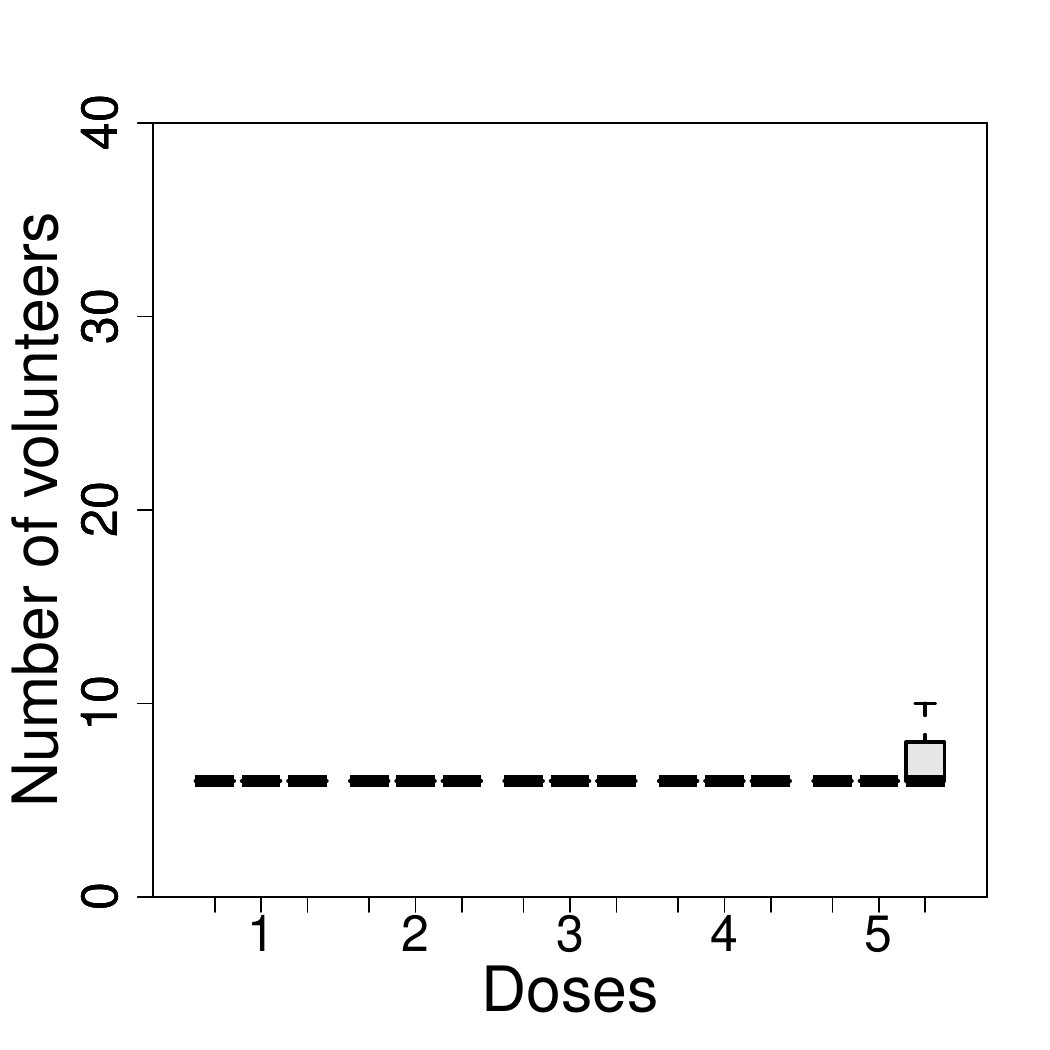}} \\ 
\subfigure[Scenario 4]{\includegraphics[scale=0.29]{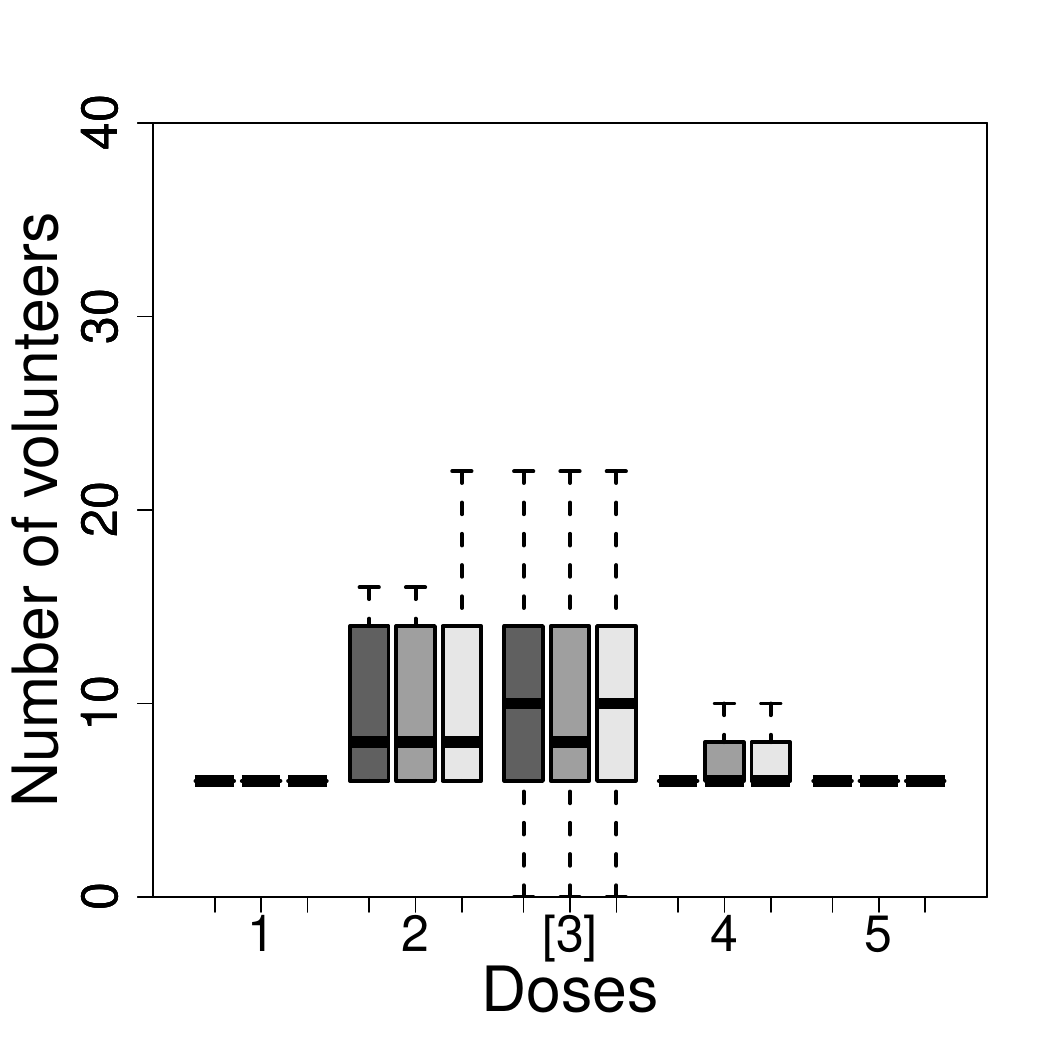}}
\subfigure[Scenario 5]{\includegraphics[scale=0.29]{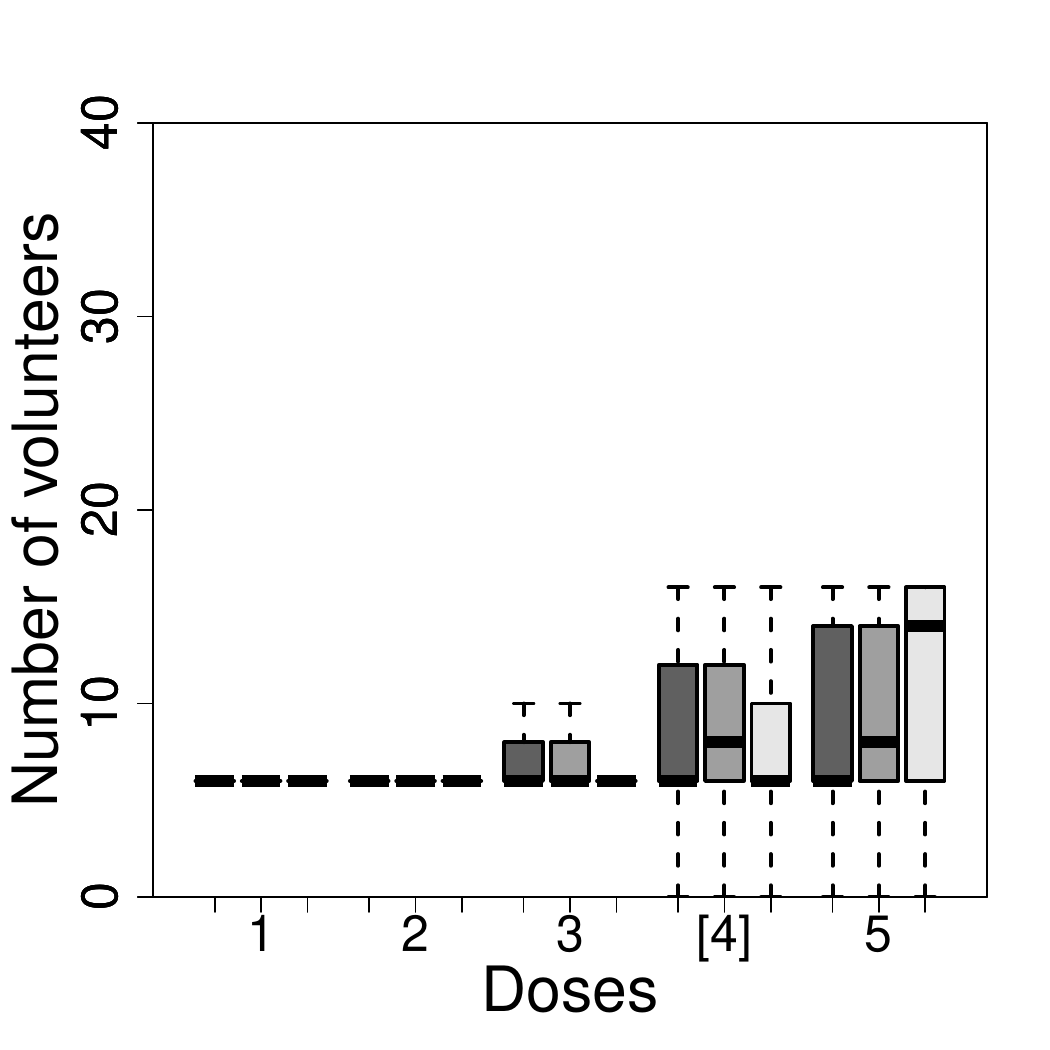}} 
\subfigure[Scenario 6]{\includegraphics[scale=0.29]{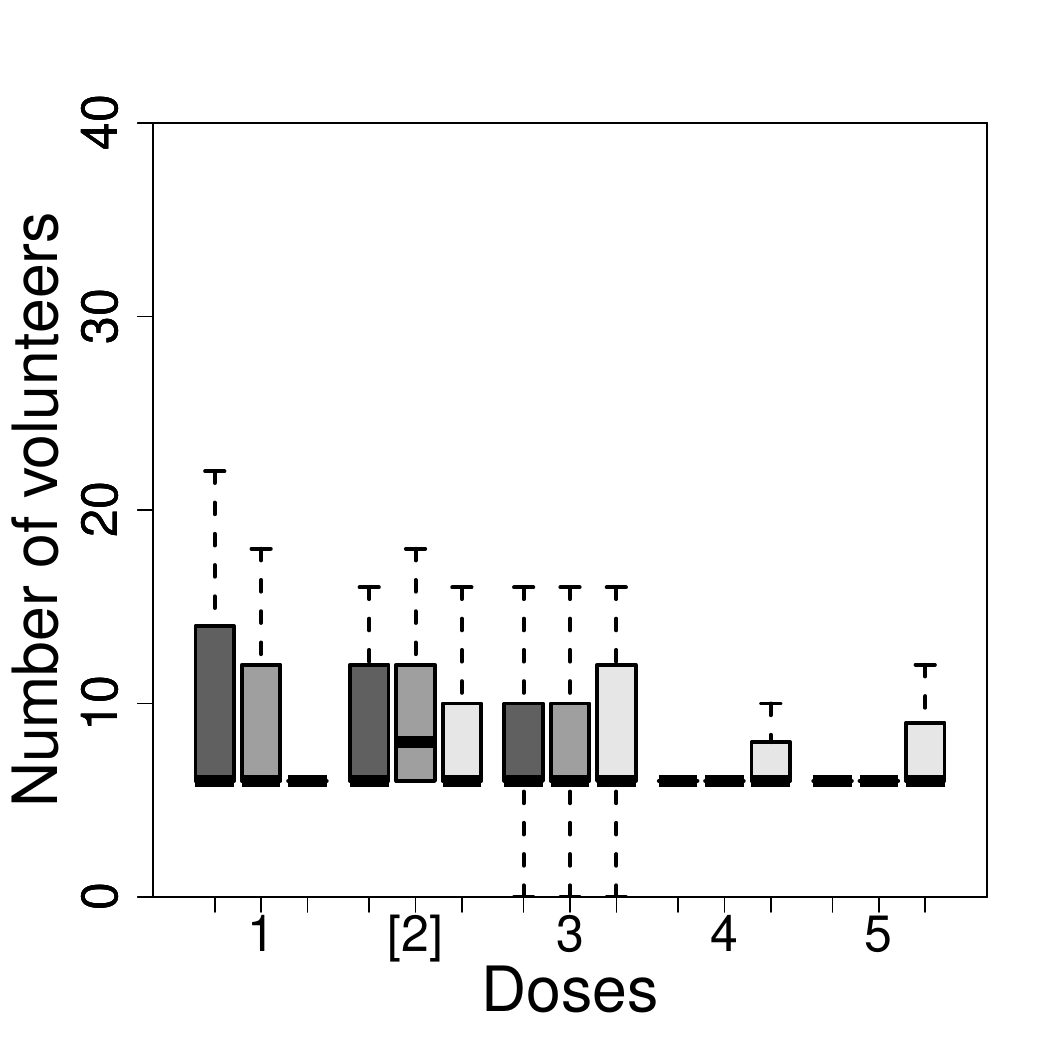}} \\ 
\subfigure[Scenario 7]{\includegraphics[scale=0.29]{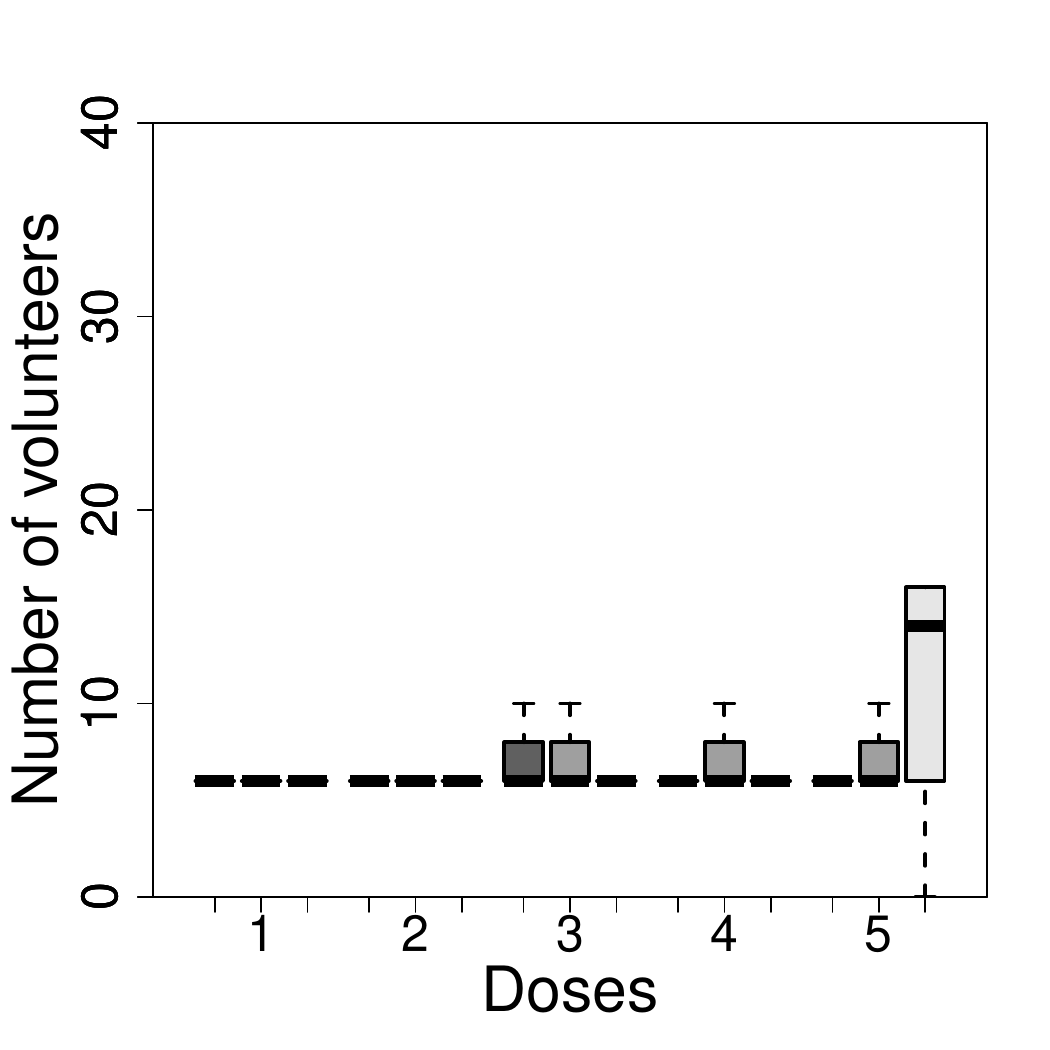}} 
\subfigure[Scenario 8]{\includegraphics[scale=0.29]{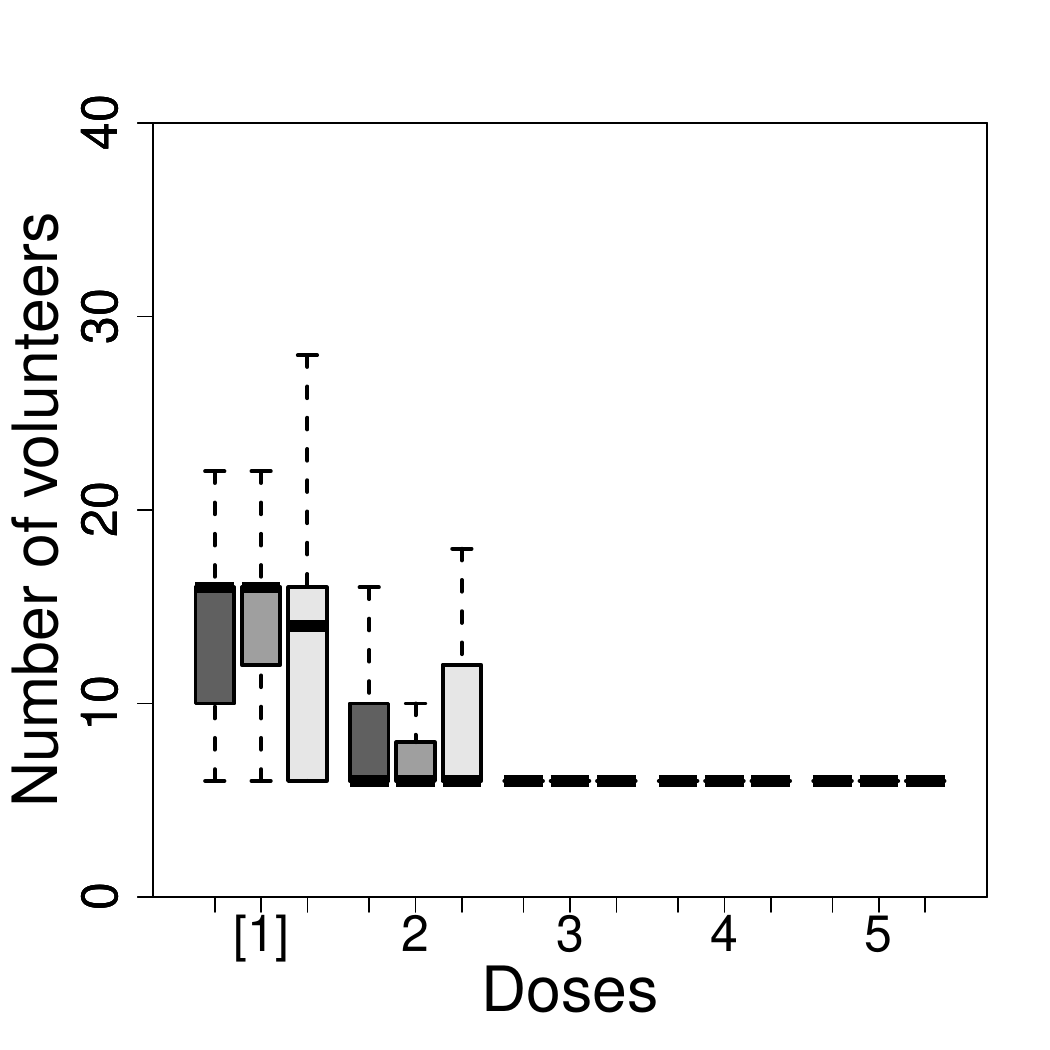}}
\subfigure{\includegraphics[scale=0.29]{legend_boxplot_nb_volunt_by_dose.pdf}}
\end{center}
\caption{Comparison of number of volunteers by dose for Selection method, BMA method  and BLRM for a maximum number of volunteers of $n = 40$, when $L = 5$ dose levels are used.  \label{fig:boxplot_nb_volunt_by_dose_n40_L5}}
\end{figure}

\clearpage

\subsection{MAD Selection}\label{subsec:app_MAD_hat_res}

\begin{figure}[h!]
\begin{center}
\subfigure[Scenario 1]{\includegraphics[scale=0.29]{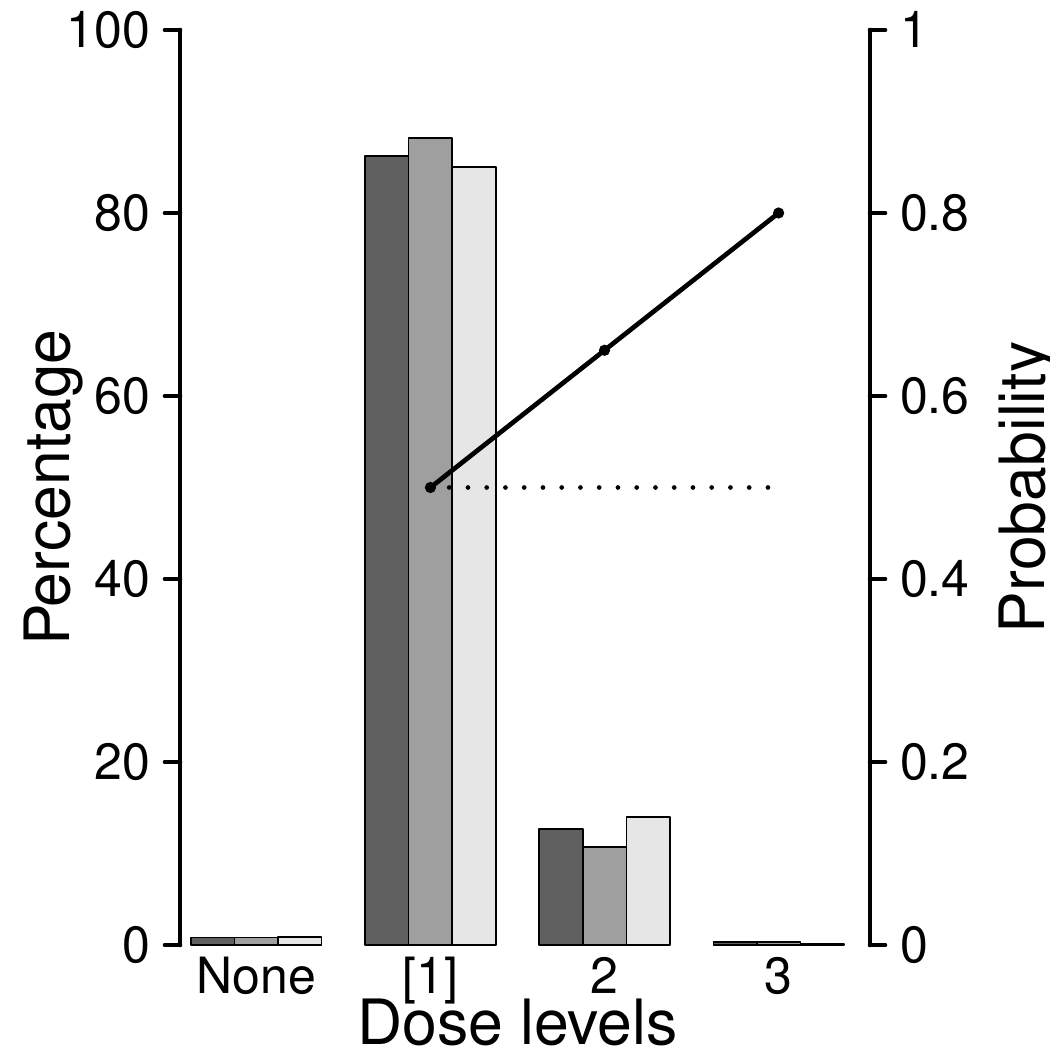}} 
\subfigure[Scenario 2]{\includegraphics[scale=0.29]{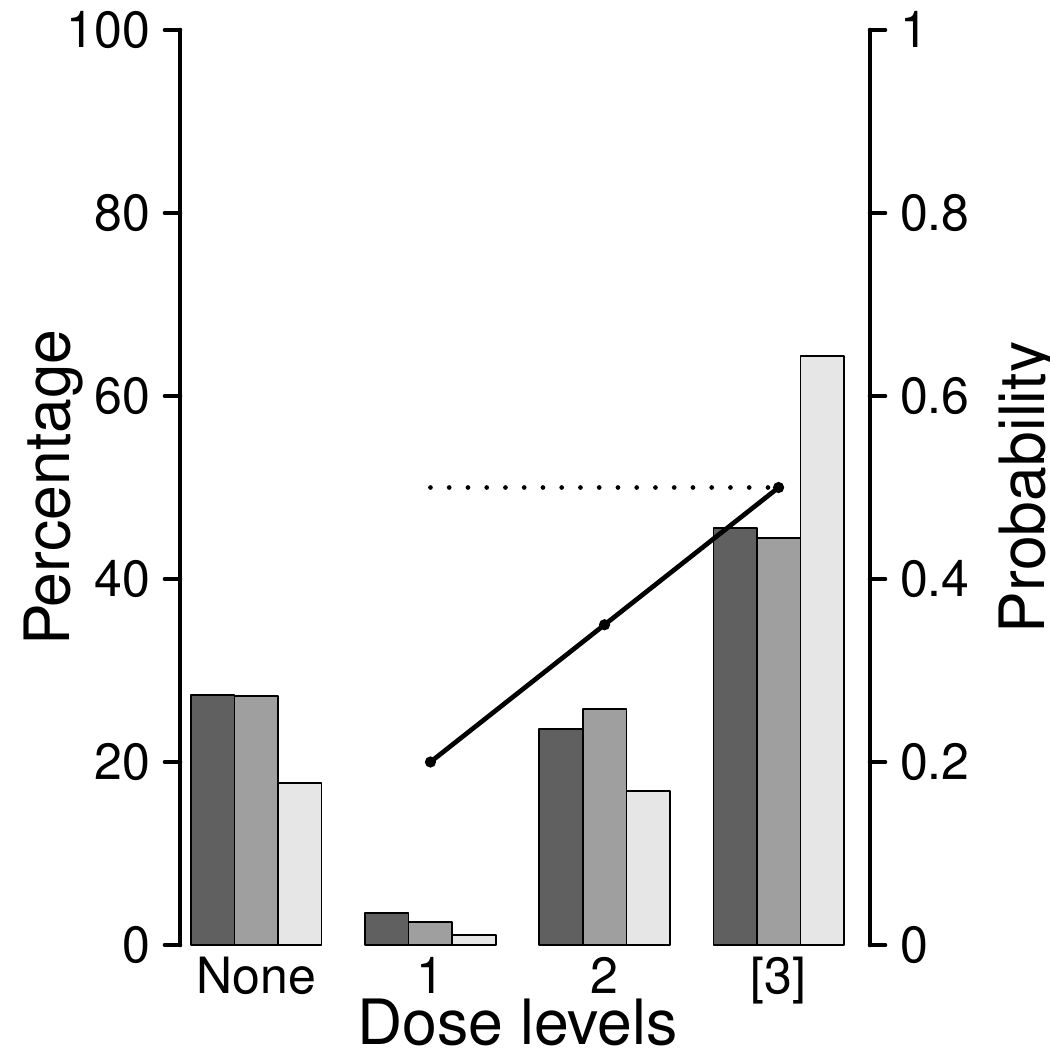}} 
\subfigure[Scenario 3]{\includegraphics[scale=0.29]{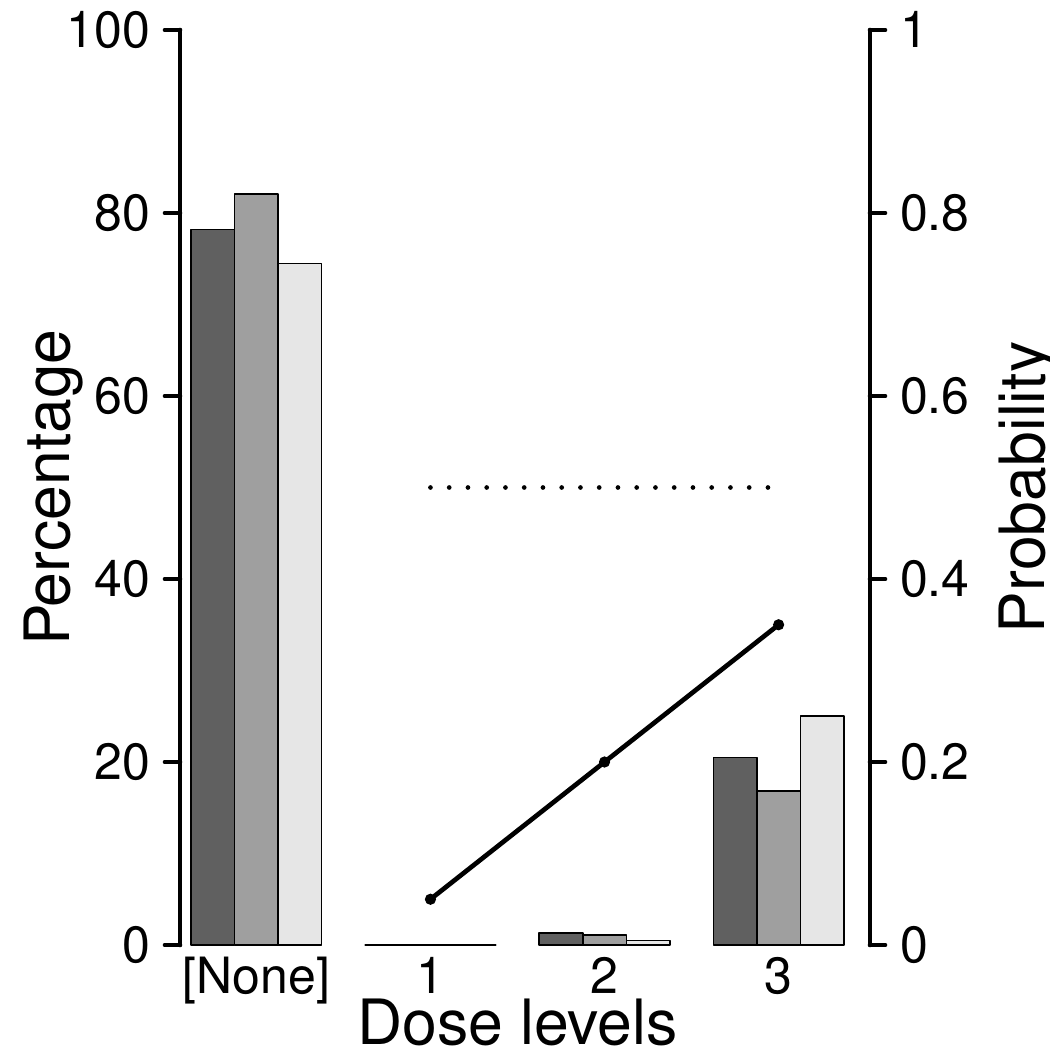}} \\ 
\subfigure[Scenario 4]{\includegraphics[scale=0.29]{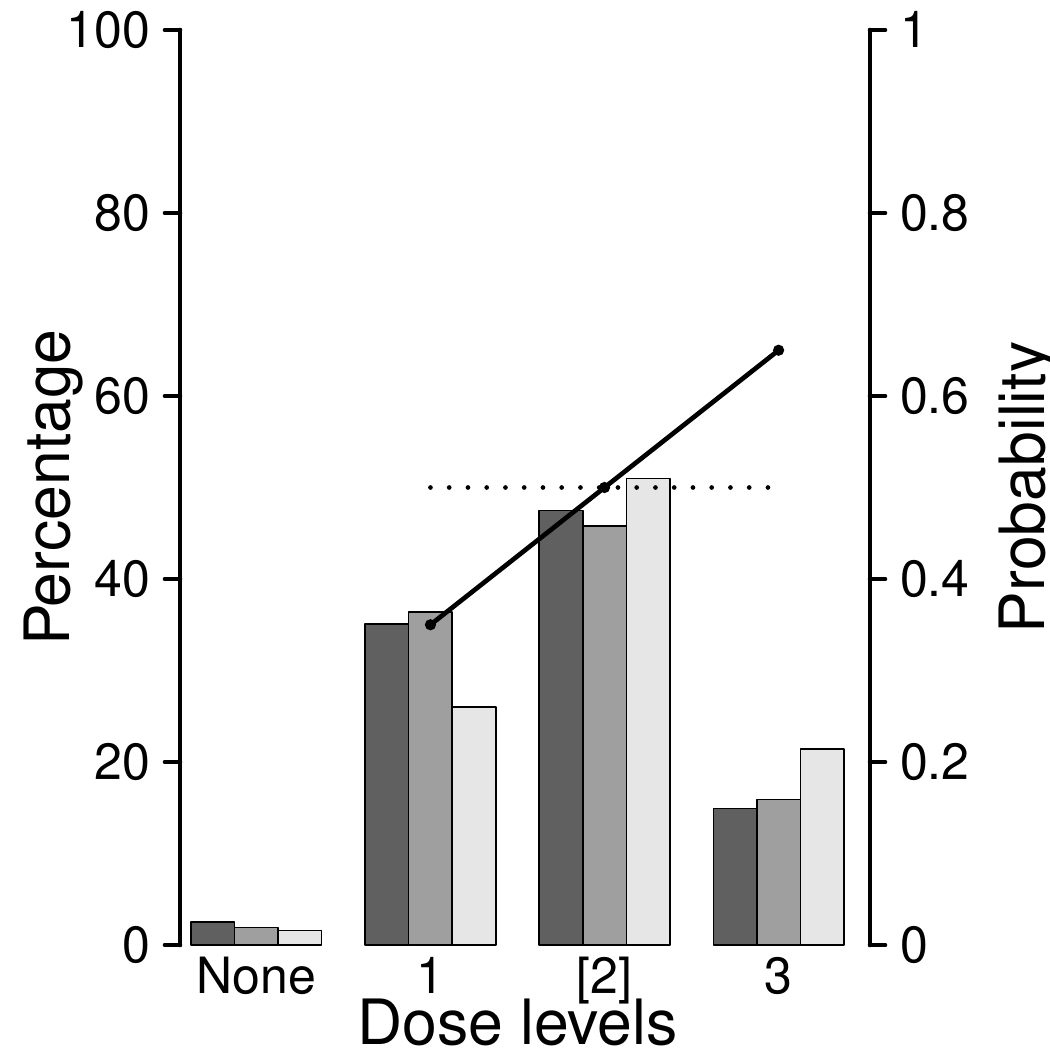}}
\subfigure[Scenario 5]{\includegraphics[scale=0.29]{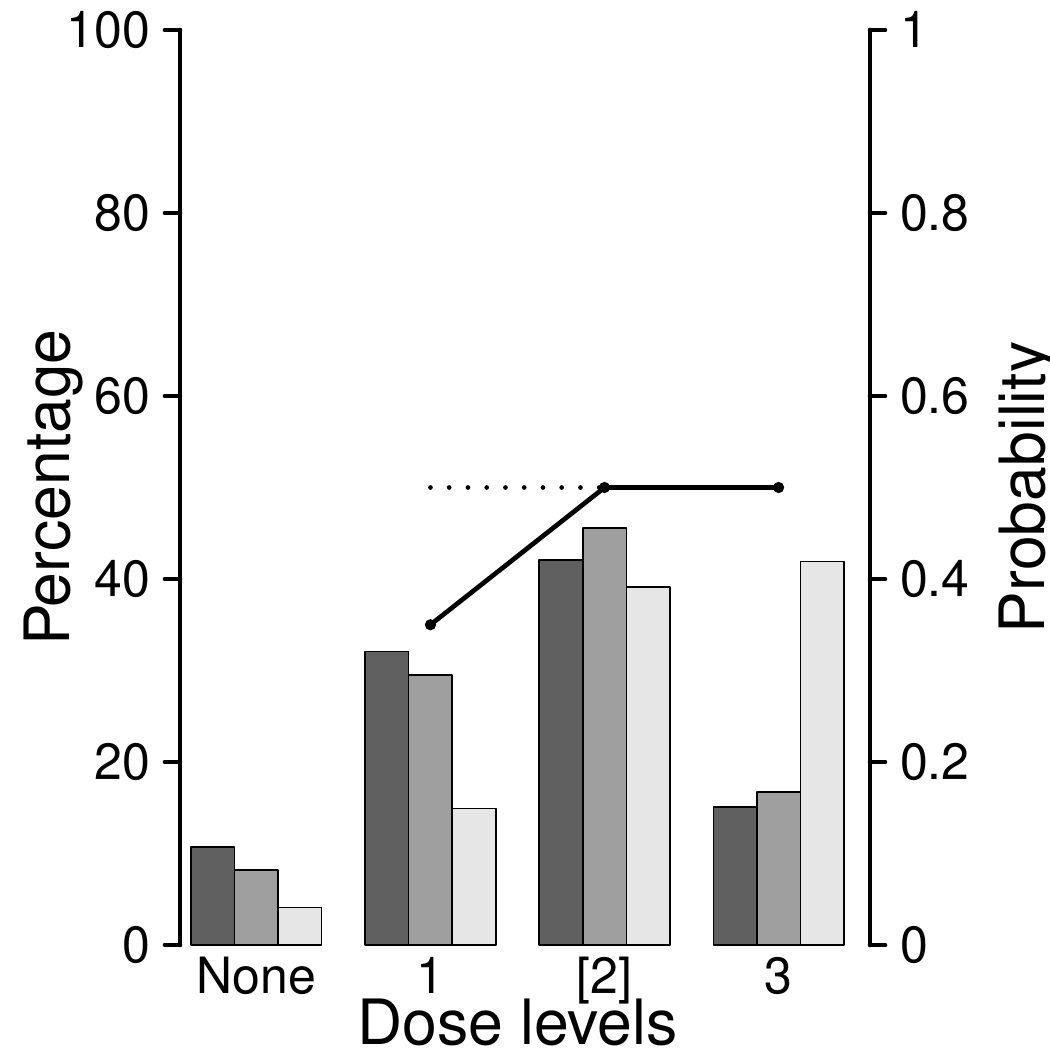}} 
\subfigure[Scenario 6]{\includegraphics[scale=0.29]{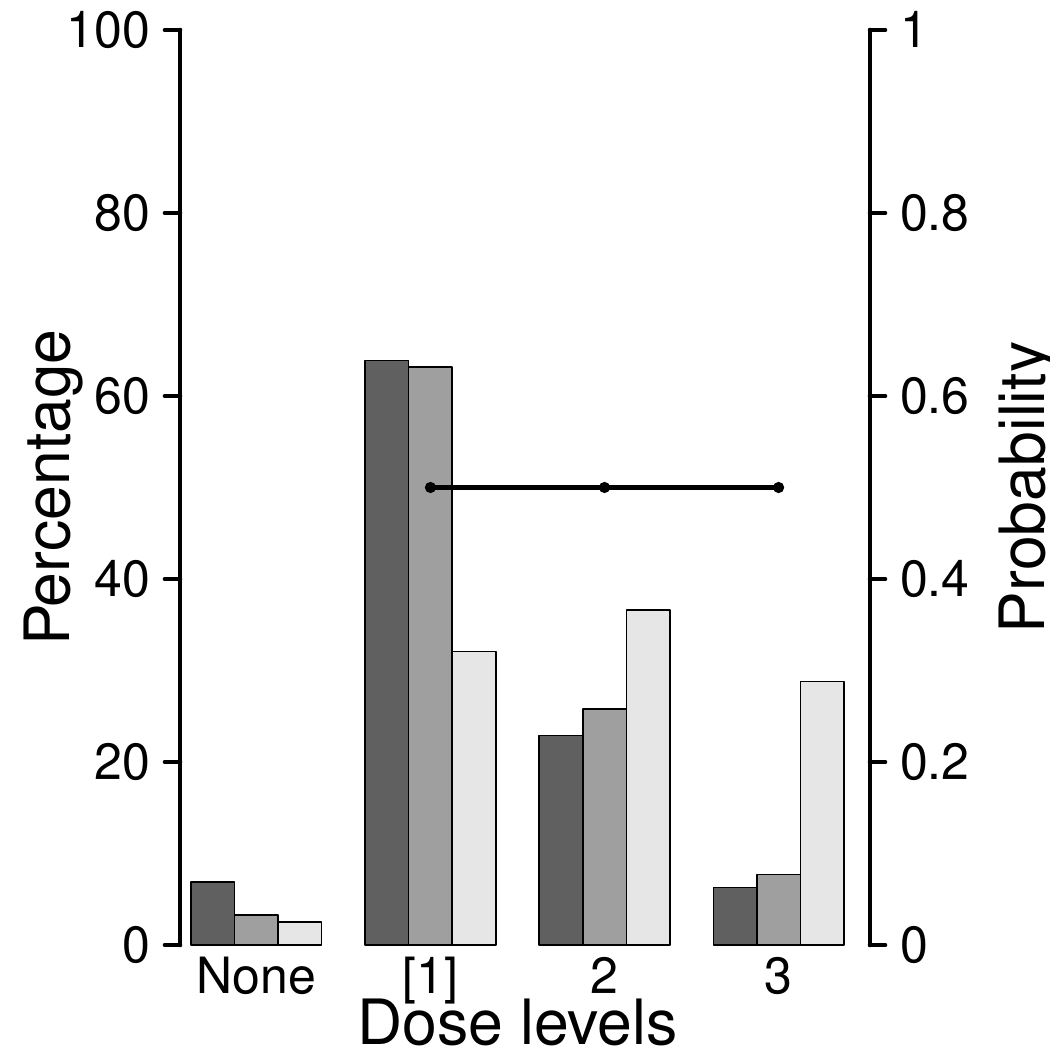}} \\ 
\subfigure[Scenario 7]{\includegraphics[scale=0.29]{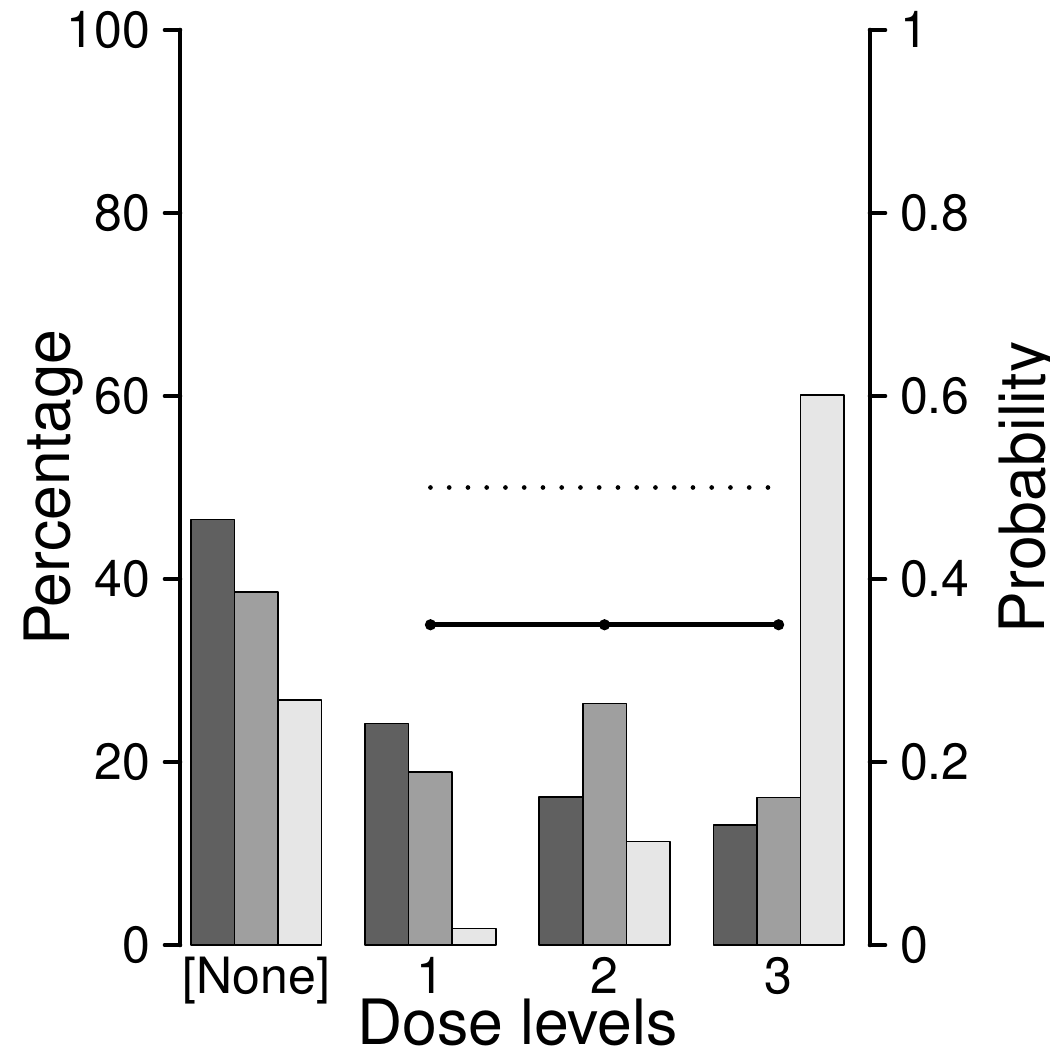}} 
\subfigure[Scenario 8]{\includegraphics[scale=0.29]{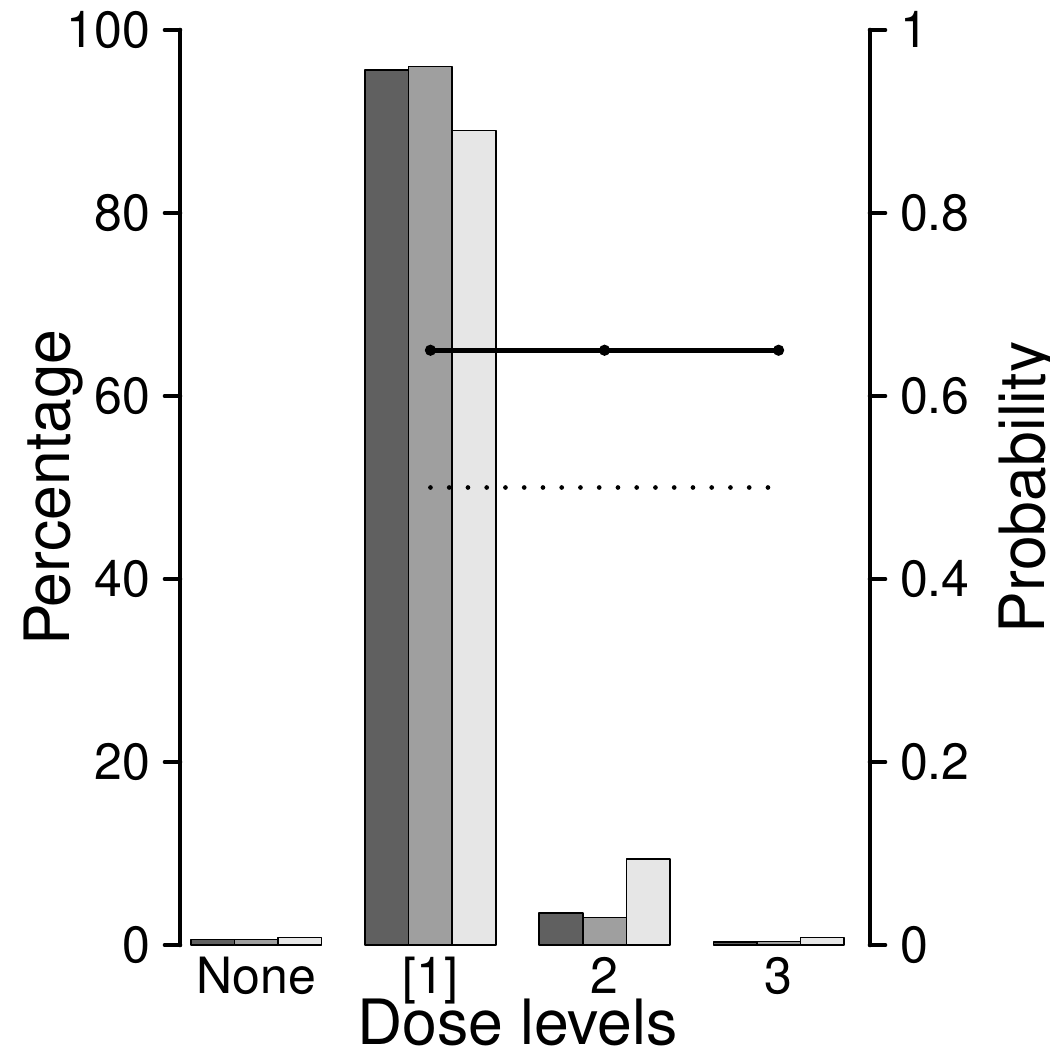}} 
\subfigure{\includegraphics[scale=0.29]{legend_barplot_OBD_hat}} 
\end{center}
\caption{Comparison of dose selection for Selection method, BMA method  and BLRM for a maximum number of volunteers of $n = 30$, when $L = 3$ dose levels are used. The solid line with circles represents the true dose-activity relationship. The horizontal dashed line is the activity probability target. \label{fig:barplot_MAD_hat_n30_L3}}
\end{figure}

\begin{figure}[h!]
\begin{center}
\subfigure[Scenario 1]{\includegraphics[scale=0.29]{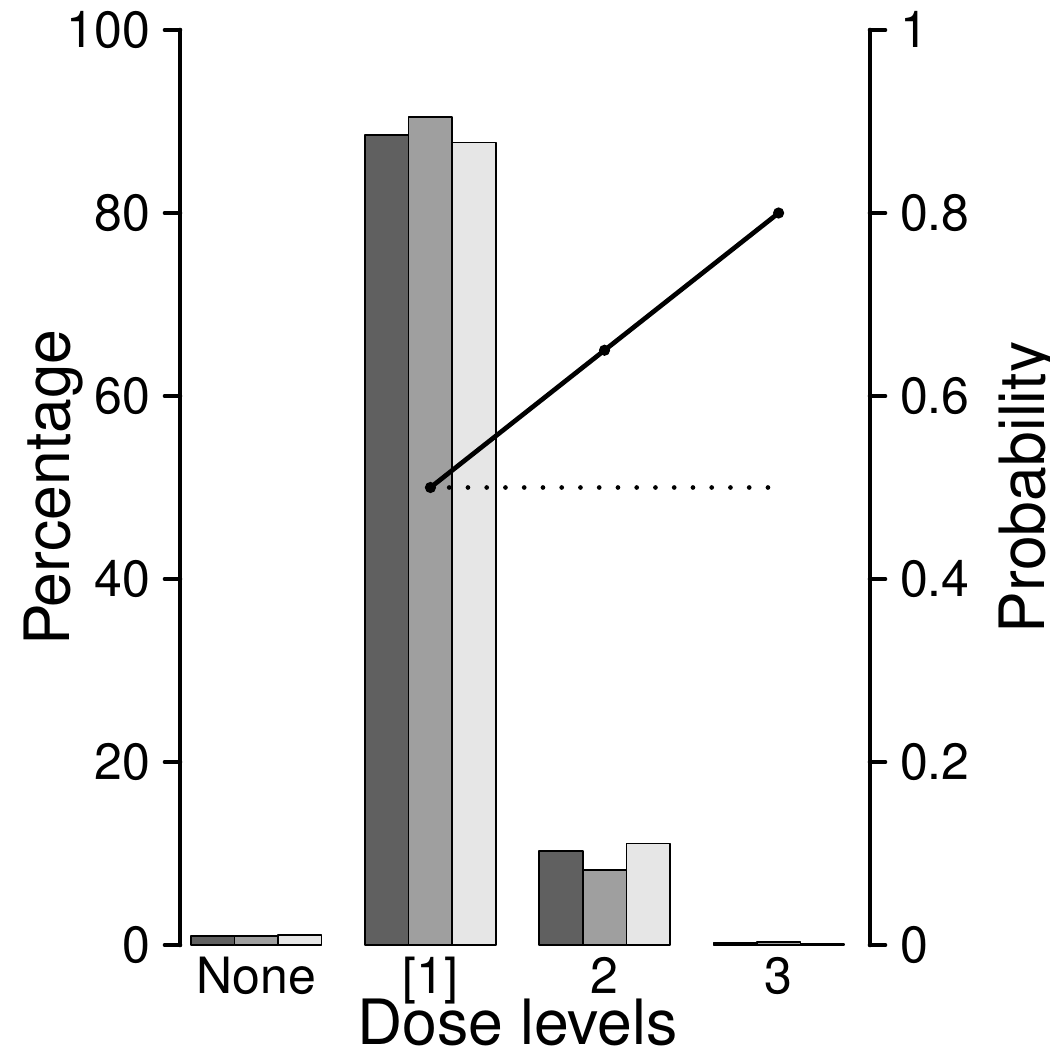}} 
\subfigure[Scenario 2]{\includegraphics[scale=0.29]{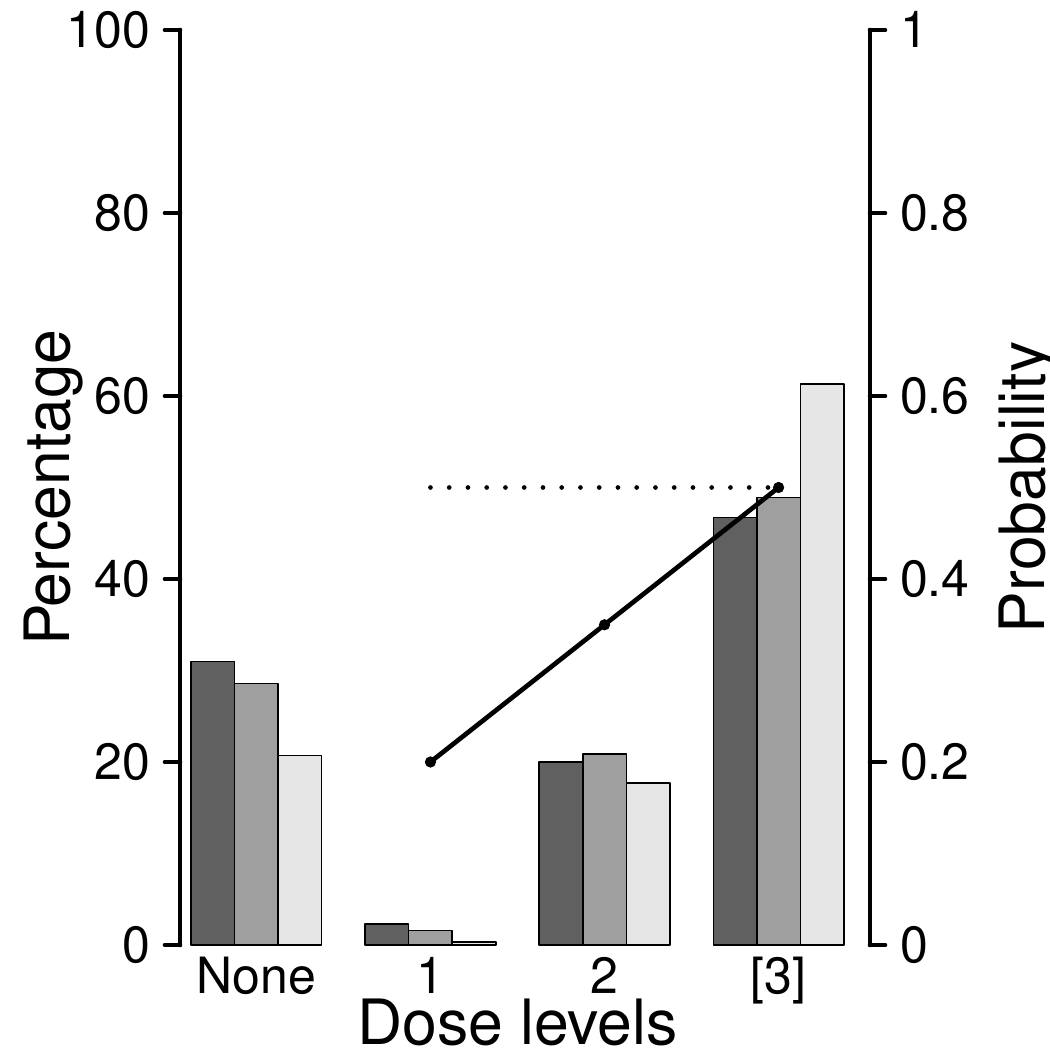}} 
\subfigure[Scenario 3]{\includegraphics[scale=0.29]{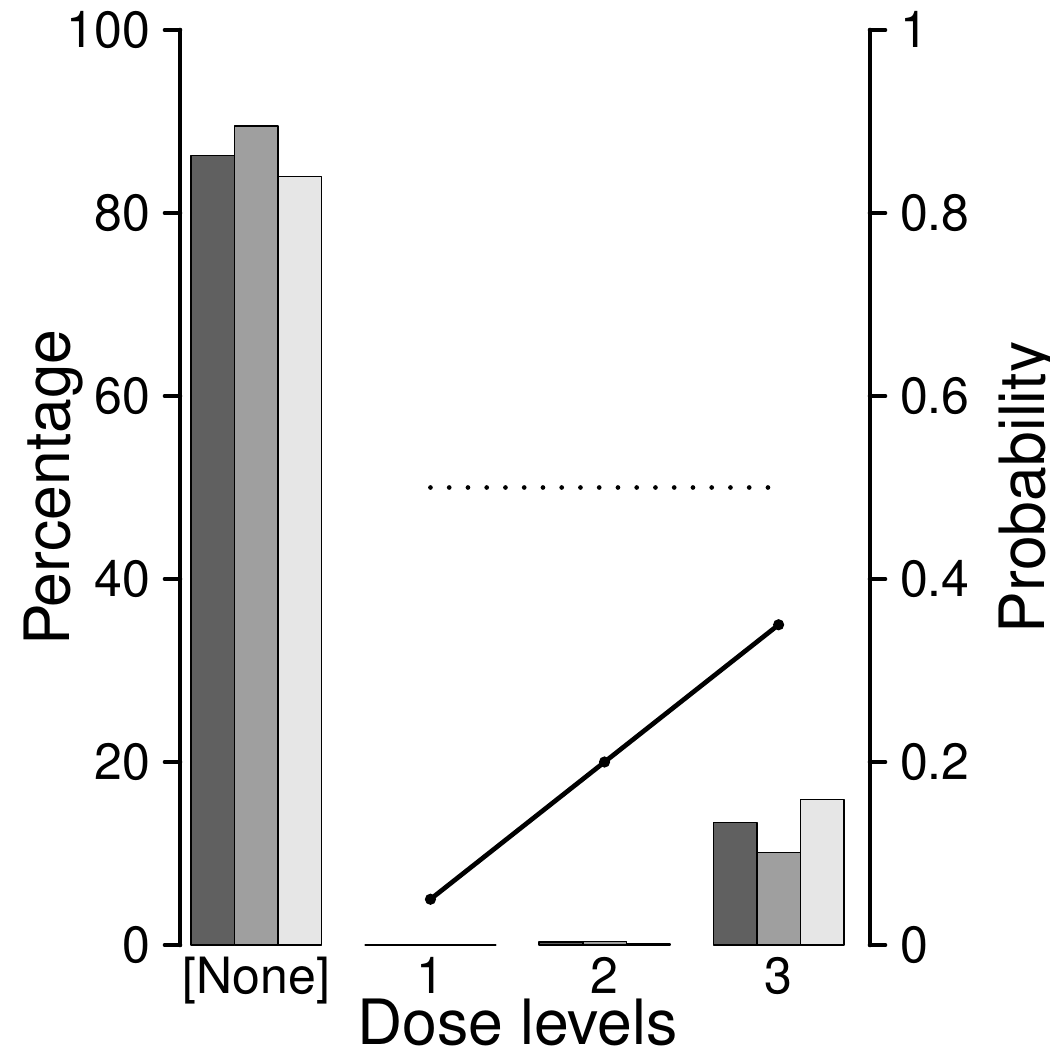}} \\ 
\subfigure[Scenario 4]{\includegraphics[scale=0.29]{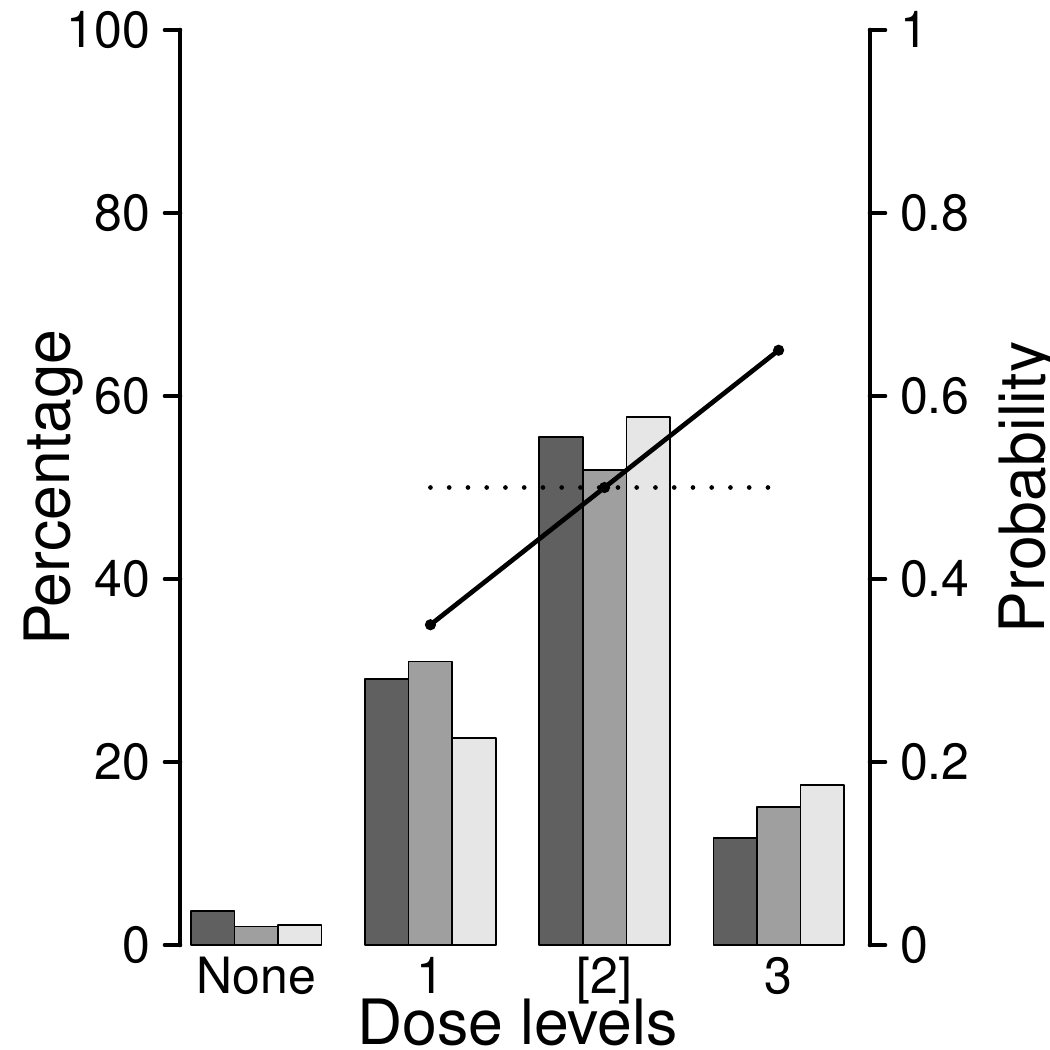}}
\subfigure[Scenario 5]{\includegraphics[scale=0.29]{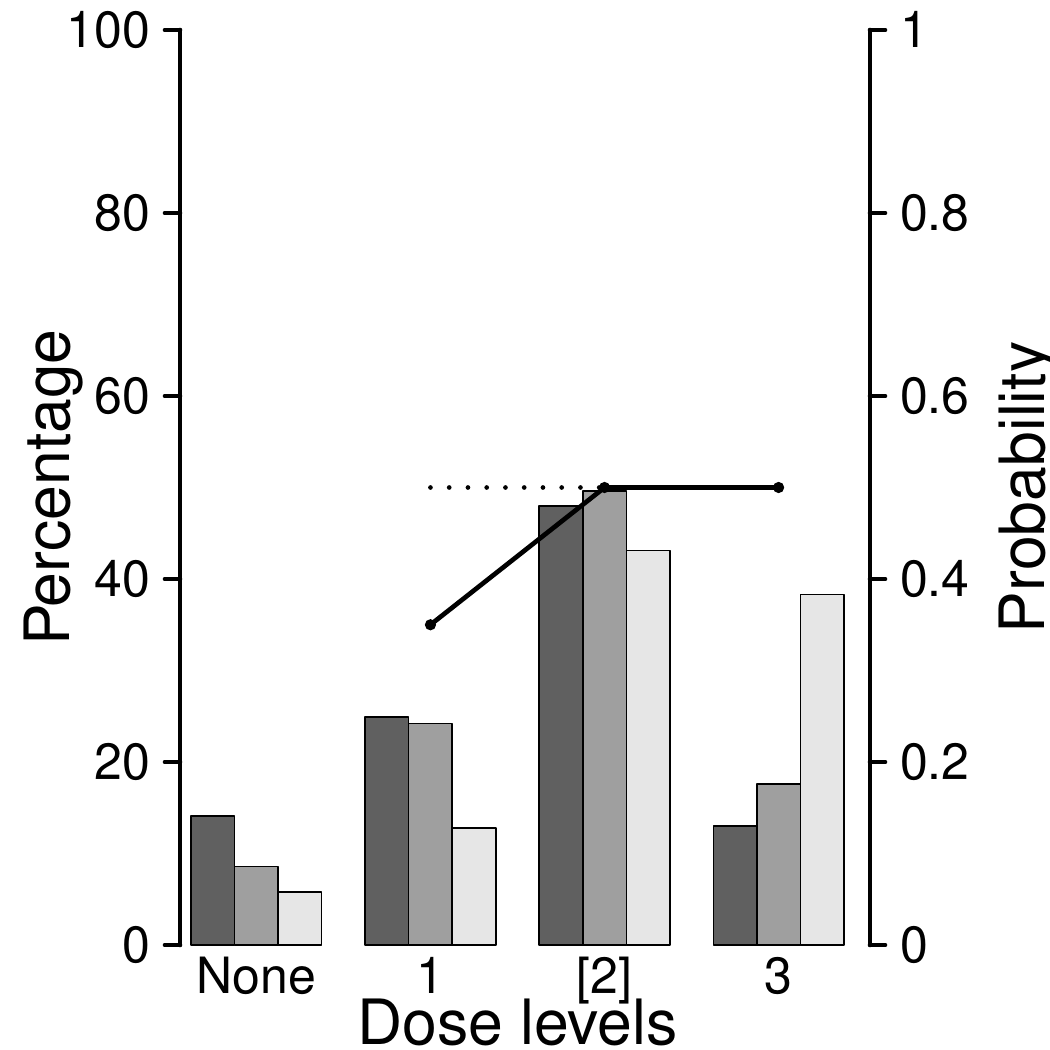}} 
\subfigure[Scenario 6]{\includegraphics[scale=0.29]{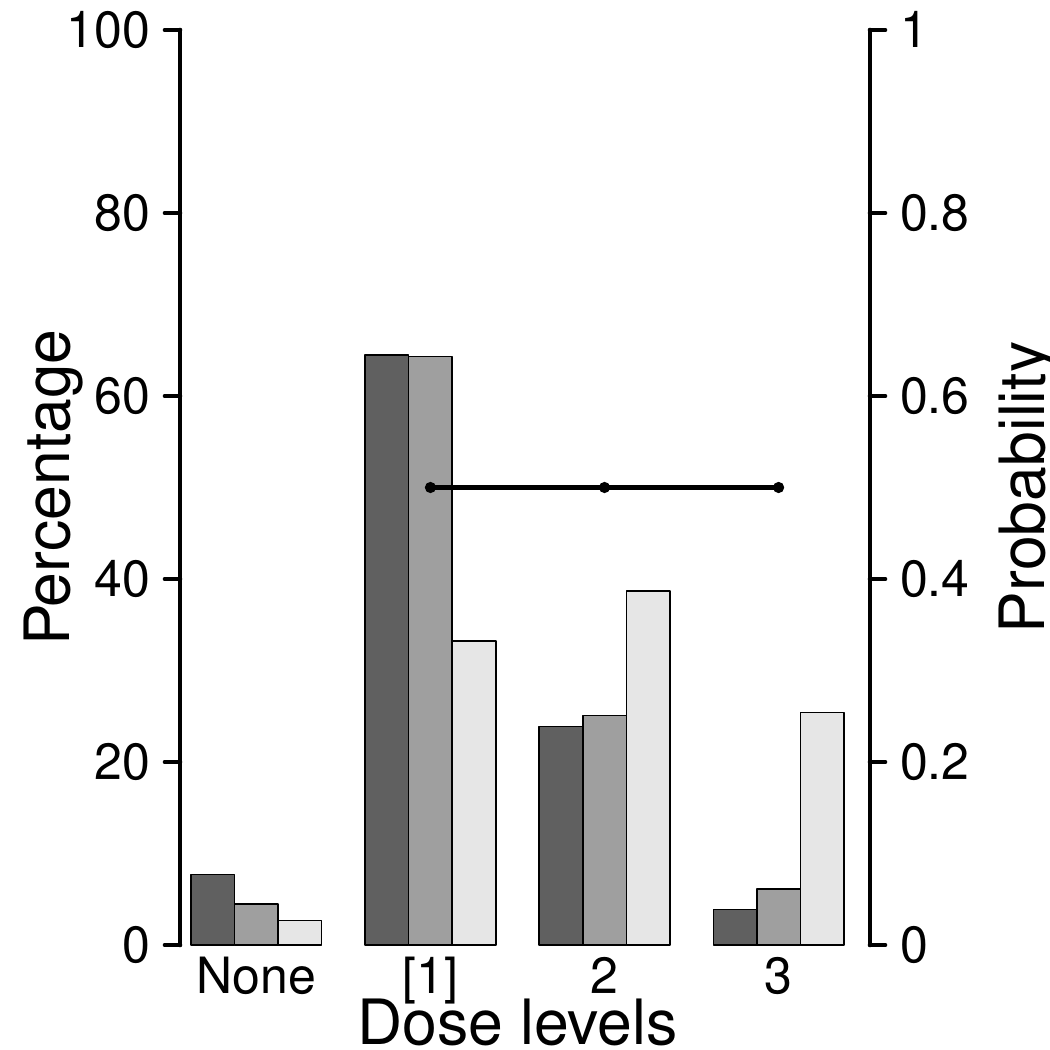}} \\ 
\subfigure[Scenario 7]{\includegraphics[scale=0.29]{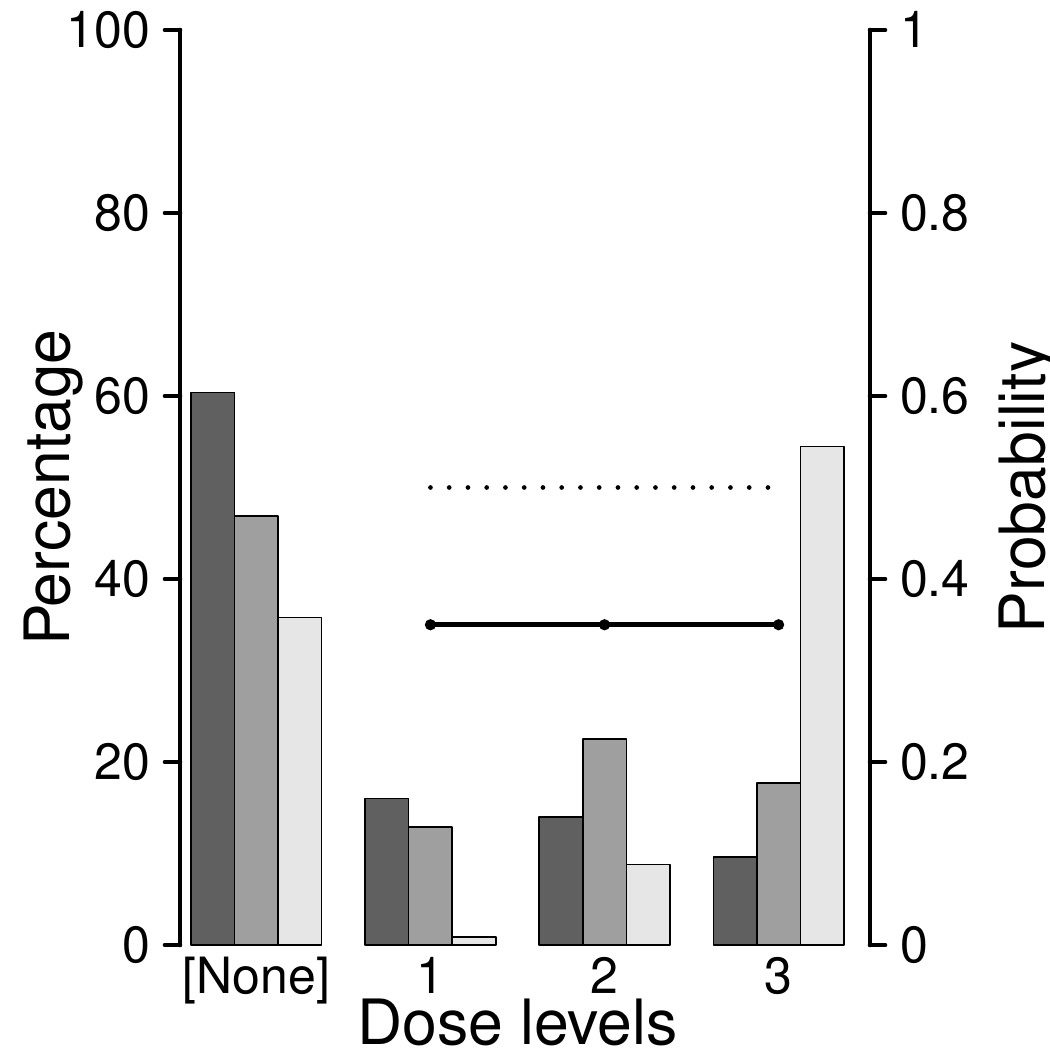}} 
\subfigure[Scenario 8]{\includegraphics[scale=0.29]{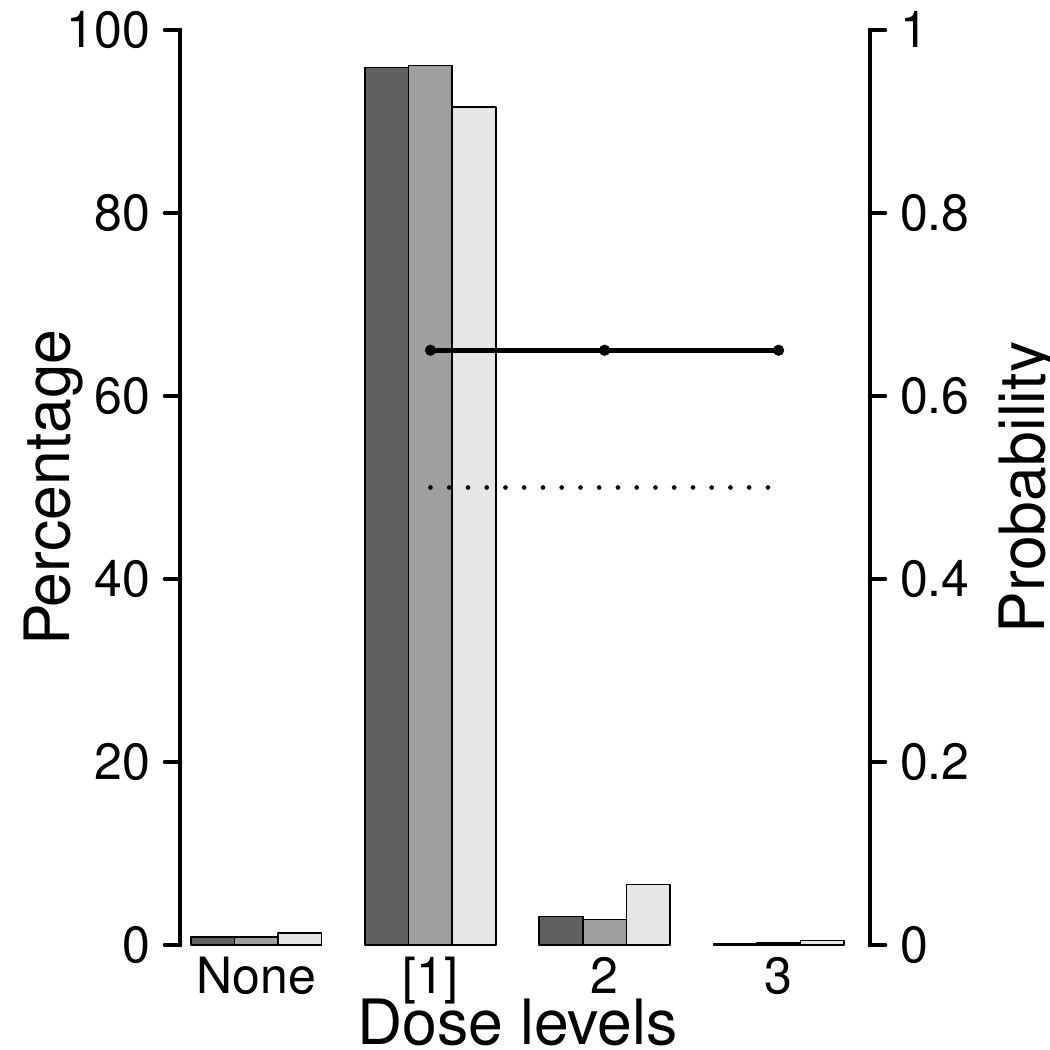}} 
\subfigure{\includegraphics[scale=0.29]{legend_barplot_OBD_hat}} 
\end{center}
\caption{Comparison of dose selection for Selection method, BMA method  and BLRM for a maximum number of volunteers of $n = 40$, when $L = 3$ dose levels are used. The solid line with circles represents the true dose-activity relationship. The horizontal dashed line is the activity probability target. \label{fig:barplot_MAD_hat_n40_L3}}
\end{figure}

\begin{figure}[h!]
\begin{center}
\subfigure[Scenario 1]{\includegraphics[scale=0.29]{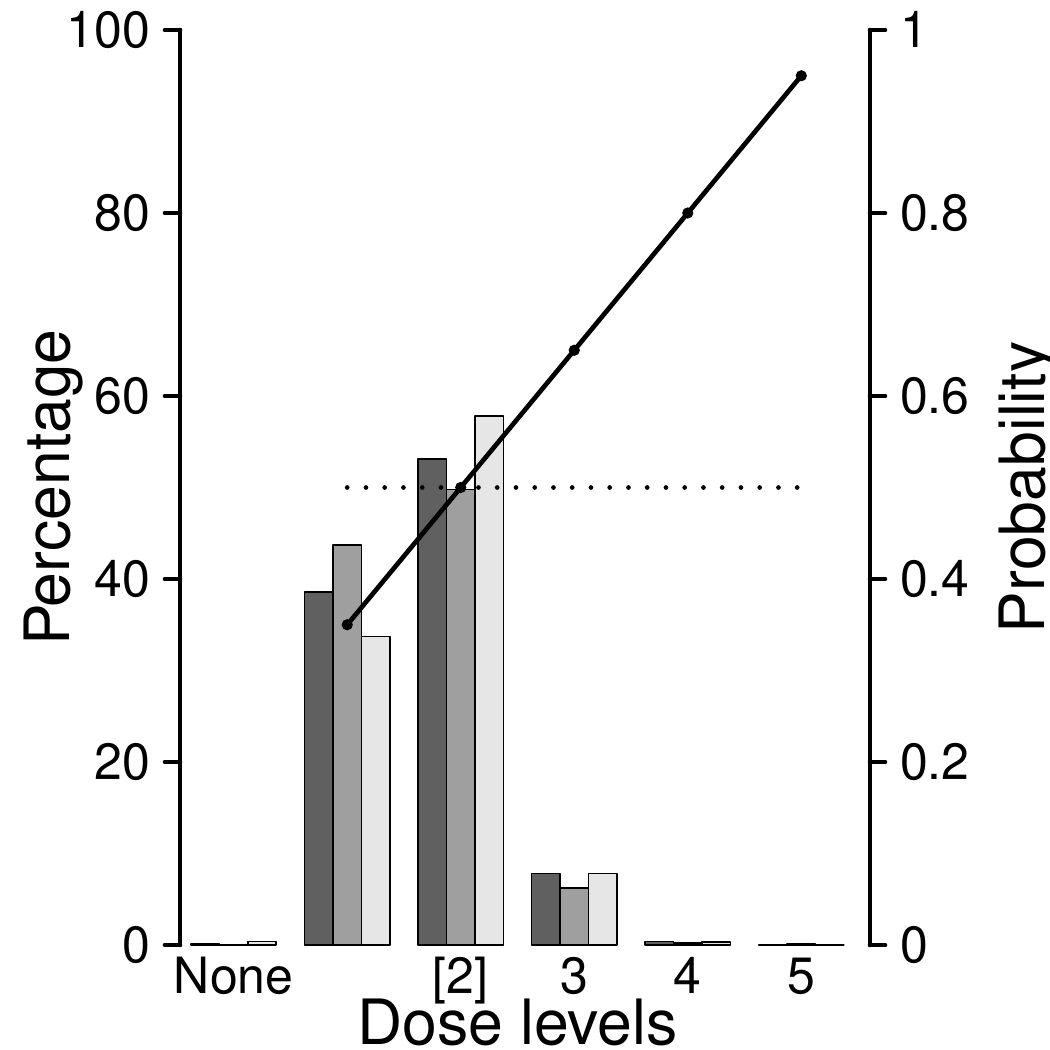}} 
\subfigure[Scenario 2]{\includegraphics[scale=0.29]{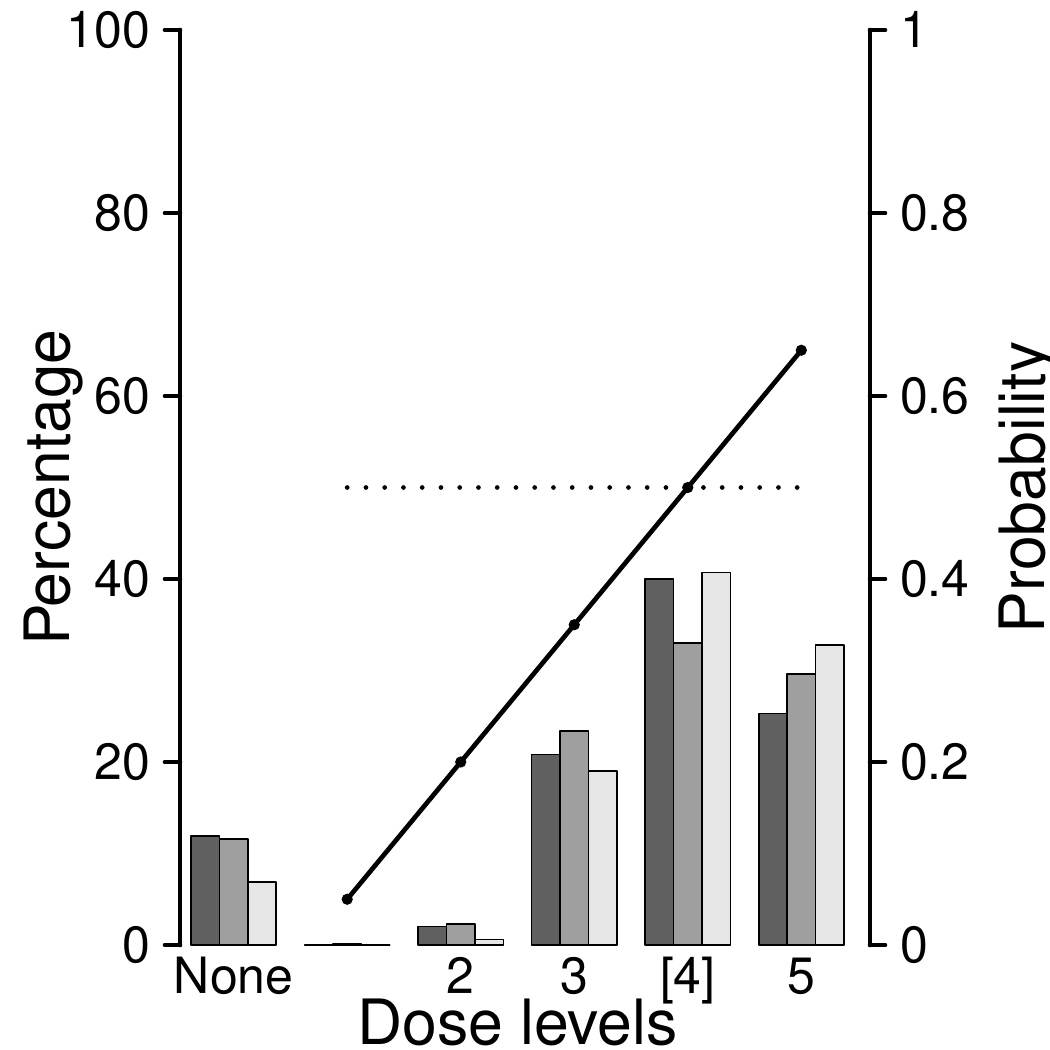}} 
\subfigure[Scenario 3]{\includegraphics[scale=0.29]{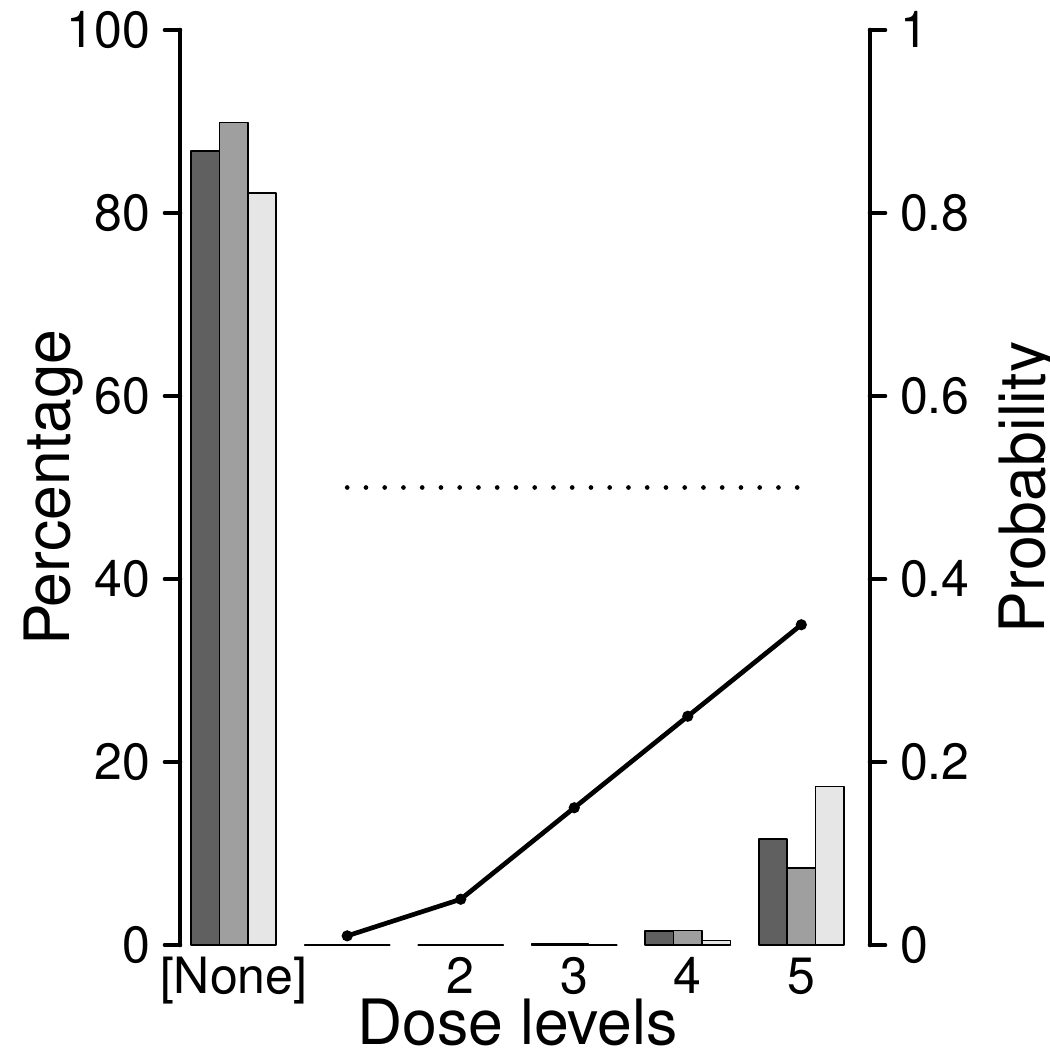}} \\ 
\subfigure[Scenario 4]{\includegraphics[scale=0.29]{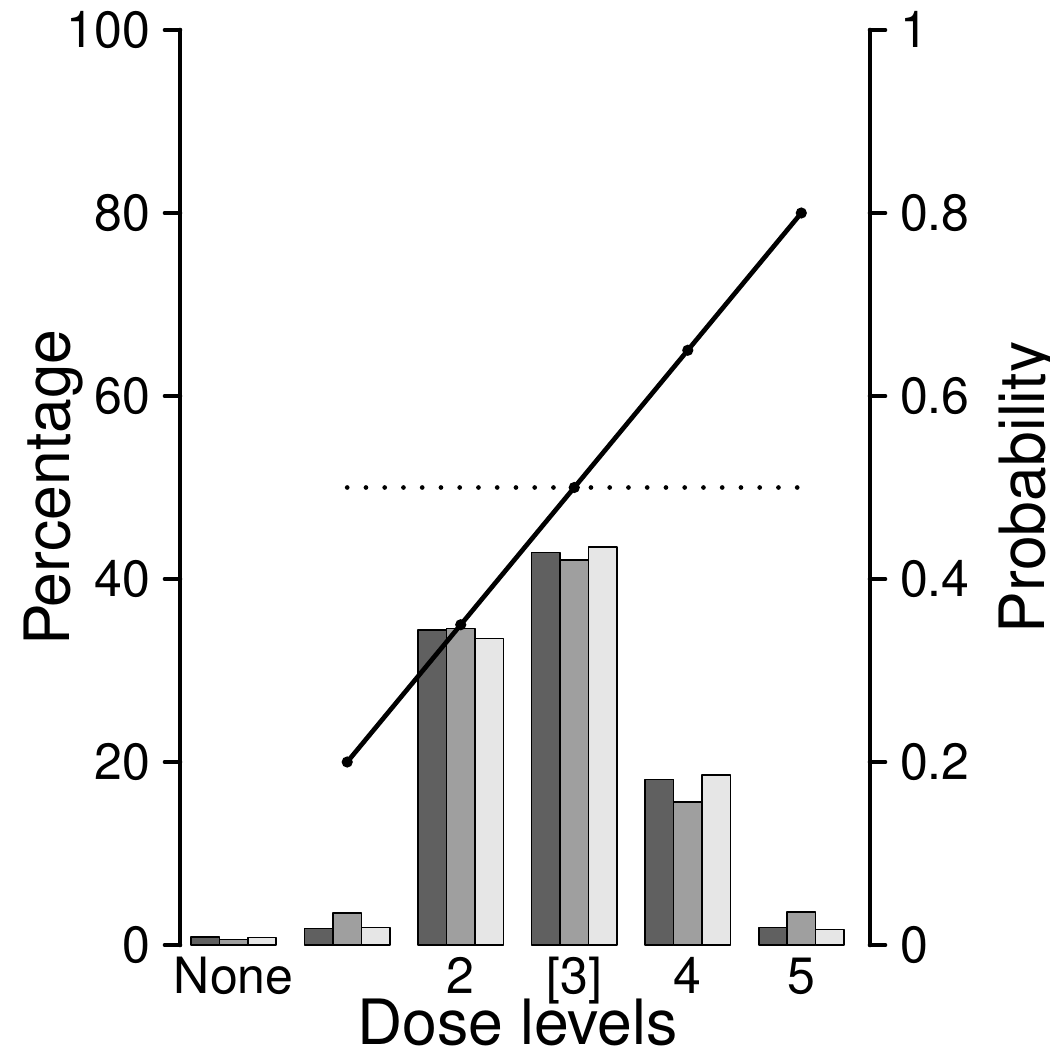}}
\subfigure[Scenario 5]{\includegraphics[scale=0.29]{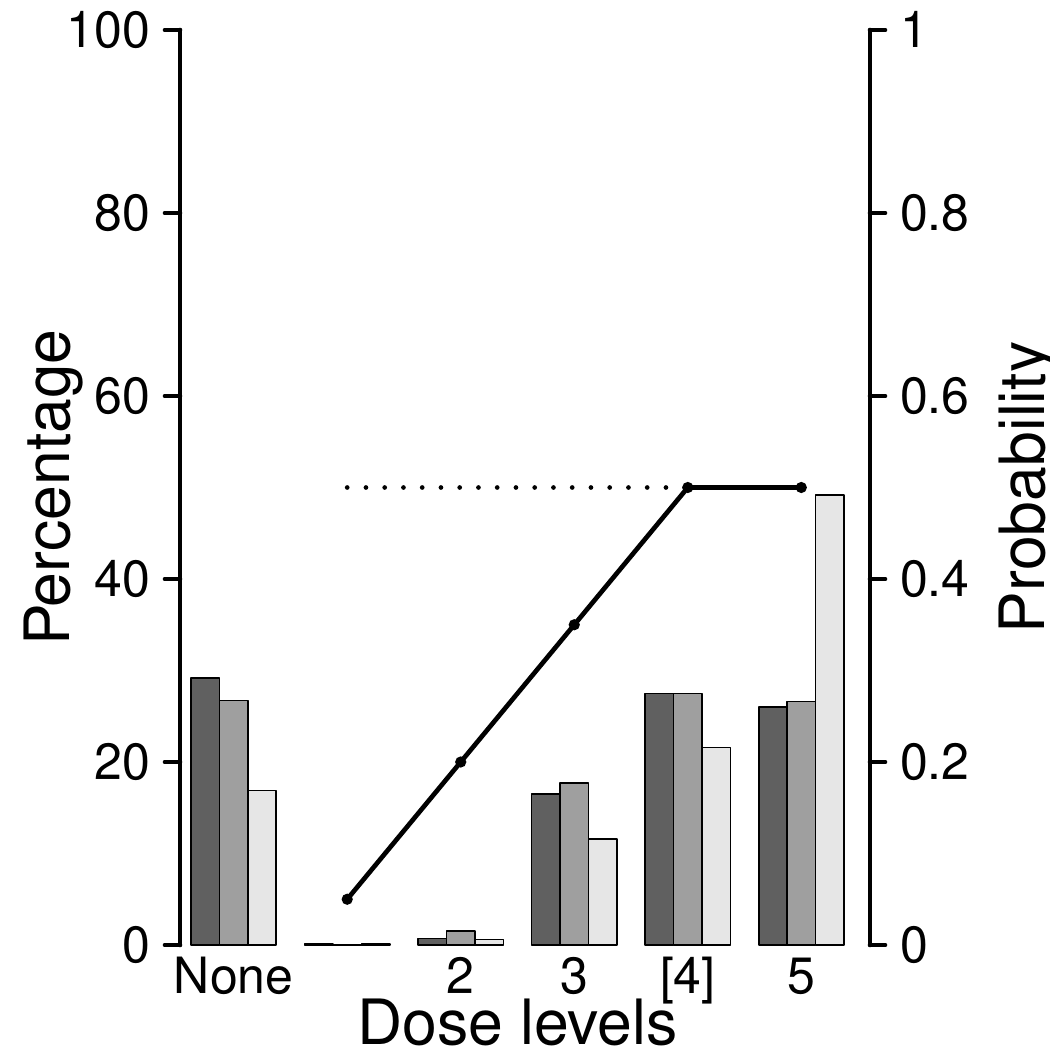}} 
\subfigure[Scenario 6]{\includegraphics[scale=0.29]{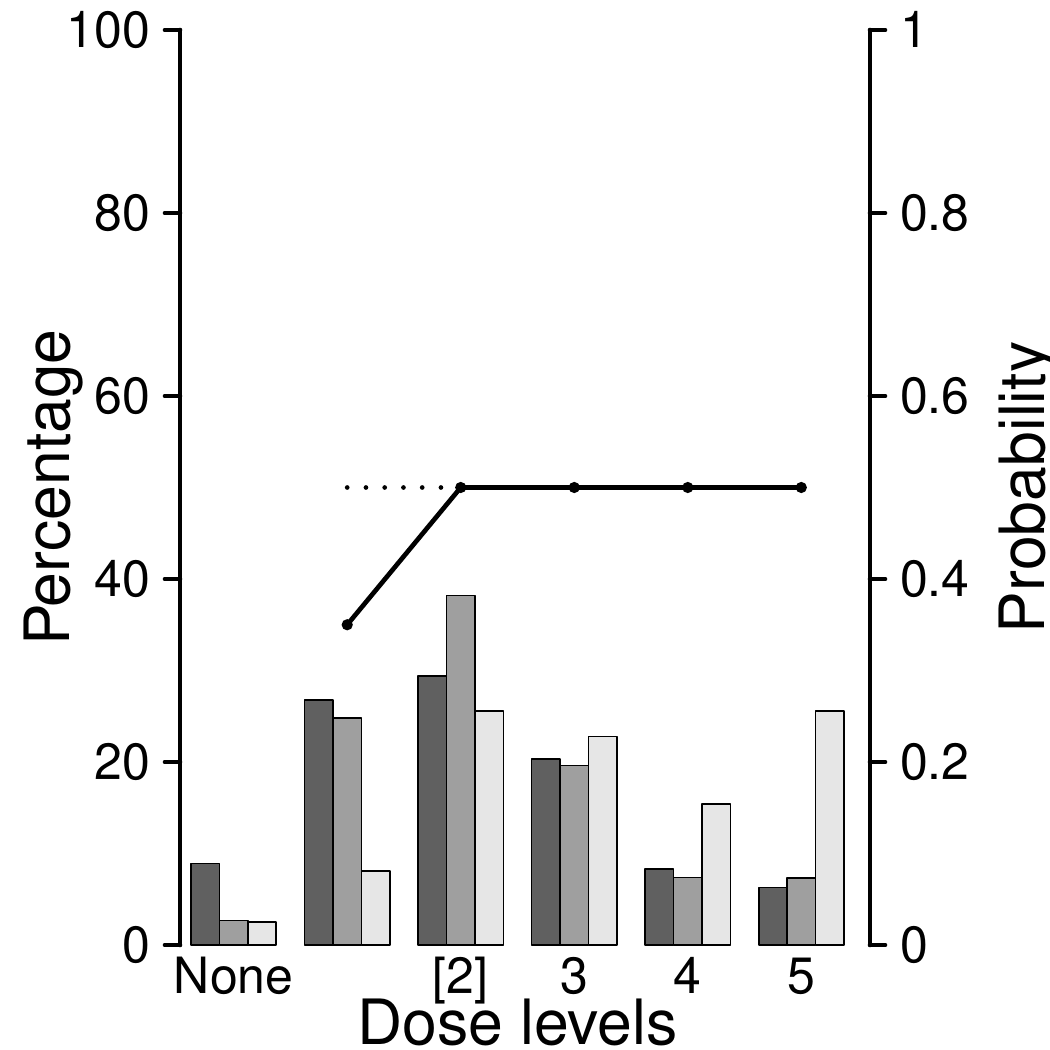}} \\ 
\subfigure[Scenario 7]{\includegraphics[scale=0.29]{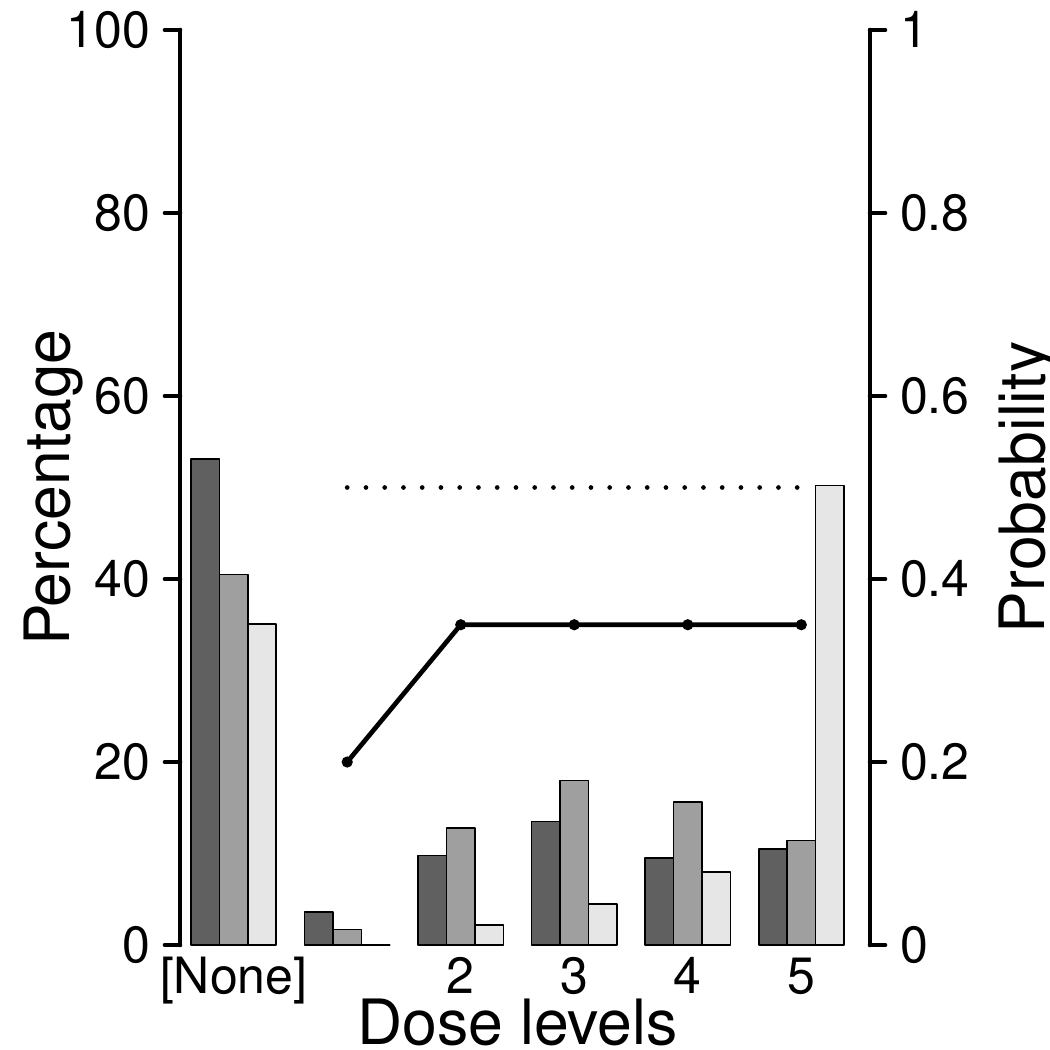}} 
\subfigure[Scenario 8]{\includegraphics[scale=0.29]{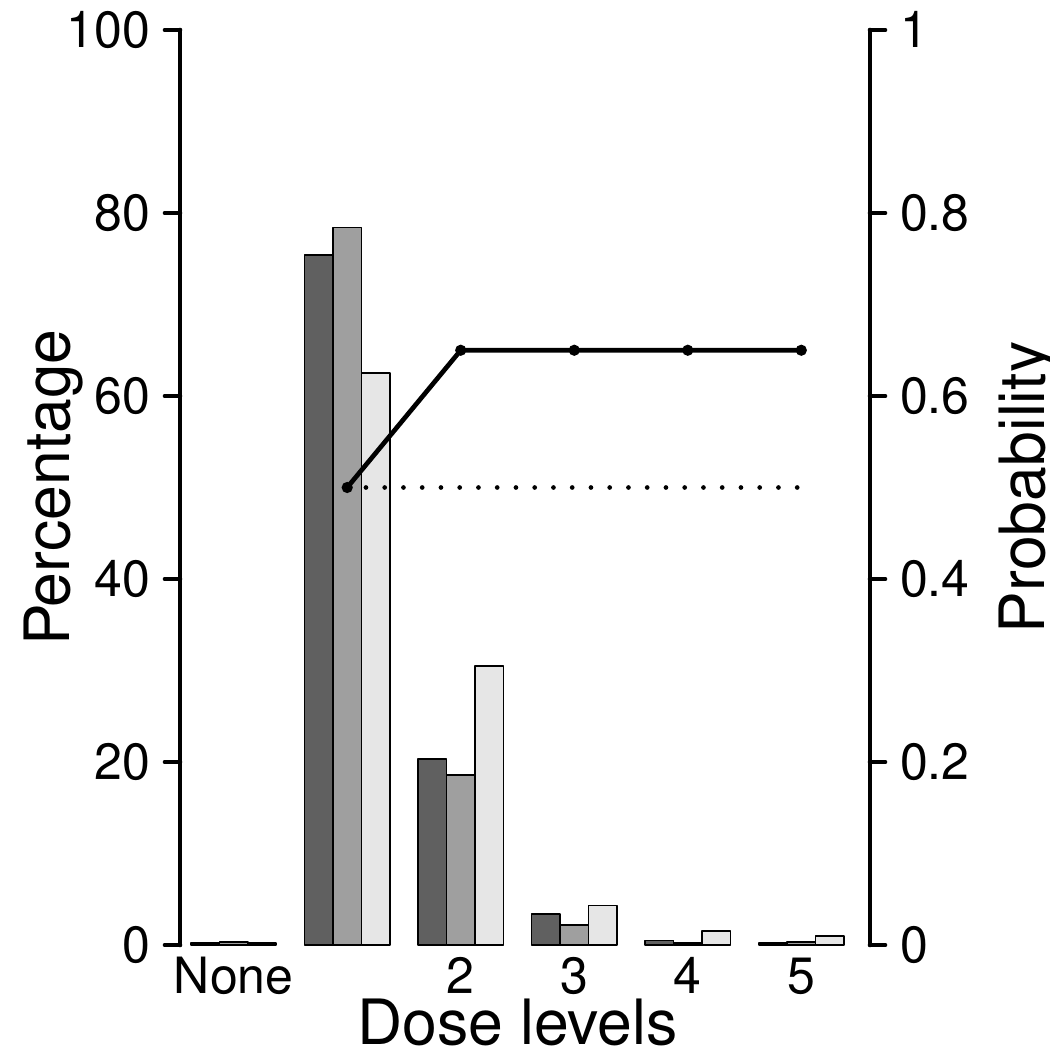}} 
\subfigure{\includegraphics[scale=0.29]{legend_barplot_OBD_hat}} 
\end{center}
\caption{Comparison of dose selection for Selection method, BMA method and BLRM for a maximum number of volunteers of $n = 30$, when $L = 5$ dose levels are used. The solid line with circles represents the true dose-activity relationship. The horizontal dashed line is the activity probability target. \label{fig:barplot_MAD_hat_n30_L5}}
\end{figure}

\begin{figure}[h!]
\begin{center}
\subfigure[Scenario 1]{\includegraphics[scale=0.29]{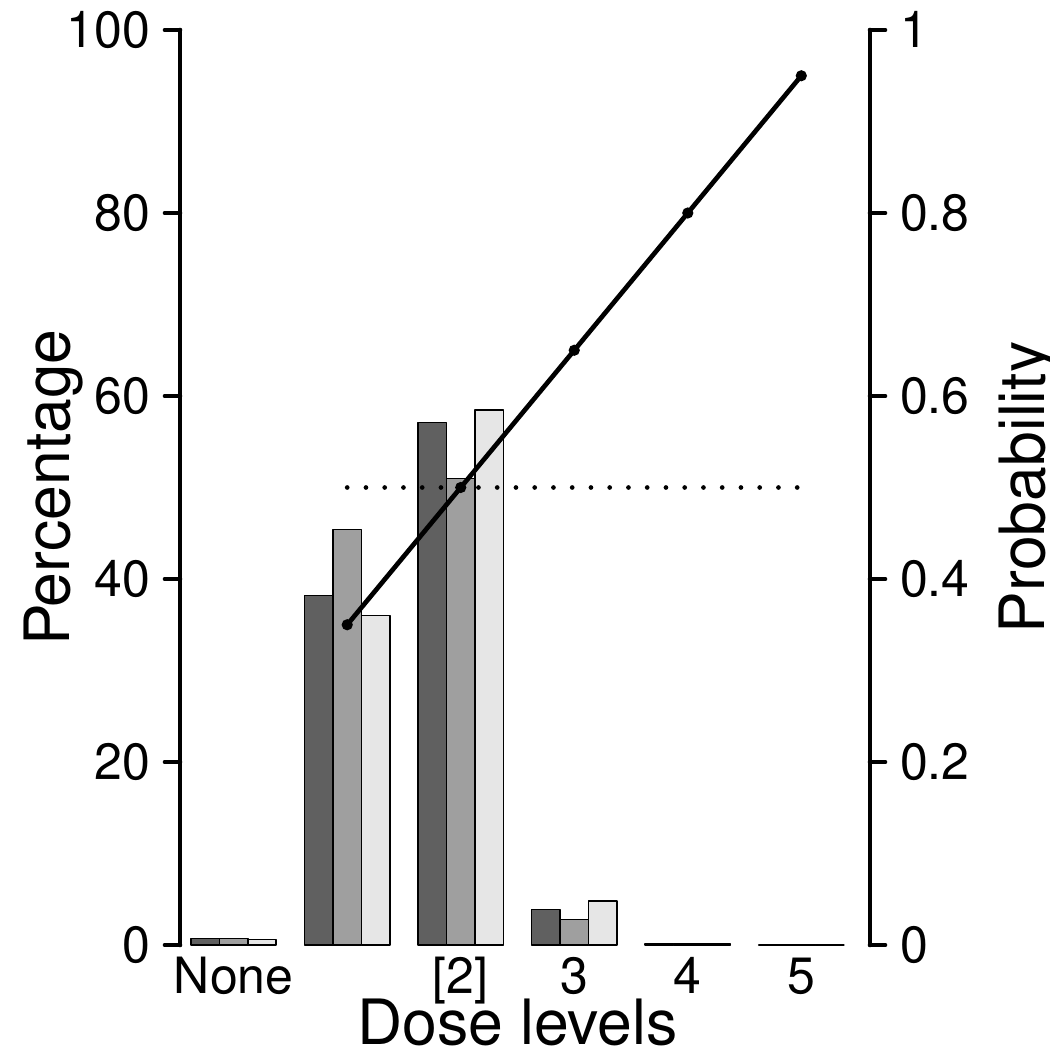}} 
\subfigure[Scenario 2]{\includegraphics[scale=0.29]{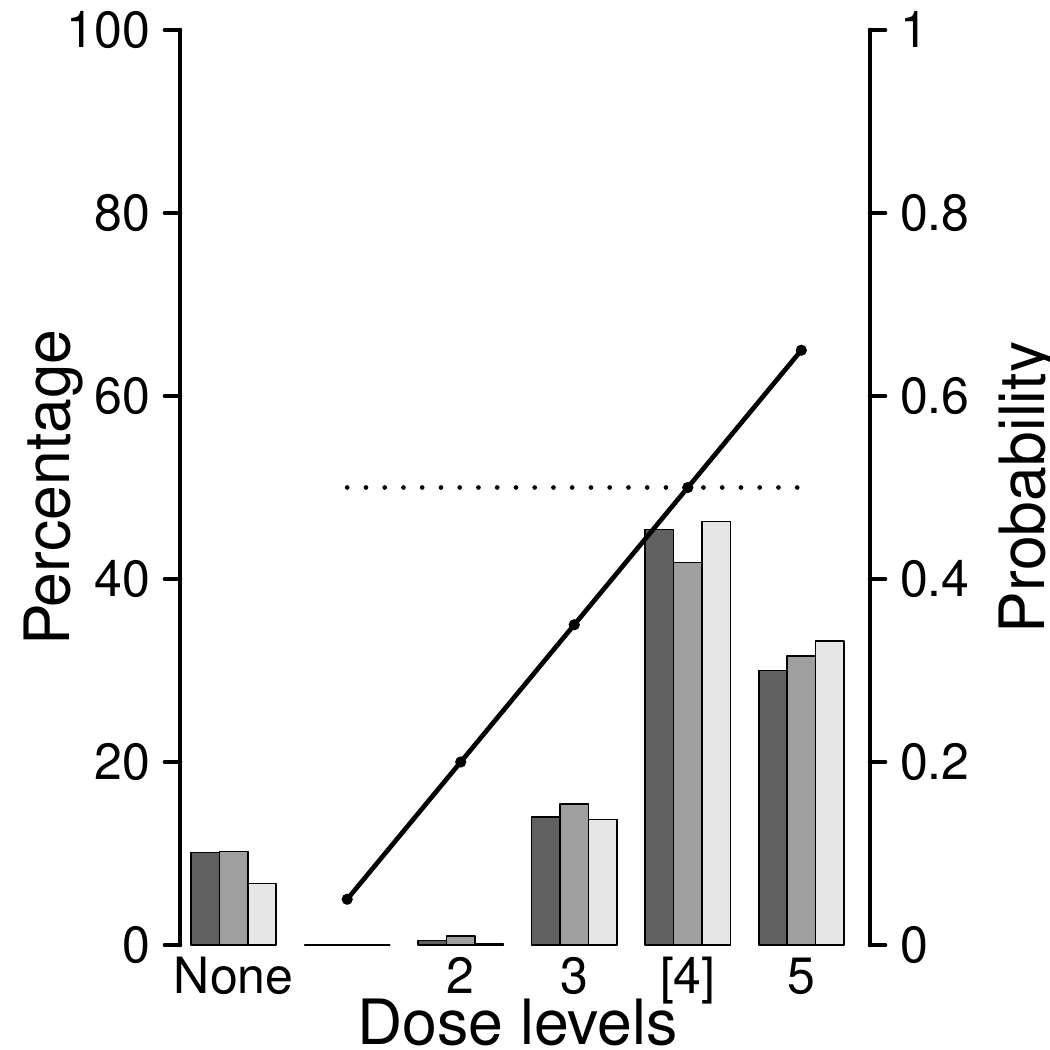}} 
\subfigure[Scenario 3]{\includegraphics[scale=0.29]{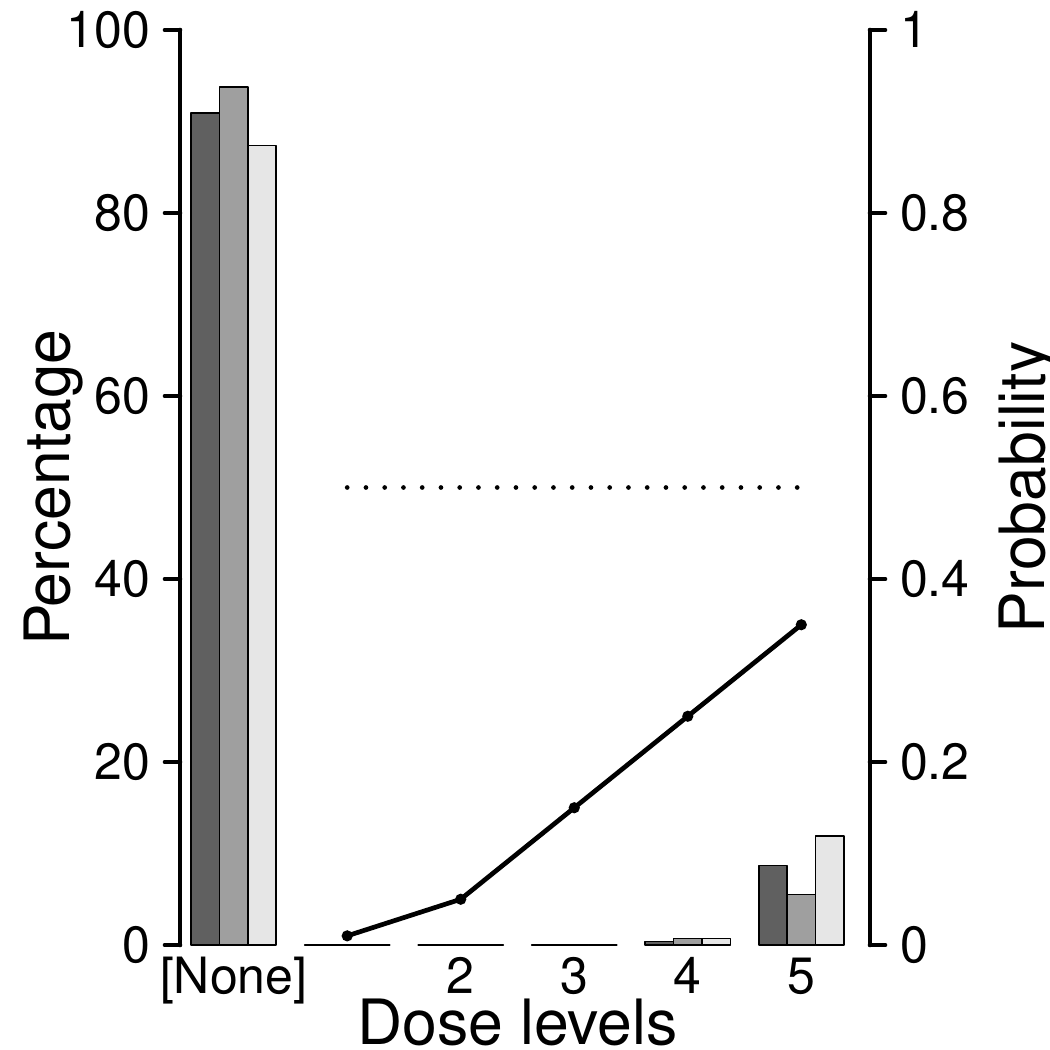}} \\ 
\subfigure[Scenario 4]{\includegraphics[scale=0.29]{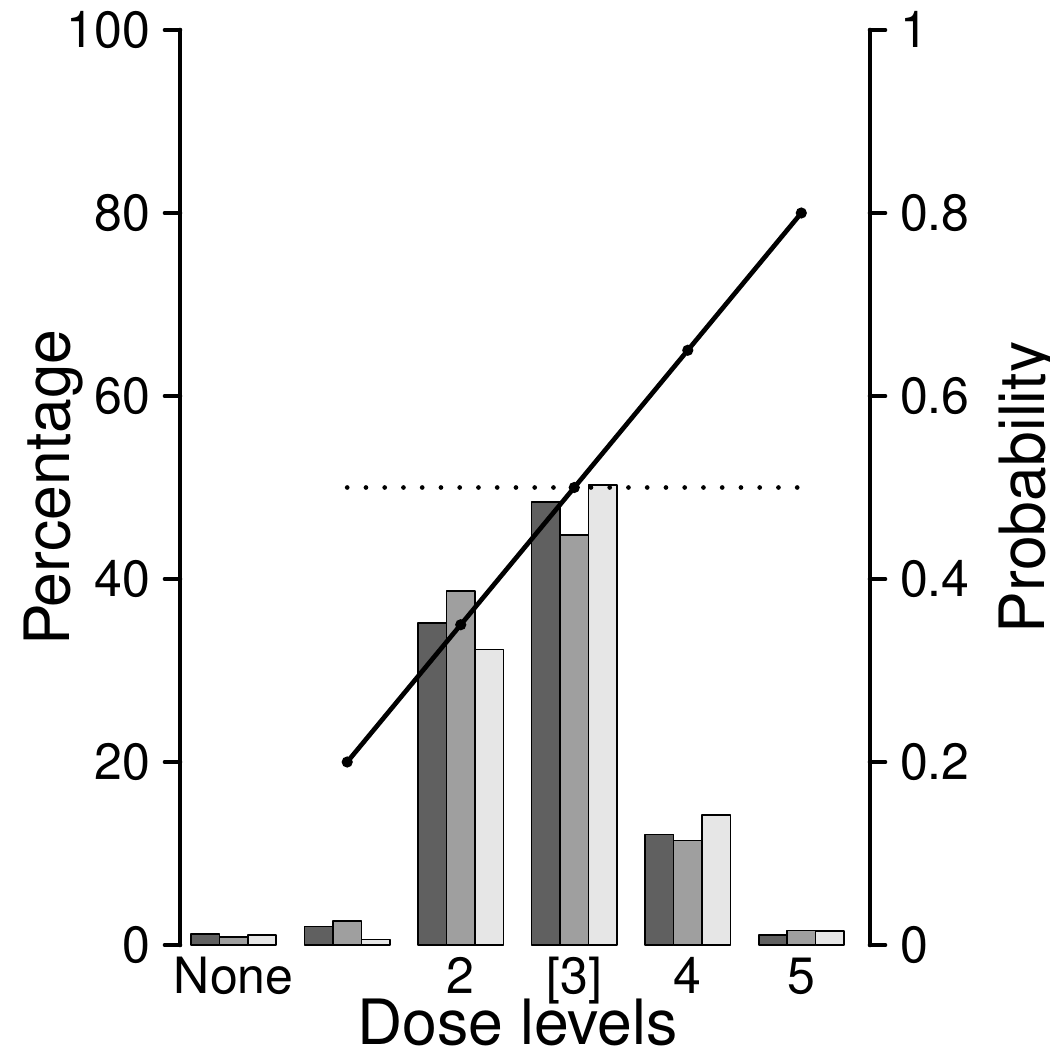}}
\subfigure[Scenario 5]{\includegraphics[scale=0.29]{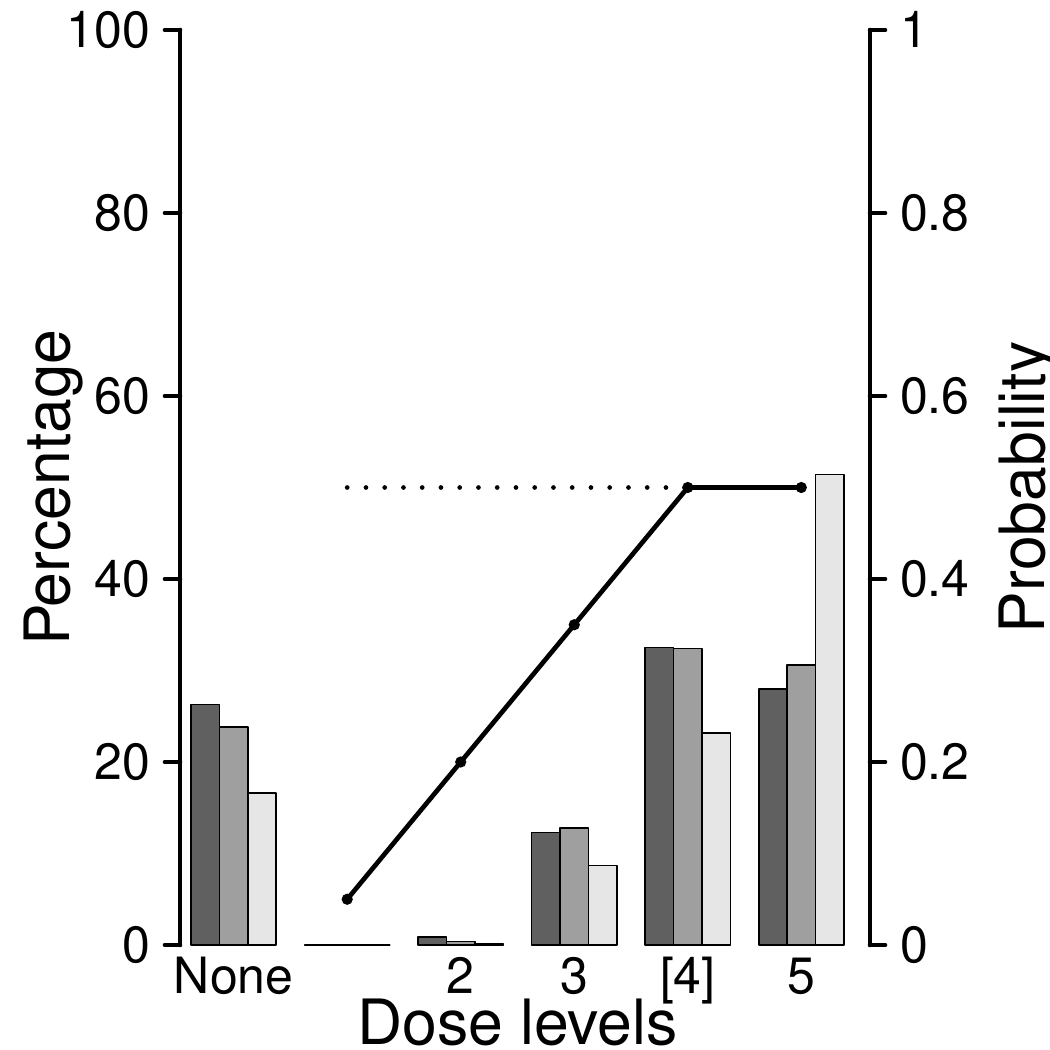}} 
\subfigure[Scenario 6]{\includegraphics[scale=0.29]{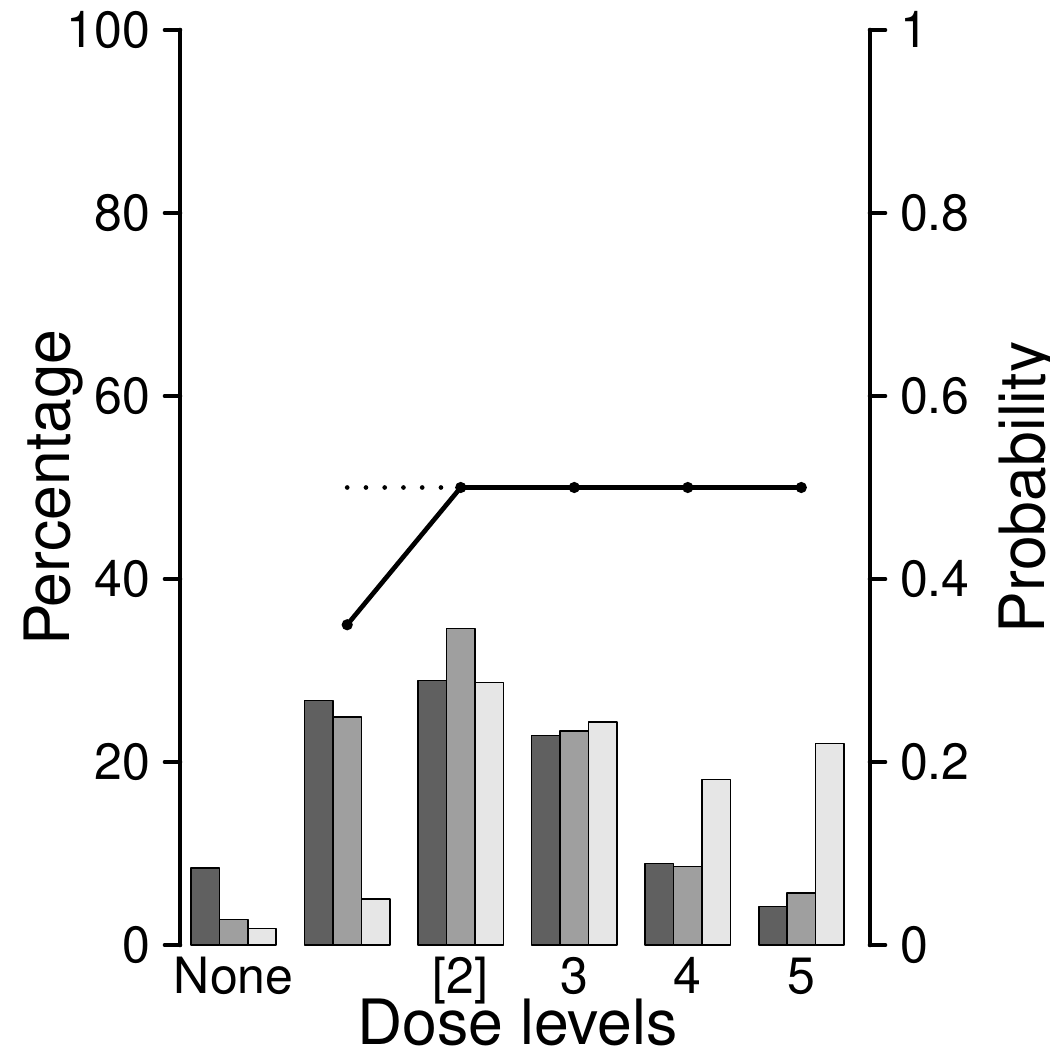}} \\ 
\subfigure[Scenario 7]{\includegraphics[scale=0.29]{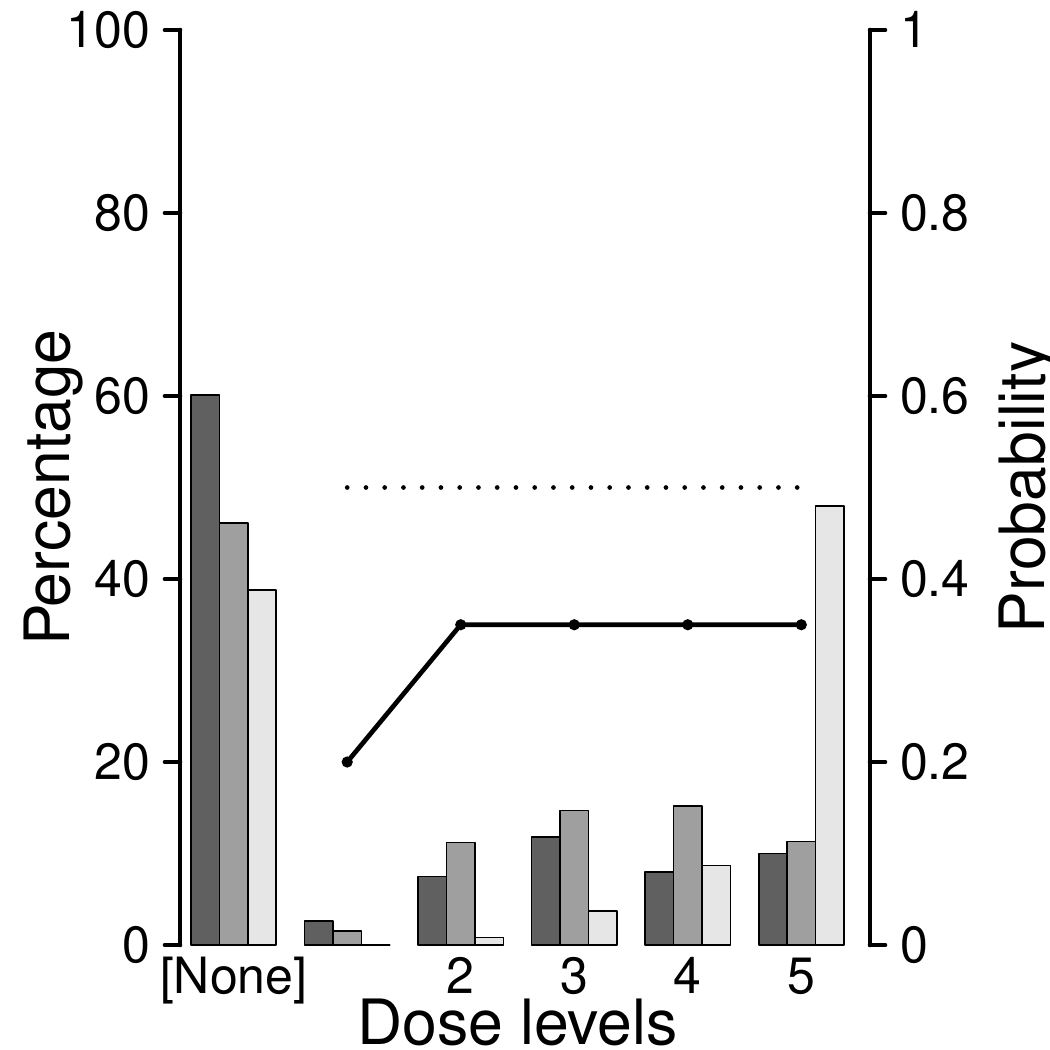}} 
\subfigure[Scenario 8]{\includegraphics[scale=0.29]{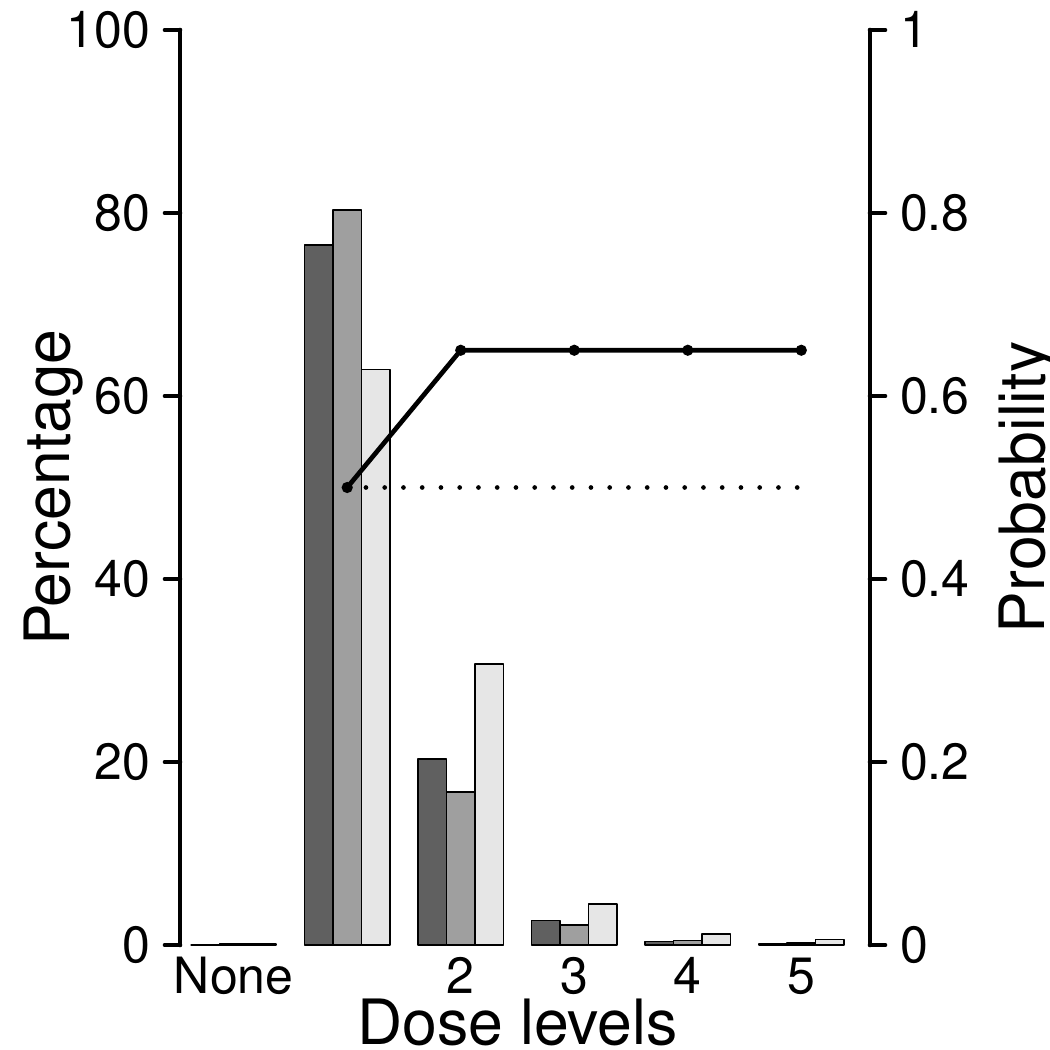}} 
\subfigure{\includegraphics[scale=0.29]{legend_barplot_OBD_hat}} 
\end{center}
\caption{Comparison of dose selection for Selection method, BMA method  and BLRM for a maximum number of volunteers of $n = 40$, when $L = 5$ dose levels are used. The solid line with circles represents the true dose-activity relationship. The horizontal dashed line is the activity probability target. \label{fig:barplot_MAD_hat_n40_L5}}
\end{figure}

\end{document}

\maketitle

\section{Simulation Results}\label{sec:supp_simu_res}

\subsection{Simulation Results Using the Selection Method}\label{subsec:supp_Selection_res}

Tables \ref{tab:Selection_L3_res}, \ref{tab:Selection_L4_res} and \ref{tab:Selection_L5_res} give the results using the selection method.

\begin{table}[h!]
\footnotesize
\caption{Selection percentage and mean number of volunteers (shown in parentheses) allocated to each dose for each scenario and for several values of maximum number of volunteers, for the selection method, when $L = 3$ dose levels are used. $\boldsymbol{\phi}$: activity probabilities; $n$: maximal number of healthy volunteers.\label{tab:Selection_L3_res}}
\begin{center}
\begin{tabular}{l|c|c|c|c|c}
\hline
& \multicolumn{3}{c}{\textbf{Dose levels}} & \textbf{Early} & \textbf{Total number}\\   
& 1 & 2 & 3  & \textbf{Termination} & \textbf{of participants} \\   \hline \hline
&\multicolumn{4}{c}{\textbf{Scenario 1}}  \\
$\boldsymbol{\phi}$ & \textbf{0.5} & 0.65 & 0.8 & \\ \hline
$n = 18$ & 81.2 (8.9) & 16.6 (4.9) & 1.4 (4.2) & 0.8 & 18.0 (0.7) \\ 
$n = 24$ & 85.2 (11.2) & 13.8 (6.8) & 0.4 (6.0) & 0.6 & 23.9 (1.1)\\ 
$n = 30$ & 86.2 (13.1) & 12.7 (8.8) & 0.3 (7.9) & 0.8 & 29.9 (1.4)\\  
$n = 40$ & 88.5 (18.9) & 10.3 (11.0) & 0.2 (9.9) & 1.0 & 39.8 (2.4) \\  \hline
&\multicolumn{4}{c}{\textbf{Scenario 2}}  \\
$\boldsymbol{\phi}$ & 0.2 & 0.35 & \textbf{0.5} & \\ \hline
$n = 18$ & 6.8 (4.8) & 31.2 (5.8) & 35.4 (6.2) & 26.6 & 16.9 (2.2)\\ 
$n = 24$ & 5.2 (6.5) & 27.6 (7.7) & 41.4 (8.5) & 25.8 & 22.8 (2.5)\\ 
$n = 30$ & 3.5 (8.4) & 23.6 (9.5) & 45.6 (10.7) & 27.3 & 28.6 (2.8)\\  
$n = 40$ & 2.3 (10.4) & 20.0 (12.3) & 46.7 (14.6) & 31.0 & 37.4 (4.7) \\  \hline
&\multicolumn{4}{c}{\textbf{Scenario 3}}  \\
$\boldsymbol{\phi}$ & 0.05 & 0.2 & 0.35 & \\ \hline
$n = 18$ & 0 (4.1) & 4.8 (4.5) & 27.9 (6.4) & 67.3 & 15.0 (2.7)\\ 
$n = 24$ & 0 (6.0) & 2.8 (6.2) & 23.3 (7.8) & 73.9 & 20.1 (2.8)\\ 
$n = 30$ & 0 (8) & 1.3 (8.1) & 20.5 (9.6) & 78.2 & 25.7 (2.9)\\  
$n = 40$ & 0 (10.0) & 0.3 (10.0) & 13.4 (11.9) & 86.3 &31.9 (4.2) \\  \hline
&\multicolumn{4}{c}{\textbf{Scenario 4}}  \\
$\boldsymbol{\phi}$ & 0.35 & \textbf{0.5} & 0.65 & \\ \hline
$n = 18$ & 38.9 (6.9) & 43.3 (6.0) & 13.7 (5.0) & 4.1 & 17.8 (1.0)\\ 
$n = 24$ & 36.2 (8.6) & 44.6 (8.3) & 15.0 (6.9) & 4.2 & 29.8 (1.6)\\ 
$n = 30$ & 35.1 (10.3) & 47.5 (10.7) & 14.9 (8.8) & 2.5 & 29.8 (1.6)\\  
$n = 40$ & 29.1 (13.5) & 55.5 (15.0) & 11.7 (11.1) & 3.7 &39.6 (2.9)\\  \hline
&\multicolumn{4}{c}{\textbf{Scenario 5}} \\
$\boldsymbol{\phi}$ & 0.35 & \textbf{0.5} & 0.5 & \\ \hline
$n = 18$ & 33.0 (6.5) & 42.0 (6.0) & 12.8 (4.9) & 12.2 & 17.5 (1.6)\\ 
$n = 24$ & 31.2 (8.4) & 40.8 (8.1) & 15.3 (7.0) & 12.7 & 23.4 (1.9)\\ 
$n = 30$ & 32.1 (10.2) & 42.1 (10.4) & 15.1 (8.9) & 10.7 & 29.5 (2.1)\\  
$n = 40$ & 24.9 (13.1) & 48.0 (14.5) & 13.0 (11.3) & 14.1 &38.8 (6.3)\\  \hline
&\multicolumn{4}{c}{\textbf{Scenario 6}} \\
$\boldsymbol{\phi}$ & \textbf{0.5} & 0.5 & 0.5 & \\ \hline
$n = 18$ & 62.8 (8.0) & 22.2 (5.1) & 7.3 (4.5) & 7.7 & 17.6 (1.4)\\ 
$n = 24$ & 63.0 (9.9) & 22.0 (7.3) & 7.9 (6.5) & 7.1 & 23.6 (1.7)\\ 
$n = 30$ & 63.9 (11.9) & 22.9 (9.4) & 6.3 (8.3) & 6.9 & 29.6 (1.9)\\  
$n = 40$ & 64.5 (16.7) & 23.9 (12.2) & 3.9 (10.3) & 7.7 &39.2 (3.3) \\  \hline
&\multicolumn{4}{c}{\textbf{Scenario 7}}  \\
$\boldsymbol{\phi}$ & 0.35 & 0.35 & 0.35 & \\ \hline
$n = 18$ & 28.7 (6.1) & 21.7 (5.2) & 12.5 (4.9) & 37.1 & 16.3 (2.6) \\ 
$n = 24$ & 26.4 (8.0) & 16.4 (7.1) & 14.5 (6.9) & 42.7 & 22.0 (2.8) \\ 
$n = 30$ & 24.2 (9.8) & 16.2 (9.1) & 13.1 (8.9) & 46.5 & 27.8 (3.0)\\  
$n = 40$ & 16.0 (12.2) & 14.0 (11.6) & 9.6 (11.0) & 60.4 &34.8 (5.1)\\  \hline
&\multicolumn{4}{c}{\textbf{Scenario 8}}  \\
$\boldsymbol{\phi}$ & \textbf{0.65} & 0.65 & 0.65 & \\ \hline
$n = 18$ & 92.4 (9.5) & 5.9 (4.4) & 1.1 (4.1) & 0.6 & 18.0 (0.6)\\ 
$n = 24$ & 94.7 (11.6) & 4.4 (6.3) & 0.2 (6.0) & 0.7 & 23.9 (1.2)\\ 
$n = 30$ & 95.6 (13.7) & 3.5 (8.3) & 0.3 (7.9) & 0.6 & 29.9 (1.4) \\  
$n = 40$ & 95.9 (19.5) & 3.1 (10.4) & 0.1 (9.9) & 0.9 &39.8 (2.3) \\  \hline
\end{tabular}
\end{center}
\end{table}

\begin{table}[h!]
\footnotesize
\caption{Selection percentage and mean number of volunteers (shown in parentheses) allocated to each dose for each scenario and for several values of maximum number of volunteers, for the selection method, when $L = 4$ dose levels are used. $\boldsymbol{\phi}$: activity probabilities; $n$: maximal number of healthy volunteers. \label{tab:Selection_L4_res}}
\begin{center}
\begin{tabular}{l|c|c|c|c|c|c}
\hline
& \multicolumn{4}{c}{\textbf{Dose levels}} & \textbf{Early} & \textbf{Total number}\\   
& 1 & 2 & 3 & 4 & \textbf{Termination} & \textbf{of participants} \\ \hline \hline
&\multicolumn{5}{c}{\textbf{Scenario 1}}  \\
$\boldsymbol{\phi}$ & 0.35 & \textbf{0.5} & 0.65 & 0.8 &\\ \hline
$n = 24$ & 34.3 (7.4) & 48.3 (7.3) & 15.5 (5.0) & 1.6 (4.3) & 0.3 & 24.0 (0.3) \\ 
$n = 30$ & 31.4 (9.0) & 52.5 (9.4) & 14.2 (6.3) & 1.1 (5.2) & 0.8 &30.0 (0.7) \\  
$n = 40$ & 30.7 (11.0) & 58.8 (12.2) & 9.1 (8.8) & 0.4 (8.0) & 1.0 & 39.9 (1.6) \\  \hline
&\multicolumn{5}{c}{\textbf{Scenario 2}}  \\
$\boldsymbol{\phi}$ & 0.05 & 0.2 & 0.35 & \textbf{0.5} & \\ \hline
$n = 24$ & 0.2 (4.0) & 2.1 (4.4) & 19.4 (5.5) & 39.9 (7.5) & 38.4 & 21.4 (3.6) \\ 
$n = 30$ & 0 (5.0) & 1.5 (5.3) & 15.1 (6.8) & 44.8 (9.9) & 38.6 & 26.9 (4.4) \\  
$n = 40$ & 0 (8.0) & 0.3 (8.1) & 15.2 (9.1) & 46.8 (12.2) & 37.7  & 37.4 (4.0) \\  \hline
&\multicolumn{5}{c}{\textbf{Scenario 3}}  \\
$\boldsymbol{\phi}$ & 0.01 & 0.05 & 0.2 & 0.35 & \\ \hline
$n = 24$ & 0 (4) & 0 (4.0) & 1.4 (4.2) & 15.1 (5.7) & 83.5 & 18.0 (3.2)\\ 
$n = 30$ & 0 (5.0) & 0 (5.0) & 0.9 (5.1) & 12.7 (7.0) & 86.4 & 22.1 (3.8)\\  
$n = 40$ & 0 (8) & 0 (8) & 0.2 (8.0) & 6.5 (8.8) & 93.3  & 32.8 (2.8) \\  \hline
&\multicolumn{5}{c}{\textbf{Scenario 4}}  \\
$\boldsymbol{\phi}$ & 0.2 & 0.35 & \textbf{0.5} & 0.65 & \\ \hline
$n = 24$ & 4.2 (4.7) & 26.6 (6.5) & 42.3 (6.3) & 19.7 (6.1) & 7.2 & 23.6 (1.7)\\ 
$n = 30$ & 3.1 (5.7) & 25.0 (7.8) & 45.4 (8.8) & 19.7 (7.3) & 6.8 & 29.5 (2.3) \\  
$n = 40$ & 1.6 (8.3) & 23.8 (10.0) & 52.0 (11.5) & 18.8 (9.8) & 3.8  & 39.7 (2.2) \\  \hline
&\multicolumn{5}{c}{\textbf{Scenario 5}}  \\
$\boldsymbol{\phi}$ & 0.2 & 0.35 & \textbf{0.5} & 0.5 & \\ \hline
$n = 24$ & 4.0 (4.8) & 22.6 (6.0) & 35.1 (6.4) & 19.1 (5.8) & 19.2 & 22.9 (2.6) \\ 
$n = 30$ & 4.3 (5.7) & 19.4 (7.3) & 36.5 (8.3) & 21.0 (7.2) & 18.8 & 28.6 (3.3) \\  
$n = 40$ & 2.0 (8.3) & 20.1 (9.6) & 42.3 (11.2) & 19.1 (9.7) & 16.5  & 38.8 (3.1) \\  \hline
&\multicolumn{5}{c}{\textbf{Scenario 6}}  \\
$\boldsymbol{\phi}$ & 0.35 & \textbf{0.5} & 0.5 & 0.5 & \\ \hline
$n = 24$ & 29.2 (6.9) & 32.2 (6.5) & 18.5 (5.3) & 9.3 (4.8) & 10.8 & 23.4 (1.8) \\ 
$n = 30$ & 27.5 (8.5) & 37.5 (8.3) & 15.9 (6.7) & 8.7 (5.8) & 10.4 & 29.3 (2.3) \\  
$n = 40$ & 28.1 (10.7) & 35.6 (10.7) & 19.4 (9.4) & 7.0 (8.5) & 9.9  & 39.3 (2.4) \\  \hline
&\multicolumn{5}{c}{\textbf{Scenario 7}}  \\
$\boldsymbol{\phi}$ & 0.2 & 0.35 & 0.35 & 0.35 & \\ \hline
$n = 24$ & 3.4 (4.8) & 16.2 (5.4) & 15.8 (5.3) & 13.9 (5.5) & 50.7 & 21.0 (3.6) \\ 
$n = 30$ & 3.3 (5.8) & 13.9 (6.7) & 15.4 (6.7) & 15.2 (6.9) & 52.2 & 26.1 (4.5) \\  
$n = 40$ & 2.2 (8.4) & 14.0 (9.2) & 15.0 (9.2) & 12.7 (9.3) & 56.1  & 36.2 (4.1) \\  \hline
&\multicolumn{5}{c}{\textbf{Scenario 8}}  \\
$\boldsymbol{\phi}$ & \textbf{0.5} & 0.65 & 0.65 & 0.65 & \\ \hline
$n = 24$ & 76.5 (9.8) & 18.0 (5.7) & 4.3 (4.3) & 0.6 (4.1) & 0.6 & 24.0 (0.6) \\ 
$n = 30$ & 78.4 (12.6) & 17.9 (6.9) & 3.1 (5.4) & 0.3 (5.1) & 0.3 & 30.0 (0.4) \\  
$n = 40$ & 80.9 (14.4) & 18.0 (9.7) & 0.9 (8.0) & 0 (7.9) & 0.2  & 40.0 (0.8) \\  \hline
\end{tabular}
\end{center}
\end{table}

\begin{table}[h!]
\footnotesize
\caption{Selection percentage and mean number of volunteers (shown in parentheses) allocated to each dose for each scenario and for several values of maximum number of volunteers, for the selection method, when $L = 5$ dose levels are used. $\boldsymbol{\phi}$: activity probabilities; $n$: maximal number of healthy volunteers. \label{tab:Selection_L5_res}}
\begin{center}
\begin{tabular}{l|c|c|c|c|c|c|c}
\hline
& \multicolumn{5}{c}{\textbf{Dose levels}} & \textbf{Early} & \textbf{Total number}\\   
& 1 & 2 & 3 & 4 & 5 & \textbf{Termination} & \textbf{of participants} \\  \hline \hline
&\multicolumn{6}{c}{\textbf{Scenario 1}}  \\
$\boldsymbol{\phi}$ & 0.35 & \textbf{0.5} & 0.65 & 0.8 & 0.95 & \\ \hline 
$n = 30$ & 38.6 (8.6) & 53.1 (8.5) & 7.8 (5.0) & 0.4 (4.0) & 0 (4.0) & 0.1 & 30.0 (0.1) \\  
$n = 40$ & 38.2 (10.6) & 57.1 (10.9) & 3.9 (6.5) & 0.1 (6.0) & 0 (5.9) & 0.7 & 39.9 (1.7)\\  \hline
&\multicolumn{6}{c}{\textbf{Scenario 2}}  \\
$\boldsymbol{\phi}$ & 0.05 & 0.2 & 0.35 & \textbf{0.5} & 0.65 & \\ \hline
$n = 30$ & 0 (4.0) & 2.0 (4.3) & 20.8 (6.1) & 40.0 (7.3) & 25.3 (7.3) & 11.9 & 28.9 (3.2) \\  
$n = 40$ & 0 (6.0) & 0.5 (6.1) & 14.0 (7.6) & 45.4 (9.7) & 30.0 (9.5) & 10.1 & 39.0 (3.5) \\  \hline
&\multicolumn{6}{c}{\textbf{Scenario 3}}  \\
$\boldsymbol{\phi}$ & 0.01 & 0.05 & 0.15 & 0.25 & 0.35 & \\ \hline
$n = 30$ & 0 (4) & 0 (4) & 0.1 (4.0) & 1.5 (4.3) & 11.6 (5.7) & 86.8 & 22.0 (3.9)\\  
$n = 40$ & 0 (6) & 0 (6) & 0 (6.0) & 0.4 (6.1) & 8.7 (7.3) & 90.9 & 31.4 (3.7) \\  \hline
&\multicolumn{6}{c}{\textbf{Scenario 4}}  \\
$\boldsymbol{\phi}$ & 0.2 & 0.35 &  \textbf{0.5} & 0.65 & 0.8 & \\ \hline
$n = 30$ & 1.8 (4.7) & 34.4 (7.3) & 42.9 (8.1) & 18.1 (5.3) & 1.9 (4.5) & 0.9 & 29.9 (1.2)\\  
$n = 40$ & 2.0 (6.3) & 35.2 (9.7) & 48.4 (10.4) & 12.1 (7.2) & 1.1 (6.2) & 1.2 & 39.8 (2.1) \\  \hline
&\multicolumn{6}{c}{\textbf{Scenario 5}}  \\
$\boldsymbol{\phi}$ & 0.05 & 0.2 & 0.35 &  \textbf{0.5} & 0.5 & \\ \hline
$n = 30$ & 0.1 (4.0) & 0.7 (4.2) & 16.5 (5.8) & 27.5 (6.8) & 26.0 (6.9) & 29.2 & 27.7 (4.2) \\  
$n = 40$ & 0 (6.0) & 0.9 (6.1) & 12.3 (7.4) & 32.5 (9.2) & 28.0 (9.1) & 26.3 & 37.8 (4.4)\\  \hline
&\multicolumn{6}{c}{\textbf{Scenario 6}}  \\
$\boldsymbol{\phi}$ & 0.35 &  \textbf{0.5} & 0.5 & 0.5 & 0.5 & \\ \hline
$n = 30$ & 26.8 (7.4) & 29.4 (6.4) & 20.3 (6.2) & 8.3 (4.8) & 6.3 (4.5) & 8.9 & 29.4 (2.3) \\  
$n = 40$ & 26.7 (9.4) & 28.9 (8.5) & 22.9 (8.3) & 8.9 (6.8) & 4.2 (6.3) & 8.4 & 39.4 (2.6) \\  \hline
&\multicolumn{6}{c}{\textbf{Scenario 7}}  \\
$\boldsymbol{\phi}$ &0.2 & 0.35 & 0.35 & 0.35 & 0.35 & \\ \hline
$n = 30$ & 3.6 (4.8) & 9.8 (4.9) & 13.5 (5.7) & 9.5 (5.1) & 10.5 (5.2) & 53.1 & 25.9 (4.6) \\  
$n = 40$ & 2.6 (6.7) & 7.5 (6.8) & 11.8 (7.5) & 8.0 (6.8) & 10.0 (7.1) & 60.1 & 35.0 (4.9) \\  \hline
&\multicolumn{6}{c}{\textbf{Scenario 8}}  \\
$\boldsymbol{\phi}$ &  \textbf{0.5} & 0.65 & 0.65 & 0.65 & 0.65 & \\ \hline
$n = 30$ & 75.4 (11.1) & 20.3 (6.2) & 3.4 (4.6) & 0.5 (4.1) & 0.2 (4.0) & 0.2 & 30.0 (0.6) \\  
$n = 40$ & 76.5 (13.5) & 20.3 (8.2) & 2.7 (6.3) & 0.4 (6.0) & 0.1 (5.9) & 0 & 40.0 (0.0)\\  \hline
\end{tabular}
\end{center}
\end{table}

\clearpage

\subsection{Simulation Results Using BMA}\label{subsec:supp_BMA_res}

Tables \ref{tab:BMA_L3_res}, \ref{tab:BMA_L4_res} and \ref{tab:BMA_L5_res} give the results using BMA.

\begin{table}[h!]
\footnotesize
\caption{Selection percentage and mean number of volunteers (shown in parentheses) allocated to each dose for each scenario and for several values of maximum number of volunteers, for BMA, when $L = 3$ dose levels are used. $\boldsymbol{\phi}$: activity probabilities; $n$: maximal number of healthy volunteers. \label{tab:BMA_L3_res}}
\begin{center}
\begin{tabular}{l|c|c|c|c|c}
\hline
& \multicolumn{3}{c}{\textbf{Dose levels}} & \textbf{Early} & \textbf{Total number}\\   
& 1 & 2 & 3  & \textbf{Termination} & \textbf{of participants} \\   \hline \hline
&\multicolumn{4}{c}{\textbf{Scenario 1}}  \\
$\boldsymbol{\phi}$ & \textbf{0.5} & 0.65 & 0.8 & \\ \hline
$n = 18$ & 84.4 (8.9) & 12.4 (4.8) & 2.6 (4.2) & 0.6 & 18.0 (0.6)\\ 
$n = 24$ & 87.6 (11.3) & 11.2 (6.6) & 0.6 (6.0) & 0.6 & 23.9 (1.1) \\ 
$n = 30$ & 88.2 (13.4) & 10.7 (8.6) & 0.3 (8.0) & 0.8 & 29.9 (1.4) \\  
$n = 40$ & 90.5 (19.2) & 8.2 (10.7) & 0.3 (10.0) & 1.0 & 39.8 (2.4)\\  \hline
&\multicolumn{4}{c}{\textbf{Scenario 2}}  \\
$\boldsymbol{\phi}$ & 0.2 & 0.35 & \textbf{0.5} & \\ \hline
$n = 18$ & 6.2 (4.8) & 31.3 (5.9) & 40.8 (6.2) & 21.7  & 16.9 (2.3) \\ 
$n = 24$ & 4.0 (6.4) & 28.2 (7.8) & 43.7 (8.6) & 24.1  & 22.8 (2.6) \\ 
$n = 30$ & 2.5 (8.4) & 25.8 (9.4) & 44.5 (10.7) & 27.2  & 28.5 (2.9) \\  
$n = 40$ & 1.6 (10.4) & 20.9 (12.2) & 48.9 (14.9) & 28.6 & 37.5 (4.7)\\  \hline
&\multicolumn{4}{c}{\textbf{Scenario 3}}  \\
$\boldsymbol{\phi}$ & 0.05 & 0.2 & 0.35 & \\ \hline
$n = 18$ & 0 (4.0) & 3.4 (4.6) & 27.4 (5.7) & 69.2  & 14.3 (2.8)\\ 
$n = 24$ & 0 (6.0) & 2.1 (6.3) & 20.8 (7.6) & 77.1  & 19.8 (2.8) \\ 
$n = 30$ & 0 (8) & 1.1 (8.1) & 16.8 (9.2) & 82.1  & 25.3 (2.7) \\  
$n = 40$ & 0 (10) & 0.4 (10) & 10.1 (11.5) & 89.5 & 31.5 (3.9)\\  \hline
&\multicolumn{4}{c}{\textbf{Scenario 4}}  \\
$\boldsymbol{\phi}$ & 0.35 & \textbf{0.5} & 0.65 & \\ \hline
$n = 18$ & 44.7 (6.9) & 34.4 (5.8) & 18.3 (5.2) & 2.6  & 17.9 (1.0) \\ 
$n = 24$ & 37.6 (8.7) & 41.9 (8.1) & 17.1 (7.0) & 3.4  & 23.8 (1.5) \\ 
$n = 30$ & 36.4 (10.7) & 45.8 (10.2) & 15.9 (8.9) & 1.9  & 29.8 (1.6) \\  
$n = 40$ & 31.0 (13.7) & 51.9 (14.5) & 15.1 (11.5) & 2.0 & 39.7 (2.7) \\  \hline
&\multicolumn{4}{c}{\textbf{Scenario 5}} \\
$\boldsymbol{\phi}$ & 0.35 & \textbf{0.5} & 0.5 & \\ \hline
$n = 18$ & 33.5 (6.4) & 42.7 (6.2) & 16.3 (4.9) & 7.5  & 17.6 (1.5) \\ 
$n = 24$ & 31.3 (8.2) & 42.3 (8.2) & 18.8 (7.1) & 7.6  & 23.6 (1.8) \\ 
$n = 30$ & 29.5 (10.2) & 45.6 (10.3) & 16.7 (9.0) & 8.2  & 29.5 (2.0) \\  
$n = 40$ & 24.2 (13.1) & 49.6 (14.4) & 17.6 (11.7) & 8.6 & 39.2 (3.2) \\  \hline
&\multicolumn{4}{c}{\textbf{Scenario 6}} \\
$\boldsymbol{\phi}$ & \textbf{0.5} & 0.5 & 0.5 & \\ \hline
$n = 18$ & 62.0 (7.9) & 23.4 (5.3) & 10.7 (4.6) & 3.9  & 17.8 (1.1) \\ 
$n = 24$ & 62.0 (9.8) & 24.1 (7.4) & 9.6 (6.6) & 4.3  & 23.7 (1.5) \\ 
$n = 30$ & 63.2 (12.0) & 25.8 (9.4) & 7.7 (8.4) & 3.3  & 29.8 (1.7) \\  
$n = 40$ & 64.3 (16.7) & 25.1 (12.3) & 6.1 (10.5) & 4.5 & 39.5 (2.9) \\  \hline
&\multicolumn{4}{c}{\textbf{Scenario 7}}  \\
$\boldsymbol{\phi}$ & 0.35 & 0.35 & 0.35 & \\ \hline
$n = 18$ & 24.9 (5.9) & 27.6 (5.9) & 20.7 (5.0) & 26.8  & 16.8 (2.3) \\ 
$n = 24$ & 22.9 (7.6) & 25.5 (7.7) & 19.5 (7.1) & 32.1  & 22.4 (2.7) \\ 
$n = 30$ & 18.9 (9.5) & 26.4 (9.4) & 16.1 (9.1) & 38.6  & 28.1 (3.0) \\  
$n = 40$ & 12.9 (11.8) & 22.5 (12.6) & 17.7 (11.7) & 46.9 & 36.2 (5.0) \\  \hline
&\multicolumn{4}{c}{\textbf{Scenario 8}}  \\
$\boldsymbol{\phi}$ & \textbf{0.65} & 0.65 & 0.65 & \\ \hline
$n = 18$ & 92.7 (9.5) & 5.3 (4.4) & 1.4 (4.1) & 0.6  & 18.0 (0.6) \\ 
$n = 24$ & 94.9 (11.6) & 3.9 (6.3) & 0.5 (6.0) & 0.7  & 23.9 (1.2) \\ 
$n = 30$ & 96.0 (13.8) & 3.0 (8.2) & 0.4 (7.9) & 0.6 & 29.9 (1.4)  \\  
$n = 40$ & 96.1 (19.6) & 2.8 (10.3) & 0.2 (9.9) & 0.9 & 39.8 (2.3) \\  \hline
\end{tabular}
\end{center}
\end{table}

\begin{table}[h!]
\footnotesize
\caption{Selection percentage and mean number of volunteers (shown in parentheses) allocated to each dose for each scenario and for several values of maximum number of volunteers, for BMA, when $L = 4$ dose levels are used. $\boldsymbol{\phi}$: activity probabilities; $n$: maximal number of healthy volunteers.\label{tab:BMA_L4_res}}
\begin{center}
\begin{tabular}{l|c|c|c|c|c|c}
\hline
& \multicolumn{4}{c}{\textbf{Dose levels}} & \textbf{Early} & \textbf{Total number}\\   
& 1 & 2 & 3 & 4 & \textbf{Termination} & \textbf{of participants} \\  \hline \hline
&\multicolumn{5}{c}{\textbf{Scenario 1}}  \\
$\boldsymbol{\phi}$ & 0.35 & \textbf{0.5} & 0.65 & 0.8 &\\ \hline
$n = 24$ & 39.6 (8.0) & 45.5 (6.8) & 12.6 (4.9) & 2.1 (4.4) & 0.2  & 24.0 (0.3)\\ 
$n = 30$ & 37.7 (9.7) & 48.2 (8.8) & 12.1 (6.1) & 1.5 (5.3) & 0.5  & 30.0 (0.6)\\  
$n = 40$ & 40.1 (11.7) & 50.2 (11.5) & 8.2 (8.6) & 0.7 (8.0) & 0.8  & 39.9 (1.7) \\  \hline
&\multicolumn{5}{c}{\textbf{Scenario 2}}  \\
$\boldsymbol{\phi}$ & 0.05 & 0.2 & 0.35 & \textbf{0.5} & \\ \hline
$n = 24$ & 0 (4.0) & 3.6 (4.4) & 18.9 (5.5) & 37.5 (7.3) & 40.0  & 21.2 (3.7)\\ 
$n = 30$ & 0 (5.0) & 1.4 (5.2) & 15.3 (6.6) & 40.2 (9.5) & 43.1  & 26.3 (4.7)\\  
$n = 40$ & 0 (8.0) & 0.2 (8.1) & 14.1 (9.0) & 42.5 (11.8) & 43.2 & 36.8 (4.2) \\  \hline
&\multicolumn{5}{c}{\textbf{Scenario 3}}  \\
$\boldsymbol{\phi}$ & 0.01 & 0.05 & 0.2 & 0.35 & \\ \hline
$n = 24$ & 0 (4) & 0 (4.0) & 1.6 (4.2) & 10.0 (5.3) & 88.4  & 17.5 (3.0) \\ 
$n = 30$ & 0 (5) & 0 (5) & 0.6 (5.1) & 8.2 (6.3) & 91.2  & 21.4 (3.3) \\  
$n = 40$ & 0 (8) & 0 (8) & 0.1 (8.0) & 5.1 (8.6) & 94.8 & 32.5 (2.5) \\  \hline
&\multicolumn{5}{c}{\textbf{Scenario 4}}  \\
$\boldsymbol{\phi}$ & 0.2 & 0.35 & \textbf{0.5} & 0.65 & \\ \hline
$n = 24$ & 4.3 (4.8) & 30.7 (6.4) & 38.0 (6.2) & 21.1 (6.1) & 5.9  & 23.6 (1.7) \\ 
$n = 30$ & 3.2 (5.7) & 28.7 (7.8) & 40.9 (8.2) & 22.1 (7.8) & 5.1  & 29.5 (2.2) \\  
$n = 40$ & 2.2 (8.4) & 24.9 (10.4) & 46.9 (10.8) & 22.9 (10.2) & 3.1 & 39.7 (2.2) \\  \hline
&\multicolumn{5}{c}{\textbf{Scenario 5}}  \\
$\boldsymbol{\phi}$ & 0.2 & 0.35 & \textbf{0.5} & 0.5 & \\ \hline
$n = 24$ & 3.5 (4.7) & 27.3 (6.1) & 34.1 (6.6) & 22.6 (5.8) & 12.5  & 23.2 (2.4) \\ 
$n = 30$ & 3.0 (5.7) & 23.4 (7.5) & 34.1 (8.0) & 24.8 (7.6) & 14.7  & 28.8 (3.2) \\  
$n = 40$ & 1.9 (8.3) & 21.8 (9.9) & 40.8 (10.9) & 23.2 (9.9) & 12.3 & 39.1 (3.0) \\  \hline
&\multicolumn{5}{c}{\textbf{Scenario 6}}  \\
$\boldsymbol{\phi}$ & 0.35 & \textbf{0.5} & 0.5 & 0.5 & \\ \hline
$n = 24$ & 29.5 (7.0) & 37.1 (6.5) & 17.9 (5.4) & 10.8 (4.8) & 4.7  & 23.7 (1.3) \\ 
$n = 30$ & 25.5 (8.4) & 43.3 (8.6) & 14.6 (6.6) & 10.2 (6.0) & 6.4  & 29.6 (1.9) \\  
$n = 40$ & 25.6 (10.5) & 42.7 (11.3) & 18.9 (9.2) & 8.5 (8.6) & 4.3 & 39.6 (2.1) \\  \hline
&\multicolumn{5}{c}{\textbf{Scenario 7}}  \\
$\boldsymbol{\phi}$ & 0.2 & 0.35 & 0.35 & 0.35 & \\ \hline
$n = 24$ & 3.2 (4.6) & 21.2 (5.7) & 20.7 (5.8) & 15.9 (5.6) & 39.0  & 21.6 (3.4) \\ 
$n = 30$ & 2.2 (5.6) & 17.9 (7.1) & 17.9 (6.9) & 19.0 (7.0) & 43.0  & 26.6 (4.4) \\  
$n = 40$ & 1.4 (8.3) & 13.9 (9.3) & 20.8 (9.6) & 15.9 (9.5) & 48.0 & 36.7 (4.1)  \\  \hline
&\multicolumn{5}{c}{\textbf{Scenario 8}}  \\
$\boldsymbol{\phi}$ & \textbf{0.5} & 0.65 & 0.65 & 0.65 & \\ \hline
$n = 24$ & 77.8 (10.2) & 17.3 (5.4) & 3.6 (4.3) & 0.9 (4.1) & 0.4  & 24.0 (0.7) \\ 
$n = 30$ & 80.1 (12.8) & 16.7 (6.7) & 2.7 (5.4) & 0.4 (5.1) & 0.1  & 30.0 (0.1) \\  
$n = 40$ & 83.1 (14.6) & 16.0 (9.4) & 0.5 (8.0) & 0.1 (7.9) & 0.3 & 39.9 (1.1) \\  \hline
\end{tabular}
\end{center}
\end{table}

\begin{table}[h!]
\footnotesize
\caption{Selection percentage and mean number of volunteers (shown in parentheses) allocated to each dose for each scenario and for several values of maximum number of volunteers, for BMA, when $L = 5$ dose levels are used. $\boldsymbol{\phi}$: activity probabilities; $n$: maximal number of healthy volunteers.\label{tab:BMA_L5_res}}
\begin{center}
\begin{tabular}{l|c|c|c|c|c|c|c}
\hline
& \multicolumn{5}{c}{\textbf{Dose levels}} & \textbf{Early} & \textbf{Total number}\\   
& 1 & 2 & 3 & 4 & 5 & \textbf{Termination} & \textbf{of participants} \\  \hline \hline
&\multicolumn{6}{c}{\textbf{Scenario 1}}  \\
$\boldsymbol{\phi}$ & 0.35 & \textbf{0.5} & 0.65 & 0.8 & 0.95 & \\ \hline 
$n = 30$ & 43.7 (9.5) & 49.8 (7.9) & 6.2 (4.6) & 0.2 (4.0) & 0.1 (4.0) & 0  & 30.0 (0.0)\\  
$n = 40$ & 45.4 (11.4) & 51.0 (10.3) & 2.8 (6.0) & 0.1 (6.0) & 0 (5.9) & 0.7 & 39.9 (1.7) \\  \hline
&\multicolumn{6}{c}{\textbf{Scenario 2}}  \\
$\boldsymbol{\phi}$ & 0.05 & 0.2 & 0.35 & \textbf{0.5} & 0.65 & \\ \hline
$n = 30$ & 0.1 (4.0) & 2.3 (4.5) & 23.4 (6.0) & 33.0 (6.8) & 29.6 (7.5) & 11.6  & 28.9 (3.3) \\  
$n = 40$ & 0 (6.0) & 1.0 (6.2) & 15.4 (7.7) & 41.8 (9.2) & 31.6 (9.9) & 10.2 & 38.9 (3.6) \\  \hline
&\multicolumn{6}{c}{\textbf{Scenario 3}}  \\
$\boldsymbol{\phi}$ & 0.01 & 0.05 & 0.15 & 0.25 & 0.35 & \\ \hline
$n = 30$ & 0 (4) & 0 (4) & 0.1 (4.0) & 1.6 (4.3) & 8.4 (5.4) & 89.9  & 21.6 (3.6) \\  
$n = 40$ & 0 (6) & 0 (6) & 0 (6.0) & 0.7 (6.1) & 5.5 (6.8) & 93.8 & 30.9 (3.3) \\  \hline
&\multicolumn{6}{c}{\textbf{Scenario 4}}  \\
$\boldsymbol{\phi}$ & 0.2 & 0.35 &  \textbf{0.5} & 0.65 & 0.8 & \\ \hline
$n = 30$ & 3.5 (4.9) & 34.6 (8.0) & 42.1 (7.2) & 15.6 (5.1) & 3.6 (4.7) & 0.6  & 30.0 (0.8) \\  
$n = 40$ & 2.6 (6.5) & 38.7 (10.2) & 44.8 (9.7) & 11.4 (7.1) & 1.6 (6.3) & 0.9 & 39.8 (2.2) \\  \hline
&\multicolumn{6}{c}{\textbf{Scenario 5}}  \\
$\boldsymbol{\phi}$ & 0.05 & 0.2 & 0.35 &  \textbf{0.5} & 0.5 & \\ \hline
$n = 30$ & 0 (4.0) & 1.5 (4.3) & 17.7 (5.8) & 27.5 (6.7) & 26.6 (7.0) & 26.7  & 27.8 (4.1)\\  
$n = 40$ & 0 (6.0) & 0.4 (6.1) & 12.8 (7.4) & 32.4 (9.0) & 30.6 (9.3) & 23.8 & 37.9 (4.4) \\  \hline
&\multicolumn{6}{c}{\textbf{Scenario 6}}  \\
$\boldsymbol{\phi}$ & 0.35 &  \textbf{0.5} & 0.5 & 0.5 & 0.5 & \\ \hline
$n = 30$ & 24.8 (7.2) & 38.2 (7.4) & 19.6 (5.9) & 7.4 (4.6) & 7.3 (4.7) & 2.7  & 29.8 (1.4)\\  
$n = 40$ & 24.9 (9.2) & 34.6 (9.1) & 23.4 (8.1) & 8.6 (6.8) & 5.7 (6.5) & 2.8 & 39.7 (2.1) \\  \hline
&\multicolumn{6}{c}{\textbf{Scenario 7}}  \\
$\boldsymbol{\phi}$ &0.2 & 0.35 & 0.35 & 0.35 & 0.35 & \\ \hline
$n = 30$ & 1.7 (4.5) & 12.8 (5.6) & 18.0 (6.0) & 15.6 (5.3) & 11.4 (5.5) & 40.5  & 27.0 (4.2) \\  
$n = 40$ & 1.5 (6.4) & 11.2 (7.3) & 14.7 (7.3) & 15.2 (7.3) & 11.3 (7.2) & 46.1 & 36.1 (4.9) \\  \hline
&\multicolumn{6}{c}{\textbf{Scenario 8}}  \\
$\boldsymbol{\phi}$ &  \textbf{0.5} & 0.65 & 0.65 & 0.65 & 0.65 & \\ \hline
$n = 30$ & 78.4 (11.5) & 18.6 (6.0) & 2.2 (4.4) & 0.2 (4.0) & 0.3 (4.0) & 0.3  & 30.0 (0.9) \\  
$n = 40$ & 80.3 (14.1) & 16.7 (7.7) & 2.2 (6.3) & 0.5 (6.0) & 0.2 (5.9) & 0.1 & 40.0 (0.8) \\  \hline
\end{tabular}
\end{center}
\end{table}

\clearpage

\subsection{Simulation Results Using a Simple BLRM}\label{subsec:supp_BLRM_res}

Tables \ref{tab:BLRM_L3_res}, \ref{tab:BLRM_L4_res} and \ref{tab:BLRM_L5_res} give the results using a simple BLRM (without plateau parameter).

\begin{table}[h!]
\footnotesize
\caption{Selection percentage and mean number of volunteers (shown in parentheses) allocated to each dose for each scenario and for several values of maximum number of volunteers, for BLRM, when $L = 3$ dose levels are used.$\boldsymbol{\phi}$: activity probabilities; $n$: maximal number of healthy volunteers. \label{tab:BLRM_L3_res}}
\begin{center}
\begin{tabular}{l|c|c|c|c|c}
\hline
& \multicolumn{3}{c}{\textbf{Dose levels}} & \textbf{Early} & \textbf{Total number}\\   
& 1 & 2 & 3  & \textbf{Termination} & \textbf{of participants} \\   \hline \hline
&\multicolumn{4}{c}{\textbf{Scenario 1}}  \\
$\boldsymbol{\phi}$ & \textbf{0.5} & 0.65 & 0.8 & \\ \hline
$n = 18$ & 78.2 (8.7) & 18.6 (5.0) & 2.6 (4.2) & 0.6 & 18.0 (0.6) \\ 
$n = 24$ & 81.6 (11.1) & 16.9 (6.8) & 0.8 (6.0) & 0.7 & 23.9 (1.1) \\ 
$n = 30$ & 85.0 (13.0) & 14.0 (9.0) & 0.1 (8.0) & 0.9 & 29.9 (1.4) \\  
$n = 40$ & 87.7 (18.8) & 11.1 (11.1) & 0.1 (9.9) & 1.1 & 39.8 (2.3) \\  \hline
&\multicolumn{4}{c}{\textbf{Scenario 2}}  \\
$\boldsymbol{\phi}$ & 0.2 & 0.35 & \textbf{0.5} & \\ \hline
$n = 18$ & 3.2 (4.4) & 22.9 (5.2) & 61.1 (7.9) & 12.8 & 17.5 (1.6) \\ 
$n = 24$ & 1.5 (6.2) & 20.5 (7.3) & 61.3 (9.7) & 16.7 & 23.2 (2.2) \\ 
$n = 30$ & 1.1 (8.1) & 16.8 (9.1) & 64.4 (12.0) & 17.7 & 29.1 (2.5) \\   
$n = 40$ & 0.3 (10.1) & 17.7 (11.8) & 61.3 (16.4) & 20.7 & 38.3 (4.2) \\  \hline
&\multicolumn{4}{c}{\textbf{Scenario 3}}  \\
$\boldsymbol{\phi}$ & 0.05 & 0.2 & 0.35 & \\ \hline
$n = 18$ & 0 (4.0) & 1.3 (4.1) & 39.7 (7.5) & 59.0 & 15.6 (2.7) \\ 
$n = 24$ & 0 (6.0) & 0.3 (6.1) & 30.9 (8.4) & 68.8 & 20.4 (3.0) \\  
$n = 30$ & 0 (8.0) & 0.5 (8.0) & 25.0 (9.9) & 74.5 & 25.9 (3.0) \\  
$n = 40$ & 0 (10) & 0.1 (10.0) & 15.9 (12.2) & 84.0 & 32.2 (4.4) \\  \hline
&\multicolumn{4}{c}{\textbf{Scenario 4}}  \\
$\boldsymbol{\phi}$ & 0.35 & \textbf{0.5} & 0.65 & \\ \hline
$n = 18$ & 30.3 (6.2) & 44.9 (6.0) & 23.1 (5.7) & 1.7 & 17.9 (0.8)\\ 
$n = 24$ & 28.8 (8.1) & 50.8 (8.5) & 20.0 (7.3) & 0.4 & 23.9 (1.3)\\  
$n = 30$ & 26.0 (9.6) & 51.0 (11.0) & 21.4 (9.3) & 1.6 & 29.8 (1.6)\\  
$n = 40$ & 22.6 (12.6) & 57.7 (15.3) & 17.5 (11.8) & 2.2 & 39.7 (2.8) \\  \hline
&\multicolumn{4}{c}{\textbf{Scenario 5}} \\
$\boldsymbol{\phi}$ & 0.35 & \textbf{0.5} & 0.5 & \\ \hline
$n = 18$ & 20.8 (5.5) & 31.6 (5.6) & 41.9 (6.7) & 5.7 & 17.8 (1.0) \\ 
$n = 24$ & 17.6 (7.3) & 38.0 (8.1) & 39.8 (8.3) & 4.6 & 23.8 (1.5) \\  
$n = 30$ & 14.9 (9.0) & 39.1 (10.2) & 41.9 (10.6) & 4.1 & 29.7 (1.8) \\ 
$n = 40$ & 12.8 (11.3) & 43.1 (14.1) & 38.3 (14.0) & 5.8 & 39.5 (3.0) \\  \hline
&\multicolumn{4}{c}{\textbf{Scenario 6}} \\
$\boldsymbol{\phi}$ & \textbf{0.5} & 0.5 & 0.5 & \\ \hline
$n = 18$ & 38.5 (6.4) & 25.5 (5.4) & 32.6 (6.1) & 3.4 & 17.9 (0.8) \\ 
$n = 24$ & 35.2 (8.3) & 32.0 (7.9) & 30.5 (7.7) & 2.3 & 23.9 (1.3) \\  
$n = 30$ & 32.1 (10.0) & 36.6 (10.2) & 28.8 (9.6) & 2.5 & 29.8 (1.6) \\ 
$n = 40$ & 33.2 (13.2) & 38.7 (14.0) & 25.4 (12.5) & 2.7 & 39.6 (2.7) \\  \hline
&\multicolumn{4}{c}{\textbf{Scenario 7}}  \\
$\boldsymbol{\phi}$ & 0.35 & 0.35 & 0.35 & \\ \hline
$n = 18$ & 4.8 (4.5) & 13.3 (4.8) & 60.1 (8.0) & 21.8 & 17.3 (1.8)\\ 
$n = 24$ & 3.8 (6.3) & 12.5 (7.1) & 59.2 (9.7) & 24.5 & 23.0 (2.3)\\  
$n = 30$ & 1.8 (8.2) & 11.3 (8.9) & 60.1 (11.8) & 26.8 & 28.9 (2.6)\\ 
$n = 40$ & 0.9 (10.2) & 8.8 (11.0) & 54.5 (16.2) & 35.8 & 37.4 (4.6) \\  \hline
&\multicolumn{4}{c}{\textbf{Scenario 8}}  \\
$\boldsymbol{\phi}$ & \textbf{0.65} & 0.65 & 0.65 & \\ \hline
$n = 18$ & 85.1 (8.9) & 10.3 (4.8) & 3.9 (4.4) & 0.7 & 18.0 (0.6) \\ 
$n = 24$ & 88.6 (11.3) & 8.4 (6.5) & 2.1 (6.1) & 0.9 & 23.9 (1.2) \\  
$n = 30$ & 89.0 (13.2) & 9.4 (8.7) & 0.8 (8.0) & 0.8 & 29.9 (1.4) \\  
$n = 40$ & 91.6 (19.0) & 6.6 (10.8) & 0.5 (10.0) & 1.3 & 39.8 (2.4) \\  \hline
\end{tabular}
\end{center}
\end{table}

\begin{table}[h!]
\footnotesize
\caption{Selection percentage and mean number of volunteers (shown in parentheses) allocated to each dose for each scenario and for several values of maximum number of volunteers, for BLRM, when $L = 4$ dose levels are used. $\boldsymbol{\phi}$: activity probabilities; $n$: maximal number of healthy volunteers.\label{tab:BLRM_L4_res}}
\begin{center}
\begin{tabular}{l|c|c|c|c|c|c}
\hline
& \multicolumn{4}{c}{\textbf{Dose levels}} & \textbf{Early} & \textbf{Total number}\\   
& 1 & 2 & 3 & 4 & \textbf{Termination} & \textbf{of participants} \\  \hline \hline
&\multicolumn{5}{c}{\textbf{Scenario 1}}  \\
$\boldsymbol{\phi}$ & 0.35 & \textbf{0.5} & 0.65 & 0.8 &\\ \hline
$n = 24$ & 33.0 (7.1) & 47.4 (7.4) & 17.8 (5.1) & 1.7 (4.4) & 0.1 & 24.0 (0.4) \\ 
$n = 30$ & 28.8 (8.6) & 55.9 (9.8) & 14.1 (6.3) & 0.9 (5.3) & 0.3 & 30.0 (0.2)\\  
$n = 40$ & 29.1 (10.8) & 59.3 (12.2) & 10.0 (8.9) & 0.6 (8.0) & 1.0 & 39.9 (1.6)  \\   \hline
&\multicolumn{5}{c}{\textbf{Scenario 2}}  \\
$\boldsymbol{\phi}$ & 0.05 & 0.2 & 0.35 & \textbf{0.5} & \\ \hline
$n = 24$ & 0.1 (4.0) & 1.5 (4.2) & 12.5 (4.7) & 60.0 (9.4) & 25.9 & 22.3 (3.1) \\ 
$n = 30$ & 0 (5.0) & 0.5 (5.1) & 12.1 (6.1) & 59.0 (11.5) & 28.4& 27.7 (4.0)  \\  
$n = 40$ & 0 (8) & 0.1 (8.0) & 8.4 (8.6) & 58.4 (13.0) & 33.1 & 37.7 (3.9) \\   \hline
&\multicolumn{5}{c}{\textbf{Scenario 3}}  \\
$\boldsymbol{\phi}$ & 0.01 & 0.05 & 0.2 & 0.35 & \\ \hline
$n = 24$ & 0 (4.0) & 0.1 (4.0) & 0.3 (4.0) & 20.1 (6.4) & 79.5 & 18.5 (3.4)\\ 
$n = 30$ & 0 (5.0) & 0 (5.0) & 0.1 (5.0) & 15.1 (7.2) & 84.8 & 22.2 (3.8) \\  
$n = 40$ & 0 (8) & 0 (8) & 0.1 (7.9) & 7.1 (8.9) & 92.8 & 32.8 (2.8)  \\   \hline
&\multicolumn{5}{c}{\textbf{Scenario 4}}  \\
$\boldsymbol{\phi}$ & 0.2 & 0.35 & \textbf{0.5} & 0.65 & \\ \hline
$n = 24$ & 1.6 (4.4) & 25.0 (6.4) & 42.8 (6.2) & 28.5 (6.9) & 2.1 & 23.9 (1.1) \\ 
$n = 30$ & 1.0 (5.2) & 23.1 (7.5) & 44.7 (8.6) & 27.8 (8.3) & 3.4 & 29.7 (2.0) \\  
$n = 40$ & 0.7 (8.1) & 20.7 (9.7) & 48.4 (11.5) & 27.9 (10.4) & 2.3 & 39.7 (2.1)  \\   \hline
&\multicolumn{5}{c}{\textbf{Scenario 5}}  \\
$\boldsymbol{\phi}$ & 0.2 & 0.35 & \textbf{0.5} & 0.5 & \\ \hline
$n = 24$ & 1.4 (4.2) & 16.6 (5.6) & 27.2 (5.6) & 45.4 (8.1) & 9.4 & 23.5 (1.8) \\ 
$n = 30$ & 0.6 (5.1) & 13.3 (6.7) & 29.4 (7.5) & 47.5 (10.0) & 9.2 & 29.3 (2.6) \\  
$n = 40$ & 0 (8.0) & 11.8 (9.0) & 33.8 (10.4) & 46.6 (12.0) & 7.8 & 39.4 (2.6)  \\   \hline
&\multicolumn{5}{c}{\textbf{Scenario 6}}  \\
$\boldsymbol{\phi}$ & 0.35 & \textbf{0.5} & 0.5 & 0.5 & \\ \hline
$n = 24$ & 13.6 (5.1) & 31.8 (6.7) & 19.0 (5.5) & 32.3 (6.6) & 3.3 & 23.9 (1.0) \\ 
$n = 30$ & 11.5 (6.3) & 29.7 (8.1) & 23.9 (7.0) & 31.5 (8.5) & 3.4 & 29.8 (1.2) \\ 
$n = 40$ & 7.1 (8.7) & 35.4 (10.7) & 26.8 (10.2) & 28.3 (10.2) & 2.4 & 39.8 (1.9)  \\   \hline
&\multicolumn{5}{c}{\textbf{Scenario 7}}  \\
$\boldsymbol{\phi}$ & 0.2 & 0.35 & 0.35 & 0.35 & \\ \hline
$n = 24$ & 0.6 (4.1) & 4.4 (4.5) & 7.2 (4.7) & 54.2 (9.0) & 33.6 & 22.3 (3.0) \\ 
$n = 30$ & 0 (5.0) & 2.6 (5.3) & 6.7 (5.9) & 53.0 (11.3) & 37.7 & 27.6 (4.0) \\  
$n = 40$ & 0 (8.0) & 1.3 (8.1) & 7.2 (8.6) & 53.6 (13.0) & 37.9 & 37.7 (3.7)  \\   \hline
&\multicolumn{5}{c}{\textbf{Scenario 8}}  \\
$\boldsymbol{\phi}$ & \textbf{0.5} & 0.65 & 0.65 & 0.65 & \\ \hline
$n = 24$ & 65.4 (8.8) & 27.1 (6.4) & 5.5 (4.6) & 1.7 (4.2) & 0.3 & 24.0 (0.6) \\ 
$n = 30$ & 67.6 (11.5) & 27.3 (7.7) & 4.0 (5.5) & 1.1 (5.2) & 0 & 30.0 (0.0) \\ 
$n = 40$ & 68.7 (13.3) & 27.5 (10.4) & 2.8 (8.2) & 0.9 (8.0) & 0.1 & 40.0 (0.8)  \\   \hline
\end{tabular}
\end{center}
\end{table}

\begin{table}[h!]
\footnotesize
\caption{Selection percentage and mean number of volunteers (shown in parentheses) allocated to each dose for each scenario and for several values of maximum number of volunteers, for BLRM, when $L = 5$ dose levels are used. $\boldsymbol{\phi}$: activity probabilities; $n$: maximal number of healthy volunteers.\label{tab:BLRM_L5_res}}
\begin{center}
\begin{tabular}{l|c|c|c|c|c|c|c}
\hline
& \multicolumn{5}{c}{\textbf{Dose levels}} & \textbf{Early} & \textbf{Total number}\\   
& 1 & 2 & 3 & 4 & 5 & \textbf{Termination} & \textbf{of participants} \\  \hline \hline
&\multicolumn{6}{c}{\textbf{Scenario 1}}  \\
$\boldsymbol{\phi}$ & 0.35 & \textbf{0.5} & 0.65 & 0.8 & 0.95 & \\ \hline
$n = 30$ & 33.7 (8.4) & 57.8 (8.7) & 7.8 (4.9) & 0.3 (4.0) & 0 (3.9) & 0.4 & 30.0 (0.6) \\   
$n = 40$ & 36.0 (10.4) & 58.5 (11.2) & 4.8 (6.5) & 0.1 (6.0) & 0 (5.9) & 0.6& 39.9 (1.4)  \\   \hline
&\multicolumn{6}{c}{\textbf{Scenario 2}}  \\
$\boldsymbol{\phi}$ & 0.05 & 0.2 & 0.35 & \textbf{0.5} & 0.65 & \\ \hline
$n = 30$ & 0 (4.0) & 0.6 (4.2) & 19.0 (5.8) & 40.7 (6.9) & 32.8 (8.3) & 6.9 &29.3 (2.8)\\  
$n = 40$ & 0 (6) & 0.1 (6.1) & 13.7 (7.4) & 46.3 (9.6) & 33.2 (10.2) & 6.7 & 39.3 (3.2) \\   \hline
&\multicolumn{6}{c}{\textbf{Scenario 3}}  \\
$\boldsymbol{\phi}$ & 0.01 & 0.05 & 0.15 & 0.25 & 0.35 & \\ \hline
$n = 30$ & 0 (4.0) & 0 (4.0) & 0 (4.0) & 0.5 (4.0) & 17.3 (6.6) & 82.2 &22.6 (4.2)\\   
$n = 40$ & 0 (6) & 0 (6) & 0 (6.0) & 0.7 (6.0) & 11.9 (7.8) & 87.4 & 31.8 (4.0) \\   \hline
&\multicolumn{6}{c}{\textbf{Scenario 4}}  \\
$\boldsymbol{\phi}$ & 0.2 & 0.35 &  \textbf{0.5} & 0.65 & 0.8 & \\ \hline
$n = 30$ & 1.9 (4.5) & 33.5 (7.4) & 43.5 (8.0) & 18.6 (5.5) & 1.7 (4.5) & 0.8 &29.9 (1.2)\\  
$n = 40$ & 0.6 (6.2) & 32.3 (9.6) & 50.3 (10.5) & 14.2 (7.3) & 1.5 (6.2) & 1.1 & 39.8 (2.0) \\   \hline
&\multicolumn{6}{c}{\textbf{Scenario 5}}  \\
$\boldsymbol{\phi}$ & 0.05 & 0.2 & 0.35 &  \textbf{0.5} & 0.5 & \\ \hline
$n = 30$ & 0.1 (4.0) & 0.6 (4.1) & 11.6 (5.1) & 21.6 (5.7) & 49.5 (9.7) & 16.9 & 28.7 (3.4)\\  
$n = 40$ &  0 (6) & 0.1 (6.0) & 8.7 (6.8) & 23.2 ( 8.0) & 51.4 (11.7) & 16.6 & 38.6 (3.9) \\   \hline
&\multicolumn{6}{c}{\textbf{Scenario 6}}  \\
$\boldsymbol{\phi}$ & 0.35 &  \textbf{0.5} & 0.5 & 0.5 & 0.5 & \\ \hline
$n = 30$ & 8.1 (4.9) & 25.6 (6.3) & 22.8 (6.6) & 15.4 (5.5) & 25.6 (6.6) & 2.5 & 29.9 (0.9)\\  
$n = 40$ & 5.0 (6.7) & 28.7 (8.5) & 24.4 (8.8) & 18.1 (7.8) & 22.0 (8.1) & 1.8 & 39.8 (1.8) \\   \hline
&\multicolumn{6}{c}{\textbf{Scenario 7}}  \\
$\boldsymbol{\phi}$ &0.2 & 0.35 & 0.35 & 0.35 & 0.35 & \\ \hline
$n = 30$ & 0 (4.0) & 2.2 (4.2) & 4.5 (4.8) & 8.0 (5.0) & 50.2 (9.9) & 35.1 & 27.9 (3.6)\\ 
$n = 40$ & 0 (6.0) & 0.8 (6.1) & 3.7 (6.5) & 8.7 (7.0) & 48.0 (11.7) & 38.8 & 37.3 (4.4) \\   \hline
&\multicolumn{6}{c}{\textbf{Scenario 8}}  \\
$\boldsymbol{\phi}$ &  \textbf{0.5} & 0.65 & 0.65 & 0.65 & 0.65 & \\ \hline
$n = 30$ & 62.5 (9.8) & 30.5 (7.0) & 4.3 (5.0) & 1.5 (4.1) & 1.0 (4.1) & 0.2 &30.0 (0.9)\\   
$n = 40$ & 62.9 (12.1) & 30.7 (9.2) & 4.5 (6.6) & 1.2 (6.1) & 0.6 (6.0) & 0.1 & 40.0 (0.2) \\   \hline
\end{tabular}
\end{center}
\end{table}

\clearpage

\section{Figures Comparing Simulation Results}\label{sec:supp_fig_res}

\subsection{Number of Volunteers by Dose}\label{subsec:supp_nb_volunt_by_dose_res}

\begin{figure}[h!]
\begin{center}
\subfigure[Scenario 1]{\includegraphics[scale=0.29]{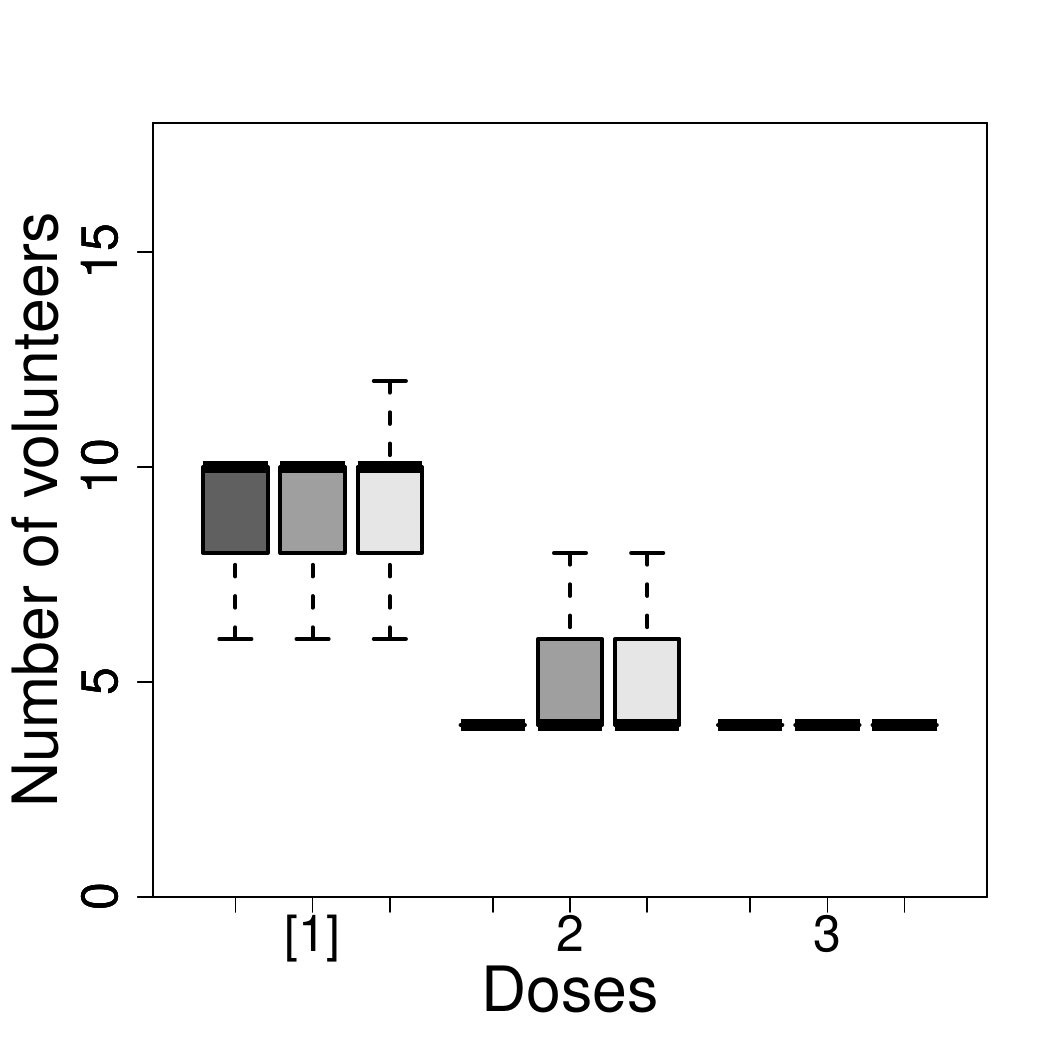}} 
\subfigure[Scenario 2]{\includegraphics[scale=0.29]{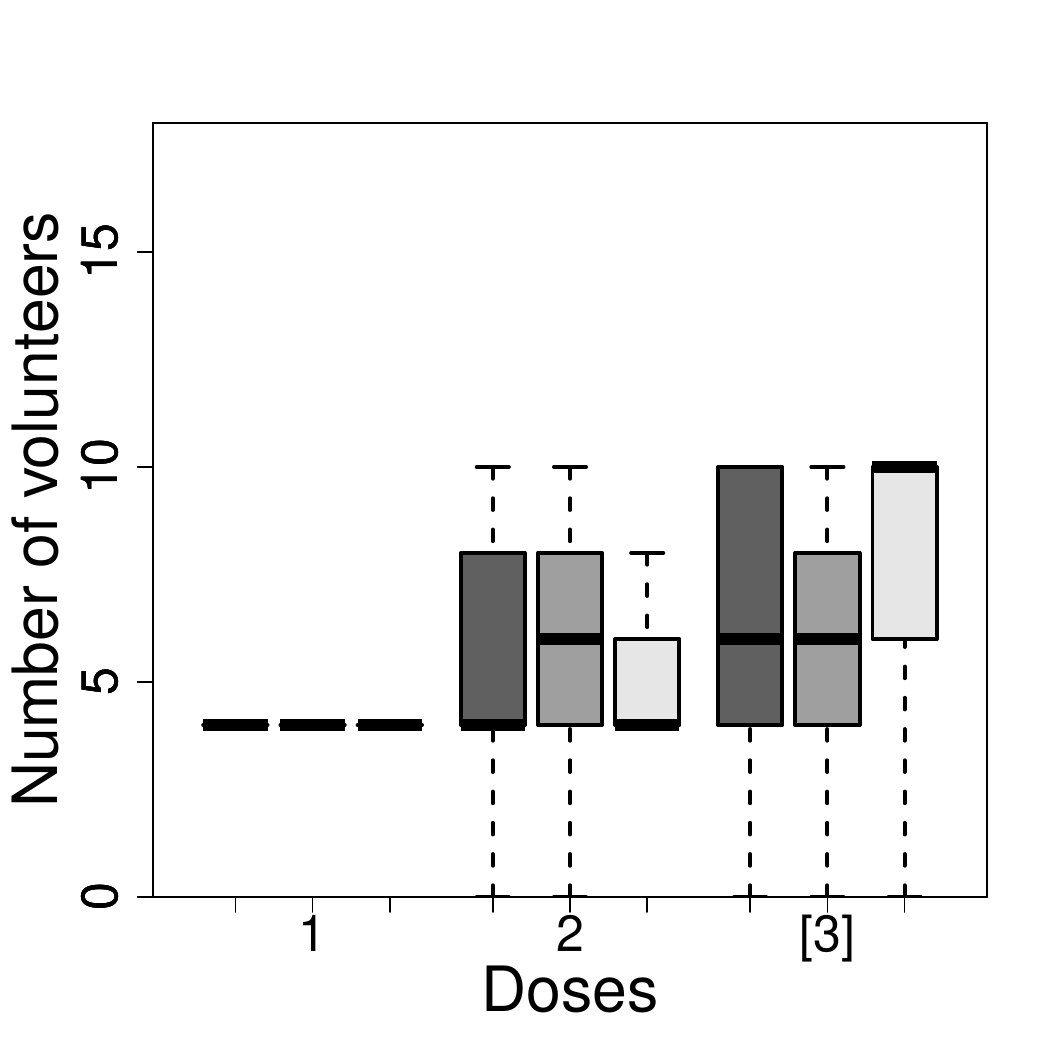}}  
\subfigure[Scenario 3]{\includegraphics[scale=0.29]{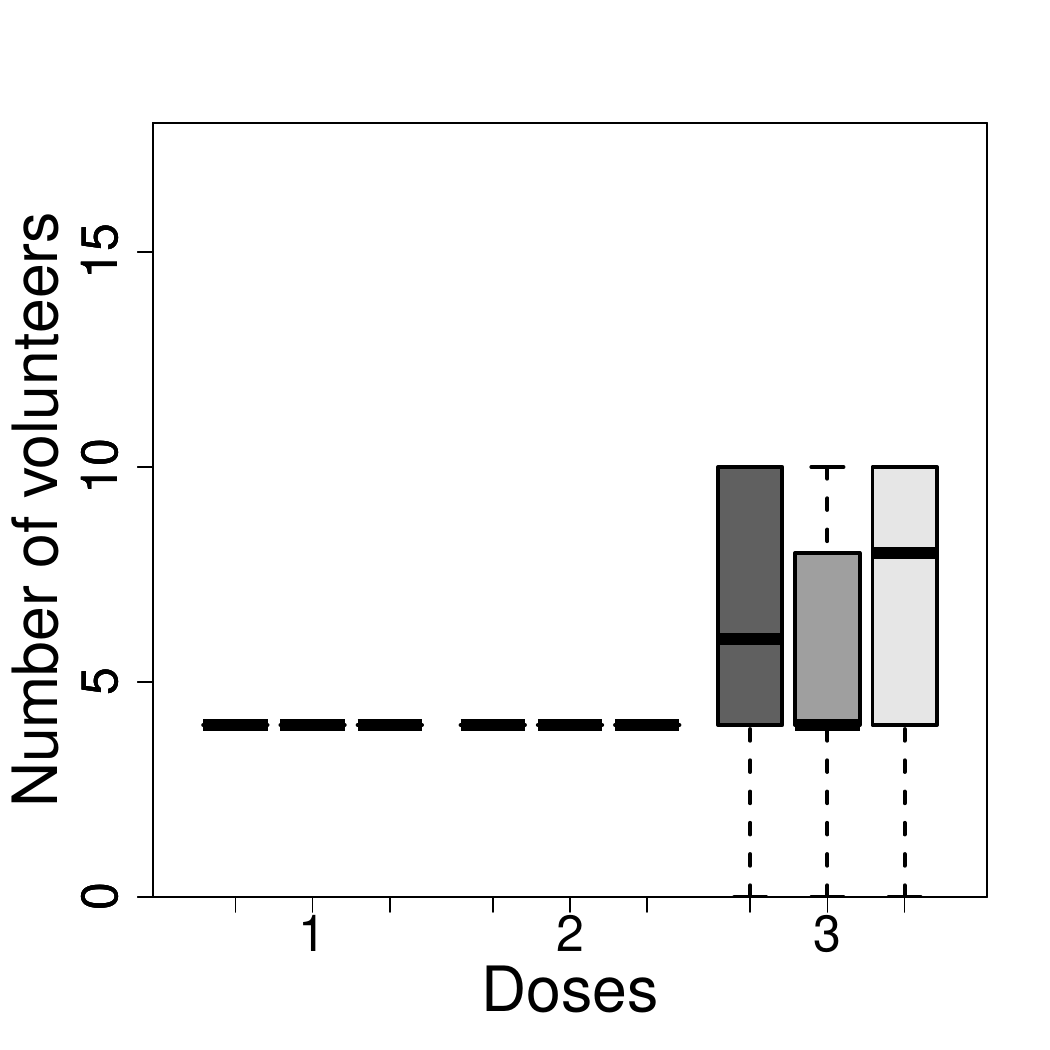}} \\
\subfigure[Scenario 4]{\includegraphics[scale=0.29]{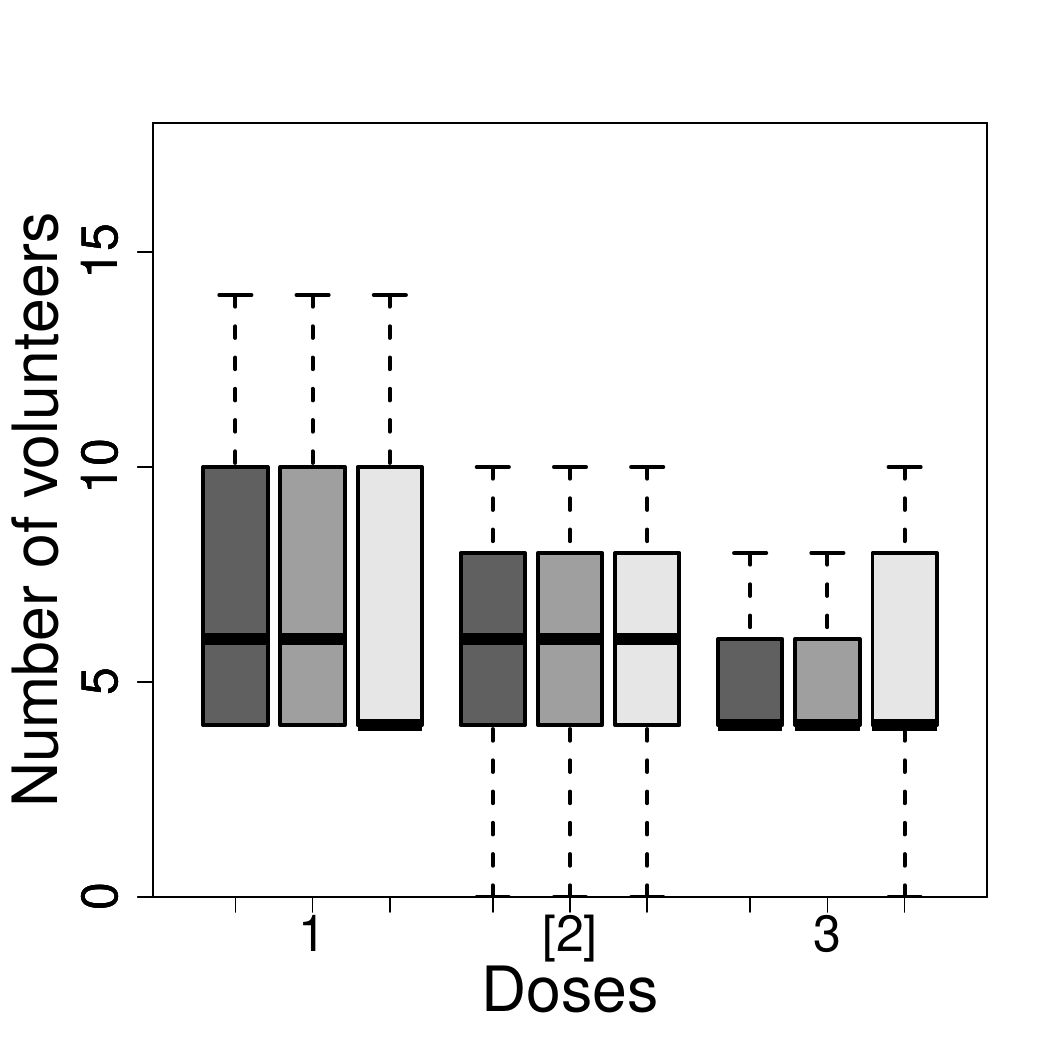}} 
\subfigure[Scenario 5]{\includegraphics[scale=0.29]{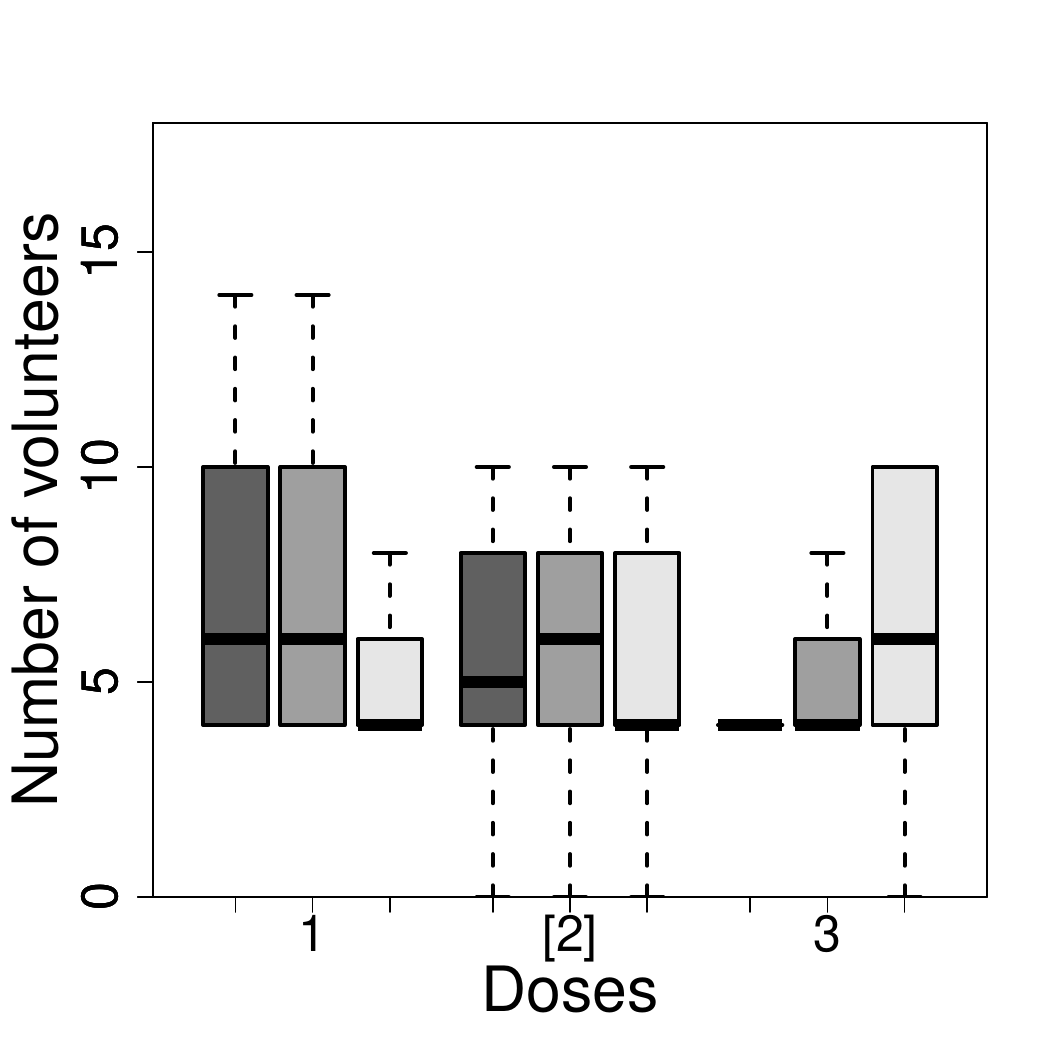}} 
\subfigure[Scenario 6]{\includegraphics[scale=0.29]{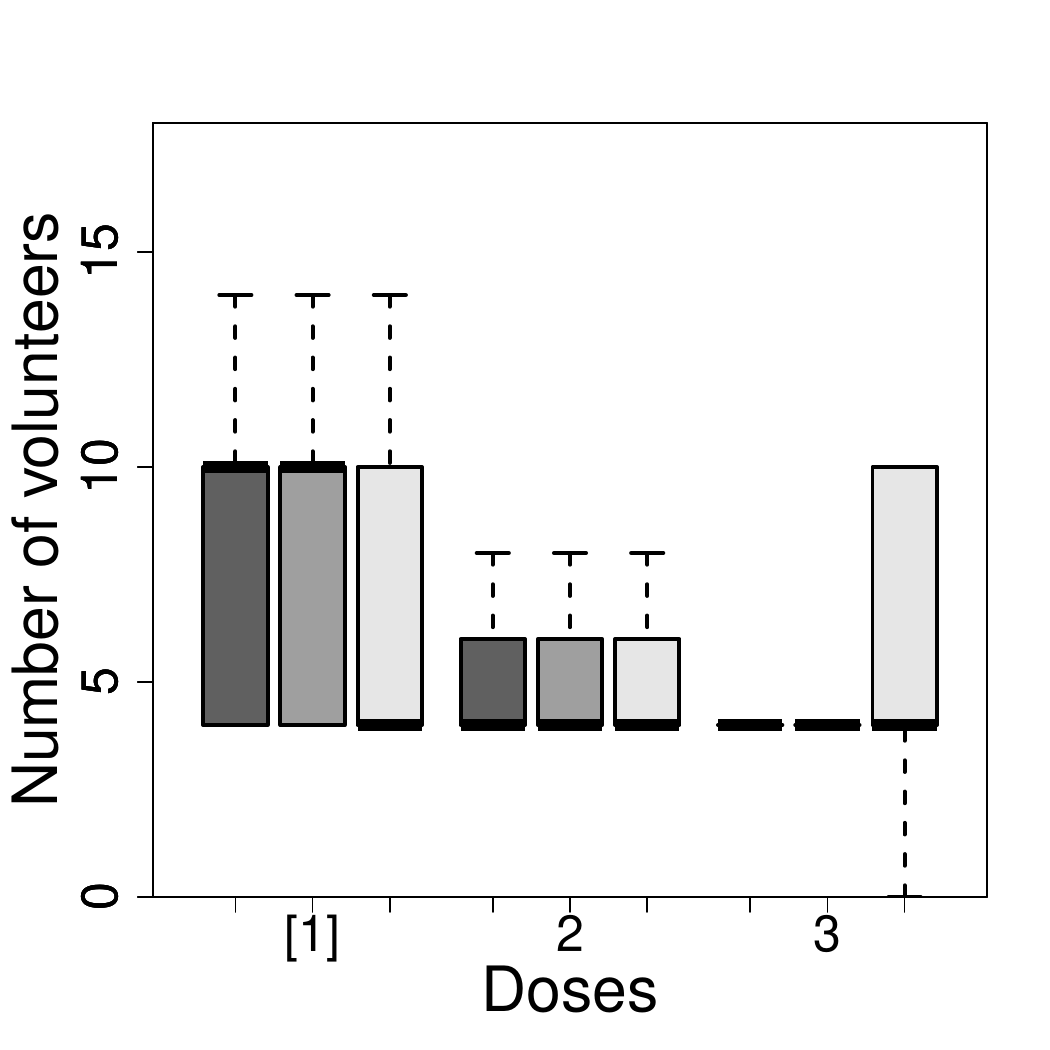}} \\
\subfigure[Scenario 7]{\includegraphics[scale=0.29]{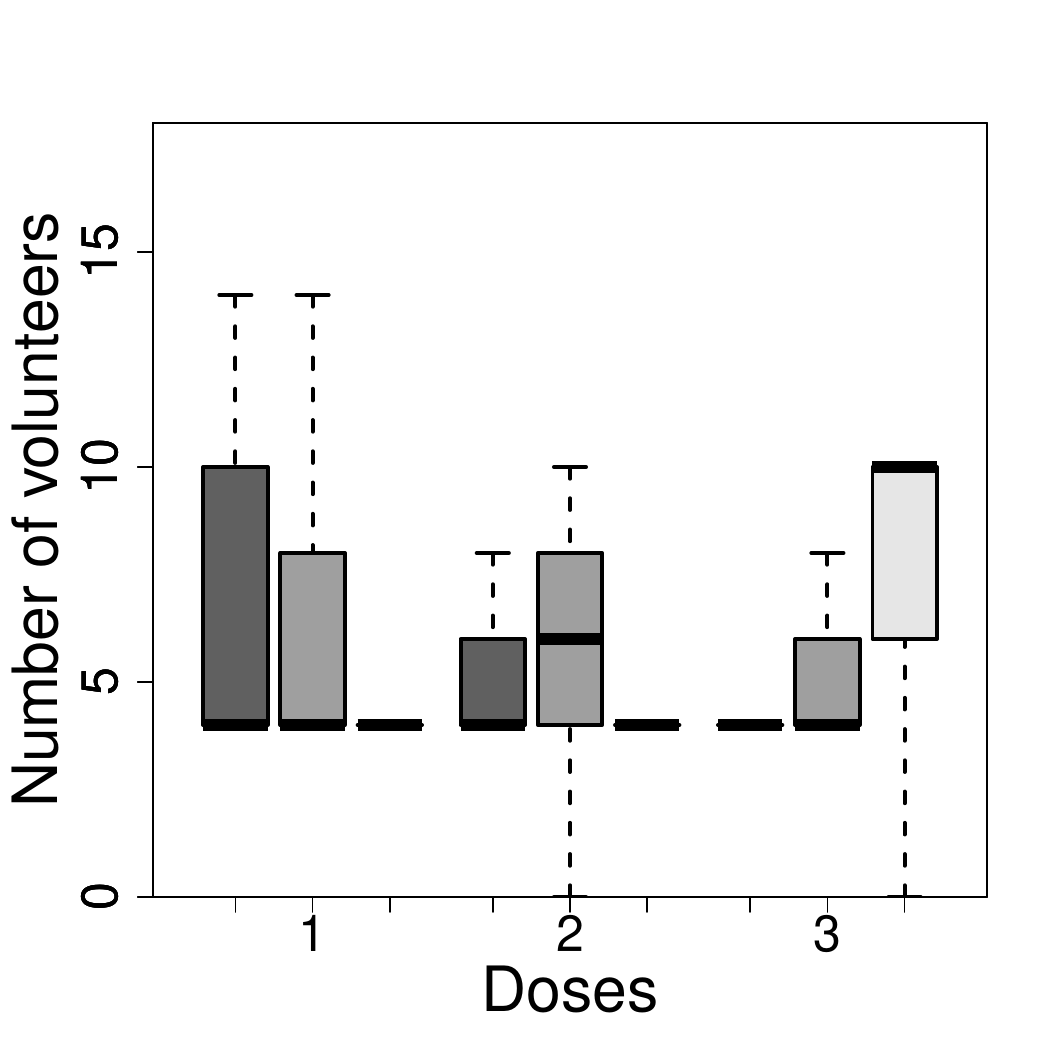}} 
\subfigure[Scenario 8]{\includegraphics[scale=0.29]{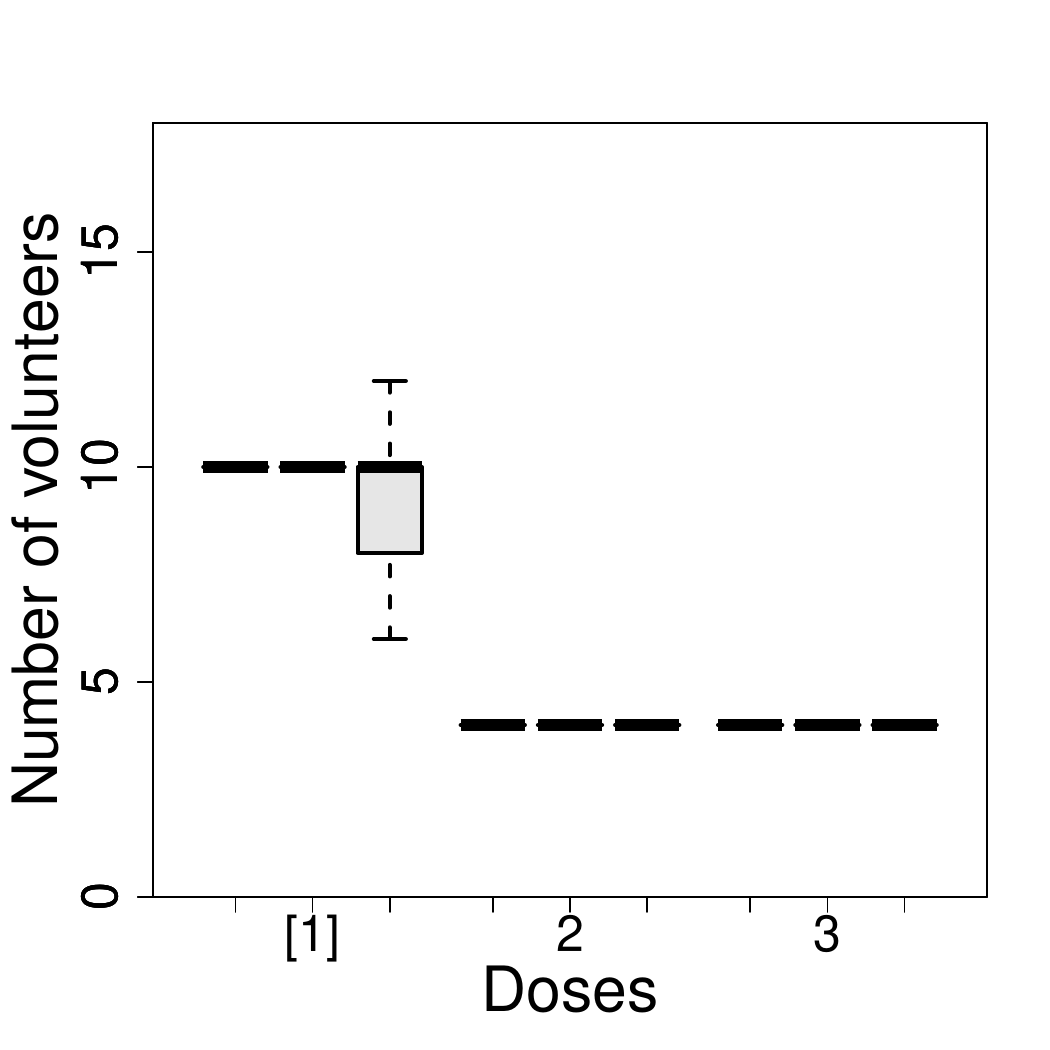}} 
\subfigure{\includegraphics[scale=0.29]{legend_boxplot_nb_volunt_by_dose.pdf}}
\end{center}
\caption{Comparison of the number of volunteers by dose for the selection method, BMA method  and BLRM for a maximum number of volunteers of $n = 18$, when $L = 3$ dose levels are used. \label{fig:boxplot_nb_volunt_by_dose_n18_L3}}
\end{figure}

\begin{figure}[h!]
\begin{center}
\subfigure[Scenario 1]{\includegraphics[scale=0.29]{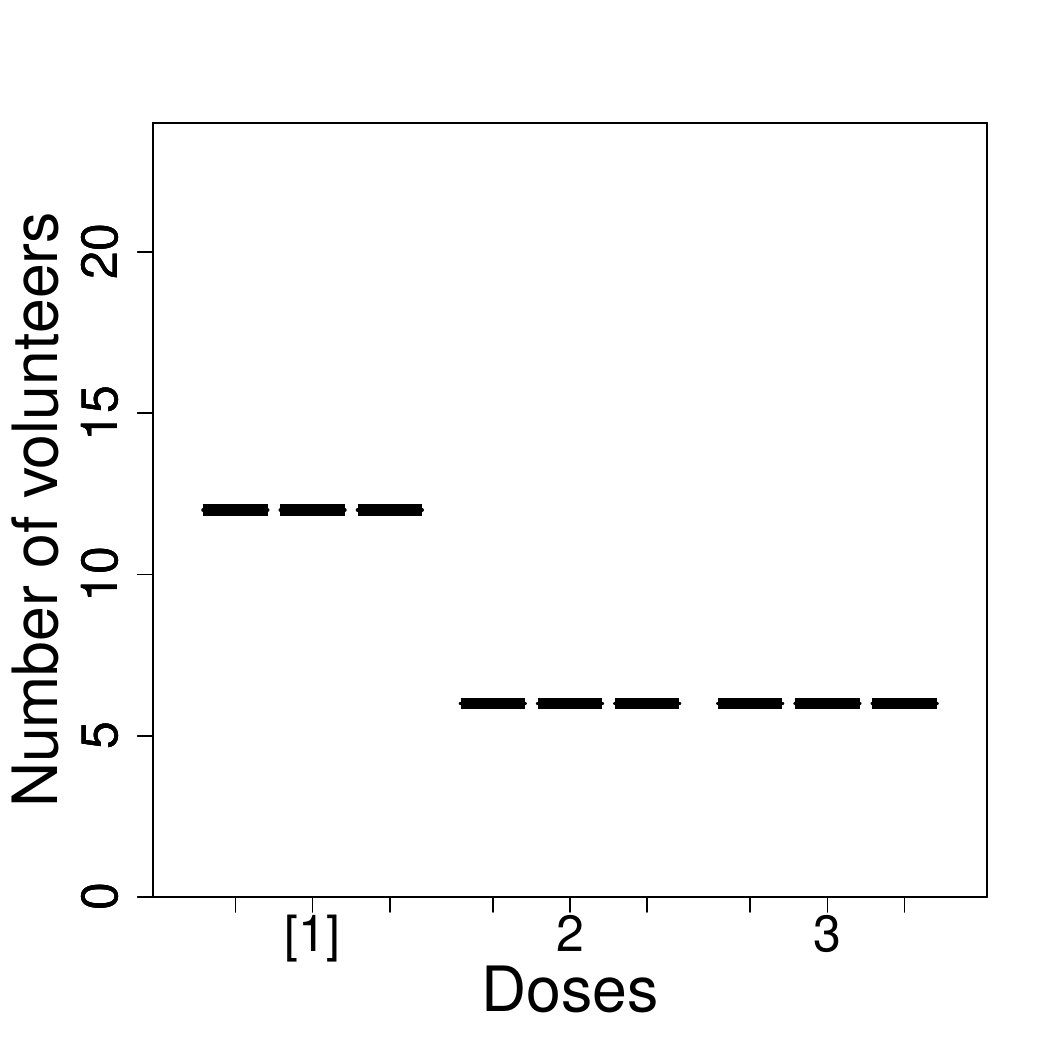}} 
\subfigure[Scenario 2]{\includegraphics[scale=0.29]{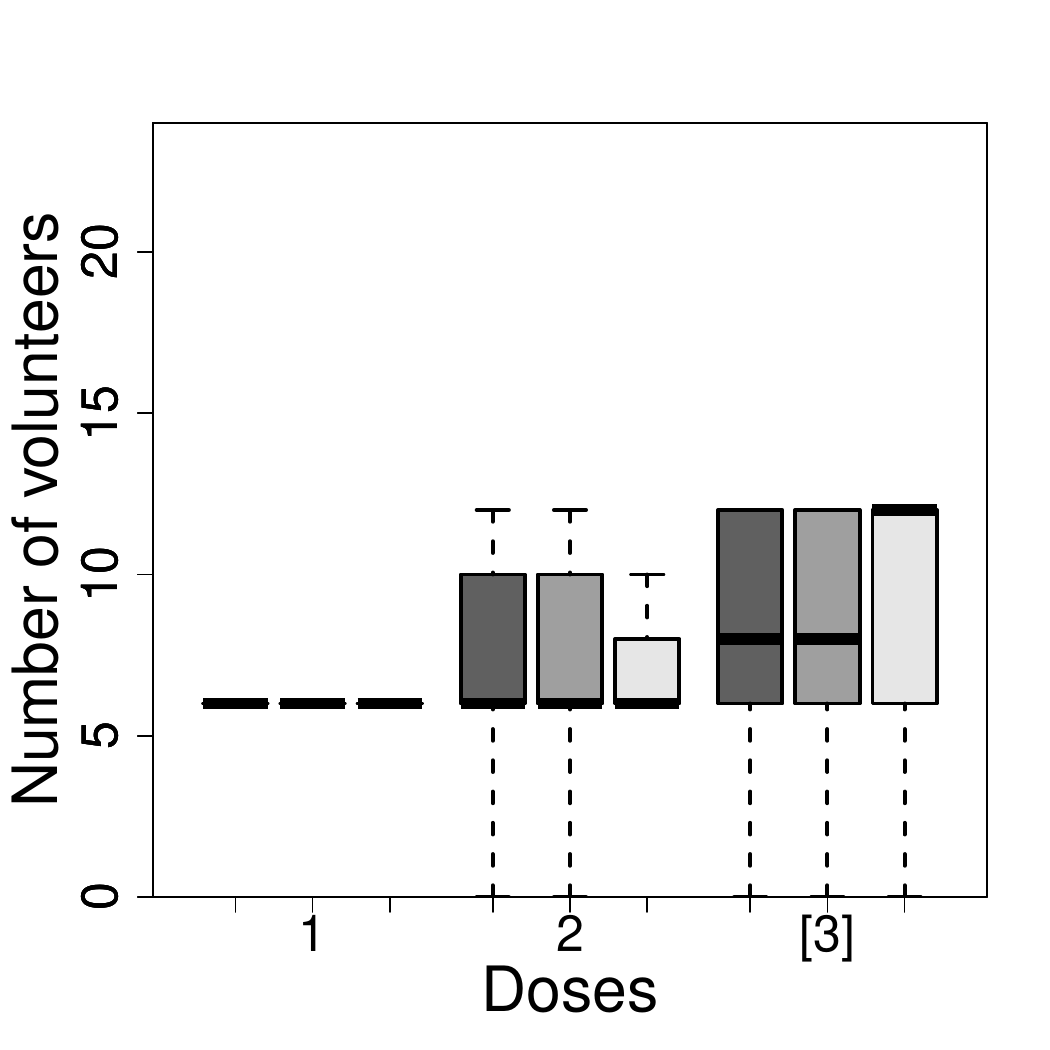}} 
\subfigure[Scenario 3]{\includegraphics[scale=0.29]{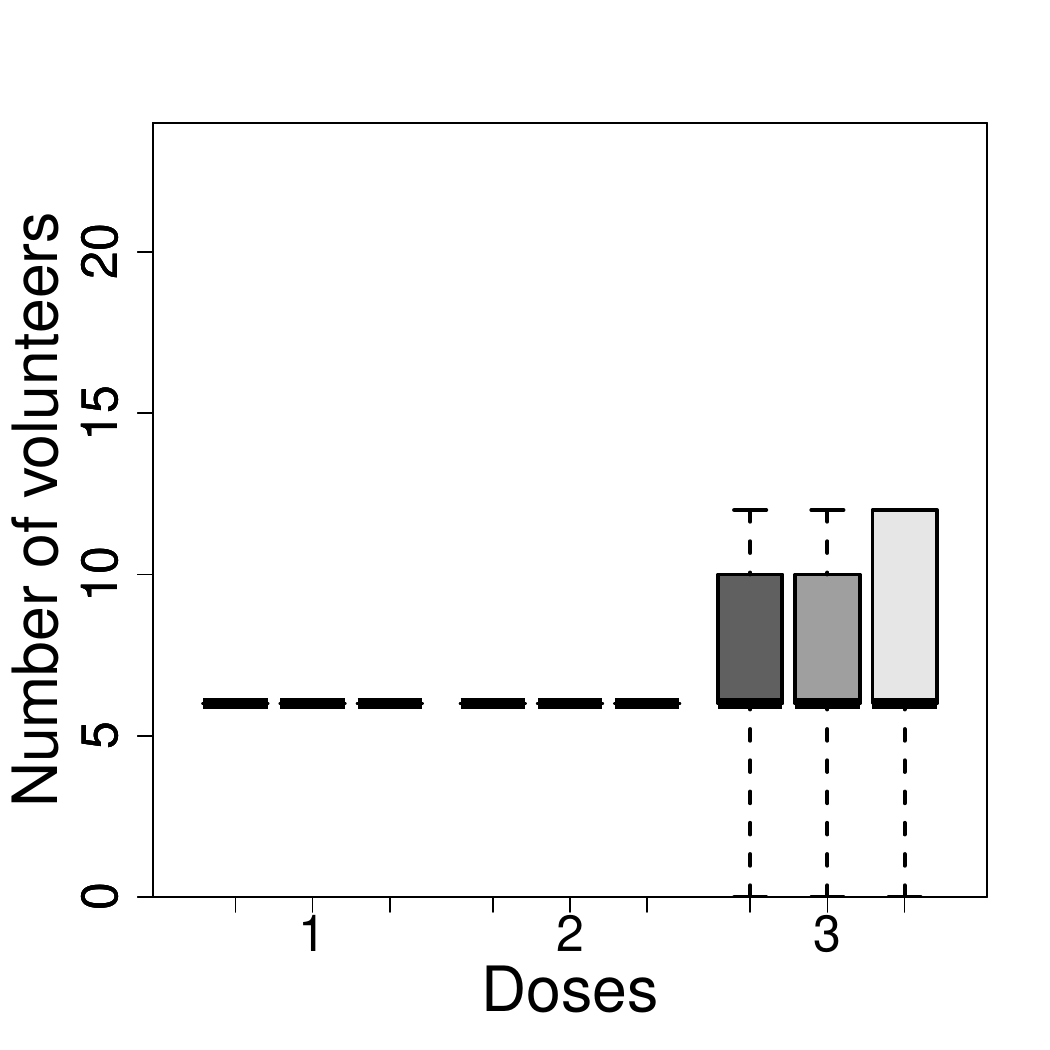}}  \\
\subfigure[Scenario 4]{\includegraphics[scale=0.29]{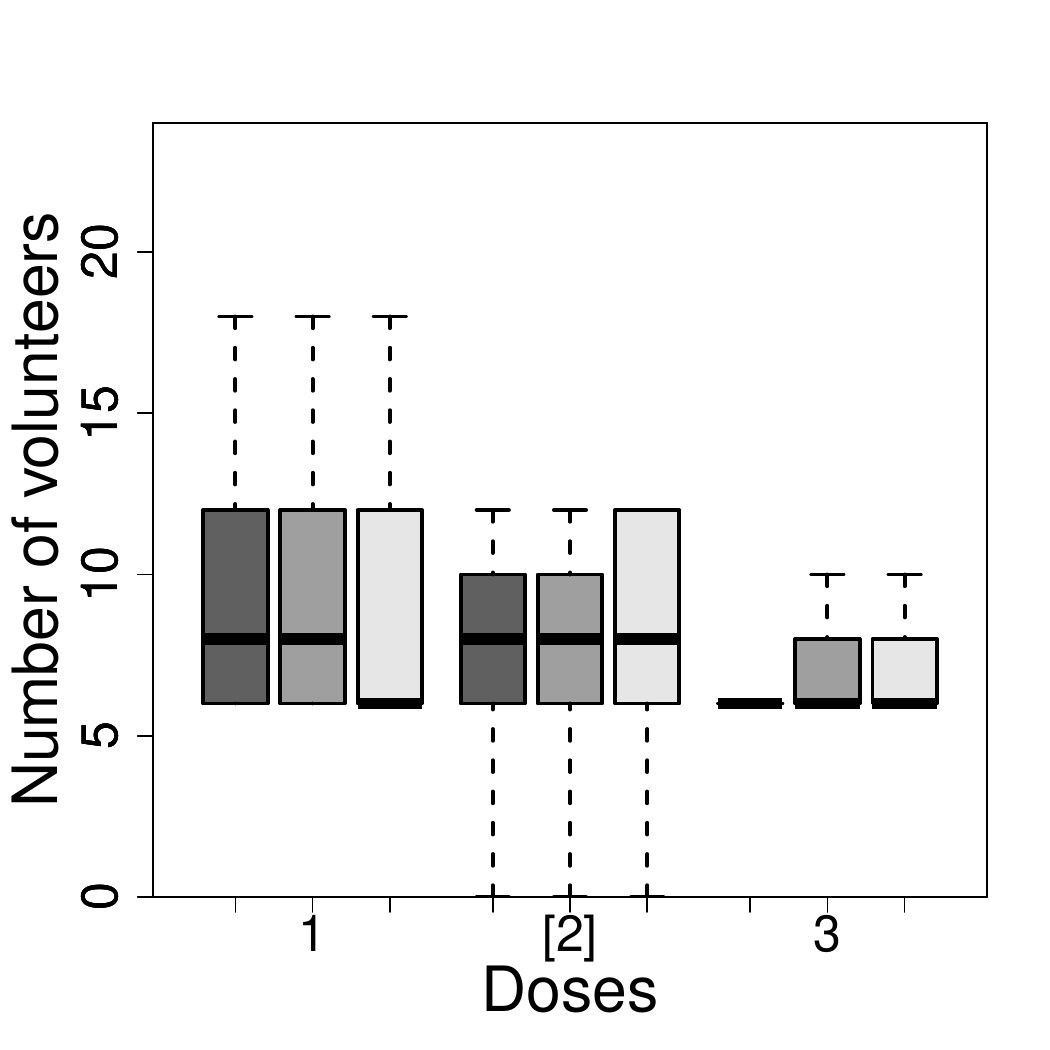}}
\subfigure[Scenario 5]{\includegraphics[scale=0.29]{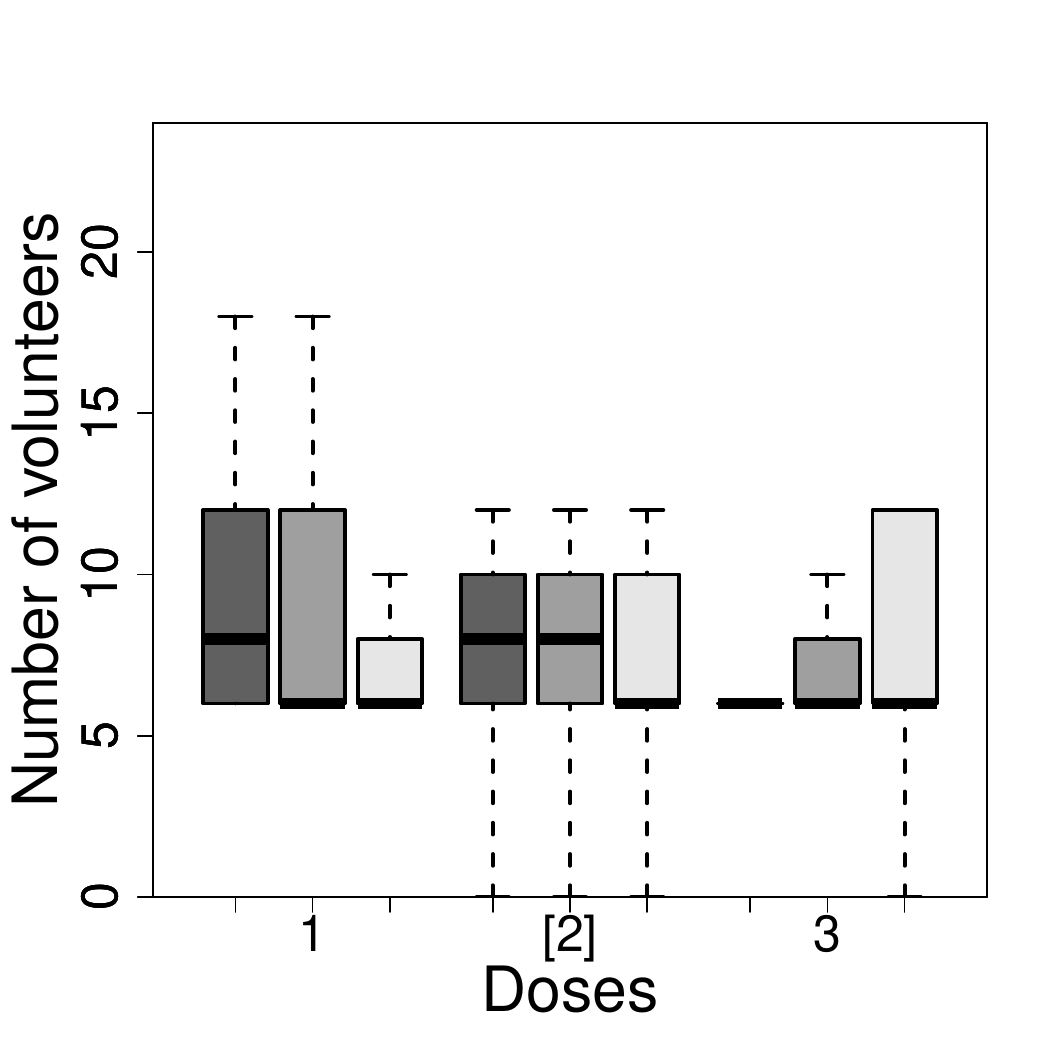}} 
\subfigure[Scenario 6]{\includegraphics[scale=0.29]{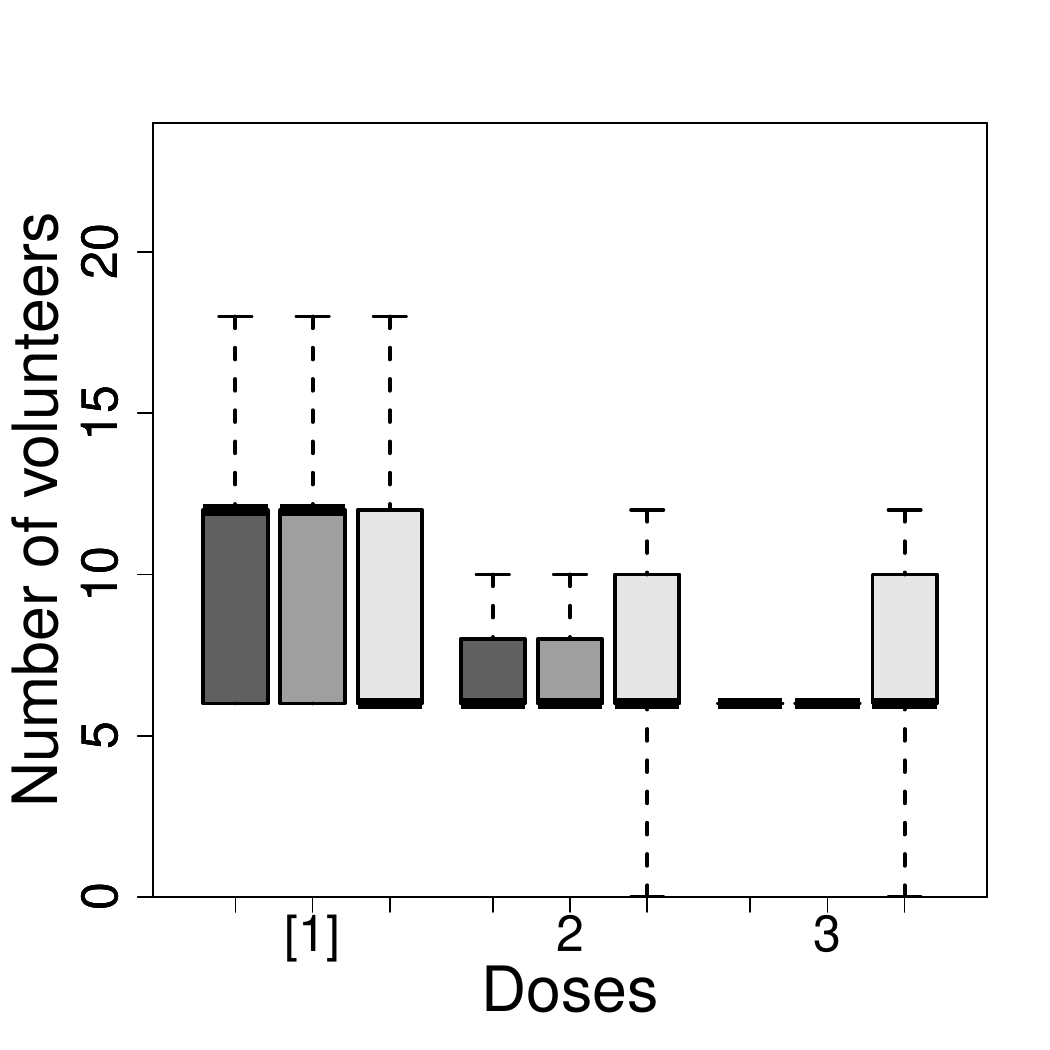}}  \\
\subfigure[Scenario 7]{\includegraphics[scale=0.29]{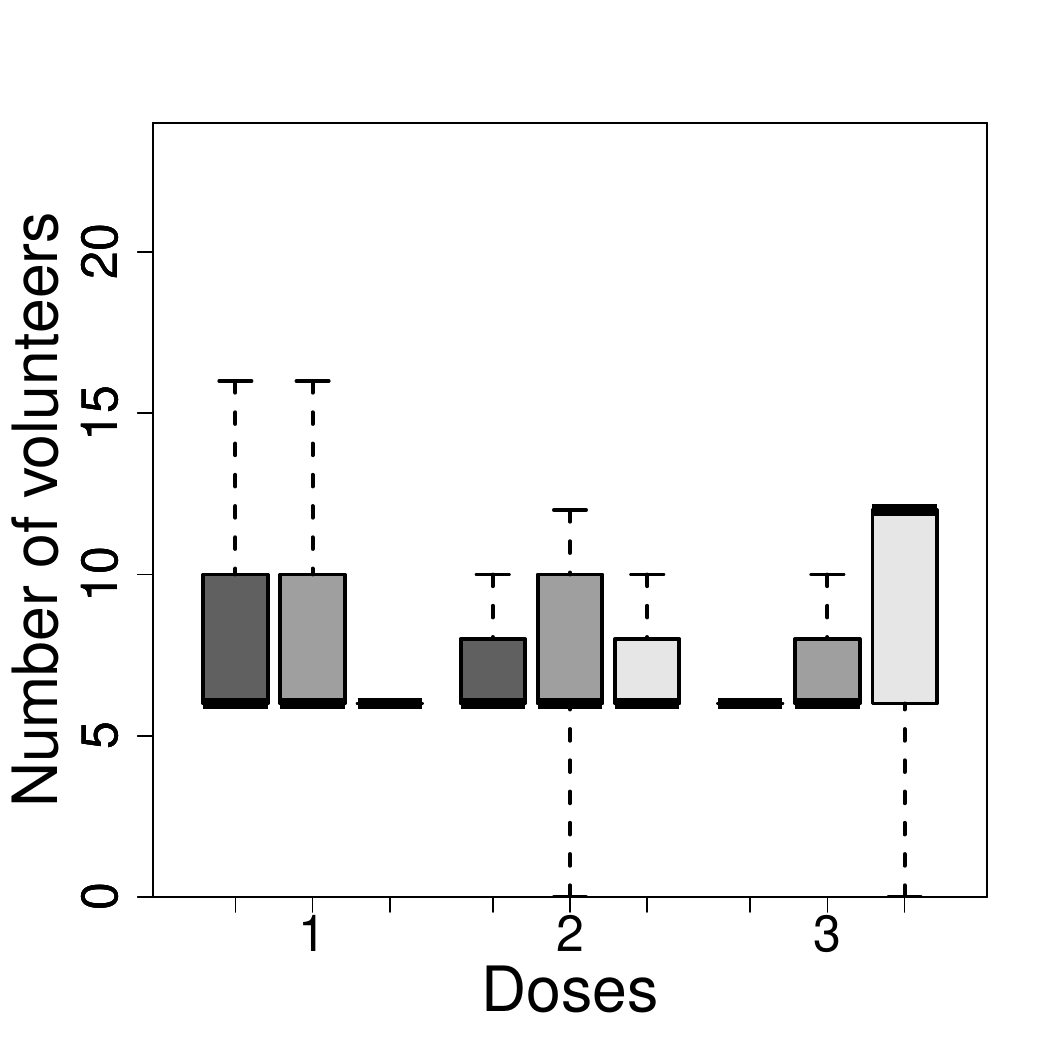}} 
\subfigure[Scenario 8]{\includegraphics[scale=0.29]{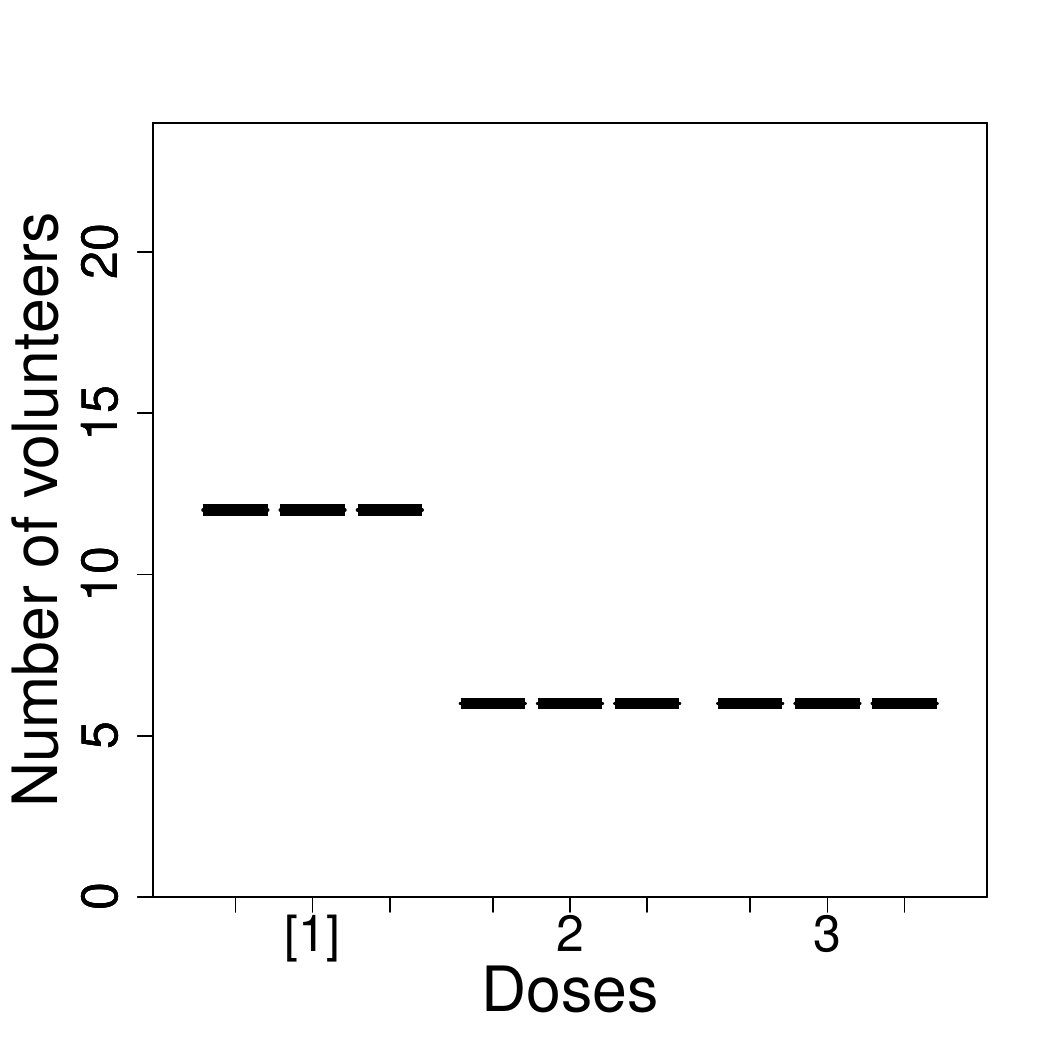}} 
\subfigure{\includegraphics[scale=0.29]{legend_boxplot_nb_volunt_by_dose.pdf}}
\end{center}
\caption{Comparison of the number of volunteers by dose for the selection method, BMA method  and BLRM for a maximum number of volunteers of $n = 24$, when $L = 3$ dose levels are used. \label{fig:boxplot_nb_volunt_by_dose_n24_L3}}
\end{figure}

\begin{figure}[h!]
\begin{center}
\subfigure[Scenario 1]{\includegraphics[scale=0.29]{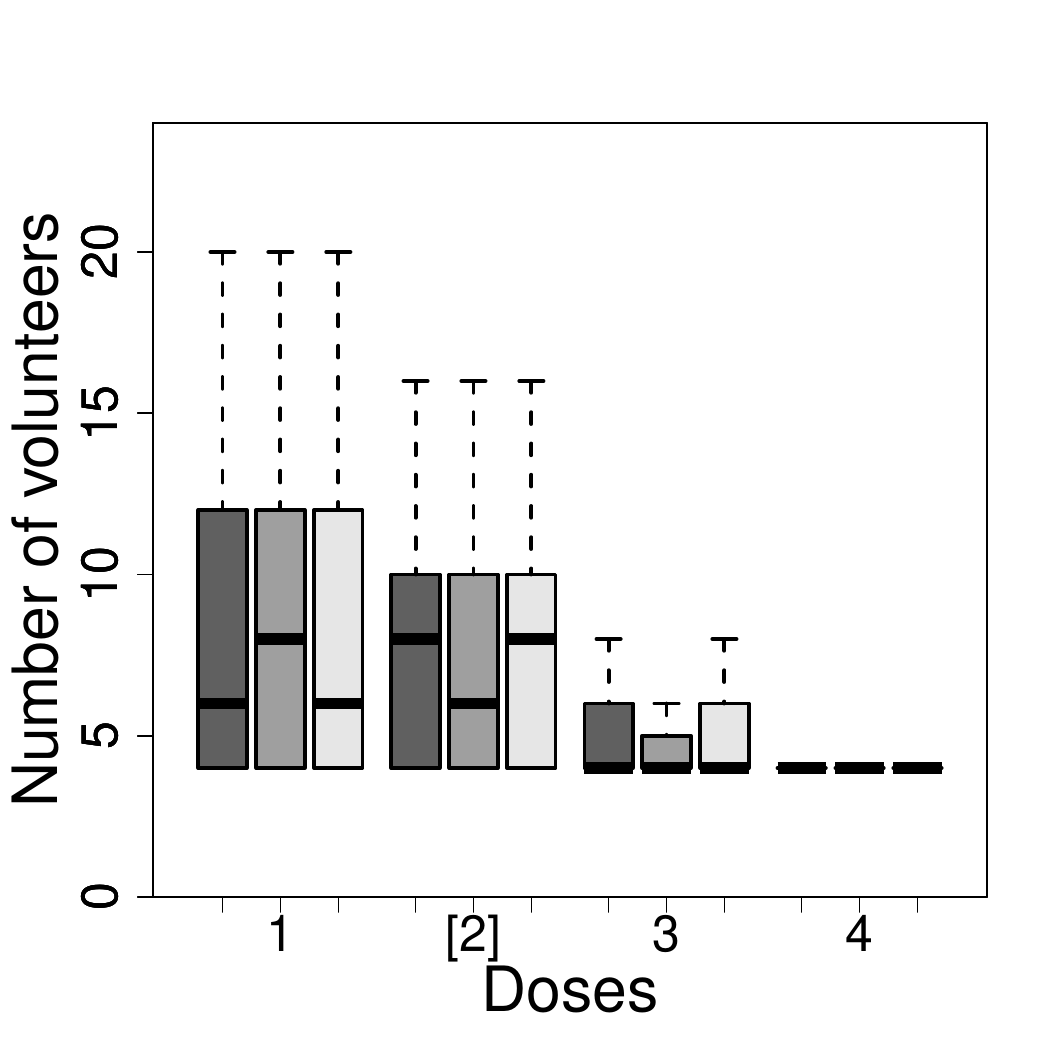}} 
\subfigure[Scenario 2]{\includegraphics[scale=0.29]{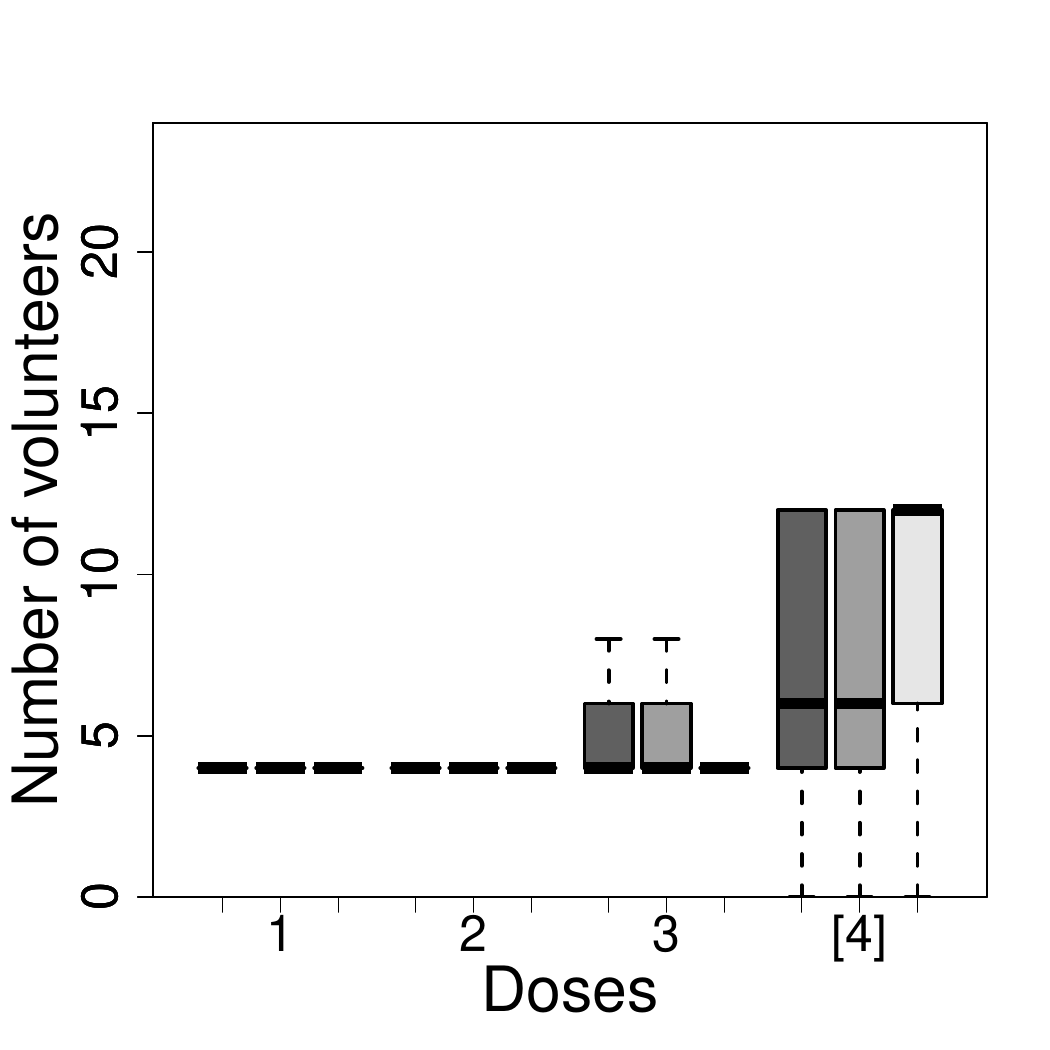}} 
\subfigure[Scenario 3]{\includegraphics[scale=0.29]{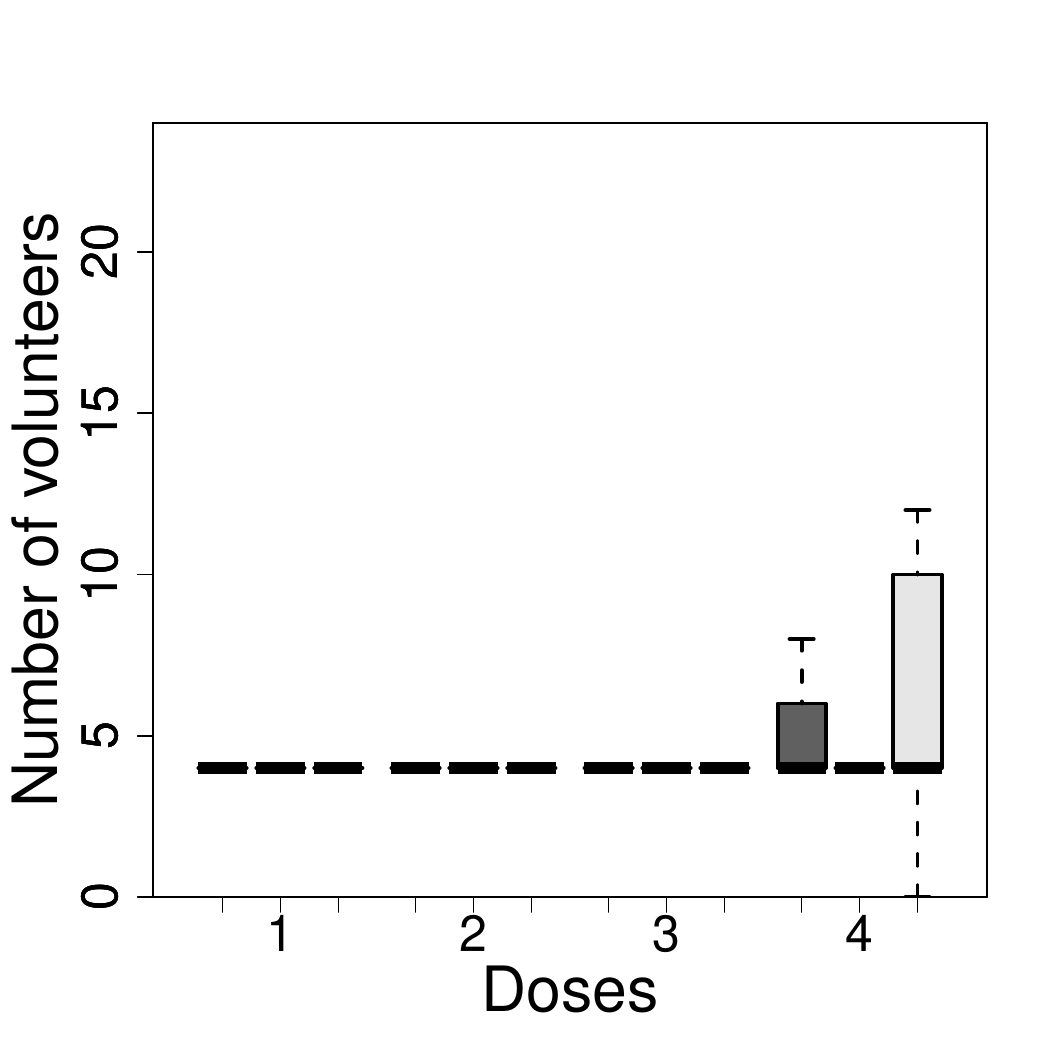}} \\
\subfigure[Scenario 4]{\includegraphics[scale=0.29]{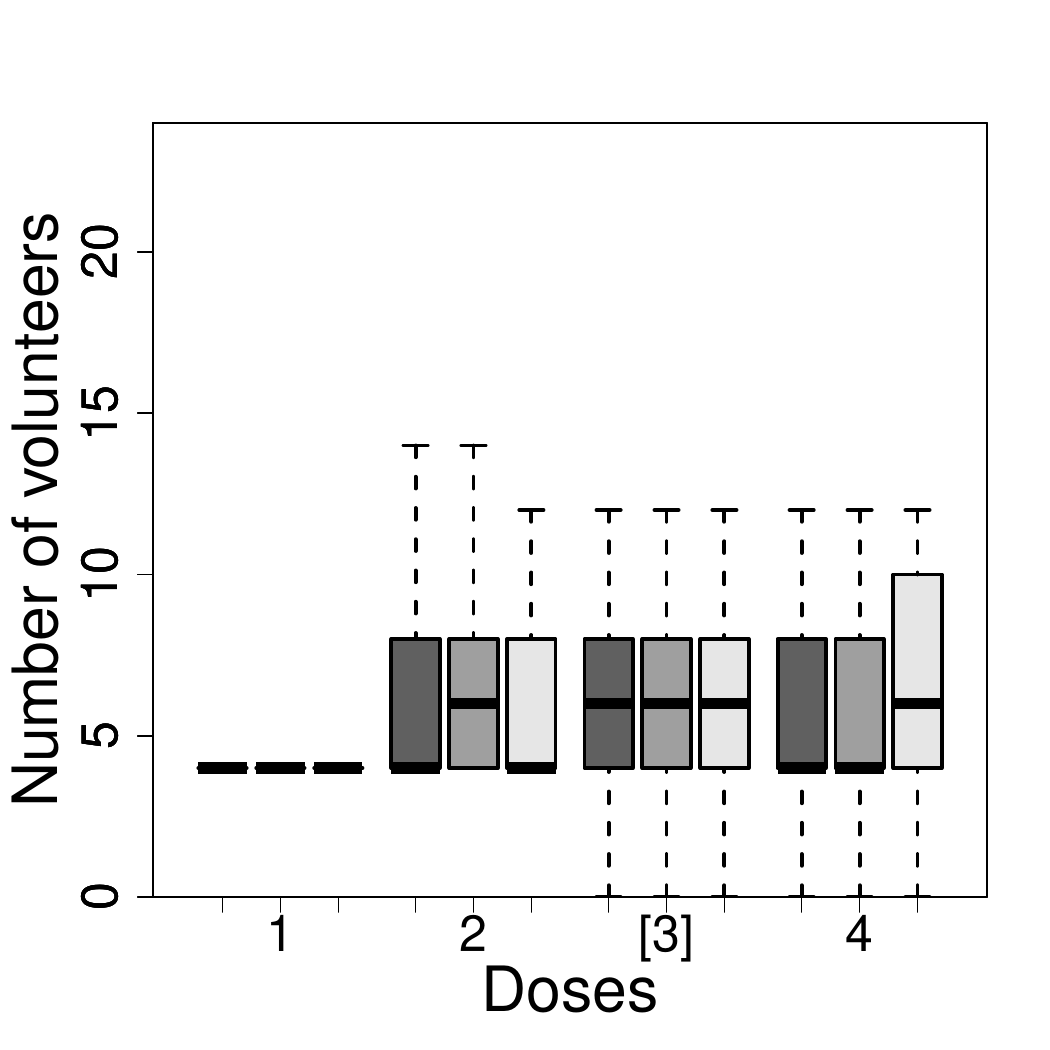}}
\subfigure[Scenario 5]{\includegraphics[scale=0.29]{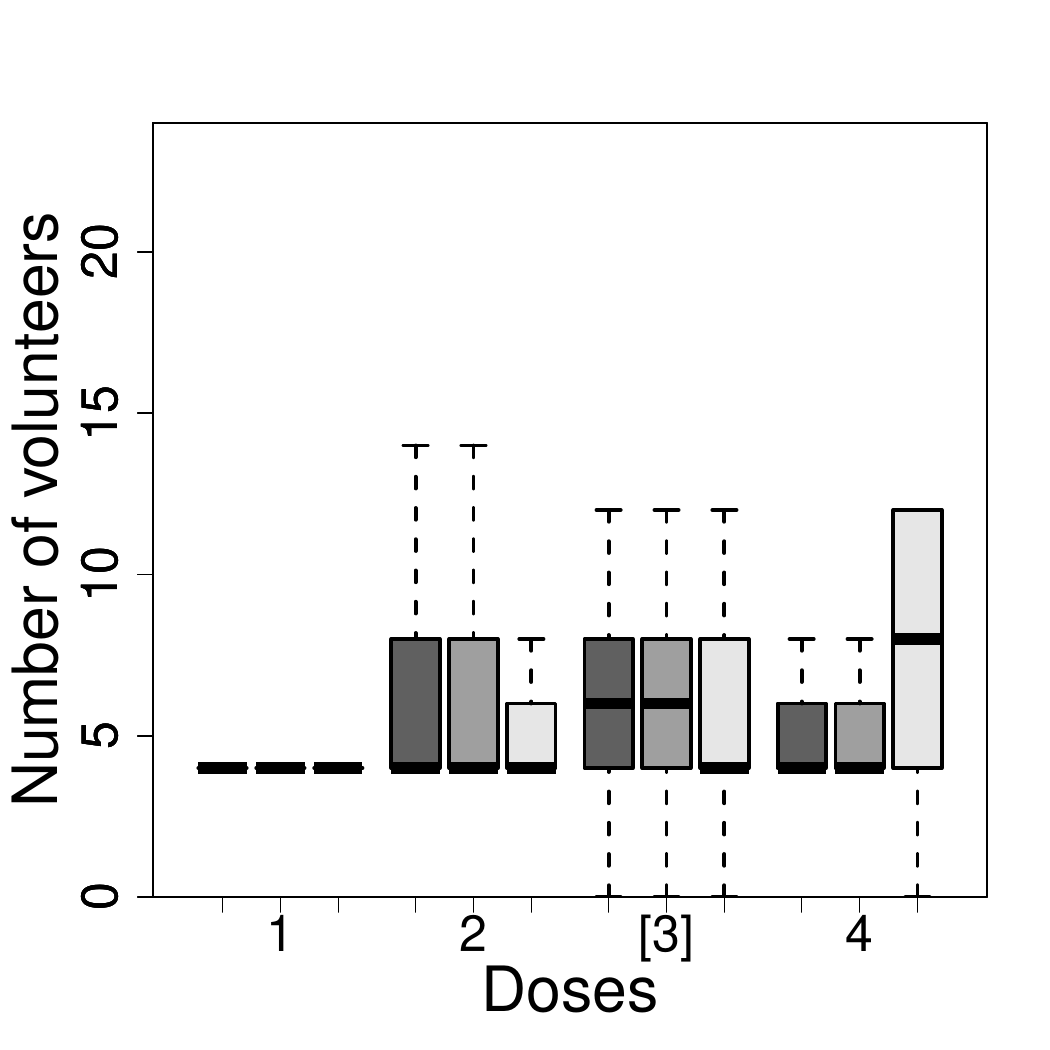}} 
\subfigure[Scenario 6]{\includegraphics[scale=0.29]{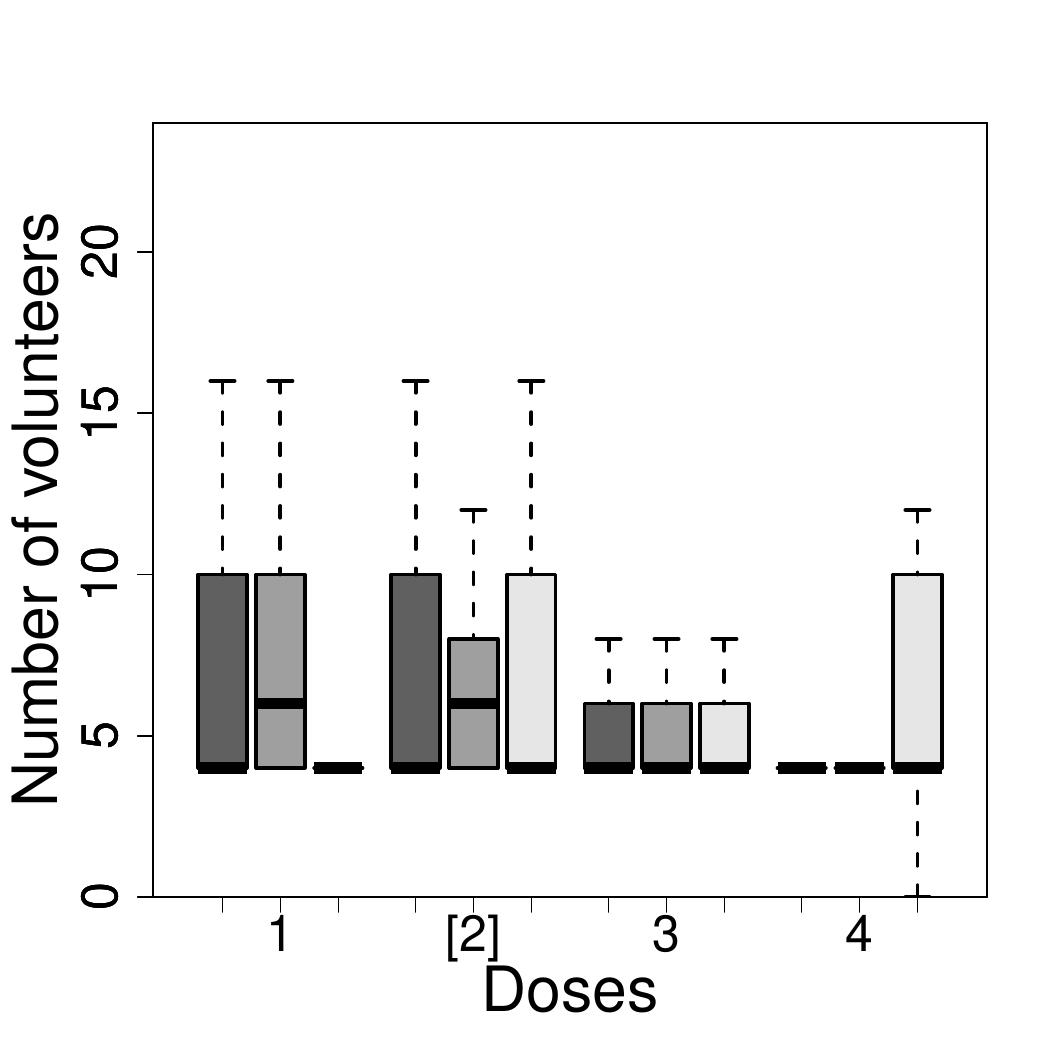}} \\ 
\subfigure[Scenario 7]{\includegraphics[scale=0.29]{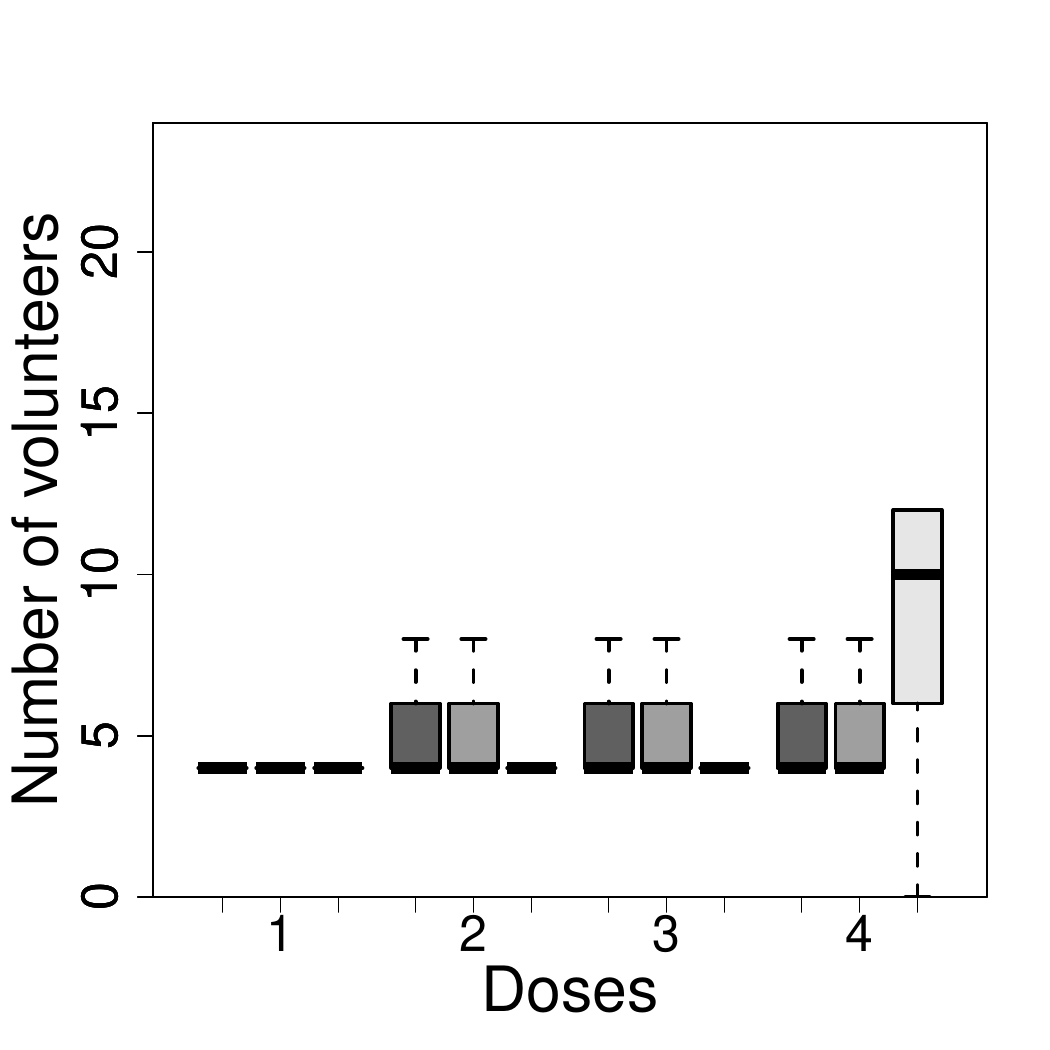}} 
\subfigure[Scenario 8]{\includegraphics[scale=0.29]{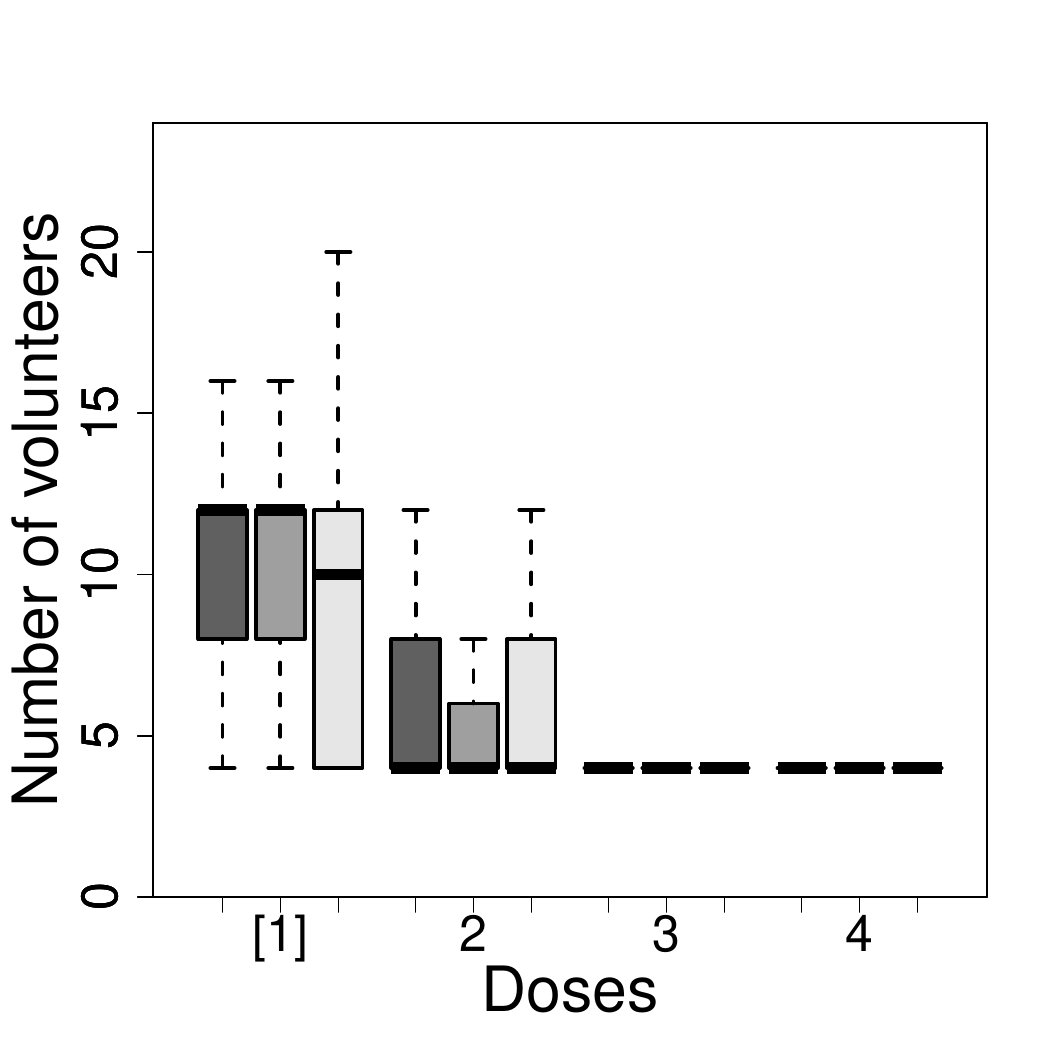}} 
\subfigure{\includegraphics[scale=0.29]{legend_boxplot_nb_volunt_by_dose.pdf}} 
\end{center}
\caption{Comparison of the number of volunteers by dose for the selection method, BMA method  and BLRM for a maximum number of volunteers of $n = 24$, when $L = 4$ dose levels are used. \label{fig:boxplot_nb_volunt_by_dose_n24_L4}}
\end{figure}

Figures \ref{fig:boxplot_nb_volunt_by_dose_n18_L3}, \ref{fig:boxplot_nb_volunt_by_dose_n24_L3} and \ref{fig:boxplot_nb_volunt_by_dose_n24_L4} respectively show the distributions of number of volunteers when $(n, L) \in \{(18, 3), (24, 3), (24, 4)\}$ are used.

\clearpage

\subsection{MAD Selection}\label{subsec:supp_MAD_hat_res}

\begin{figure}[h!]
\begin{center}
\subfigure[Scenario 1]{\includegraphics[scale=0.29]{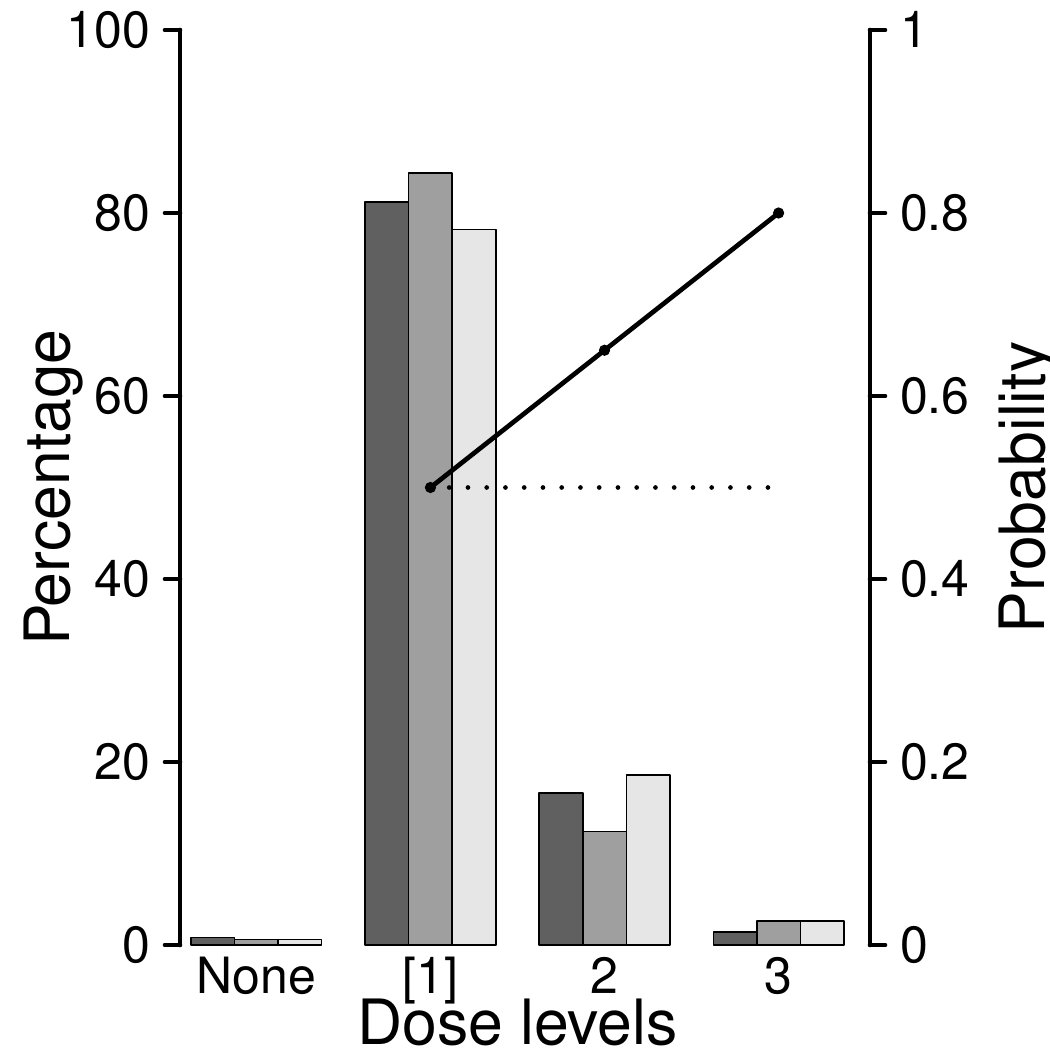}} 
\subfigure[Scenario 2]{\includegraphics[scale=0.29]{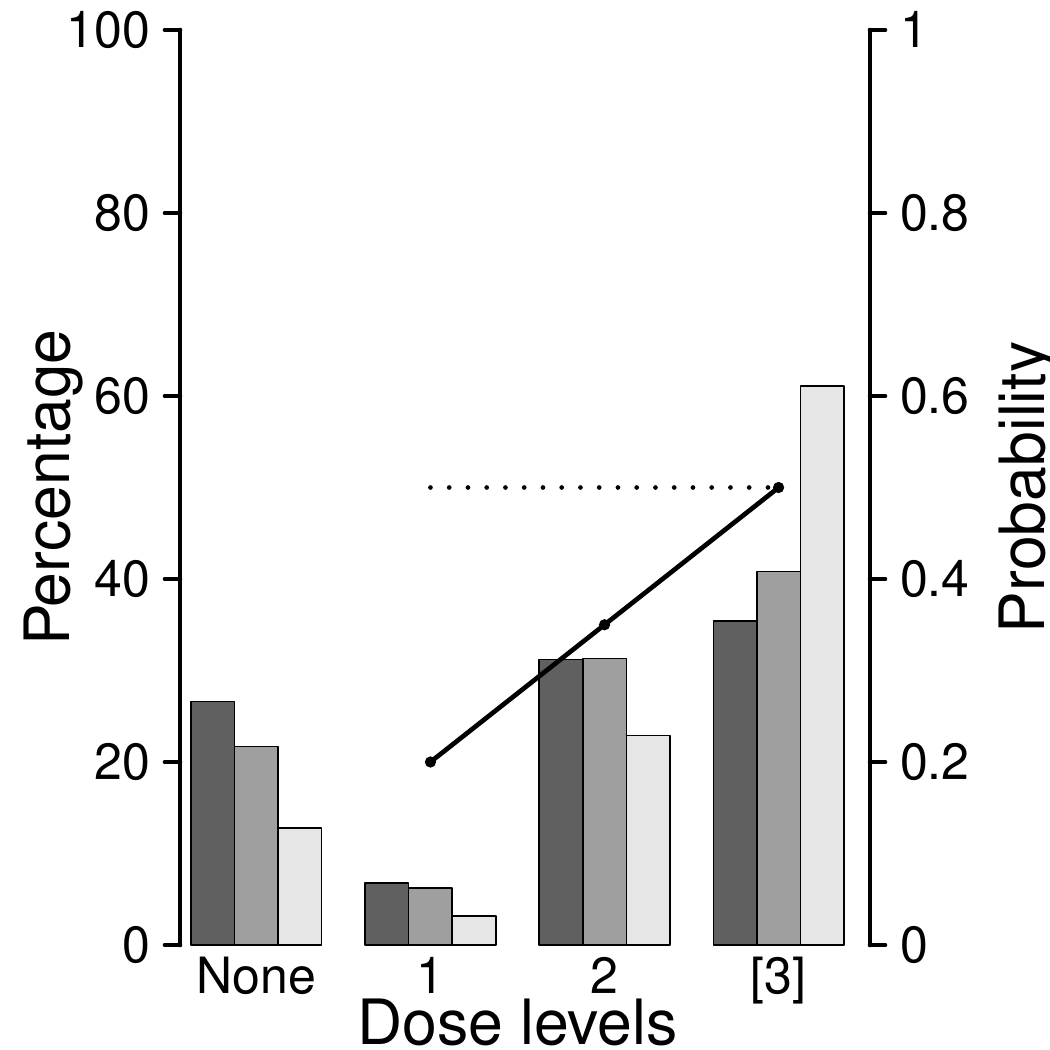}} 
\subfigure[Scenario 3]{\includegraphics[scale=0.29]{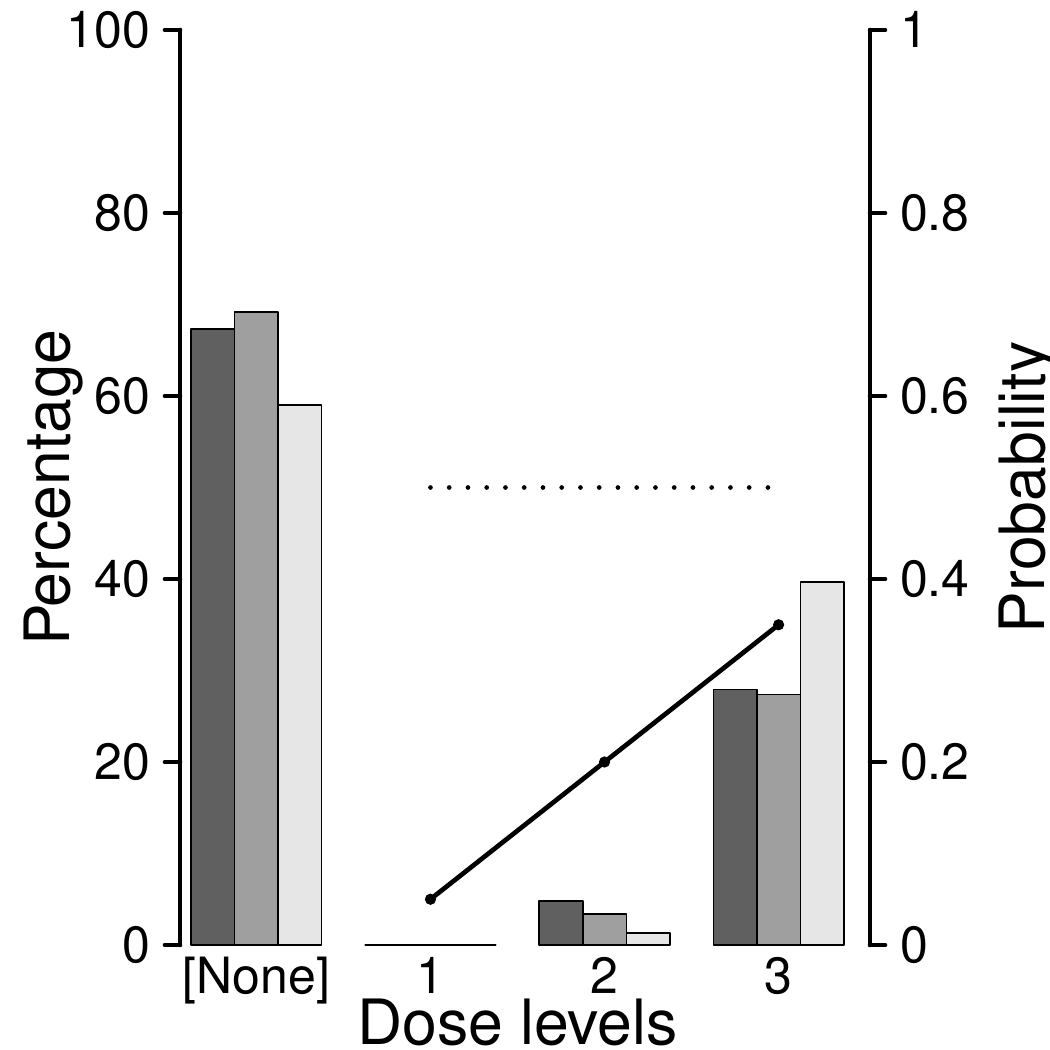}} \\ 
\subfigure[Scenario 4]{\includegraphics[scale=0.29]{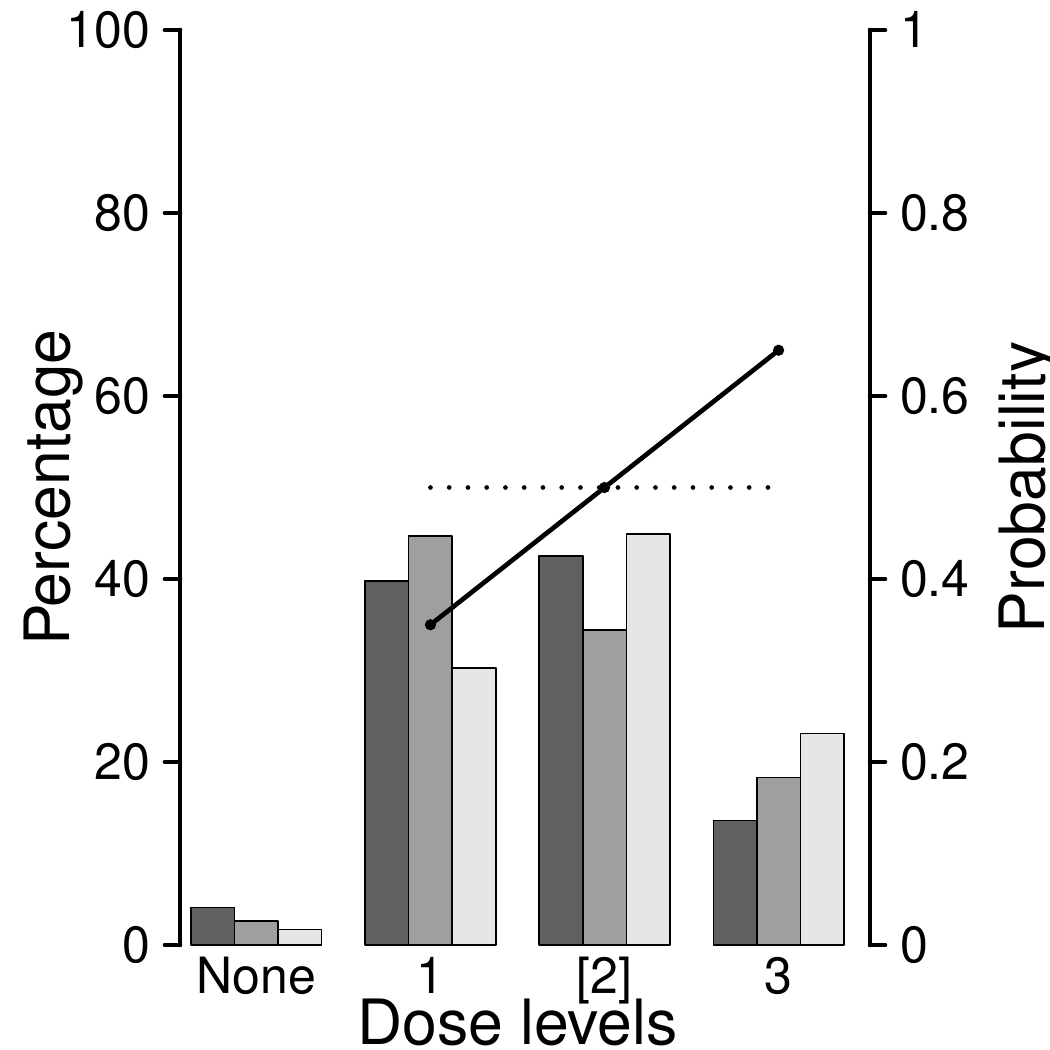}}
\subfigure[Scenario 5]{\includegraphics[scale=0.29]{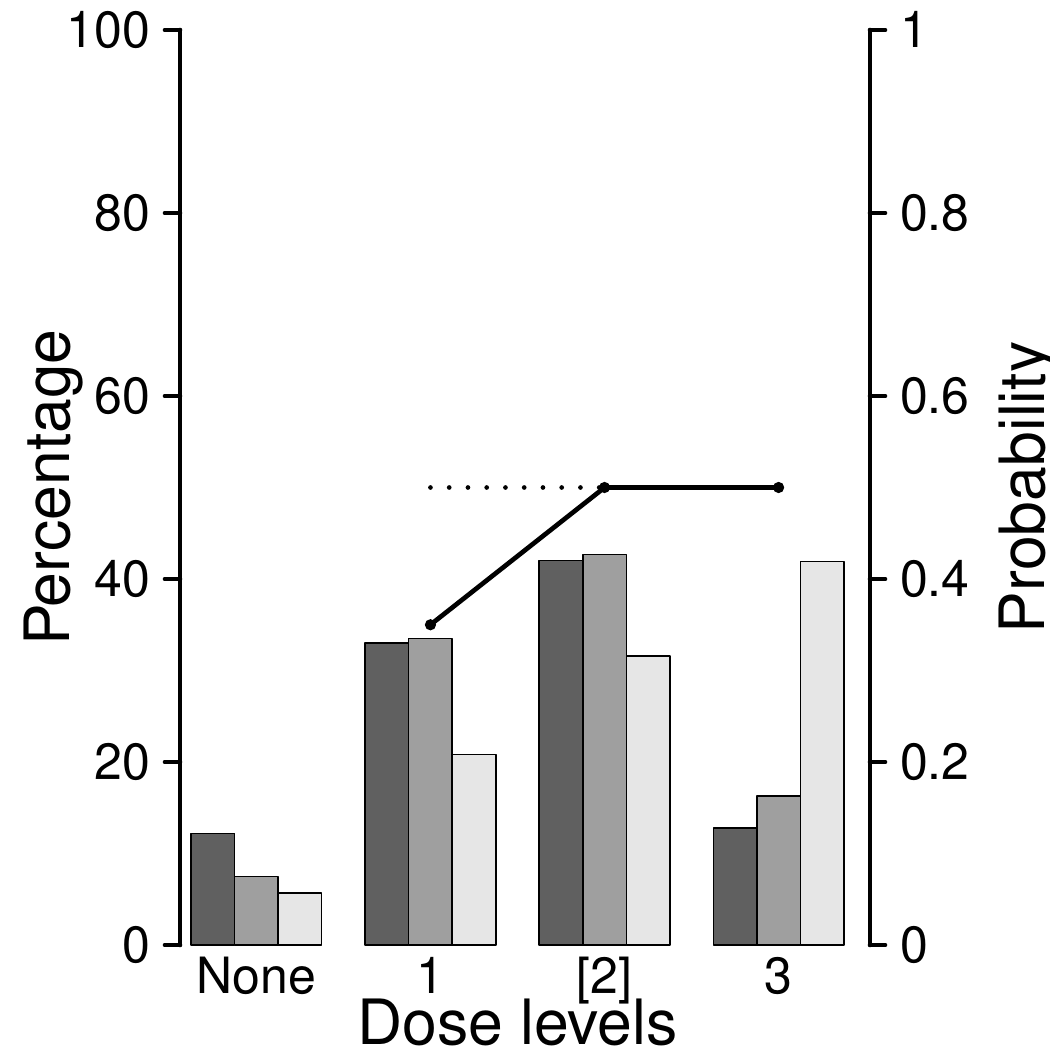}} 
\subfigure[Scenario 6]{\includegraphics[scale=0.29]{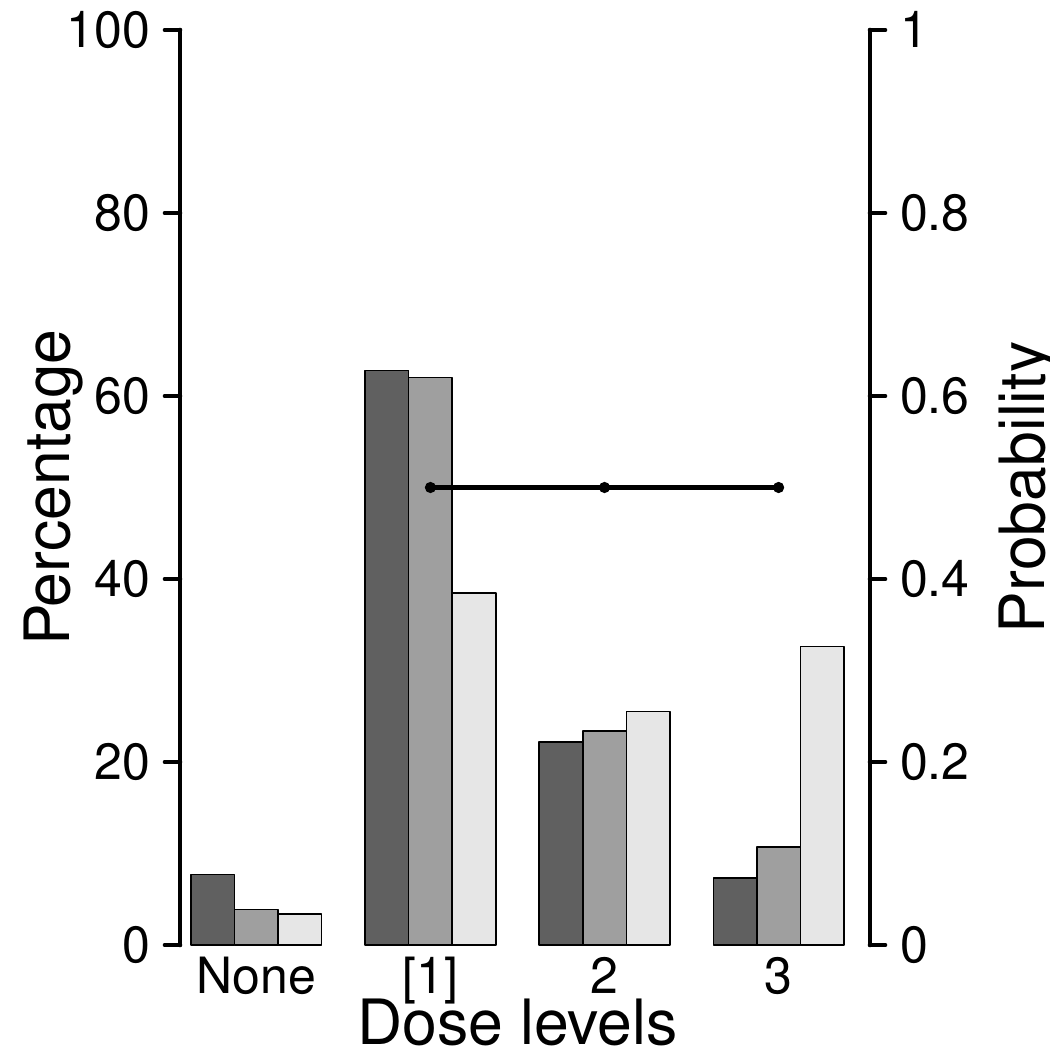}} \\ 
\subfigure[Scenario 7]{\includegraphics[scale=0.29]{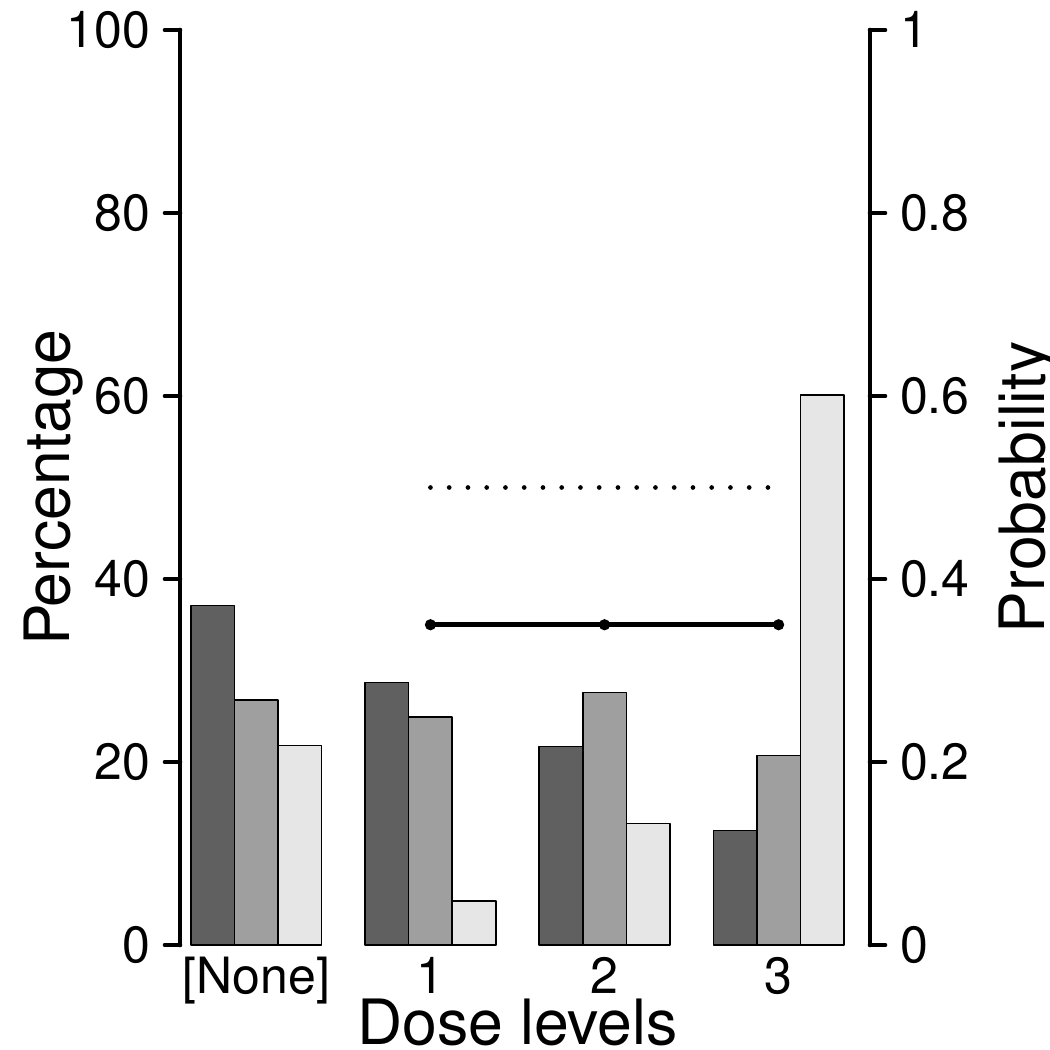}} 
\subfigure[Scenario 8]{\includegraphics[scale=0.29]{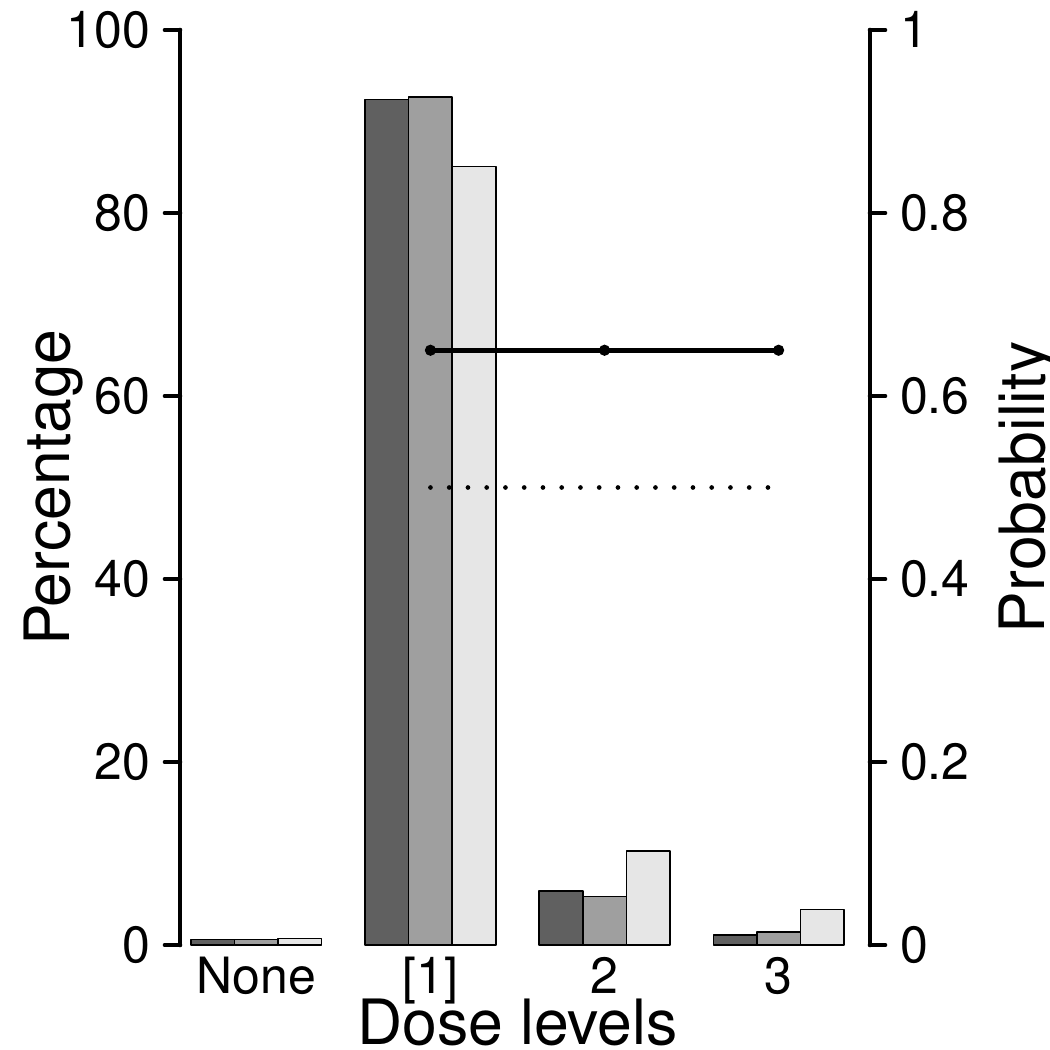}} 
\subfigure{\includegraphics[scale=0.29]{legend_barplot_OBD_hat}} 
\end{center}
\caption{Comparison of dose selection for the selection method, BMA method  and BLRM for a maximum number of volunteers of $n = 18$, when $L = 3$ dose levels are used. The solid line with circles represents the true dose-activity relationship. The horizontal dashed line is the activity probability target. \label{fig:barplot_MAD_hat_n18_L3}}
\end{figure}

\begin{figure}[h!]
\begin{center}
\subfigure[Scenario 1]{\includegraphics[scale=0.29]{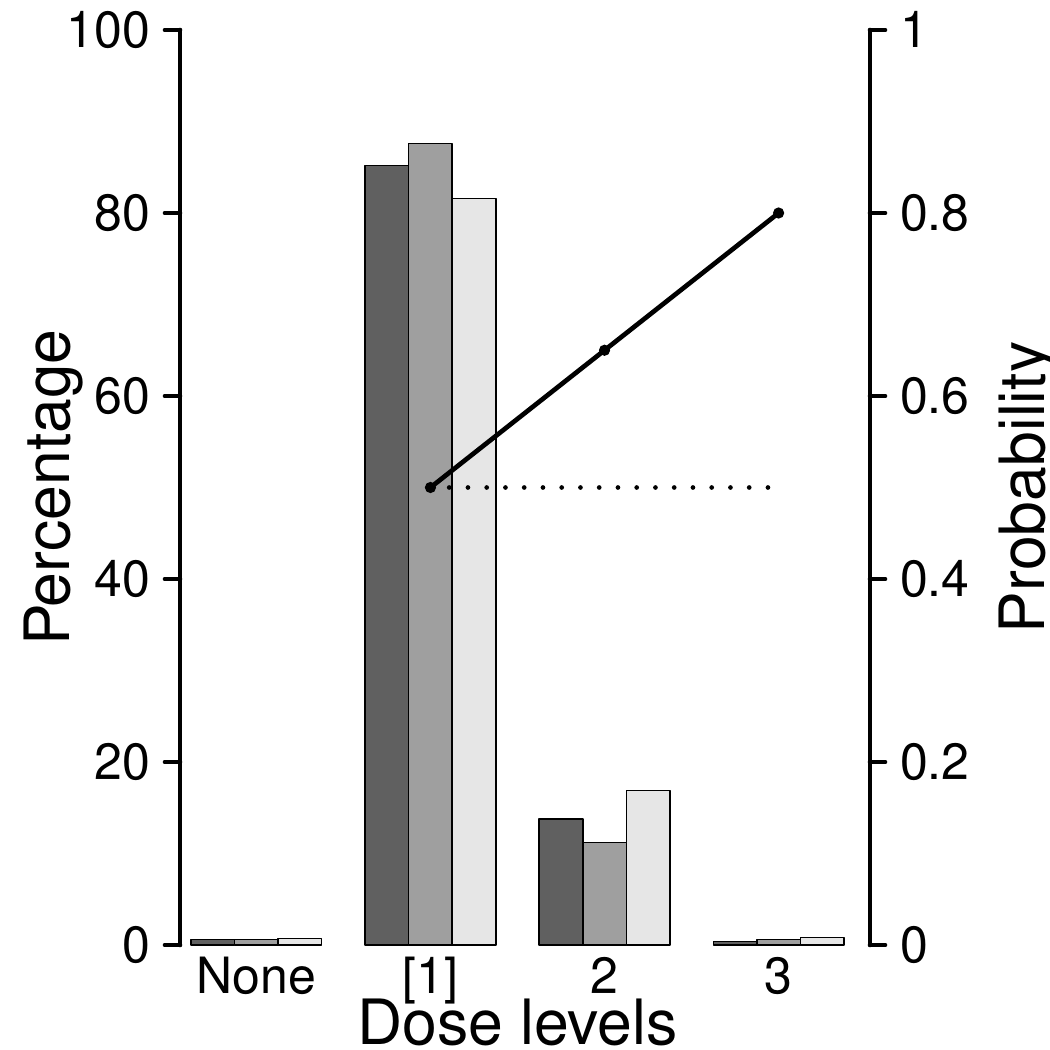}} 
\subfigure[Scenario 2]{\includegraphics[scale=0.29]{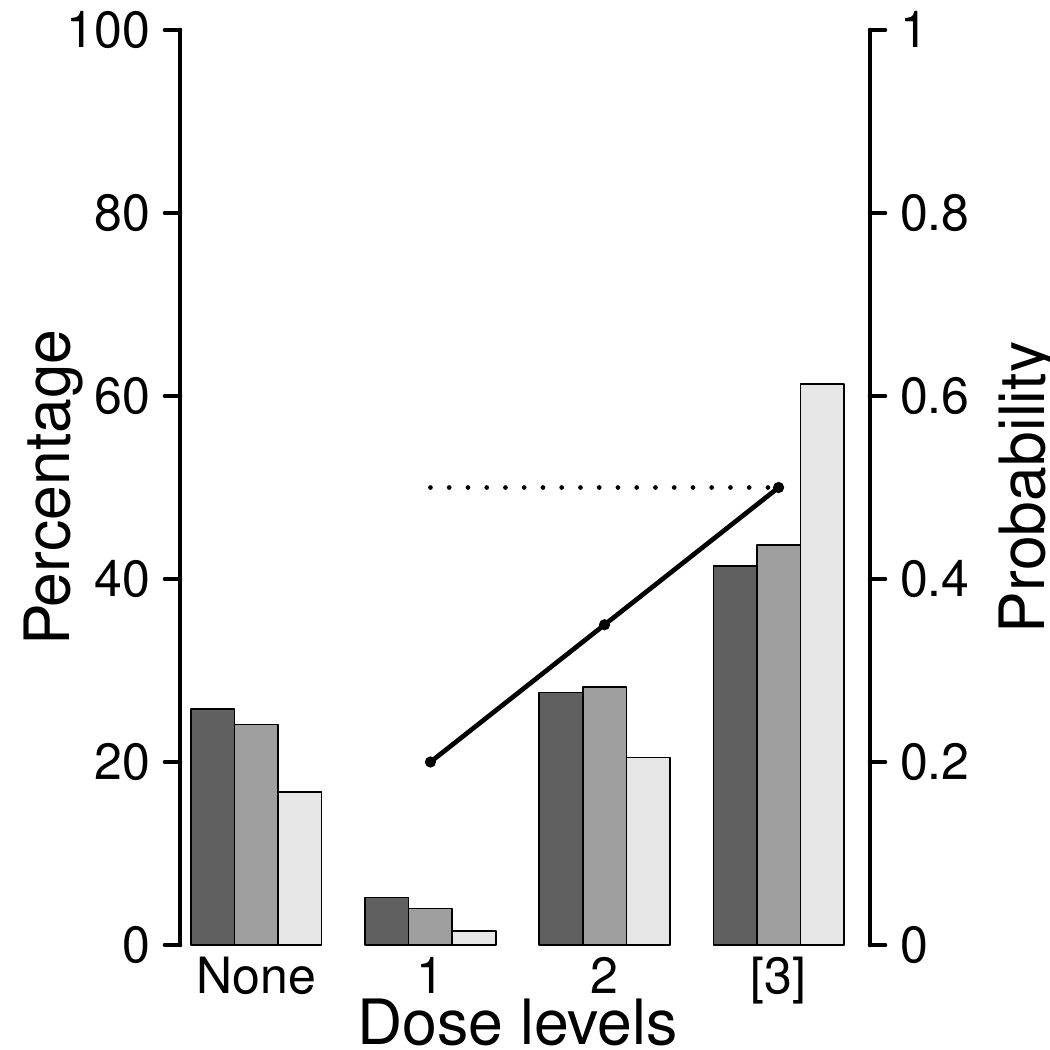}} 
\subfigure[Scenario 3]{\includegraphics[scale=0.29]{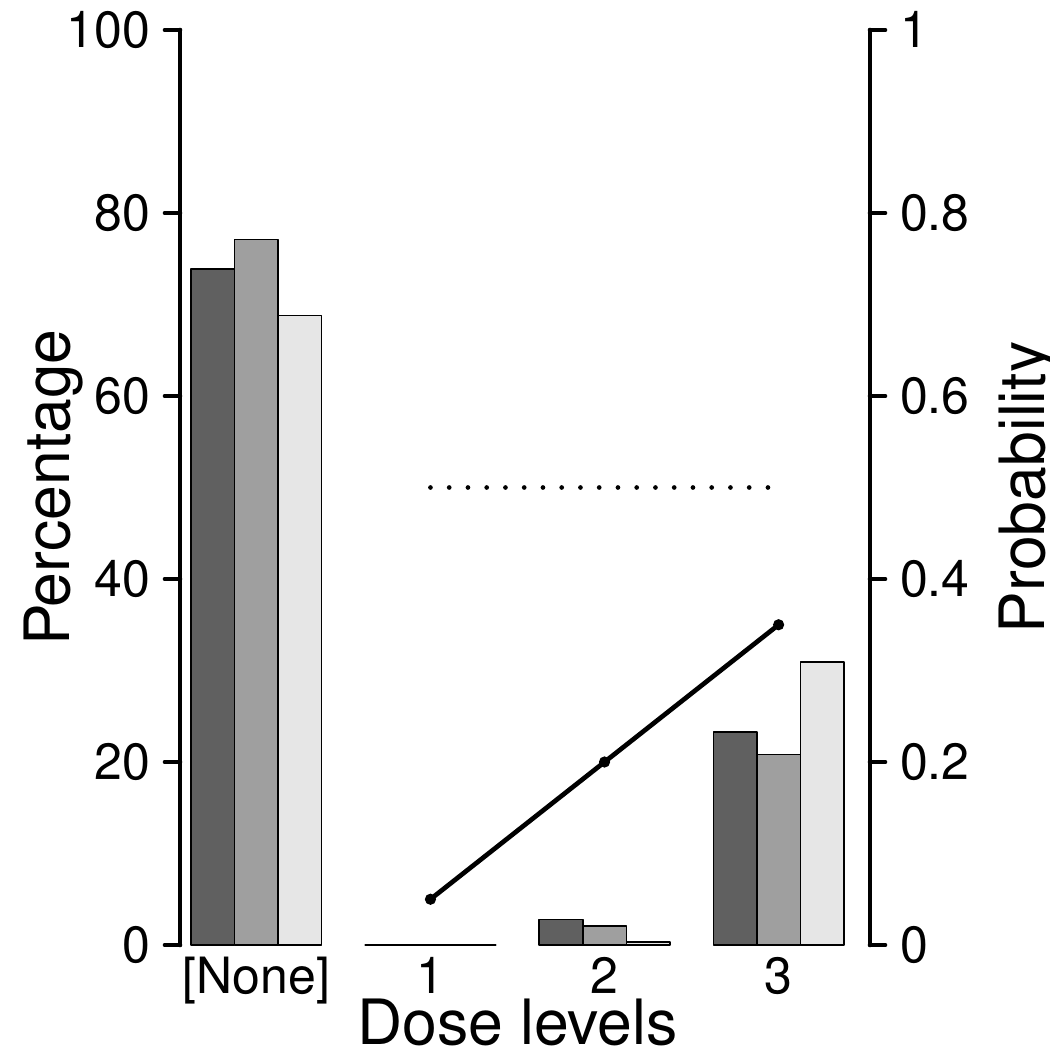}} \\ 
\subfigure[Scenario 4]{\includegraphics[scale=0.29]{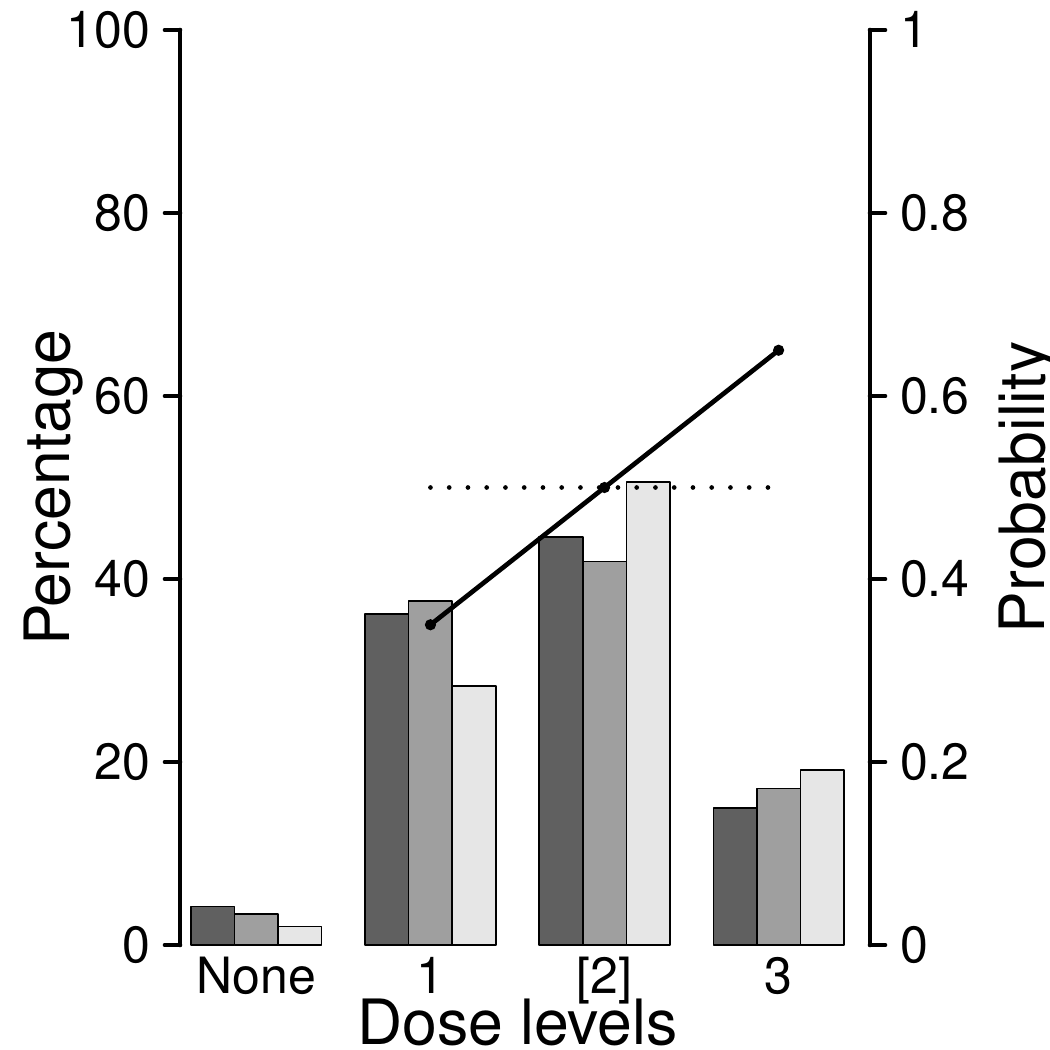}}
\subfigure[Scenario 5]{\includegraphics[scale=0.29]{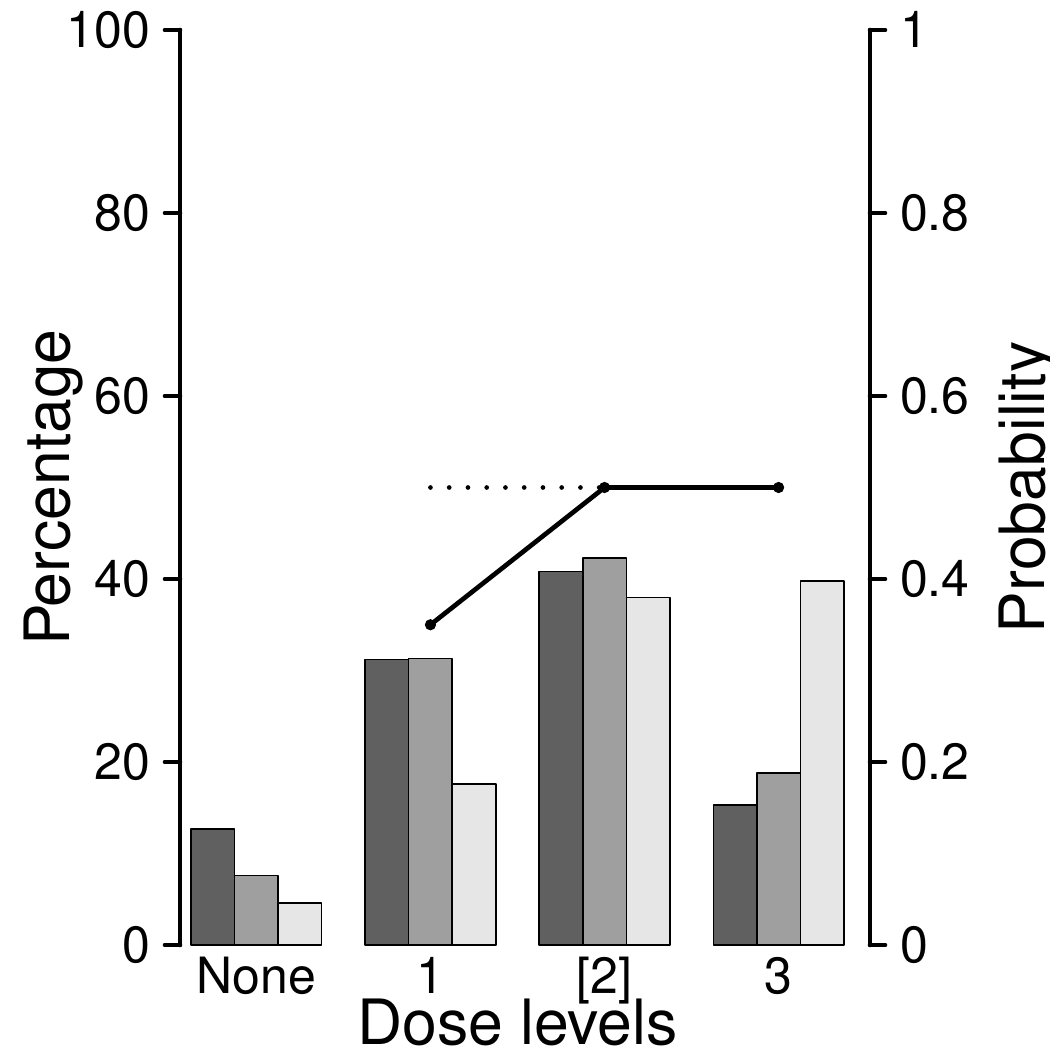}} 
\subfigure[Scenario 6]{\includegraphics[scale=0.29]{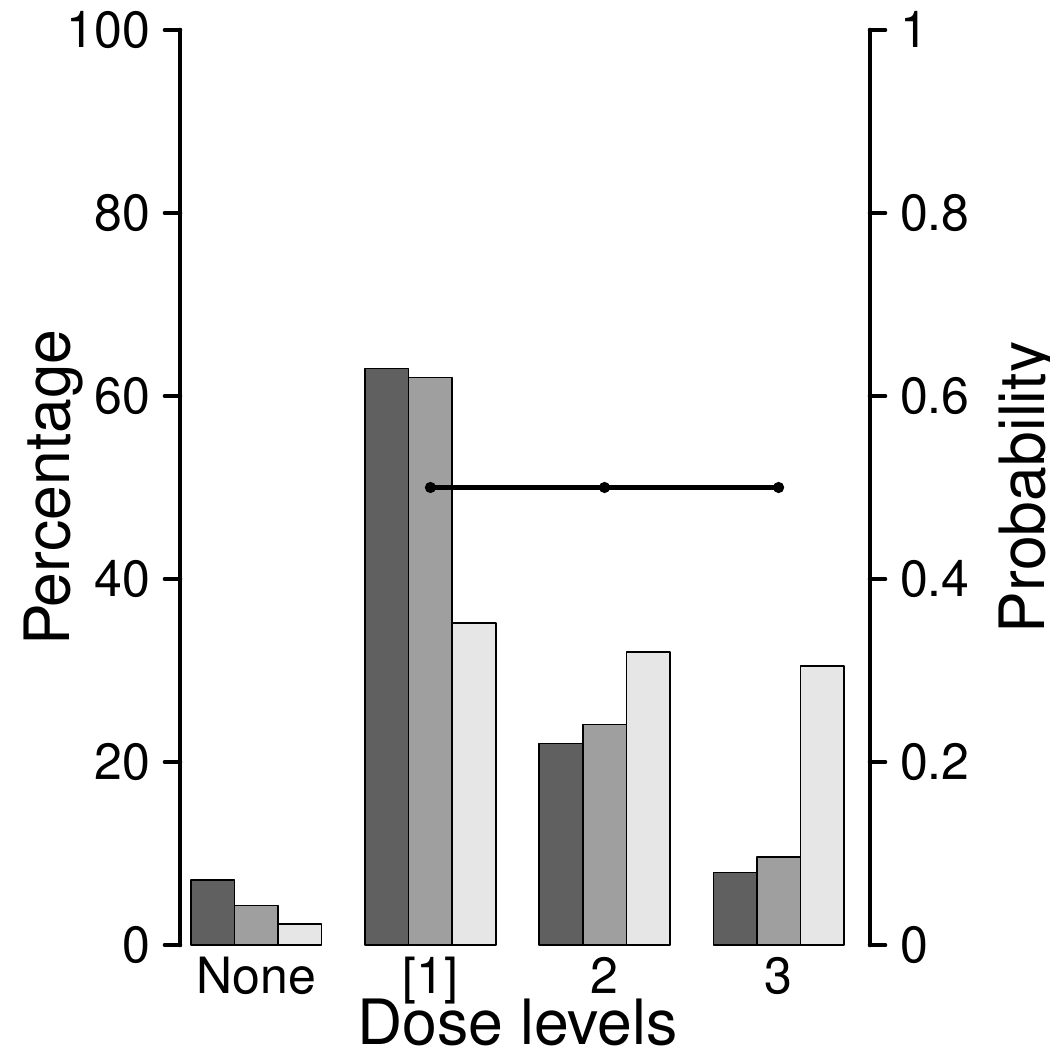}} \\ 
\subfigure[Scenario 7]{\includegraphics[scale=0.29]{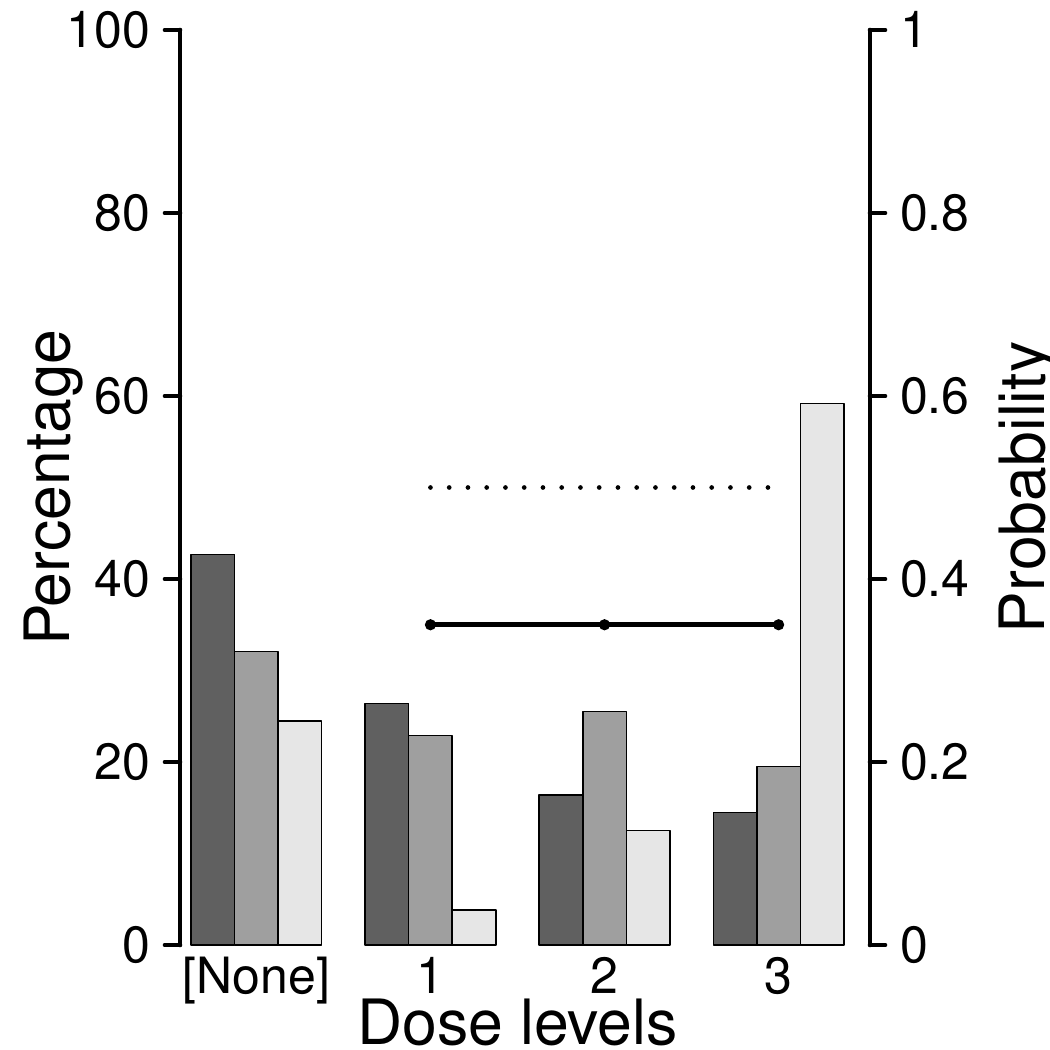}} 
\subfigure[Scenario 8]{\includegraphics[scale=0.29]{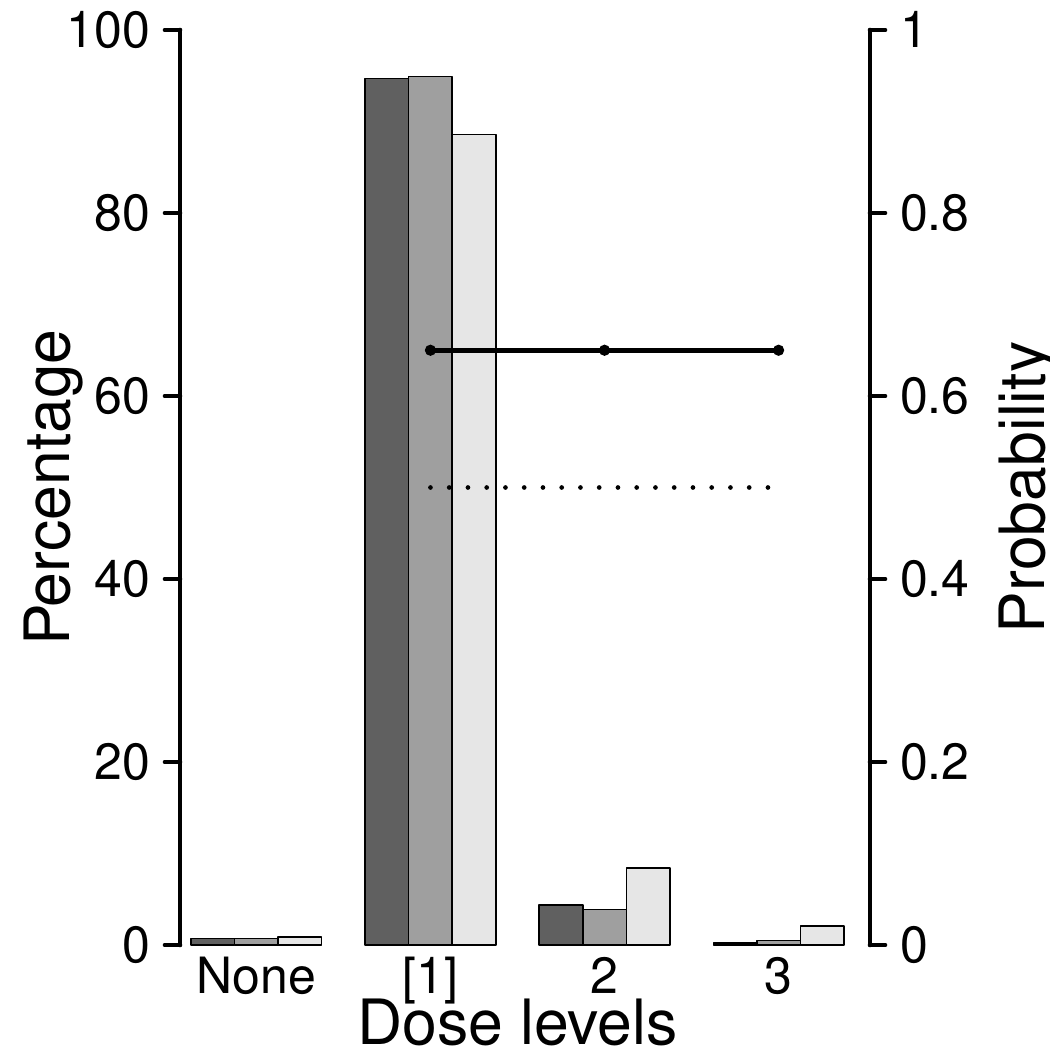}} 
\subfigure{\includegraphics[scale=0.29]{legend_barplot_OBD_hat}} 
\end{center}
\caption{Comparison of dose selection for the selection method, BMA method  and BLRM for a maximum number of volunteers of $n = 24$, when $L = 3$ dose levels are used. The solid line with circles represents the true dose-activity relationship. The horizontal dashed line is the activity probability target. \label{fig:barplot_MAD_hat_n24_L3}}
\end{figure}

\begin{figure}[h!]
\begin{center}
\subfigure[Scenario 1]{\includegraphics[scale=0.29]{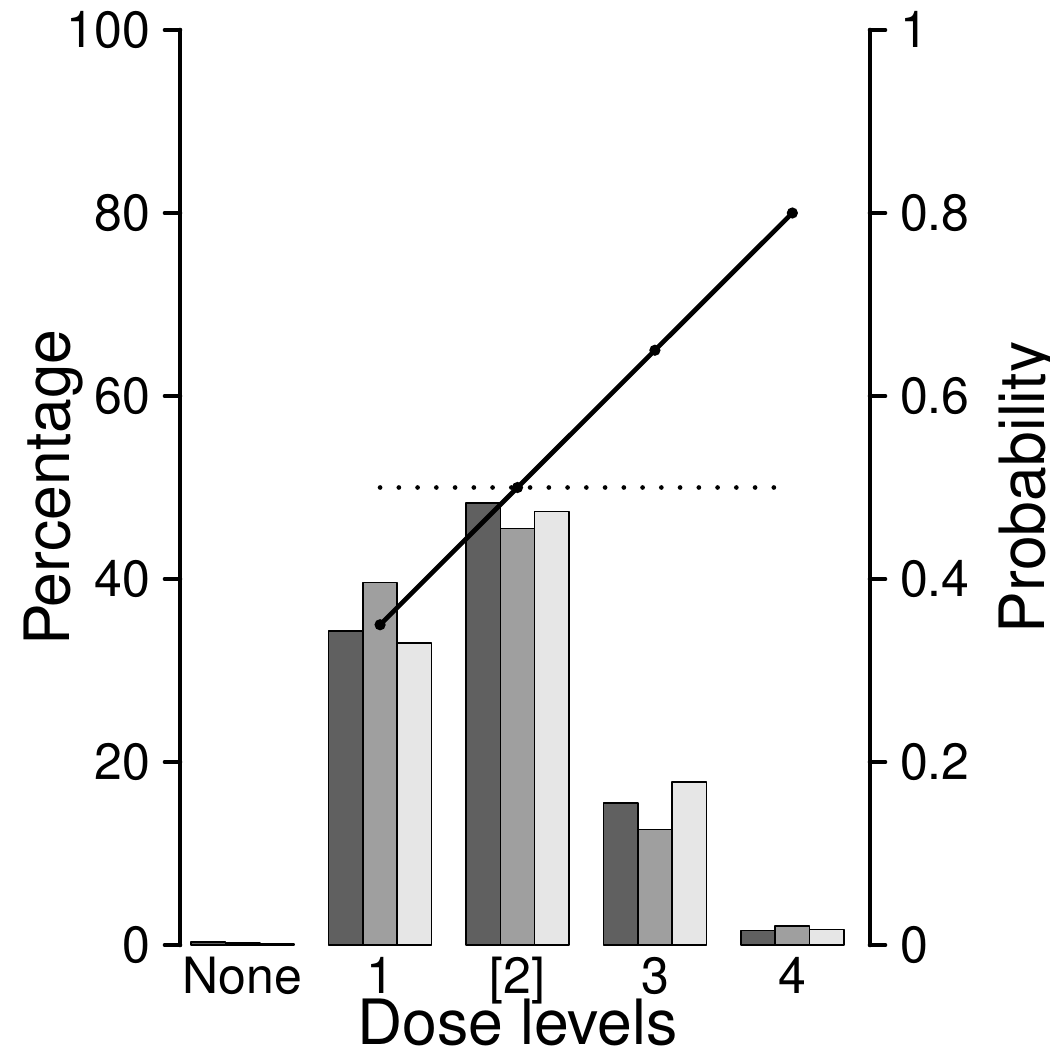}} 
\subfigure[Scenario 2]{\includegraphics[scale=0.29]{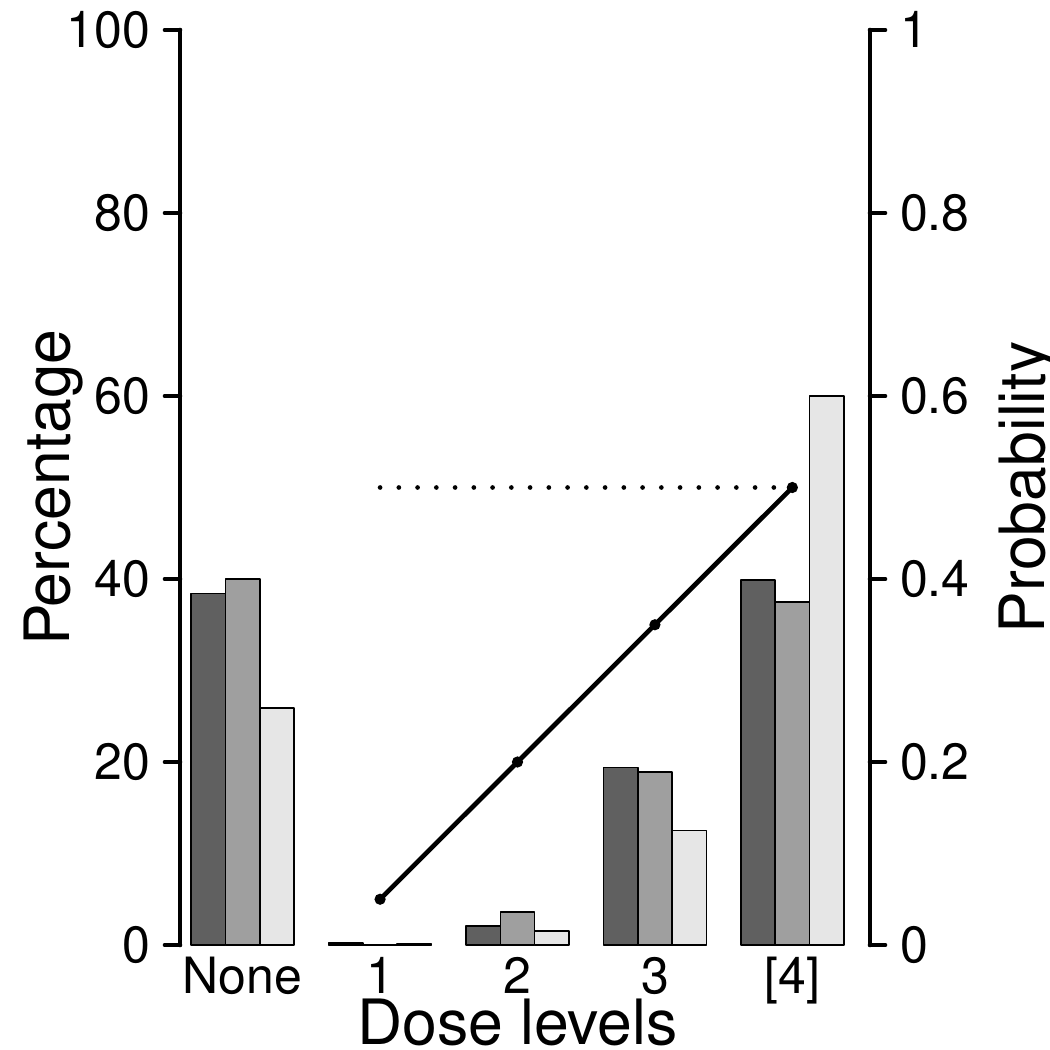}} 
\subfigure[Scenario 3]{\includegraphics[scale=0.29]{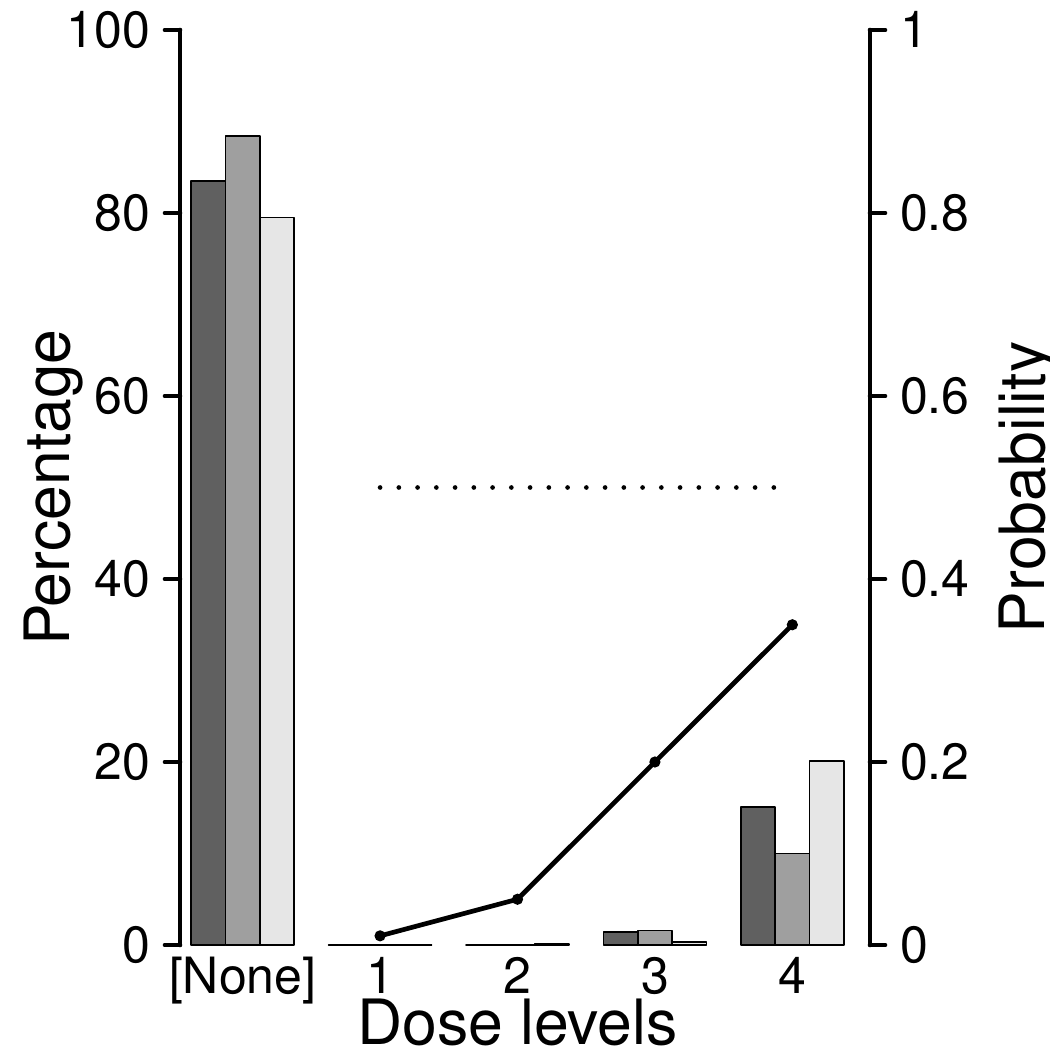}} \\ 
\subfigure[Scenario 4]{\includegraphics[scale=0.29]{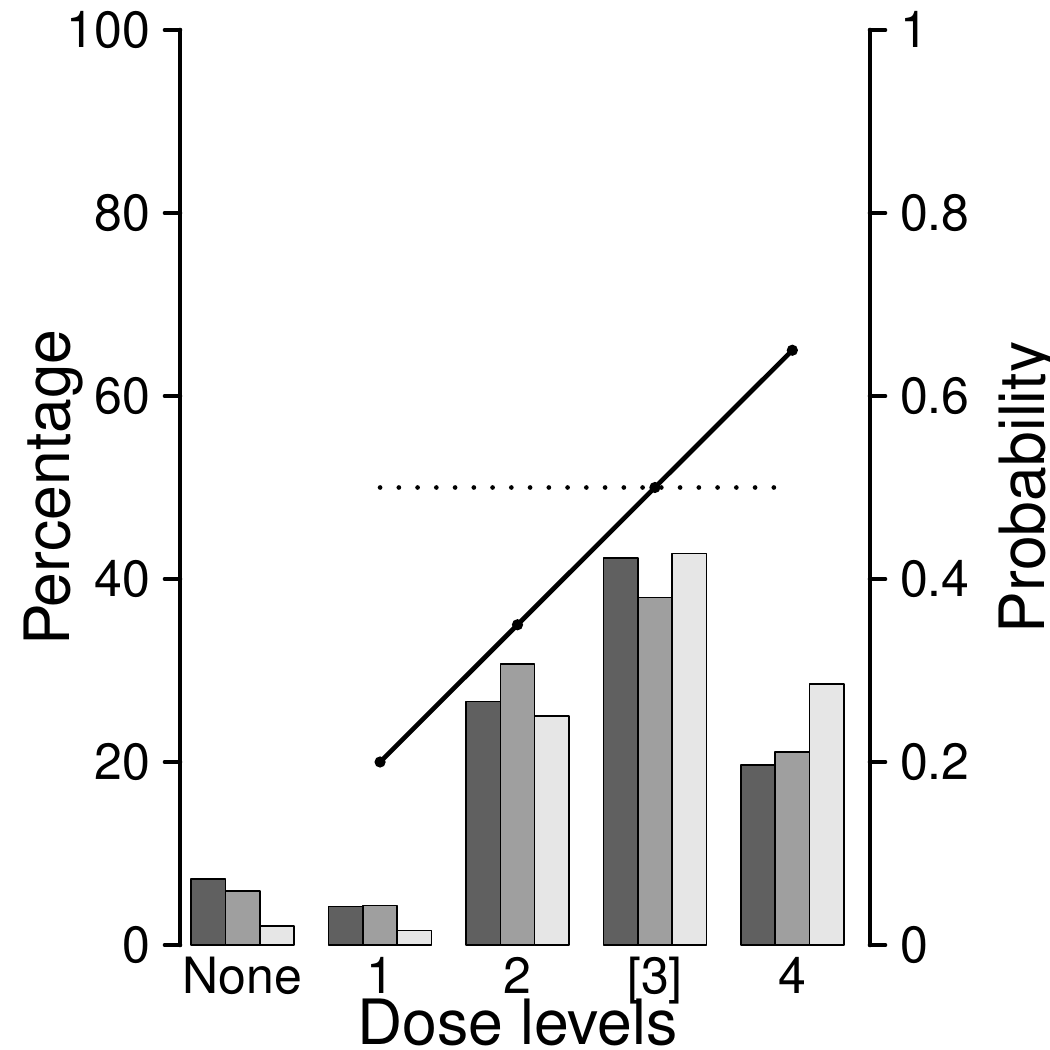}}
\subfigure[Scenario 5]{\includegraphics[scale=0.29]{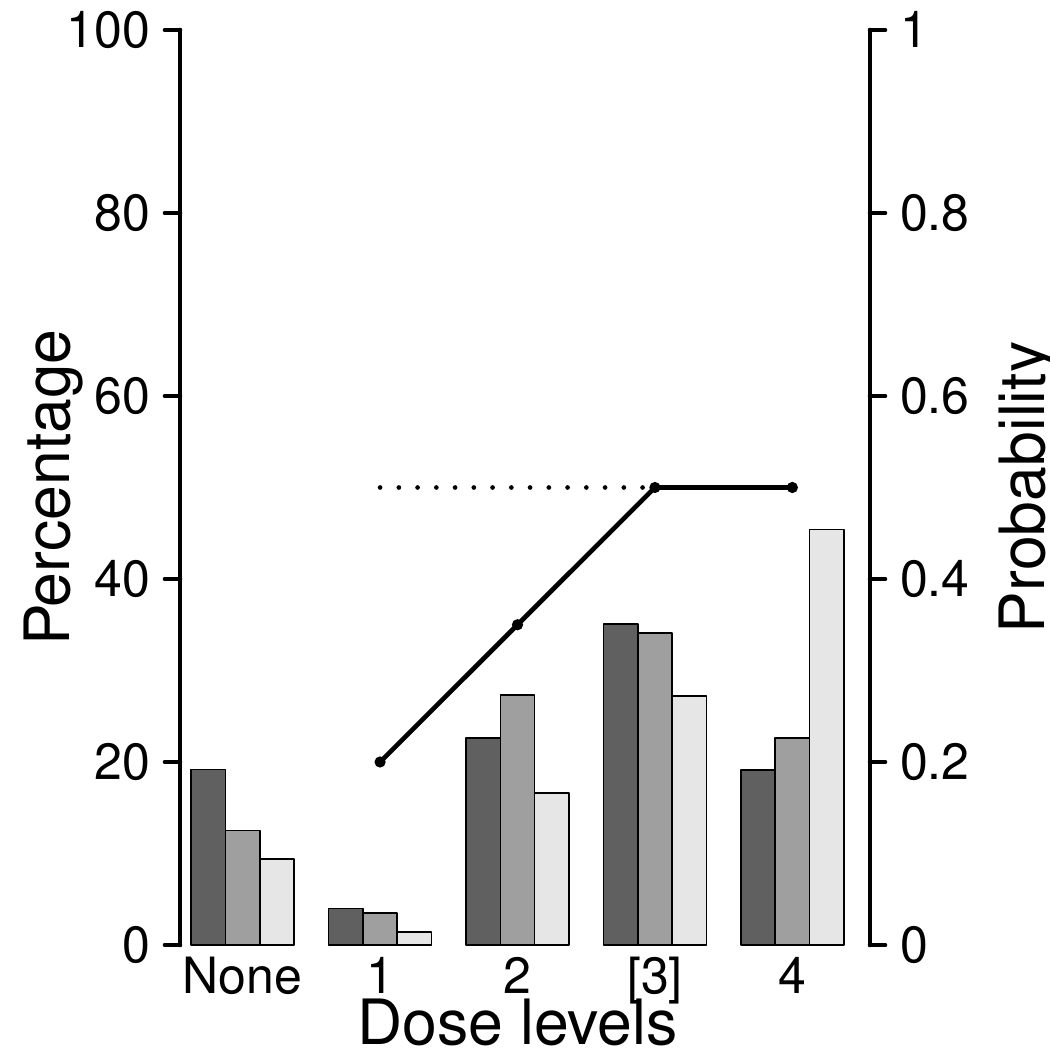}} 
\subfigure[Scenario 6]{\includegraphics[scale=0.29]{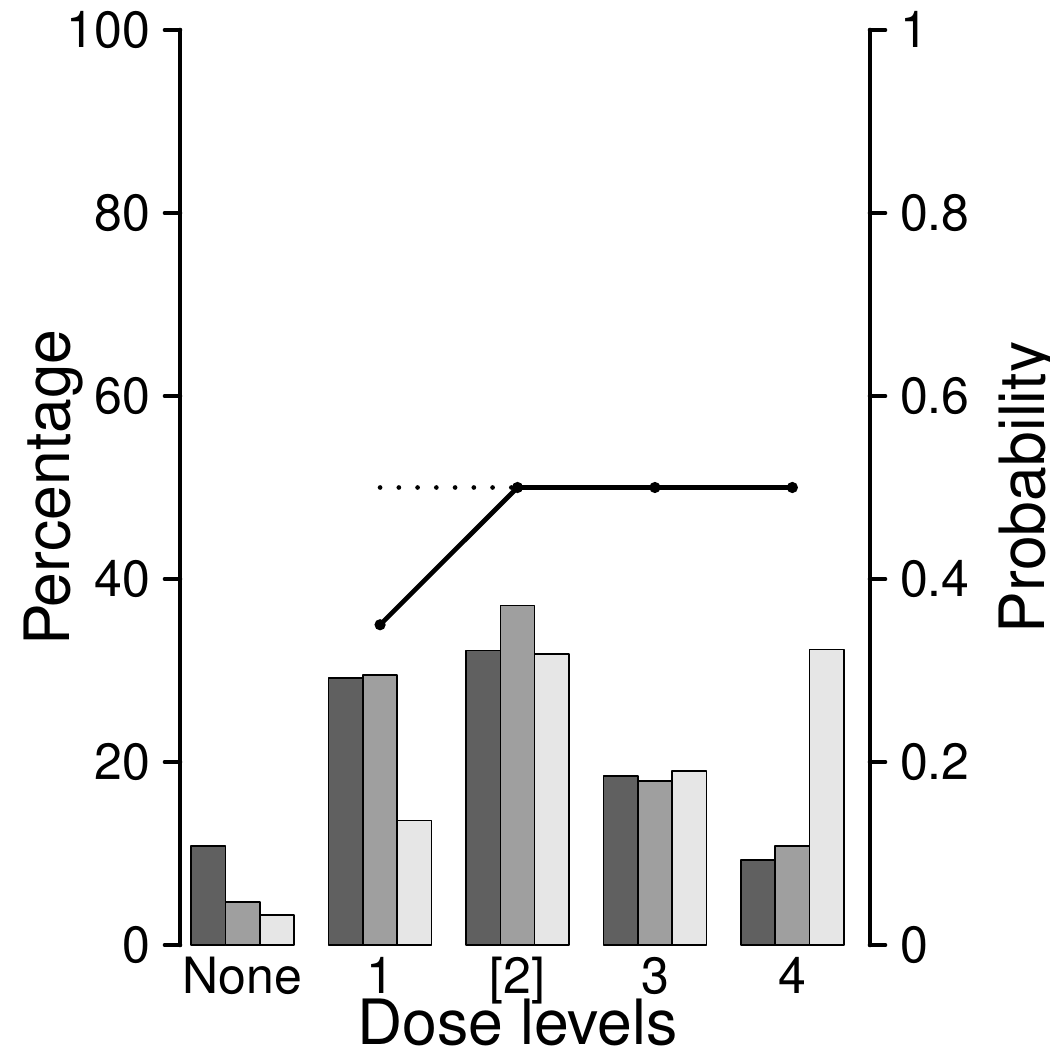}} \\ 
\subfigure[Scenario 7]{\includegraphics[scale=0.29]{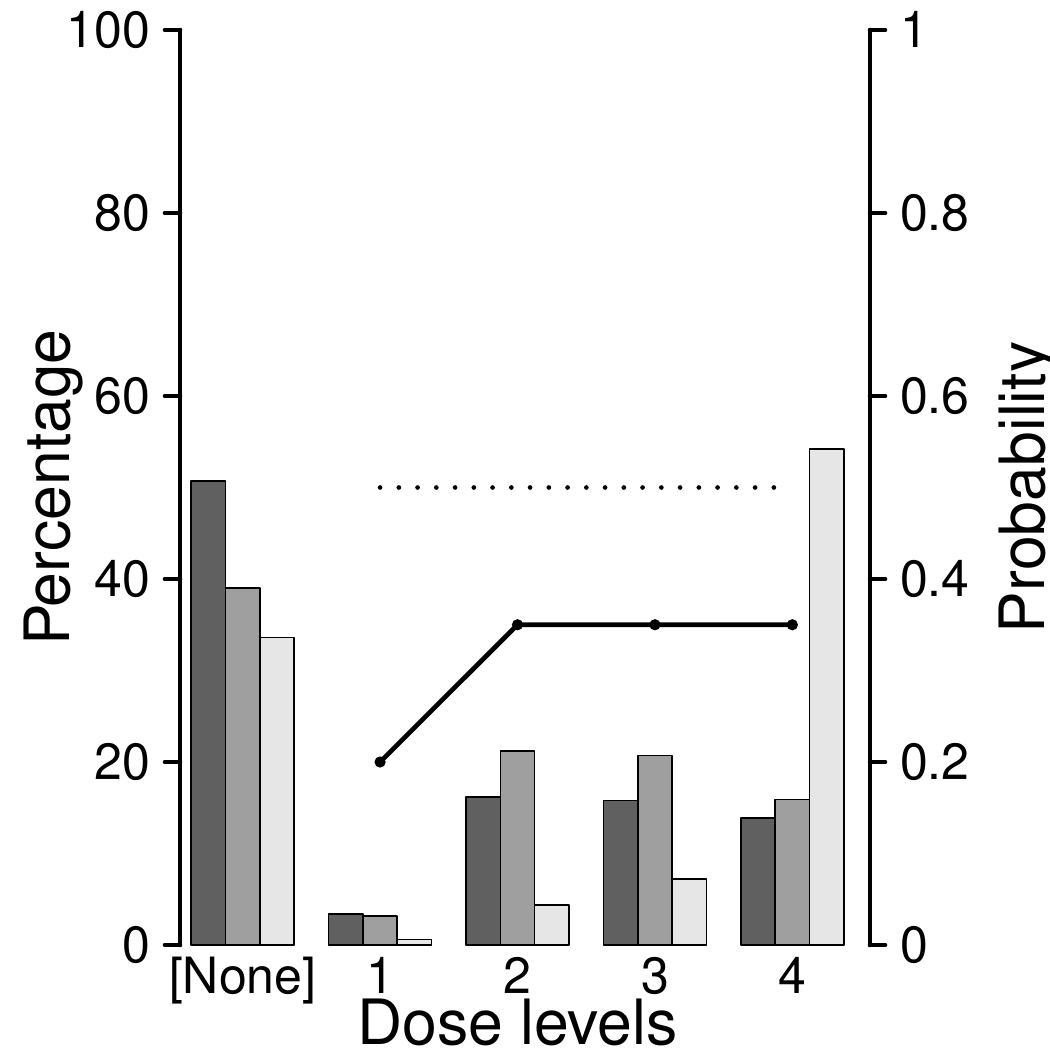}} 
\subfigure[Scenario 8]{\includegraphics[scale=0.29]{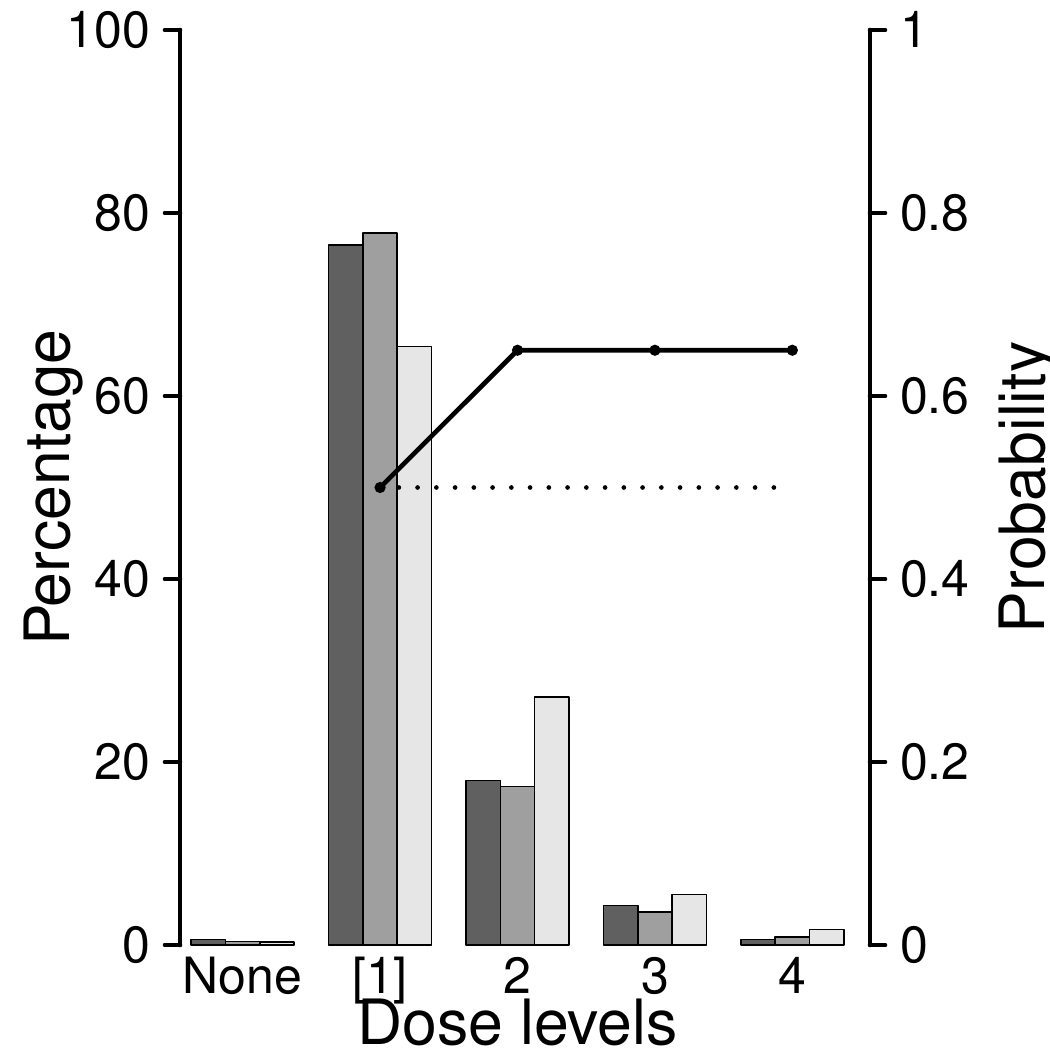}} 
\subfigure{\includegraphics[scale=0.29]{legend_barplot_OBD_hat}} 
\end{center}
\caption{Comparison of dose selection for the selection method, BMA method  and BLRM for a maximum number of volunteers of $n = 24$, when $L = 4$ dose levels are used. The solid line with circles represents the true dose-activity relationship. The horizontal dashed line is the activity probability target. \label{fig:barplot_MAD_hat_n24_L4}}
\end{figure}

Figures \ref{fig:barplot_MAD_hat_n18_L3}, \ref{fig:barplot_MAD_hat_n24_L3} and \ref{fig:barplot_MAD_hat_n24_L4} respectively show the percentages of doses selection when $(n, L) \in \{(18, 3), (24, 3), (24, 4)\}$ are used.

